\documentclass[
	fontsize=12pt, 
	twoside=true, 
	numbers=noenddot, 
]{kaobook}

\usepackage[english]{babel} 
\usepackage[english=british]{csquotes} 

\usepackage[style=alphabetic,backref=true,maxbibnames=99]{biblatex}
\usepackage{blindtext}
\usepackage{glossaries}
\usepackage{xcolor}
\usepackage{amsmath}
\usepackage{epigraph}

\hypersetup{
	citecolor = RedOrange,
	linkcolor = MidnightBlue,
	urlcolor = ForestGreen,
	hyperindex=true,
}

\usepackage{braket}
\DeclareUnicodeCharacter{0301}{\'{e}}

\definecolor{codegreen}{rgb}{0,0.6,0}
\definecolor{codegray}{rgb}{0.5,0.5,0.5}
\definecolor{codepurple}{rgb}{0.58,0,0.82}
\definecolor{backcolour}{rgb}{0.95,0.95,0.92}

\lstdefinestyle{mystyle}{
    backgroundcolor=\color{backcolour},   
    commentstyle=\color{codegreen},
    keywordstyle=\color{magenta},
    numberstyle=\tiny\color{codegray},
    stringstyle=\color{codepurple},
    basicstyle=\ttfamily\footnotesize,
    breakatwhitespace=false,         
    breaklines=true,                 
    captionpos=b,                    
    keepspaces=true,                 
    numbers=left,                    
    numbersep=5pt,                  
    showspaces=false,                
    showstringspaces=false,
    showtabs=false,                  
    tabsize=2
}

\newcommand\mydots{\hbox to 1em{.\,.\,.}}
\newcommand\Floor[1]{\lfloor#1\rfloor}
\newcommand\Ceil[1]{\lceil#1\rceil}
\newcommand{\tw}{\ensuremath{t_\mathrm{w}}\xspace}
\newcommand{\Tc}{\ensuremath{T_\mathrm{c}}\xspace}
\newcommand{\NRep}{\ensuremath{N_{\text{Rep}}}\xspace}
\newcommand{\NS}{\ensuremath{N_{\text{S}}}\xspace}
\newcommand{\NTherm}{\ensuremath{N_{\text{thermal}}}\xspace}
\newcommand{\Nov}{\ensuremath{N_{\text{ov}}}\xspace}
\newcommand{\Tiso}{\ensuremath{T^{\text{iso}}}\xspace}

\newcommand \texp{{\tau_{\exp}}}
\newcommand \texpf{{\tau_{\exp,f}}}
\newcommand \tint{{\tau_{\mathrm{int}}}}
\newcommand \tintvar{{\tau_{\mathrm{int,var}}}}
\newcommand \tintold{{\tau_{\mathrm{int,old}}}}
\newcommand \tintf{{\tau_{\mathrm{int,}f}}}
\newcommand{\tintd}{{\tau_{\mathrm{int,}16}}}
\newcommand{\tintt}{{\tau_{\mathrm{int,}13}}}
\newcommand \tintfTlblo{{\tau_{\mathrm{int,}f,T^*,\mathrm{lblo}}}}
\newcommand \mcav[1]{\braket{{#1}}_{\text{MC}}}
\newcommand \abs[1]{\lvert {#1} \rvert}
\newcommand \Nrep{{N_{\mathrm{rep}}}}
\newcommand \nmet{{n_{\mathrm{Met}}}}
\newcommand \lblo{l_{\mathrm{blo}}}
\newcommand \xchaos{X^J_{T_1,T_2}}
\newcommand \xchaosmin{X^J_{T_{\min},T_2}}
\newcommand \Tmin{T_{\min}}
\newcommand{\zpt}{z^{\mathrm{PT}}}
\newcommand{\ee}{{\mathrm{e}}}
\newcommand{\dd}{{\mathrm{d}}}
\newcommand{\ecvj}{{\varepsilon_{\mathrm{cv}}(\mathcal{J})}}
\newcommand{\TA}{T_{\mathrm{A}}}
\newcommand{\TB}{T_{\mathrm{B}}}
\newcommand{\xiA}{\xi_{\mathrm{A}}}
\newcommand{\xiB}{\xi_{\mathrm{B}}}
\newcommand{\xjk}{x^{\mathrm{JK}}}
\newcommand{\fjk}{f^{\mathrm{JK}}}
\newcommand{\fb}{f^{B}}
\newcommand{\Nb}{N_{\mathrm{boots}}}
\newcommand{\Cov}{\mathrm{Cov}}

\newcommand\gobbleone[1]{}

\newcommand{\Also}[2]{\emph{See also} #1}

\usepackage{styles/kaobiblio}
\usepackage{styles/mdftheorems}
\DeclareMathAlphabet{\mathcal}{OMS}{cmsy}{m}{n}

\graphicspath{{examples/documentation/images/}{images/}} 

\makeglossaries 
\newacronym{CuMn}{CuMn}{Copper-Manganese alloy}
\newacronym{AuFe}{AuFe}{Gold-Iron alloy}
\newacronym{RKKY}{RKKY}{Ruderman-Kittel-Kasuya-Yosida}
\newacronym{EA}{EA}{Edwards-Anderson}
\newacronym{SG}{SG}{Spin Glass}
\newacronym{RFIM}{RFIM}{Random Field Ising Model}
\newacronym{SQUID}{SQUID}{Superconducting Quantum Interference Device}
\newacronym{AgMn}{AgMn}{Silver-Manganese alloy}
\newacronym{CuFe}{CuFe}{Copper-Iron alloy}
\newacronym{YDy}{YDy}{Yttrium-Dysprosium alloy}
\newacronym{YEr}{YEr}{Yttrium-Erbium alloy}
\newacronym{ScTb}{ScTb}{Scandium-Terbium alloy}
\newacronym{LaGdAl}{La$_{1-x}$Gd$_x$Al$_2$}{Lanthanum-Gadolinium-Aluminum ternary compound}
\newacronym{RbCuCoF}{Rb$_2$Cu$_{1-x}$Co$_x$F$_4$}{Rubidium and copper (II) tetraflouride with cobalt impurities}
\newacronym{FeMnTiO}{Fe$_{1-x}$Mn$_x$TiO$_3$}{Iron (II) and titanium (IV) trioxide with manganese impurities}
\newacronym{DM}{DM}{Dzyaloshinkskii-Moriya}
\newacronym{TRM}{TRM}{thermo-remanent magnetization}
\newacronym{ZFC}{ZFC}{zero-field cooled}
\newacronym{CdCrInS}{CdCr$_{1.7}$In$_{0.3}$S$_4$}{Cadmium and chromium (III) tetrasulfide with indium impurities}
\newacronym{CLT}{CLT}{Central Limit Theorem}
\newacronym{SK}{SK}{Sherrington-Kirkpatrick}
\newacronym{ISG}{ISG}{Ising spin glass}
\newacronym{RSB}{RSB}{Replica symmetry breaking}
\newacronym{pdf}{pdf}{probability density function}
\newacronym{PBC}{PBC}{Periodic Boundary Conditions}
\newacronym{DLR}{DLR}{Dobrushin-Lanford-Ruelle}
\newacronym{CSD}{CSD}{Chaotic Size Dependence}
\newacronym{AW}{AW}{Aizenman-Wehr}
\newacronym{NS}{NS}{Newman-Stein}
\newacronym{MAS}{MAS}{Metastate-Averaged State}
\newacronym{PT}{PT}{Parallel Tempering}
\newacronym{TC}{TC}{Temperature Chaos}
\newacronym{ME}{ME}{Mpemba Effect}
\newacronym{CPU}{CPU}{Central Processing Unit}
\newacronym{GPU}{GPU}{Graphics Processing Unit}
\newacronym{FPGA}{FPGA}{Field-programmable Gate Array}
\newacronym{MSC}{MSC}{Multispin Coding}
\newacronym{MUSA}{MUSA-MSC}{Multisample Multispin Coding}
\newacronym{MUSI}{MUSI-MSC}{Multisite Multispin Coding}

\newglossaryentry{maths}
{
    name=mathematics,
    description={Mathematics is what mathematicians do}
}

\makenomenclature 

\begin{document}

\title{}
\author{}
\date{}

\makeatletter
\thispagestyle{empty}
\begin{center}
{\includegraphics[width=0.3\textwidth]{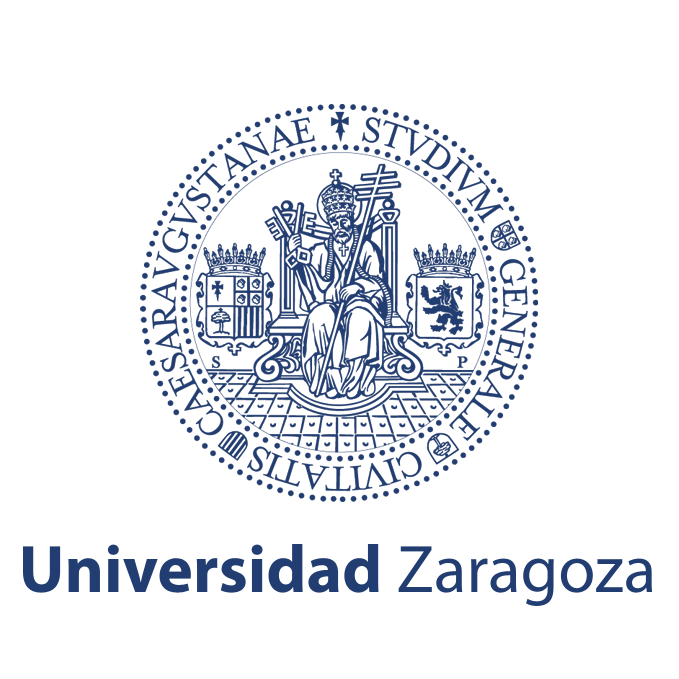}\par}
\vspace{0.2cm}
{\bfseries\LARGE Universidad de Zaragoza \par}
\vspace{1cm}
{\scshape\Large Facultad de Ciencias \par}
\vspace{2cm}
{\scshape\Huge Deepening the knowledge of Spin Glasses: Metastate, Off-equilibrium phenomena and Temperature Chaos \par}
\vspace{0.2cm}
{\itshape\Large Ph.D Thesis \par}
\vspace{2cm}
{\Large Autor: \par}
{\Large Javier Moreno-Gordo \par}
\vspace{0.5cm}
{\Large Supervisores: \par}
\begin{minipage}[l]{0.49\linewidth}
\begin{center}
{\Large Víctor Martín Mayor \par}
{\normalsize Universidad Complutense de Madrid}
\end{center}
\end{minipage}
\begin{minipage}[r]{0.49\linewidth}
\begin{center}
{\Large David Íñiguez Dieste \par}
{\normalsize Universidad de Zaragoza}
\end{center}
\end{minipage}
\end{center}
\makeatother

%


\frontmatter 




\makeatletter
\uppertitleback{\@titlehead} 

\lowertitleback{
	
	\medskip
	
	\textbf{Creative Commons License}\\
	\ccbyncsa\ This license allows reusers to distribute, remix, adapt, and build upon the material in any medium or format for noncommercial purposes only, and only so long as attribution is given to the creator. If you remix, adapt, or build upon the material, you must license the modified material under identical terms. 
	
	To view a copy of the CC BY-NC-SA code, visit: \\\url{https://creativecommons.org/licenses/by-nc-sa/4.0/}
	
%
%
%
%
	
}
\makeatother


\dedication{
\flushright \small
	Y nana ne enano ne naba en e nianno.\\
	A la yaya Paula, a mamá, a Sara y a la tuanita...
}



\maketitle
\tableofcontents 




\setchapterstyle{kao}{} 
\pagelayout{wide}

\section*{Agradecimientos}

Esta tesis ha sido posible gracias a la ayuda de muchas personas que han puesto su granito de arena para que yo pudiera avanzar en la consecución de los objetivos (no sólo académicos) que me he impuesto a lo largo de los cuatro últimos años.

A pesar de que mi tesis se ha desarrollado principalmente en la Universidad de Zaragoza, he tenido la suerte de poder colaborar estrechamente con la Universidad Complutense de Madrid. Quiero agradecer en primer lugar a mis directores de tesis, David Íñiguez Dieste de la Universidad de Zaragoza y Víctor Martín Mayor de la Universidad Complutense de Madrid, su ayuda y dedicación a lo largo de estos cuatro años.

David Íñiguez me recibió con los brazos abiertos en mi primera toma de contacto con Zaragoza y siempre me ha dado todas las facilidades del mundo, tanto al principio para establecerme en mi puesto de trabajo y en la ciudad (nueva para mí) como a lo largo de toda la tesis donde sabía que cruzando el pasillo y llamando a su puerta siempre lo encontraría dispuesto a echar una mano.

A Víctor Martín Mayor lo conocí en la realización del máster de Física Teórica en la Universidad Complutense de Madrid. Con Víctor ha sido con la persona que más he aprendido a lo largo de esta tesis, ya fuera en sesudas sesiones de trabajo en su despacho, por videoconferencias o escapándonos en algún hueco entre charlas en una conferencia en Argentina para hacer una reunión improvisada. Su labor como director ha ido mucho más allá de sus obligaciones, y su dedicación y esfuerzo ha sido todavía mayor teniendo en cuenta que la mayoría de veces tuvimos que trabajar a distancia.

Quiero agradecer también toda la ayuda que me han prestado a Alfonso Tarancón y a Luis Antonio Fernández. Alfonso Tarancón oficialmente figura como mi ``Tutor de Tesis'', un título que no hace justicia a toda la ayuda que he recibido de su parte. Al igual que me ocurría con David, sabía que cruzando el pasillo y llamando a su puerta encontraría su ayuda para cualquier problema que se me presentase. Fue la persona que me introdujo a la docencia y creo que eso no se olvida. Podría decirse que ha sido un director no oficial de mi tesis. Y si Alfonso Tarancón ha sido mi director de tesis no oficial en Zaragoza, sin duda, Luis Antonio ha sido mi director de tesis no oficial en Madrid. Desde luego, Luis Antonio ha tenido la dedicación de un director tesis sin tener ninguna obligación a ello. Siempre ha estado dispuesto a ayudarme con cualquier cosa y recuerdo con especial cariño las largas tardes delante de su ordenador, explicándome las tripas de algún programa que me serviría de base para desarrollar uno nuevo, en las que yo muchas veces salía asustado con un montón de notas y con las cosas no del todo claras, pero que más tarde comenzaban a cobrar sentido cuando me metía en harina y descubría que esa extraña línea de código era uno de los checks exhaustivos que ahora son también imprescindibles en todos mis programas.

A mis directores de tesis y también a Alfonso y a Luis Antonio a quienes considero mis ``directores no oficiales'', muchas gracias. Sin vuestra ayuda esta tesis no habría sido posible.

También me gustaría mencionar en estos agradecimientos a la gente con la que realicé mis estancias de investigación fuera de España. Mi primera estancia de investigación la realicé en Roma, en la Universidad La Sapienza, donde fui recibido por Enzo Marinari y Andrea Maiorano. En el trabajo del día a día con quien más tiempo pasé fue con Andrea, que en aquella época estaba preparándose unas oposiciones y aún así sacaba tiempo de donde no lo había para echarme una mano e interesarse por mi trabajo. A Enzo, Federico, Giorgio, y especialmente a Andrea, me gustaría agradecerles su cálida acogida durante mi estancia en Roma. 

Mi segunda estancia de investigación la realicé en Paris, en la Universidad de Paris Sud. Debido a la pandemia las condiciones fueron mucho peores por razones obvias, sin embargo agradezco enormemente a Cyril y a Aurélien tanto su ayuda (y más teniendo en cuenta la situación tan delicada que había) como la posibilidad que me brindaron de comenzar a investigar en algo tan apasionante como el Machine Learning.

También quiero agradecer a todos los investigadores de los cuales he aprendido mucho en estos cuatro años. Especialmente me gustaría mencionar a Juan Jesús Ruiz que me introdujo en el mundo de los espines en mi TFG, que siempre me prestó su ayuda desde Badajoz y que incluso organizó un curso intensivo sobre el grupo de renormalización con el único objetivo de ayudarnos en nuestra tesis, sacrificando
muchas horas sin tener ninguna obligación a ello. También a David Yllanes, que siempre está al pie del cañón, y al que le tengo en gran estima. A Sergio Pérez Gaviro, que me ayudó muchísimo en mi primera toma de contacto en Zaragoza enseñándome todos los detalles técnicos, dándome programas para ayudarme en mi tarea hasta abrumarme y bajando conmigo al CPD para verle las tripas a los Janus. Y también a Ilaria e Isidoro, con los que he compartido director de tesis y a los que espero que les vaya realmente bien en su carrera investigadora y en su vida.

Por supuesto, no puedo olvidarme de toda la gente de Zaragoza con la que he trabajado estos últimos cuatro años. Isabel Vidal me ayudó con todo el tema administrativo. Nunca he visto a una persona hacer su trabajo tan bien como lo hacía ella, con tanto cuidado, cariño y siempre con una sonrisa. El grupo de Yamir Moreno, que me acogió desde el primer día con los brazos abiertos y con los que he vivido grandes momentos trabajando juntos. Toda la gente del BIFI, especialmente Pedro, Daniel, Sergio y Rubén que me ayudaron cada vez que tuve algún problema con \textit{Cierzo}.

Dejo para la parte final a aquellos a los que conocí mucho antes de iniciar el doctorado y que sé que seguirán formando parte de mi vida mucho después de completarlo. A mis amigos Kike, Álvaro, Pilar, Carlos, Helena, Cienfu, Gordillo, Kubicki, Luiso y Vázquez, muchas gracias por estar ahí estos años y los que vendrán. Por supuesto, muchas gracias a mi familia, a Tito y a Ariadna. También a Toni, Candela, Lolo y Ana. Sin vuestra ayuda, que va mucho más allá de la tesis, nada de esto hubiera salido adelante.

No me olvido de Sarita, con la que llevo media vida y con la que espero que me queden muchos años por delante. La persona que ha sido mi apoyo emocional durante toda esta tesis y a la que le debo tanto que me sabe a poco dedicarle solo estas pocas líneas. Sin ti, no sólo no hubiera sido posible esta tesis, tampoco muchas otras cosas en mi vida. Te quiero. A la Juani y al Ramón, los perros más fantásticos del mundo. Que comparten habitación conmigo (son perros investigadores) y cuya alegría al verme siempre me levanta el ánimo (te echaré muchísimo de menos Juanita). Y por último a mamá, que lo ha hecho todo por mí. Probablemente la persona con la que más haya discutido, y también la única persona que me ha acompañado en cada paso que he dado y que no me ha dejado caerme nunca. Sin ella no es que no hubiese sido posible esta tesis, es que no hubiera podido hacer nada en esta vida. Muchísimas gracias mamá, te quiero.

Estos agradecimientos pretendían ser unas pocas líneas pero sois muchos los que me habéis ayudado y me siento afortunado por teneros. A todas las personas que he nombrado, a todas aquellas que mi terrible memoria no me haya permitido incluir y a mi yaya Paula, a la que le dedico esta tesis, muchas gracias. Habéis dejado huella en mí y habéis contribuido a que esta tesis llegase a buen puerto.

Muchas gracias a todos.

\checkoddpage
\ifoddpage{
	\afterpage{\blankpage}
}\fi

\oldmainmatter 

\setlength{\overflowingheadlen}{\linewidth}
\addtolength{\overflowingheadlen}{\marginparsep}
\addtolength{\overflowingheadlen}{\marginparwidth}

\pagestyle{scrheadings} 
\setcounter{chapter}{-1}
\chapter{General Introduction}
\labch{general_introduction}
As usually happens in science, we have just used the light of our predecessors to open the door in front of us and we expect that our work provides enough light to those who are to come.

This thesis pretends to be another step in the development of numerical research in disordered\index{disorder!systems} systems. Specifically, we will focus on spin glasses which have demonstrated to be a fertile field from both, experimental and theoretical approaches. Throughout this thesis, we will discuss a variety of interesting phenomenons and we will also open new avenues to previously unexplored effects in the context of spin glasses. However, without a doubt, the leitmotiv conducting this thesis is the role of numerical simulations as a valuable tool to explore spin-glass physics.

Indeed, the development of this thesis has been possible due to the high-quality data obtained from a huge numerical effort. In particular, the role of the FPGA\index{FPGA}-based supercomputer Janus\index{Janus} II has been determinant in this thesis. Unprecedented large off-equilibrium simulations have allowed obtaining high-accuracy data that are the basis of several chapters in the present document. The Janus\index{Janus} II success has been possible due to the Janus\index{Janus} Collaboration, an international project involving five different universities from Italy and Spain that have been joining in their efforts to design and program the spin-glass dedicated computer which is at the forefront of numerical research. In addition to Janus\index{Janus} II, thousands of hours of computational time in CPUs have been also fundamental in the analysis of data. Specifically I would like to mention Cierzo supercomputer in BIFI and the Madrid's cluster in the UCM.

This thesis is organized into five different parts. The first part, containing the~\refch{def_spinglass}, is focused on introducing the spin glasses to the reader. It is impossible to account for all the work developed in the spin-glass field. Nonetheless, we try to properly introduce the reader into the spin-glass context, starting from the very definition of the studied system. We also discuss relevant experimental results, trying to focus on those that will come up throughout the thesis. The most relevant theoretical results are also provided in this introduction. Finally, numerical simulations, the main research-resource in this thesis, are introduced.

The second part, containing the~\refch{metastate}, is dedicated to discussing the metastate\index{metastate}. The theoretical development of the spin glasses in the thermodynamic limit\index{thermodynamic limit} has historically suffered mathematical inconsistencies. The irruption of the physical-mathematics in the context of spin glasses brought a solution to this problem introducing the concept of metastate\index{metastate}. However, for finite dimensions, there exist different metastate\index{metastate} pictures with different predictions. Since the invention of the metastate\index{metastate} in 1990, this concept has been in the theoretical \textit{world}, separated from experiments and numerical simulations, nonetheless, the current state of the art in numerical simulations have allowed us to numerically construct the metastate\index{metastate} and to (partially) elucidate between the competing metastate\index{metastate} pictures.

The third part, shaped by the~\refch{aging_rate} and \refch{mpemba}, is devoted to studying the off-equilibrium dynamics in spin glasses. This topic is of glaring importance in the spin glasses since experiments are (almost) always conducted in off-equilibrium conditions. Moreover recently, quantitative relevant relations have been found between equilibrium and out-equilibrium spin glasses. Learning about off-equilibrium dynamics might be useful to unveil the equilibrium properties of spin glasses concerning their very nature. Specifically, \refch{aging_rate} is focused on discussing the growth of the coherence length\index{coherence length} in spin glasses, a key quantity that characterizes the off-equilibrium evolution of those systems. In that chapter, we will witness a \textit{tour de force} of Janus\index{Janus} II, and we will be able to solve a discrepancy between numerical simulations and experiments by extrapolating our results to the relevant experimental time-scales. In~\refch{mpemba} we will discuss an interesting phenomenon: the Mpemba effect. It has been observed that, under some circumstances, if two beakers of water, one hotter than the other, are put in contact with a thermal reservoir at low temperatures, the hot water freezes faster than the cold water. We translate this off-equilibrium phenomenon to the spin-glass context for the first time and provide a satisfactory explanation.

The fourth part, containing the~\refch{Introduction_chaos}, \refch{equilibrium_chaos} and \refch{out-eq_chaos} is devoted to study the Temperature Chaos phenomenon in spin glasses. In~\refch{Introduction_chaos} we introduce the main historical results on Temperature Chaos, from its origins to the last steps. In~\refch{equilibrium_chaos} we study equilibrated spin glasses and we characterize the Temperature Chaos from a static and a dynamical point of view. We also reproduce previous results by relating both approaches through an exploration of the correlations between them and proposing new observables to improve those correlations. Moreover, in this chapter, we develop a numerical method to improve the dynamical estimation of Temperature Chaos. In~\refch{out-eq_chaos} we tackle the problem of characterizing Temperature Chaos in off-equilibrium dynamics. The very definition of Temperature Chaos refers to the reorganization of the \textit{equilibrium} configurations\index{configuration} of a spin glass upon small changes in the temperature. Then, the existence of the phenomenon in out-equilibrium systems needs to be established. In that chapter, we find a phenomenon that closely mimics the equilibrium Temperature Chaos. Indeed, we observe a strong relationship between the equilibrium and the off-equilibrium Temperature Chaos.

The fifth and last part of the main body of the thesis corresponds to the conclusions (\refch{conclusions}). There, we recall the relevant results of the thesis and we also discuss possible future works.

Finally, some appendixes can be found in the last part of the document. In~\refch{AP_statistics} we explain the relevant statistical tools that we have used in the analysis of this thesis. In~\refch{AP_technical_details_aging} and~\refch{AP_technical_details_out-eq_chaos} we provide technical details on the~\refch{aging_rate} and~\refch{out-eq_chaos} respectively. \refch{AP_PT} is devoted to explaining technical details on the use of the parallel\index{parallel!tempering} tempering method to compute the characteristic time-scales in Markov\index{Markov chain} chains. \refch{AP_multispin_coding} provides technical details on general methods of parallelization\index{parallel!computation} that are useful in our simulations and analysis.

The original results of this thesis are published in the following articles
\begin{itemize}
\item A. Billoire, L. A. Fernandez, A. Maiorano, E. Marinari, V. Martin-Mayor, J. Moreno-Gordo, G. Parisi, F. Ricci-Tersenghi, and J. J. Ruiz-Lorenzo. ‘Numerical Construction of the Aizenman-Wehr Metastate’. In: Phys. Rev. Lett. 119 (3 July 2017), p. 037203. doi: 10.1103/PhysRevLett.119.037203 \cite{billoire:17}

\item A Billoire, L A Fernandez, A Maiorano, E Marinari, V Martin-Mayor, J Moreno-Gordo, G Parisi, F Ricci-Tersenghi, and J J Ruiz-Lorenzo. ‘Dynamic variational study of chaos: spin glasses in three dimensions’. In: Journal of Statistical Mechanics: Theory and Experiment 2018.3 (2018), p. 033302. doi: 10.1088/1742-5468/aaa387 \cite{billoire:18}

\item M. Baity-Jesi, E. Calore, A. Cruz, L. A. Fernandez, J. M. Gil-Narvion, A. Gordillo-Guerrero, D. Iñiguez, A. Maiorano, E. Marinari, V. Martin-Mayor, J. Moreno-Gordo, A. Muñoz-Sudupe, D. Navarro, G. Parisi, S. Perez-Gaviro, F. Ricci-Tersenghi, J. J. Ruiz-Lorenzo, S. F. Schifano, B. Seoane, A. Tarancon, R. Tripiccione, and D. Yllanes. ‘Aging Rate of Spin Glasses from Simulations Matches Experiments’. In: Phys. Rev. Lett. 120 (26 June 2018), p. 267203. doi: 10.1103/PhysRevLett.120.267203 \cite{janus:18}

\item M. Baity-Jesi, E. Calore, A. Cruz, L. A. Fernandez, J. M. Gil-Narvión, A. Gordillo-Guerrero, D. Iñiguez, A. Lasanta, A. Maiorano, E. Marinari, V. Martin-Mayor, J. Moreno-Gordo, A. Muñoz-Sudupe, D. Navarro, G. Parisi, S. Perez-Gaviro, F. Ricci-Tersenghi, J. J. Ruiz-Lorenzo, S. F. Schifano, B. Seoane, A. Tarancon, R. Tripiccione, and D. Yllanes. ‘The Mpemba effect in spin glasses is a persistent memory effect’. In: Proceedings of the National Academy of Sciences 116.31 (2019), pp. 15350–15355 \cite{janus:19}

\item M. Baity-Jesi, E. Calore, A. Cruz, L. A. Fernandez, J. M. Gil-Narvión, I. Gonzalez-Adalid Pemartin, A. Gordillo-Guerrero, D. Iñiguez, A. Maiorano, E. Marinari, V. Martin-Mayor, J. Moreno-Gordo, A. Muñoz-Sudupe, D. Navarro, I. Paga, G. Parisi, S. Perez-Gaviro, F. Ricci-Tersenghi, J. J. Ruiz-Lorenzo, S. F. Schifano, B. Seoane, A. Tarancon, R. Tripiccione, and D. Yllanes. ‘Temperature chaos is present in off-equilibrium spin-glass dynamics’. In: Communications Physics 4.1 (Apr. 2021), p. 74 \cite{janus:21}
\end{itemize}

\setcounter{chapter}{-1}
\chapter{Introducción General}
\labch{introduccion_general}

Como habitualmente ocurre en ciencias, nuestro cometido ha sido usar la luz de nuestros predecesores para poder abrir la puerta que se encontraba delante de nosotros. Esperamos que nuestro trabajo dé la suficiente luz a aquellos que están por venir.

Esta tesis pretende ser otro paso más en el desarrollo de la investigación numérica en los sistemas desordenados. En concreto, nos centraremos en los vidrios de espín, que han demostrado ser un campo de conocimiento fértil tanto para los experimentos como para el desarrollo teórico. A lo largo de esta tesis discutiremos una gran variedad de fenómenos interesantes y también abriremos nuevos caminos hacia efectos que no habían sido explorados previamente en el contexto de los vidrios de espín. Sin embargo, sin género de duda, el leitmotiv de esta tesis es el papel de las simulaciones numéricas como una valiosa herramienta para explorar la física de los vidrios de espín.

Ciertamente, el desarrollo de esta tesis ha sido posible gracias a la alta calidad de los datos obtenidos con un inmenso esfuerzo numérico. En concreto, el rol de la supercomputadora, basada en FPGA, Janus II ha sido determinante en esta tesis. Simulaciones masivas fuera del equilibrio sin precedentes han permitido obtener datos altamente precisos que son la base de varios capítulos en este texto. El éxito de Janus II ha sido posible gracias a la colaboración Janus, un projecto internacional que involucra cinco universidades distintas de Italia y España que han unido sus fuerzas para diseñar y programar un hardware dedicado a la simulación de vidrios de espín en la vanguardia de la investigación numérica. Además de Janus II, miles de hora de tiempo computacional en CPU han tenido también un papel fundamental en el análisis de datos. En concreto, me gustaría mencionar la supercomputadora Cierzo, en el BIFI y el cluster de Madrid, en la UCM.

Esta tesis está organizada en cinco partes diferentes. La primera parte, formada por el Capítulo 1, está centrada en introducir los vidrios de espín al lector. Es imposible dar cuenta de todo el trabajo desarrollado en el campo de los vidrios de espín. No obstante, intentamos introducir apropiadamente al lector al contexto de los vidrios de espín, empezando por la definición del sistema en cuestión. En este capítulo también discutimos los resultados experimentales más importantes, tratando de dar más relevancia a aquellos que guardan relación directa con los temas tratados en esta tesis. Los resultados teóricos más relevantes también se plasman en este capítulo introductorio. Finalmente, las simulaciones numéricas, el principal recurso para el desarrollo de la investigación en esta tesis, son introducidas en este capítulo.

La segunda parte, que incluye el Capítulo 2, está dedicada a discutir el metaestado. El desarrollo teórico de los vidrios de espín en el límite termodinámico ha sufrido históricamente de inconsistencias matemáticas. La irrupción de la física-matemática en el contexto de los vidrios de espín trajo la solución a este problema introduciendo el concepto del metaestado. Sin embargo, para dimensión finita, existen diferentes teorías del metaestado con diferentes predicciones. Desde la invención del metaestado en 1990, el concepto ha estado restringido al mundo teórico, separado de experimentos y de simulaciones numéricas, no obstante, el actual estado del arte en las simulaciones numéricas ha permitido construir numéricamente el metaestado y (al menos parcialmente) dilucidar entre las distintas teorías que lo describen.

La tercera parte, formada por los Capítulos 3 y 4, está dedicada al estudio de la dinámica de no equilibrio de los vidrios de espín. Este tema es de palmaria importancia en estos sistemas puesto que los experimentos (casi) siempre son llevados a cabo fuera del equilibrio. Además, relaciones cuantitativas relevantes han sido halladas entre los vidrios de espín en equilibrio y fuera del equilibrio. Aprender sobre la dinámica fuera del equilibrio puede ser útil para desvelas las propiedades de equilibrio de los vidrios de espín, relacionadas con la misma naturaleza de estos sistemas. Específicamente, el Capítulo 3 está centrado en discutir el crecimiento de la longitud de coherencia en los vidrios de espín, una cantidad clave que caracteriza la evolución fuera del equilibrio de estos sistemas. En este capítulo, presenciaremos un auténtico \textit{tour de force} de Janus II y seremos capaces de resolver una discrepacia entre experimentos y simulaciones numéricas extrapolando nuestros resultados hasta las escalas de tiempo relevantes para los experimentos. En el Capítulo 4 discutiremos un fenómeno interesante: el efecto Mpemba. Se ha observado que, bajo ciertas circunstancias, si dos recipientes de agua, uno caliente y otro frío, son puestos en contacto con un baño térmico a baja temperatura, el agua caliente se congela antes que el agua fría. Nosotros traducimos este fenómeno que se da fuera del equilibrio al contexto de los vidrios de espín por primera vez y damos una primera explicación del mismo.

La cuarta parte, formada por los Capítulos 5, 6 y 7, está dedicada a estudiar el fenómeno del Caos en Temperatura en vidrios de espín. En el Capítulo 5 introducimos los principales resultados históricos sobre el Caos en Temperatura, desde sus orígenes hasta los últimos pasos. En el Capítulo 6 estudiamos vidrios de espín en equilibrio y caracterizamos el Caos en Temperatura desde un punto de vista estático y dinámico. Además, relacionamos ambos enfoques explorando las correlaciones entre ambos. Además, en este capítulo desarrollamos un método numérico para mejorar la estimación dinámica del Caos en Temperatura. En el Capítulo 7, atacamos el problema de caracterizar el Caos en Temperatura en una dinámica fuera del equilibrio. La propia definición del Caos en Temperatura se refiere a las configuraciones de \textit{equilibrio} del vidrio de espín bajo un cambio arbitrariamente pequeño de temperatura. Por lo tanto, la existencia del fenómeno fuera del equilibrio debe ser establecida. En este capítulo, encontramos un fenómeno que imita a la perfección al fenómeno del Caos en Temperatura en equilibrio. De hecho, observamos una fuerte relación entre el Caos en Temperatura en equilibrio y fuera del equilibrio.

La quinta y última parte del cuerpo principal de la tesis corresponde a las conclusiones. En el Capítulo 8 revisitamos los resultados más relevantes de esta tesis y también discutimos sobre posibles futuros trabajos.

Finalmente, el lector puede encontrar algunos apéndices en la parte final de este texto. En el apéndice A explicamos cuáles son las herramientas estadísticas más relevantes que hemos usado a lo largo de esta tesis. En los apéndices B y D se dan detalles técnicos de los capítulos 3 y 7 respectivamente. En el apéndice C se explican los detalles técnicos del uso del método de \textit{Parallel Tempering} para calcular las escala de tiempo características en las cadenas de Markov. Por último, en el apéndice E se dan los detalles técnicos de métodos generales de paralelización que han sido útiles en nuestras simulaciones y análisis.

Los resultados originales de esta tesis están publicados en los siguientes artículos
\begin{itemize}
\item A. Billoire, L. A. Fernandez, A. Maiorano, E. Marinari, V. Martin-Mayor, J. Moreno-Gordo, G. Parisi, F. Ricci-Tersenghi, and J. J. Ruiz-Lorenzo. ‘Numerical Construction of the Aizenman-Wehr Metastate’. In: Phys. Rev. Lett. 119 (3 July 2017), p. 037203. doi: 10.1103/PhysRevLett.119.037203 \cite{billoire:17}

\item A Billoire, L A Fernandez, A Maiorano, E Marinari, V Martin-Mayor, J Moreno-Gordo, G Parisi, F Ricci-Tersenghi, and J J Ruiz-Lorenzo. ‘Dynamic variational study of chaos: spin glasses in three dimensions’. In: Journal of Statistical Mechanics: Theory and Experiment 2018.3 (2018), p. 033302. doi: 10.1088/1742-5468/aaa387 \cite{billoire:18}

\item M. Baity-Jesi, E. Calore, A. Cruz, L. A. Fernandez, J. M. Gil-Narvion, A. Gordillo-Guerrero, D. Iñiguez, A. Maiorano, E. Marinari, V. Martin-Mayor, J. Moreno-Gordo, A. Muñoz-Sudupe, D. Navarro, G. Parisi, S. Perez-Gaviro, F. Ricci-Tersenghi, J. J. Ruiz-Lorenzo, S. F. Schifano, B. Seoane, A. Tarancon, R. Tripiccione, and D. Yllanes. ‘Aging Rate of Spin Glasses from Simulations Matches Experiments’. In: Phys. Rev. Lett. 120 (26 June 2018), p. 267203. doi: 10.1103/PhysRevLett.120.267203 \cite{janus:18}

\item M. Baity-Jesi, E. Calore, A. Cruz, L. A. Fernandez, J. M. Gil-Narvión, A. Gordillo-Guerrero, D. Iñiguez, A. Lasanta, A. Maiorano, E. Marinari, V. Martin-Mayor, J. Moreno-Gordo, A. Muñoz-Sudupe, D. Navarro, G. Parisi, S. Perez-Gaviro, F. Ricci-Tersenghi, J. J. Ruiz-Lorenzo, S. F. Schifano, B. Seoane, A. Tarancon, R. Tripiccione, and D. Yllanes. ‘The Mpemba effect in spin glasses is a persistent memory effect’. In: Proceedings of the National Academy of Sciences 116.31 (2019), pp. 15350–15355 \cite{janus:19}

\item M. Baity-Jesi, E. Calore, A. Cruz, L. A. Fernandez, J. M. Gil-Narvión, I. Gonzalez-Adalid Pemartin, A. Gordillo-Guerrero, D. Iñiguez, A. Maiorano, E. Marinari, V. Martin-Mayor, J. Moreno-Gordo, A. Muñoz-Sudupe, D. Navarro, I. Paga, G. Parisi, S. Perez-Gaviro, F. Ricci-Tersenghi, J. J. Ruiz-Lorenzo, S. F. Schifano, B. Seoane, A. Tarancon, R. Tripiccione, and D. Yllanes. ‘Temperature chaos is present in off-equilibrium spin-glass dynamics’. In: Communications Physics 4.1 (Apr. 2021), p. 74 \cite{janus:21}
\end{itemize}

\addpart{A brief introduction to Spin Glasses}
\chapter[A brief introduction to spin glasses]{A brief introduction \\ to spin glasses}
\labch{def_spinglass}
\setlength\epigraphwidth{.5\textwidth}

\epigraph{\textit{Todo empezó aquel fatídico día \dots}}{-- Jonathan Stroud, \textit{El amuleto de Samarkanda} }

The first signs of unusual properties of alloys between noble metals and transition metals such as \gls{AuFe} or \gls{CuMn}, which later become the paradigmatic examples of spin glasses, were discovered in the second half of the 50's and the 60's through measurement of the specific heat\index{specific heat} at low temperatures \cite{denobel:59,zimmerman:60}, observation of time-dependent remanent magnetization\index{magnetization} \cite{kouvel:60,kouvel:61} and measurements of the ESR spectra \cite{owen:56}. In parallel, the first interpretations of those experimental results from a theoretical point of view came out \cite{marshall:60,klein:63} through \gls{RKKY}\index{RKKY interaction} interactions \cite{ruderman:54,kasuya:56,yosida:57}.

The experiments and the first developed theories suggested that those systems present a low-temperature magnetic order\index{magnetic order} which exhibits different properties from the known condensed-matter phases. Therefore, a natural question arose: there exists a sharp phase transition\index{phase transition} from a high-temperature phase\index{phase!high-temperature/paramagnetic} to a low-temperature phase\index{phase!low-temperature/spin-glass} capable of describing the observed phenomena? 

The very existence of this sharp phase transition\index{phase transition} was discussed during the 1960s and the first half of the 1970s \cite{violet:66,violet:67,beck:71,tholence:74}. However, the experiments of Canella, Mydosh, and Budnick \cite{cannella:71,cannella:72} unveiled a sharp cusp in the ac susceptibility\index{susceptibility!ac} of spin glasses such as \gls{CuMn}. Those experiments were strong support for the sharp transition\index{phase transition} hypothesis between a high-temperature paramagnetic phase\index{phase!high-temperature/paramagnetic} and a, at that moment new, low-temperature phase\index{phase!low-temperature/spin-glass}. Indeed, subsequent works in the following years brought evidence against this sharp transition\index{phase transition}~\cite{mulder:81,mulder:82} unveiling the rounded nature of the susceptibility\index{susceptibility} curve. It was in 1991 when the community got convinced about the existence of the phase transition\index{phase transition} in experimental spin-glasses \cite{gunnarsson:91}. Nonetheless, at that time (the early 1970s), although still debated, the experimental results were supporting the phase transition\index{phase transition} to occur.

Moreover, previous experiments of neutron scattering\index{neutron scattering}, indicated that there was not an order of the constitutive elements responsible for the magnetization\index{magnetization}, the spins, in the low-temperature phase\index{phase!low-temperature/spin-glass} \cite{arrott:65}. 

The evidence of no order in the low-temperature phase\index{phase!low-temperature/spin-glass}, together with the results that suggested a sharp phase transition\index{phase transition}, led Edwards and Anderson to propose their theory \cite{edwards:75}. This theory was a breakthrough in the study of the spin glasses because, although it was a mean-field\index{Mean-Field!theory} theory, proposed a model that put aside the \gls{RKKY}\index{RKKY interaction} interactions in favor of random interactions following probability distribution. This simple conceptual change opened the door to a large class of systems to be studied in the same terms that spin glasses did.

After the Edwards-Anderson theory\index{Edwards-Anderson!theory} not only the spin glasses became a hot topic in the condensed matter\index{condensed matter} research, but also the divergence between the theoretical and the experimental researches was considerable. Some effort was indeed done to establish some interplay, for example, scaling theories of domain\index{magnetic domain} growth \cite{fisher:88,koper:88,ocio:90,bouchaud:01}, however, the mainstream of both, theory and experiments, were quite apart from each other, as we briefly discuss below.

Due to the glassy nature of the spin glasses in low temperatures, experiments always were developed in off-equilibrium conditions while the theoretical work put its efforts on capture the nature of spin glasses at low temperatures, with results that require access to the spin configurations\index{configuration} for equilibrated systems in the thermodynamic limit\index{thermodynamic limit}.

Traditionally, numerical work has been a powerful tool of theoretical research. Numerical studies were, almost always, developed in small systems at thermal equilibrium, very far from the experiments, most of them performed out of equilibrium in very large systems. Moreover, the numerical simulations performed in off-equilibrium conditions could only reach time-scales much smaller than the experimental ones. However, the continuous increase of the computational power and the development of special-purpose computers\index{special-purpose computer} have allowed the study of the spin glasses at low temperatures out of equilibrium in the relevant experimental scales.

An exciting new venue for numerical research is bridging the experimental approach to spin glasses and the theoretical one. This thesis pretends to be a step forward in this sense.

In this chapter, the reader may find a general introduction to spin glasses. First of all, a first definition of the concept of spin glass will be introduced (see \refsec{definition_spinglass}). Then, some experimental background will be provided in \refsec{experimental_spinglass}. The main theoretical pictures, with their different quantitative predictions, will be exposed in \refsec{theoretical_spinglass}. Finally, a brief introduction to the numerical work performed in this thesis together with some useful definitions can be found in \refsec{numerical_spinglass}.

All the results presented in this chapter are not original from this thesis. In addition to the cited papers that will appear in this chapter, the reader may find useful general references about spin glasses in \cite{binder:86,mezard:87,mydosh:93,fischer:93,young:98,dotsenko:01,dedominicis:06}.

\section{A first definition of spin glass} \labsec{definition_spinglass}
Spin glasses are magnetic systems whose low-temperature phase\index{phase!low-temperature/spin-glass} is frozen and disordered\index{disorder}. From a practical point of view, we consider the spins (i.e. the magnetic moments) as the constitutive elements of the spin glass. We denote $\vec{s}_{\vec{r}}$ as the spin at the position $\vec{r}$, and $\braket{\dots}_t$ as the mean over the experimental time.

A direct consequence of the low-temperature phase\index{phase!low-temperature/spin-glass} to be frozen is $\braket{\vec{s}_{\vec{r}}}_t \neq 0$. Moreover, the direction in which the spins freeze are random, which has a direct implication in the magnetization\index{magnetization} $\sum_{\vec{r}} \braket{\vec{s}_{\vec{r}}}_t = 0$. Actually, no long-range magnetic order\index{magnetic order} will appear in the low-temperature phase\index{phase!low-temperature/spin-glass}
\begin{equation}
 \vec{M}_{\vec{k}} = \dfrac{1}{N} \sum_i e^{-i\vec{k}\cdot\vec{r}}\braket{\vec{s}_{\vec{r}}}_t = 0 \quad \,\, \forall \, \vec{k} .  \labeq{no_magnetic_order}
\end{equation} 
This general expression includes the ferromagnetic order parameter\index{order parameter} $\vec{k} = \vec{0}$ and the antiferromagnetic one $\vec{k} = (\pi,\pi,\pi)$.

These two characteristics, freezing and disorder\index{disorder}, give us the key to understand the name \gls{SG} which was proposed for the first time by Anderson \cite{anderson:70}. The sluggish evolution of the system together with the disorder\index{disorder} reminds the structural glass\index{structural glass} main properties trading the magnetic moments by the position of the particles. In structural glasses\index{structural glass}, particles occupy random positions with no time-evolution (disorder\index{disorder} and freezing).

There exists a far well establish consensus about the principal ingredients that one model which aspires to exhibit the \gls{SG}'s phenomenology must include: \textit{randomness}\index{randomness} in the interactions and \textit{frustration}\index{frustration} \cite{toulouse:77,blandin:78}. 

\subsection{A key property: frustration}
Suppose that a pair of spins $\vec{s}_{\vec{r}_i}$ and $\vec{s}_{\vec{r}_j}$ interact with each other through the Hamiltonian\index{Hamiltonian} $\mathcal{H}_{ij} = -J_{ij}\vec{s}_{\vec{r}_i}\cdot \vec{s}_{\vec{r}_j}$, being $J_{ij}$ the \textit{coupling}\index{couplings} interaction between the spins at the positions $\vec{r}_i$ and $\vec{r}_j$. For $J_{ij}>0$ the more energetically favorable values for the pair of spins $\vec{s}$ i.e. the more energetically favorable \textit{configurations},\index{configuration} are those in which the spins are parallel to each other. In the same way, for $J_{ij}<0$ the spins tends to align in an anti-parallel way. For a collection of spins, the Hamiltonian\index{Hamiltonian} can be easily generalized
\begin{equation}
\mathcal{H} = -\sum_{i,j}J_{ij}\vec{s}_{\vec{r}_i}\cdot \vec{s}_{\vec{r}_j} \labeq{first_Hamiltonian} \, .
\end{equation}
We will say that a system is frustrated\index{frustration} when it is not possible to satisfy simultaneously all the pairwise interactions i.e. there is no way to maximize simultaneously all the summands of \refeq{first_Hamiltonian}.

\begin{figure}[h]
\centering
\includegraphics[width=0.8\textwidth]{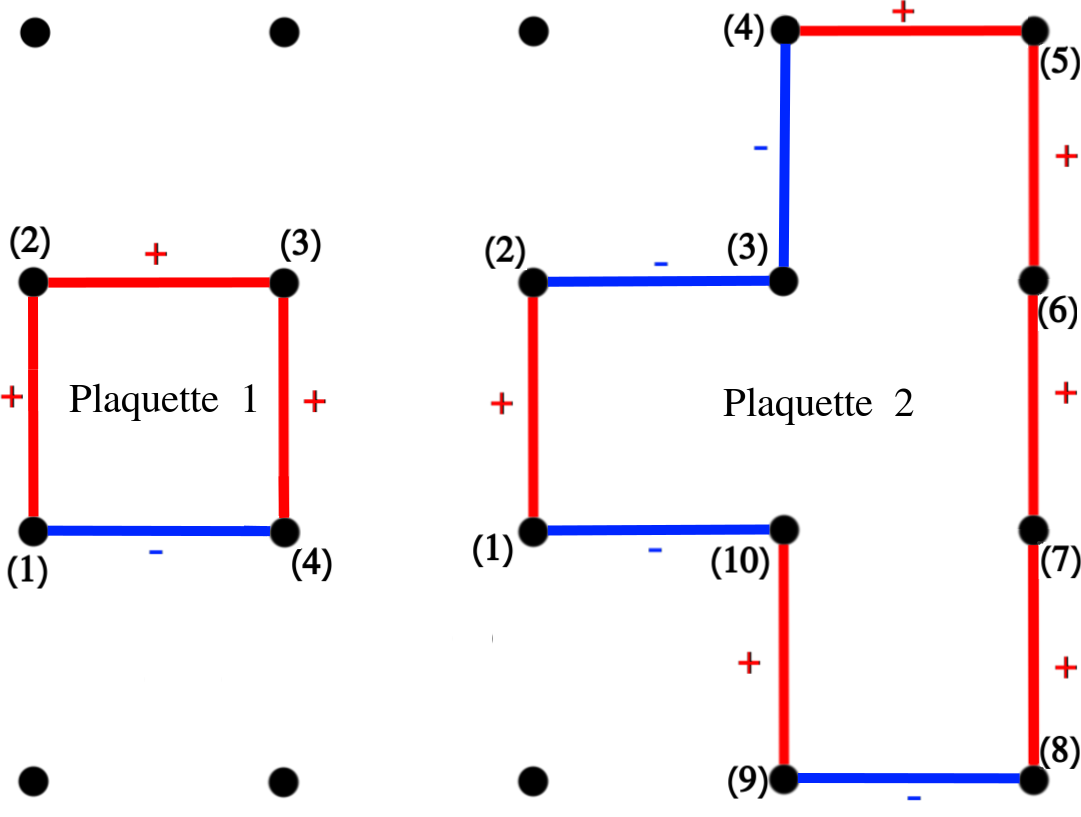}
\caption[\textbf{A graphical example of frustration.}]{\textbf{A graphical example of frustration\index{frustration}.} Plaquette\index{plaquette} 1 is said to be frustrated\index{frustration} because the spins, that tend to align in a parallel or anti-parallel way depending on the couplings\index{couplings} (+ or - respectively), can not satisfy all the interactions simultaneously. However, in the unfrustrated plaquette\index{plaquette} 2, we observe that neither the size of the plaquette\index{plaquette} nor the mixture of positive and negative interactions is responsible for the frustration\index{frustration}. The key is the number of negative couplings\index{couplings} (respectively positive): if the number of, say, negative couplings\index{couplings} is odd, then we will have a frustrated\index{frustration} plaquette\index{plaquette}, otherwise, we will have an unfrustrated plaquette\index{plaquette}.}
\labfig{frustration}
\end{figure}

The canonical example of a frustrated\index{frustration} system can be found in \reffig{frustration} where the spins lie in the nodes of a square-regular\index{regular lattice!square} lattice\footnote{As far as we are just dealing with the sign of the couplings\index{couplings}, that particular disposition of the spins only try to simplify the visualization without any generality loss.} in which the edges represent the coupling\index{couplings} interactions between the spins. We firstly focus our attention in the closed-loop, also called \textit{plaquette}\index{plaquette}, labeled with the number 1. As long as we are discussing here just if the spins are parallel or anti-parallel, let us take for the sake of simplicity spins with values $\uparrow$ (up) and $\downarrow$ (down). Suppose that the spin (1) is $\uparrow$, following to spin (2) through a positive coupling\index{couplings} ($+$) which favors the parallel interactions, the value of spin (2) also should be $\uparrow$. The same reasoning can be applied to the spin (3). Finally, the spin (4) is at the end of the loop and, therefore, has to satisfy two couplings\index{couplings}: the positive ($+$) coupling\index{couplings} with the spin (3) and the negative ($-$) coupling\index{couplings} with the spin (1). Any value that the spin (4) takes will lead to one unsatisfied interaction. We say that the plaquette\index{plaquette} 1 is frustrated\index{frustration}. 

However, if we focus now on plaquette\index{plaquette} 2 and we apply the same mental exercise, we can easily find a configuration\index{configuration} of spins that satisfies all the interactions, for example, the sequence (1)$\uparrow$, (2)$\uparrow$, (3)$\downarrow$, (4) $\uparrow$, (5) $\uparrow$, (6) $\uparrow$, (7) $\uparrow$, (8) $\uparrow$, (9) $\downarrow$ and (10) $\downarrow$. The length of the plaquette\index{plaquette} is irrelevant and one can infer from these examples that the key is the number of negatives (or equivalently positives) interactions; if the number of negative (positive) interactions that favor anti-parallel (respectively parallel) spin-alignment is odd, then the plaquette\index{plaquette} will be frustrated\index{frustration}, otherwise, we will say that the plaquette\index{plaquette} is unfrustrated.

In frustrated\index{frustration} systems there exist many configurations\index{configuration} with several unsatisfied interactions in which any local change would lead to an increase of the energy\index{energy}, thus, frustration\index{frustration} draws a rugged free-energy\index{free energy!landscape} landscape with many metastable\index{metastability} states and high-energy barriers. Indeed, the rugged free-energy\index{free energy!landscape} landscape is directly related to the frozen nature of \gls{SG}s and other characteristic properties as the slow time-evolution.

Two clarifications should be made at this point. First, the mixture of positive and negative interactions do not guarantee the system to be frustrated\index{frustration},\index{frustration} see \reffig{frustration}, plaquette\index{plaquette} 2, or Mattis model\index{Mattis model} \cite{mattis:76} which can be mapped to a uniform ferromagnet\index{ferromagnet}. Lastly, frustration\index{frustration} without randomness\index{randomness} (or vice versa) does not lead to \gls{SG} behavior, the most simple example is the regular\index{regular lattice!triangular} triangular lattice with antiferromagnetic\index{antiferromagnetic} interactions\footnote{Exactly solvable, see \cite{wannier:50}}, which is a fully frustrated\index{frustration} system with a large ground-state\index{ground-state} degeneracy but without phase transition\index{phase transition} at finite temperature\footnote{More examples of frustrated\index{frustration} systems without randomness\index{randomness} can be found in \cite{villain:77,villain:77b,wolff:82,wolff:83,wolff:83b,mackenzie:81}}.

We conclude that frustration\index{frustration} and randomness\index{frustration}\index{randomness} are necessary conditions to have a \gls{SG} but not sufficient ones. An illustrative example with this respect can be found in the Ising\index{Ising} ferromagnet in a small random magnetic field, where frustration\index{frustration} and randomness\index{randomness} appear in a very weak way and it is possible to find long-range magnetic order\index{magnetic order}.

\subsection{Why spin glasses?}
A natural question is, if the \gls{SG}s reproduce the glassy behavior at low temperatures, why should we focus on them instead on the structural\index{structural glass} glasses? The main reason to study the glassy behavior through \gls{SG}s is their simplicity. This simplicity allows the development of theoretical tools in \gls{SG}s that can be later applied to other fields of the complex systems \cite{mezard:85c,mezard:86b,amit:85,amit:85b,goldstein:92} (paradoxically, including the structural\index{structural glass} glasses \cite{charbonneau:14}).

Moreover, \gls{SG}s exhibit a wide set of characteristic phenomenons of glassy behavior with several advantages from the theoretical and experimental points of view. 

First of all, it is worthy to note that, unlike the structural\index{structural glass} glasses, the phase transition\index{phase transition} is well-known in \gls{SG}s, both in experiments through the study of the susceptibility\index{susceptibility} \cite{gunnarsson:91} and in the theoretical models \cite{palassini:99,ballesteros:00}. The physicists can greatly benefit from this fact by establishing quantitative criteria to determine whether or not they are studying the glassy phase\index{phase!low-temperature/spin-glass}.

Besides, from the experimental point of view, \gls{SQUID}\index{Superconducting Quantum Interference Device (SQUID)} allows for very precise measures, much more difficult to perform in structural\index{structural glass} glasses.

Finally, a technical reason is that \gls{SG}s are much simpler to simulate than structural\index{structural glass} glasses. The lattice models are very easy to simulate numerically.

\subsubsection{Further motivations in the study of spin glasses}
In addition to the above-exposed advantages of studying \gls{SG}s in order to explore the glassy behavior, there exist other fields in which the study of these systems is interesting and prolific. The study of complexity in optimization problems constitutes a paradigmatic example in this regard~\cite{barahona:82b}. 

The Turing machine\index{Turing machine} is a theoretical machine widely used in computation theory introduced by Turing~\cite{turing:37}. The deterministic version of the machine is able to give at most one result for every situation while the nondeterministic one is able to provide more than one result in each situation.

The set of problems that can be solved by a deterministic Turing machine\index{Turing machine} in polynomial time belongs to the set P\index{complexity!P} of problems. In the same way, the set of problems that can be solved by a non-deterministic Turing machine in polynomial time belong to the set of NP problems. It is trivial to see that P\index{complexity!P} problems are a subset of NP\index{complexity!NP} problems.

Specifically, there exists a subset of NP\index{complexity!NP} which is called NP-complete\index{complexity!NP-complete}\footnote{The concept of NP-completeness was introduced in~\cite{cook:71}.}, which is of special interest. We say that a problem is NP-complete if it belongs to the complexity class NP and all the NP problems are reducible (in polynomial time) to that problem. From a computational point of view, the NP-complete\index{complexity!NP-complete} problems are the hardest in the set NP\index{complexity!NP} and are equivalent to each other (see Cook-Levin theorem~\cite[p.38]{garey:79}), in the sense that founding a polynomial-bounded algorithm for any one of them would effectively yield a polynomial-bounded algorithm for all.

There exist many problems in the NP-complete\index{complexity!NP-complete} set (see, for instance ~\cite{karp:72}) and, specifically, the problem of finding the ground-state\index{ground-state} for a three-dimensional Ising \gls{SG} \footnote{See~\refsubsec{source_randomness} for the concept of Ising \gls{SG}.} is NP-complete\index{complexity!NP-complete}~\cite{barahona:82}. 

Although the three-dimensional case is of special interest in this thesis, there exist several models of \gls{SG}s that have been studied with great detail from the complexity point of view. In particular, the question of the \textit{planarity}\index{planar graph} of the \gls{SG} lattice has proven to be central in the complexity discussion~\cite{istrail:00}.

The concept of planar graph\index{planar graph} is rather intuitive. A graph is said to be planar\index{planar graph} if it can be drawn in a plane without edge-crossing\footnote{Although the concept is rather intuitive, in general, prove that a graph is planar\index{planar graph} requires non-trivial criteria like Kuratowski's theorem.}. On the contrary, the fact that the problem of finding the ground-state\index{ground-state} in a non-planar graph\index{non-planar graph} is NP-complete\index{complexity!NP-complete} is certainly not intuitive. The reader may find an interesting study of this problem for several graphs in~\cite{istrail:00}.

An interesting fact in the case of \gls{SG}s is the two-dimensional case. The typical case in the study of spin systems is to consider the spins placed in the vertex of a lattice and only take into account nearest-neighbors interactions. In this case, for the two-dimensional case, finding the ground-state is a P problem. However, if one considers next-nearest-neighbors interactions, the problem of finding the ground-state becomes NP-complete. This case might be surprising at first sight since the basic elements building the lattice are complete graphs of fourth order\footnote{A complete graph of $n^{\mathrm{th}}$ order, also knows as $K_n$ is a graph with $n$ nodes where every node is connected to the rest of them. In other contexts are also known as \textit{fully-connected networks}.}, that fulfill Kuratowski's criteria and, therefore, are planar. The reader may find a deep discussion in this respect in~\cite{istrail:00}.

Therefore, the study of \gls{SG}s is not only interesting from the statistical physics or the solid-state physics point of view but also from the complexity point of view.

\subsection{Beyond spin glasses. Weakly disordered versus strongly disordered systems}
Throughout this first approach to \gls{SG}s, we have set that the disorder is one of the essential characteristics that a system must have in order to exhibit the \gls{SG} behavior. However, there exist a variety of systems exhibiting disorder that cannot be identified as \gls{SG}s.

We have already introduced the Hamiltonian for spin systems in~\refeq{first_Hamiltonian}. The simplest case, with $J_{ij}=J$ constant, corresponds to the Ising ferromagnet. Therefore, the introduction of disorder can be regarded as an additional random term
\begin{equation}
J_{ij} = J + \delta J_{ij} \, .
\end{equation}

The limiting cases $\delta J_{ij} \ll J$ and $\delta J_{ij} \gg J$ correspond to weak and strong disorder respectively. Specifically, in this thesis we will focus on systems with a strong disorder\index{disorder!systems}, which is the case of \gls{SG}s. However, there exist a variety of systems exhibiting weak disorder with a very different but rich phenomenology.

The study of weak-disorder systems\index{disorder!systems} is a natural generalization of the study of pure systems\footnote{In this context, pure systems refer to the absence of impurities in their composition.} by the introduction of impurities that are unavoidable in real systems. In these systems, the ground-state\index{ground-state} and the equilibrium properties keep a close relationship with the pure system obtained by removing its impurities. However, the presence of these impurities may affect the behavior of the system in the neighborhood of the critical temperature\index{critical temperature} \cite{mccoy:70,balagurov:73,harris:74,harris:74b,khmelnitskii:75,grinstein:76}. 

Moreover, for systems undergoing a first-order phase transition, the effects of the weak disorder have been widely studied. Particularly, in two-dimensional systems, any arbitrarily small amount of disorder makes the first-order transition to become a second-order one~\cite{hui:89,aizenman:90,cardy:97,jacobsen:98,chatelain:98}. The three-dimensional case is more difficult to study but there exist results suggesting that a change of the order of the transition from first- to second-order occurs when the disorder increases~\cite{fernandez:12}. This is known as the Cardy-Jacobsen conjecture\index{Cardy-Jacobsen conjecture}~\cite{cardy:97,jacobsen:98}.

On the contrary, in strong-disorder systems, like \gls{SG}s, the ground-state, the equilibrium properties of the system, and the phase transition completely differs from the pure system, as we will illustrate throughout this thesis.

The disorder can also be present in the form of an external random field $h_i$. This type of disorder, with the absence of the coupling disorder, leads to the well-known and widely-studied \gls{RFIM}\index{Random Field Ising Model}~\cite{imry:75,nattermann:98,belanger:98}. This model is paradigmatic in the study of disordered systems and can be regarded as an intermediate case between weak and strong disorder.

In the two-dimensional case, the ordered phase is destroyed by the random field~\cite{binder:84,aizenman:90}, i.e. the effects in the transition are strong and the properties of the pure system give no clue to understand the disordered system. However, in the three-dimensional case, the critical behavior of the system is changed by the presence of the random fields but still exhibits an ordered phase~\cite{imbrie:84,bricmont:87}, similarly to what occur in the weak-disorder systems.

The importance of the \gls{RFIM} in the statistical physics literature yields over several reasons. Standing as one of the simplest disordered systems with a rich phenomenology is one. Moreover, this model has numerous representatives in nature, for example, the diluted antiferromagnets in a homogeneous external field~\cite{fishman:79} and it has been intensively studied, from both the experimental~\cite{belanger:98} and the theoretical~\cite{nattermann:98} point of view. Besides, contrary to the \gls{SG} case (as we will discuss in subsequent sections), the theoretical and experimental \gls{RFIM} have been developed in parallel for many years.

The number of weak-disorder systems and the results characterizing them is large and of central importance in the statistical physics field, as we have illustrated above. Nonetheless, the presence of weak disorder is not always affecting the physics of the pure system. We discuss this issue next.

\subsubsection{The Harris criterion}
We have stated above that the presence of impurities may affect the critical behavior of a system. There exists a quantitative criterion to know if the presence of the disorder is going to be relevant: the Harris criterion\index{Harris criterion}~\cite{harris:74}.

The argument behind the original formulation of the Harris criterion is rather simple and a useful discussion can be found in~\cite{brooks:16}. The main idea is that the sign of the critical exponent associated with the specific heat\index{specific heat} ($\alpha$) for the pure system, is the stability condition itself.

If $\alpha>0$, the disorder will change the critical behavior. On the contrary, the disorder will be irrelevant and the critical behavior of the system will not change.

\section{Experimental spin glasses} \labsec{experimental_spinglass}
As we said in the introduction of the chapter, the paradigmatic examples of \gls{SG}s are transition metal impurities in noble metal hosts like \gls{CuMn}, \gls{AuFe}, \gls{AgMn}, \gls{CuFe}, etc. However, there exist other types of materials that exhibit the \gls{SG} phenomenology. Is the case of rare earth constituents in metallic host \gls{YDy}, \gls{YEr}, \gls{ScTb}, etc. Also the same holds for ternary systems, e.g \gls{LaGdAl}. We will discuss the interactions that are the source of the randomness\index{randomness} and frustration\index{frustration} necessary to have a \gls{SG} low-temperature phase\index{phase!low-temperature/spin-glass} in \refsubsec{source_randomness} and we will expose relevant experiments that had a historical impact in the development of the \gls{SG}s field and that are somehow related to the results presented in this thesis in \refsubsec{aging_memory_rejuvenation}

\subsection{Internal structure and magnetic interactions: the source of randomness}
\labsubsec{source_randomness}
We have set that one of the main ingredients to have a \gls{SG} is the randomness\index{randomness}. How the real systems achieve that randomness\index{randomness} in the interactions? There exist two main ways to obtain it: \textit{bond randomness}\index{randomness!bond} and \textit{site randomness}\index{randomness!site}.

Bond randomness\index{randomness!bond} is a type of disorder\index{disorder} present in real systems like \gls{RbCuCoF} and \gls{FeMnTiO}. These systems present regular\index{regular lattice} lattices where the dominant magnetic interactions are short-ranged, then the impurities (cobalt and manganese respectively) are introduced. This procedure mimics what is called \textit{ideal spin-glass} i.e. a regular\index{regular lattice} lattice of spins interacting with their nearest-neighbors in a ferromagnetic or antiferromagnetic random way.

Site randomness\index{randomness!site} is the type of disorder\index{disorder} that is present in the most commonly studied \gls{SG}s such as \gls{CuMn}. In this system, the substitution of the magnetic solute for the non-magnetic solvent should occur completely randomly\footnote{There are other procedures used to create this type of disorder\index{disorder} by destroying the crystal structure of the materials, making them amorphous.}. However, this type of disorder\index{disorder} needs something else to generate randomness\index{randomness} in the interactions. We need a kind of magnetic interaction that depending on the distance between the magnetic impurities generates antiferromagnetic or ferromagnetic couplings\index{couplings}.																				
The dominant interaction in those systems is the so-called \gls{RKKY}\index{RKKY interaction} interaction \cite{ruderman:54,kasuya:56,yosida:57}. This interaction is long-ranged and the underlying mechanism is the conduction electrons of the host metal acting as intermediaries between the magnetic moments of the magnetic solute. 

A magnetic impurity placed at $\vec{r}_i$ changes the susceptibility\index{susceptibility} of the conduction electrons surrounding it through hyperfine interaction. A second magnetic impurity placed at $\vec{r}_j$ will behave in the same way, thus the two \gls{RKKY}\index{RKKY interaction} polarization will overlap, establishing an effective interaction between the two spins of the magnetic impurities. This interaction is given by
\begin{equation}
J(r_{ij}) = 6\pi Z J^2 N(E_F) \left[ \dfrac{\sin (2k_F r_{ij})}{(2k_F r_{ij})^4} - \dfrac{\cos (2k_F r_{ij})}{(2k_Fr_{ij})^3}\right] \quad , \labeq{RKKY}
\end{equation}
where $Z$ is the number of conduction electrons per atom, $J$ is the s-d exchange constant, $N(E_F)$ is the density of states at the Fermi level, $k_F$ is the Fermi momentum and $r_{ij} = \lvert \vec{r}_i - \vec{r}_j \rvert$. At large distance the quartic term of the distance becomes irrelevant with respect to the cubic one, therefore
\begin{equation}
J(r_{ij}) \approx \dfrac{J_0 \cos (2k_F r_{ij} + \phi)}{(2k_F r_{ij})^3} \quad ,\labeq{RKKY_reduced}
\end{equation}
being $J_0$ a constant which agglutinates all the constant terms of \refeq{RKKY} and $\phi$ a phase that takes into account the charge difference between impurity and host. The reader may wonder whether the decaying-sinusoidal behavior of \refeq{RKKY} or \refeq{RKKY_reduced} would be enough to generate random interactions. We would like to stress the fact that the Fermi moment $k_F$ is, actually, quite large (of the order of the inverse of the interatomic spacing). That makes the sinusoidal oscillations to be very sensitive to any change of the distance $r_{ij}$.

Of course, there exist other types of interactions between spins capable to generate randomness\index{randomness}\footnote{For example, superexchange interaction is relevant in insulating and semiconducting materials due to the lack of conduction electrons. Moreover, there exist weaker interactions like dipolar\index{dipolar} interaction that play an important role because they introduce anisotropies\index{anisotropy}.}, however, we only stop to explain \gls{RKKY}\index{RKKY interaction} for historical and practical reasons. From the historical point of view, \gls{RKKY}\index{RKKY interaction} interactions are bounded to the birth of the \gls{SG} research. Moreover, along this thesis experiments with \gls{CuMn} will have an important role and the dominant interaction in \gls{CuMn} turn out to be the \gls{RKKY}\index{RKKY interaction} interaction.

Last, we have to keep in mind that the computation of \refeq{RKKY} involves several approximations. The assumption of the free electron and the random impurities are the stronger ones. On the one hand, the consideration of an electronic band structure leads to considerable modifications of the \gls{RKKY}\index{RKKY interaction} interaction, see for example \cite{narita:84} or, for computations in specific materials like graphene \cite{annica:10,sherafati:11}. On the other hand, the positions of the impurities are not truly random as we assumed before. Experimentally it is possible to find significant correlations in the position of the impurities through neutron-scattering\index{neutron scattering} techniques. Actually, the knowledge of those correlations allows the experimental computation for the couplings\index{couplings} in different \gls{SG}s, see \reffig{RKKY} extracted from \cite{morgownik:83}.

\begin{figure}[h]
\centering
\includegraphics[width=0.7\textwidth]{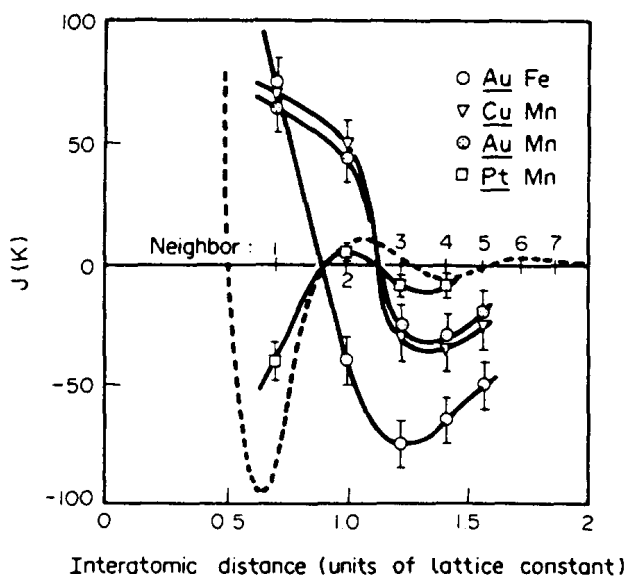}
\caption[\textbf{RKKY interaction for real Spin Glasses.}]{\textbf{RKKY interaction for real Spin Glasses.} Dotted line represents the original computation for the \gls{RKKY}\index{RKKY interaction} coupling\index{couplings} as a function of the interatomic distance. Computations over real \gls{SG}s shows significant differences with the theoretical results due to the correlations between the position of the impurities in the host metal. Figure from \cite{morgownik:83}.}
\labfig{RKKY}
\end{figure}

The naive computation may not be quantitatively accurate but this, \textit{a priori}, unfortunate fact turns out to be a hope for the study of \gls{SG}s. The \gls{RKKY}\index{RKKY interaction} model still captures the fundamental requirements to find glassy behavior and thus, open the door to theoretical models which we will discuss in future sections that do not reproduce the couplings\index{couplings} distribution of the real systems but which still contains the two main ingredients needed for finding \gls{SG} behavior: randomness\index{randomness} and frustration\index{frustration}.

\subsubsection{Subtle but fundamental: anisotropies}
Up to now, the magnetic interaction that we have attributed to \gls{CuMn} real systems, the \gls{RKKY}\index{RKKY interaction} interaction, presents isotropic behavior, thus, there is no reason to restrict the value of the spin $\vec{s}$ of the impurities in any dimension. This three-dimensional spin leads to the so-called \textit{Heisenberg\index{Heisenberg} spin glass}.

However, even the purest real system presents some anisotropies\index{anisotropy}. The role of those anisotropies\index{anisotropy} is fundamental because they can restrict the degrees of freedom\index{degree of freedom} of the spins to a plane (resulting in the $XY$ \textit{spin glass}) or to a single dimension, leading to the known as \textit{Ising\index{Ising} spin glass}.

Throughout this thesis, several results will be compared with real \gls{CuMn} systems which are, essentially, Heisenberg\index{Heisenberg}-like \gls{SG}. A lot has been said about the effect of the anisotropies\index{anisotropy} in the \gls{CuMn} \gls{SG}s \cite{prejean:80,fert:80,levy:81,bray:82,mendels:87,chu:94,petit:02,bert:04,zhai-janus:21} where the main ones that we can found are the dipolar\index{dipolar} anisotropies\index{anisotropy}, weak but present in every spin system, and the \gls{DM} anisotropies\index{anisotropy}\index{Dzyaloshinksii-Moriya}, whose origin is a large spin-orbit coupling\index{couplings} of the conduction electron with the impurities, acting the conduction electron as an intermediary (similar to \gls{RKKY}\index{RKKY interaction}). 

From the computational point of view, Ising\index{Ising} \gls{SG}s are very convenient since they are much easier to simulate than Heisenberg\index{Heisenberg} \gls{SG}s and the research developed in the context of this thesis is focused on the former. Differences between Heisenberg\index{Heisenberg} and Ising\index{Ising} \gls{SG}s are numerous\footnote{For example, the very existence of a phase transition\index{phase transition} in 3D systems is not clear in the Heisenberg\index{Heisenberg} models while is commonly accepted in 3D Ising\index{Ising} \gls{SG}s.}, therefore, a natural question is whether or not the results obtained in this thesis are general.

This question is positively answered in \cite{baityjesi:14}, actually, we know now that the ruling universality\index{universality} class in presence of coupling\index{couplings} anisotropies\index{anisotropy} is Ising\index{Ising} and even the purest real \gls{SG} will present some anisotropies\index{anisotropy}.

\subsection{Aging, memory and rejuvenation}\labsubsec{experiments_introduction}
\labsubsec{aging_memory_rejuvenation}
Here, we present emblematic experiments showing that \gls{SG}s are out of equilibrium in the experimental time-scales. We also take the opportunity to expose those experiments that will be, at least conceptually, related in some way to the original results presented throughout this thesis.

\subsubsection{Aging}\index{aging}
We have said that \gls{SG}s, in absence of an external magnetic field, have null magnetization\index{magnetization!zero}\footnote{Not only, but also no magnetic order\index{magnetic order} can be found, see \refeq{no_magnetic_order}} $M=\sum_{\vec{r}} \braket{\vec{s}_{\vec{r}}}=0$, however, when a magnetic field is applied, a magnetization\index{magnetization} $M \neq 0$ can be measured. When the external magnetic field is turned off, the system evolves from $M=M(t_0)\neq 0 $ to $M(t_f)=0$. This process is called \textit{relaxation}\index{relaxation}.

\gls{SG}s exhibit in the low-temperature phase\index{phase!low-temperature/spin-glass} extremely-large relaxation\index{relaxation} times, remaining out of equilibrium during the whole experiments. Still, the most striking feature is the emergence of a second time-scale: the relaxation\index{relaxation} process strongly depends on the time that the system spent in the low-temperature phase\index{phase!low-temperature/spin-glass} before turning off the external field. We say that the system \textit{ages}.

There exist two mirror experimental setups that are proven to be equivalent \cite{nordblad:86}: the relaxation\index{relaxation} of the \gls{TRM} and the relaxation\index{relaxation} of the \gls{ZFC} magnetization\index{magnetization!thermo-remanent}.

The typical protocol to study the \gls{TRM}\index{magnetization!thermo-remanent} is the following. First, we set a small external magnetic field and the system is cooled in its presence from a temperature $T_0$ above the critical\index{critical temperature} temperature\footnote{The temperature that separates the low-temperature phase\index{phase!low-temperature/spin-glass} and the high-temperature one\index{phase!high-temperature/paramagnetic}.} $\Tc$ to a temperature $T_1<\Tc$. We let the system age at temperature $T_1$ for a waiting time $t_w$ and then, the external magnetic field is switched off. At that moment, the decreasing magnetization\index{magnetization} is recorded as a function of time $t$ where $t=0$ corresponds to the moment in which we turned off the field. One part of the total magnetization\index{magnetization} falls immediately, the so-called reversible magnetization\index{magnetization}. The other part is the so-called remanent magnetization\index{magnetization!thermo-remanent} and falls slowly with the time $t$. The slow fall of the remanent magnetization\index{magnetization!thermo-remanent} is shown in \reffig{TRM_relaxation} from \cite{vincent:97}. The reader may find similar experiments of \gls{TRM}\index{magnetization!thermo-remanent} relaxation\index{relaxation} processes in \cite{chamberlin:84,ocio:85,nordblad:86,alba:86}.

\begin{figure}[t]
\centering
\includegraphics[width=0.8\textwidth]{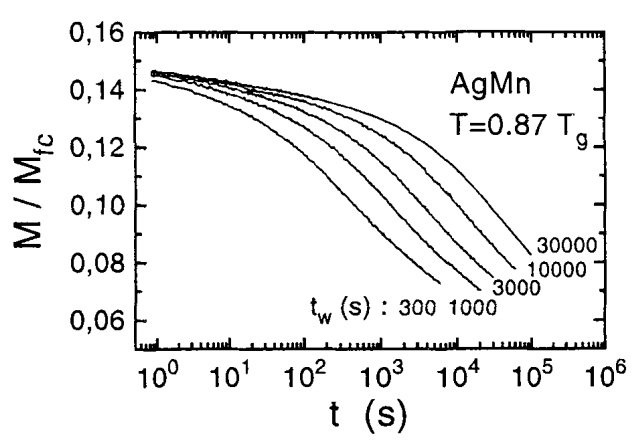}
\caption[\textbf{Thermo-remanent magnetization in spin glass.}]{\textbf{Thermo-remanent magnetization in spin glass.} Thermo-remanent magnetization\index{magnetization!thermo-remanent} $M$ normalized by the field-cooled value $M_{fc}$ is plotted against the time $t$ in a semi-log scale. \gls{AgMn} with the $2.6 \%$ of impurities is cooled from $T_0 >\Tc=10.4$ K to $T_1=9$ K$ = 0.87 \Tc$ in the presence of an external field of $0.1$ Oe. The system stays at $T_1$ with the field for a time $t_w$, then, the field is cut and the decaying magnetization\index{magnetization} is recorded as a function of time $t$ where $t=0$ corresponds to the moment of turning off the field. Different curves corresponds to different $t_w$. Figure from \cite{vincent:97}.}
\labfig{TRM_relaxation}
\end{figure}

\begin{figure}[h]
\centering
\includegraphics[width=0.8\textwidth]{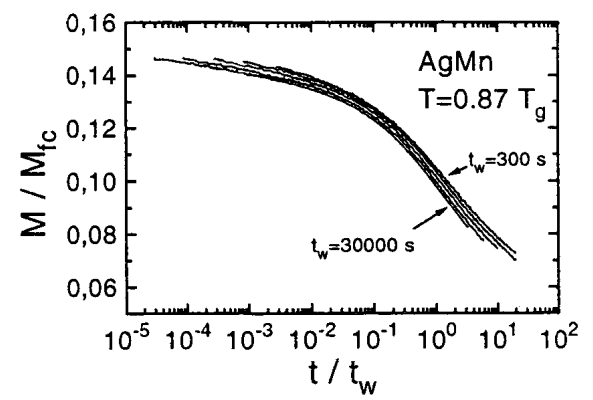}
\caption[\textbf{The relevant time-scale of the thermo-remanent magnetization relaxation.}]{\textbf{The relevant time-scale of the thermo-remanent magnetization relaxation\index{relaxation}.} Thermo-remanent magnetization\index{magnetization!thermo-remanent} $M$ normalized by the field-cooled value $M_{fc}$ is plotted against $t/t_w$ in a semi-log scale in the same experimental conditions that in \reffig{TRM_relaxation}. An almost perfect collapse is observed. Figure from \cite{vincent:97}.}
\labfig{TRM_relaxation_collapse}
\end{figure}

In fact, one can observe in \reffig{TRM_relaxation} for every curve an inflection point which roughly corresponds to $t=t_w$. The natural representation is, therefore, the one showed in \reffig{TRM_relaxation_collapse} where the abscissa axis corresponds to the time normalized as $t/t_w$. A perfect collapse of the curves under this representation is known as \textit{full aging\index{aging}}. However, the collapse is only approximated. The scaling variable\footnote{The original symbol associated with this quantity, and the most used is $\xi$, however, we use here $\tau$ to avoid confusion with the coherence length\index{coherence length} $\xi$.} $\tau$, firstly used in structural\index{structural glass} glasses \cite[p.~129]{struik:80} and lately introduced in \gls{SG} by \cite{ocio:85} as quoted by \cite{rodriguez:03}, solves the problem and achieves a much better collapse
\begin{equation}
\tau=\dfrac{t_w^{1-\mu}}{1-\mu}\left[ \left( 1 + \dfrac{t}{t_w} \right)^{1-\mu} -1 \right] \quad \mu < 1 \,\, . \labeq{not_full_aging}
\end{equation}

Putting the subtleties aside, we now know that the relevant time-scale of the aging\index{aging} processes is $t_w$ and this is fundamental, as we will discuss in the following chapters because it has a deep relation with the coherence length\index{coherence length} $\xi$ acting as the key quantity governing the non-equilibrium phenomena, see~\refch{aging_rate}.

The mirror protocol is the \gls{ZFC}. In that protocol, the sample\index{sample} is cooled from a temperature $T_0 > \Tc$ to a temperature $T_1<\Tc$ in zero-field. After a time $t_w$ a small field is turned on and the magnetization\index{magnetization} is recorded as a function of time $t$, analogously to the previous experiment, $t=0$ corresponds to the moment in which the field is applied. This protocol is equivalent to the \gls{TRM}\index{magnetization!thermo-remanent} but with an increasing magnetization\index{magnetization}. Furthermore, the sum of the \gls{ZFC}-magnetization\index{magnetization!zero-field-cooled} plus the \gls{TRM}\index{magnetization!thermo-remanent} is equal to the field-cooled magnetization\index{magnetization}\footnote{This is not true in general, the reader should note that we are in the small field limit where the only relevant term is the linear one. The equality is not guaranteed for larger fields where the non-linear responses are sizable, see for example recent relevant works~\cite{zhai-janus:20,zhai-janus:21}}. For experimental results of this protocol, see \cite{lundgren:83,nordblad:86}.

An experiment that can be regarded as a generalization (if that word can be used in the context of experiments) was performed, actually, earlier by Nagata \textit{et al.}, see \cite{nagata:79}. The main results of their research can be summarized in \reffig{nagata_dcsuscept}

\begin{figure}[h]
\centering
\includegraphics[width=0.8\textwidth]{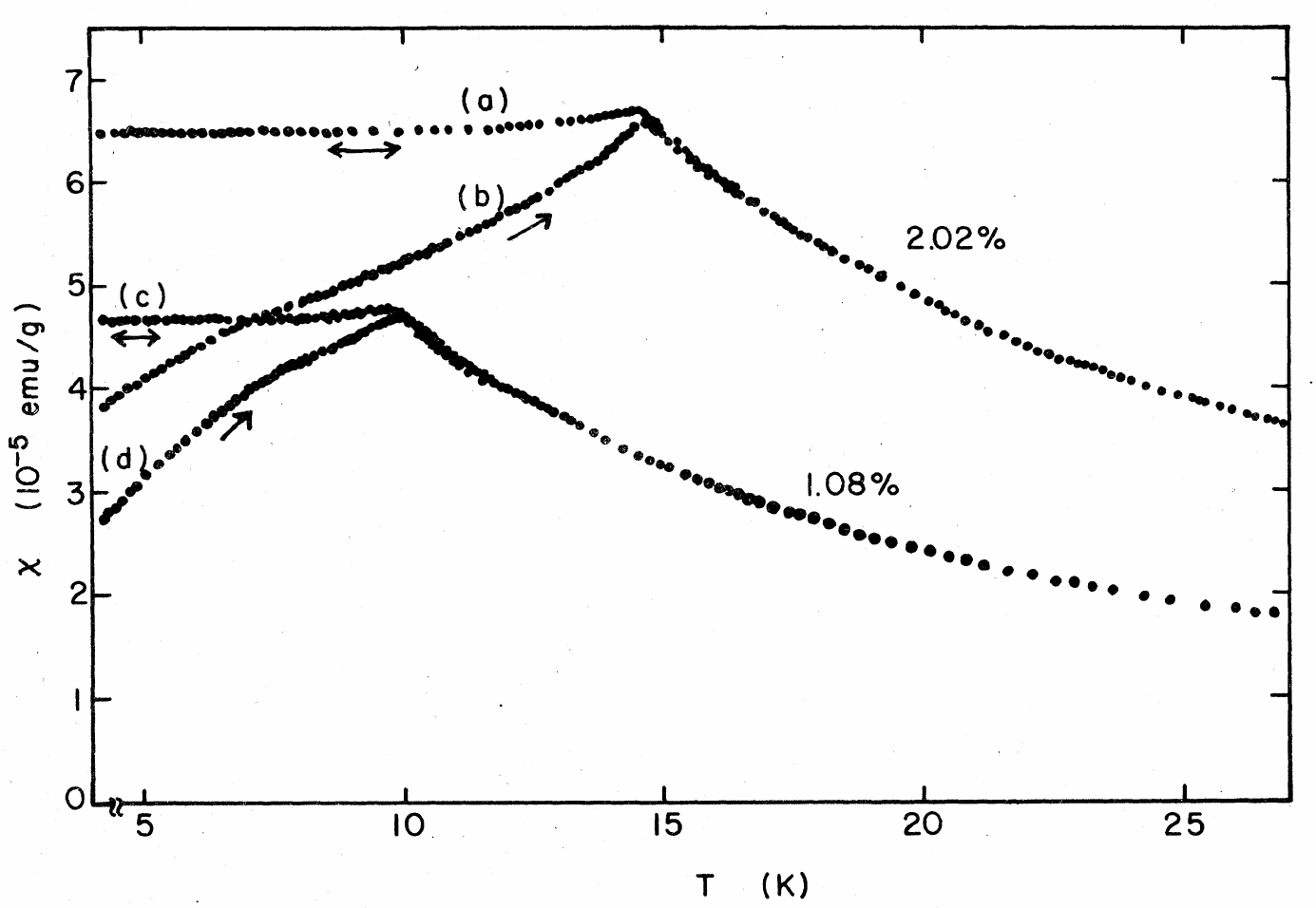}
\caption[\textbf{DC-Susceptibility in CuMn spin glass.}]{\textbf{DC-Susceptibility\index{susceptibility!dc} in CuMn spin glass.}  Susceptibility\index{susceptibility!dc} of a DC applied-field $h=5.9$ G is measure in a $1.08 \%$ and a $2.02 \%$ CuMn spin-glass and plotted against the temperature. Curves (a) and (c) corresponds to a field cooled protocol while curves (b) and (d) corresponds to a zero-field cooled protocol. Arrows indicate the reversibility ($\leftrightarrow$) or irreversibility ($\rightarrow$) of the protocol. Figure from \cite{nagata:79}.}
\labfig{nagata_dcsuscept}
\end{figure}

In this experiment, the authors measure the magnetic response to an applied magnetic field i.e. \textit{the susceptibility}\index{susceptibility} $\chi = \lvert \vec{M} \lvert / \lvert \vec{H} \lvert$.

Two different protocols are performed in this experiment. In the protocol corresponding to the curves (a) and (c), the sample\index{sample} is cooled in the presence of a constant field $h = 5.9$ G. Below the critical\index{critical temperature} temperature $\Tc$ the susceptibility\index{susceptibility} remains almost constant and the process is reversible.

On the contrary, in the protocol corresponding to the curves (b) and (d), the sample\index{sample} is cooled in absence of any field. Then, the sample\index{sample} is heated and the susceptibility\index{susceptibility} increases monotonically until it reaches the critical\index{critical temperature} temperature. Moreover, cooling again the system in presence of the field leads to an irreversible behavior. Above $\Tc$ both protocols are identical.

Moreover, if we switch on the field after \gls{ZFC} and let the system evolve at a fixed temperature, we observe that the susceptibility\index{susceptibility} grows towards the field-cooled value but without reaching it in the experimental time-scales. This is the connection to the aging\index{aging}-experiments showed above, where the temperature cycle\index{temperature cycle} was performed only between two temperatures, but the aging\index{aging} and the non-equilibrium behavior were captured the same.

The experiments exposed above are stressing us two main things:
\begin{enumerate}
\item Experimental \gls{SG}s are out of equilibrium in the experimental time-scales.
\item The system ages, i.e. the time that the system expends below $\Tc$ is a key quantity to understand its non-equilibrium behavior in the low-temperature phase\index{phase!low-temperature/spin-glass}.
\end{enumerate}

\subsubsection{Memory and rejuvenation}

\begin{figure}[h]
\centering
\hspace{-3cm}
\includegraphics[width=0.8\textwidth]{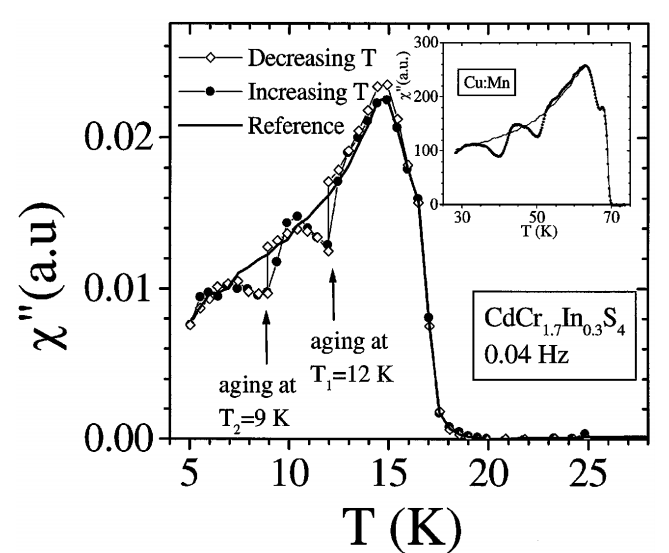}
\caption[\textbf{Memory and rejuvenation experiment in spin glasses.}]{\textbf{Memory\index{memory effects} and rejuvenation\index{rejuvenation} experiment in spin glasses.} Out-of-phase susceptibility\index{susceptibility!ac} is recorded when temperature vary in the presence of a sinusoidal field $h_\mathrm{ac}$ of frequency $f=0.04$ Hz and peak amplitude of $0.3$ Oe. Main plot corresponds to CdCr$_{1.7}$In$_{0.3}$S$_4$ and the inset corresponds to the same plot for CuMn. Figure from \cite{jonason:98}.}
\labfig{memory_rejuvenation}
\end{figure}

In previous experiments, when a constant field was applied, we defined the so-called \textit{dc-susceptibility}\index{susceptibility!dc}, therefore, a straightforward generalization is the \textit{ac-susceptibility}\index{susceptibility!ac}, which is none but the magnetic response of the system when a sinusoidal field is applied to it. 

In \gls{SG}s, the susceptibility\index{susceptibility!ac} $\chi_\mathrm{ac}$ measured from a sinusoidal applied field $h_{\mathrm{ac}}$ is also a sinusoidal quantity that presents a phase-delay $\varphi$ with respect to the field $h_{\mathrm{ac}}$. This delay lead to the definition of two quantities: the in-phase susceptibility\index{susceptibility!ac} $\chi'$ and the out-of-phase susceptibility\index{susceptibility!ac} $\chi''$
\begin{equation}
\begin{split}
\chi' & = \chi \cos \varphi \\
\chi'' & = \chi \sin \varphi
\end{split}
\end{equation}

In \cite{jonason:98} \footnote{Experiments of memory\index{memory effects} and rejuvenation\index{rejuvenation} in temperature cycles\index{temperature cycle} of two temperatures were performed earlier, see for example \cite{lefloch:92} however, the richer phenomenology of \cite{jonason:98} make us to focus on this experiment for the sake of clarity.}, the authors measured the out-of-phase susceptibility\index{susceptibility!ac} when a $h_\mathrm{ac}$ field of frequency $f=0.04$ Hz and peak amplitude of $0.3$ Oe was applied to a \gls{CdCrInS} \gls{SG}. The main result of this research is summarized in \reffig{memory_rejuvenation}

The system was cooled from $T>\Tc=16.7$ K down to $T_f=5$ K at a constant cooling-rate of $0.1$ K/min. Then, the system was heated back continuously at the same rate.

$\chi''$ was recorded during the cooling and heating protocol, resulting in two close curves (still slightly different) where the heated one was used as a reference curve (continuous line in \reffig{memory_rejuvenation}). Next, the experiment was repeated but the system was let age at temperatures $T_1=12$ K and $T_2=9$ K for a waiting time $t_w=7$ h for $T_1$ and $t_w=40$ h for $T_2$. After the age, the cooling was restarted at the same rate and the out-of-phase susceptibility\index{susceptibility!ac} merged back with the reference curve, it is said that the system \textit{rejuvenates}. The measurements of this protocol correspond to the white squares in \reffig{memory_rejuvenation}. 

Last, the system was reheated at a constant heating rate. When the temperature approached the age temperatures $T_1$ and $T_2$, the system, somehow, ``remembered'' the age and followed the susceptibility\index{susceptibility} curve, reproducing the data of the cooling protocol. In the inset, the authors included the same experiment appearing in \cite{djurberg:99} in a \gls{CuMn} \gls{SG}, where the behavior was completely similar.

This experiment sends a very clear message: the aging\index{aging} at a temperature $T$ does not affect the susceptibility\index{susceptibility} value at lower temperatures. This temperature independence has been commonly related to the \textit{temperature chaos}\index{temperature chaos} phenomenon that we extensively treat in \refch{Introduction_chaos}, \refch{equilibrium_chaos} and \refch{out-eq_chaos} and, in particular, the \refch{out-eq_chaos} sums up the spirit of this thesis: separated researches from the experimental and the theoretical point of view that can be related by numerical simulations.

\section{Theoretical spin glasses} \labsec{theoretical_spinglass}
In this section, we will briefly expose the theoretical development of the \gls{SG}'s theory. The aim of this section is not to deeply review all the important results on \gls{SG} but to provide some context of its history and to highlight the main theoretical predictions which we will deal with by means of numerical simulations.

The tools of statistical mechanics that we use in this section will be introduced in \refsubsec{tools_statistical_mechanics}. Then, the most popular theoretical models will be shown in \refsubsec{theoretical_models}. Finally, we will present the main theoretical pictures which predict different results in the \gls{SG} low-temperature phase\index{phase!low-temperature/spin-glass} in \refsubsec{theoretical_pictures}. One of the tasks of our numerical simulations will be partially to discriminate between them.

\subsection{The tools of statistical mechanics}
\labsubsec{tools_statistical_mechanics}
We will present some basic results on statistical mechanics that will be useful in \gls{SG} theory. The reader may find general references in \cite{landau:80,lebellac:91,amit:05,greiner:12}.

Although most of the things we are saying in this part are general for statistical mechanics, the notation and the language will be always focused on spin systems.

We consider a system described by $N$ microscopic variables $\{s_i\}$. The most basic quantity describing the system is the \textit{energy}\index{energy}
\begin{equation}
E(\{s_i\}) \equiv \mathcal{H}(s_0,s_1,\dots,s_n) \, , \labeq{Hamiltonian}
\end{equation}
a function (usually called \textit{Hamiltonian}\index{Hamiltonian}) of the microscopic variables $\{s_i\}$ that tends to be minimized when the system approach the thermal equilibrium.

However, the energy\index{energy} is not the only quantity describing the system, nor the only quantity we are interested in. A general quantity $\mathcal{O}$ depending on the concrete values of the microscopic variables is called \textit{observable}: $\mathcal{O}(\{s_i\})$. 

In general, the fluctuation of the observables due to the fluctuation of the microscopic variables makes desirable to consider averaged quantities
\begin{equation}
\braket{\mathcal{O}} = \lim_{t\to \infty} \dfrac{1}{t} \int_0^t \mathcal{O}(\{s_i\}(\tau)) d\tau \, . \labeq{time_average}
\end{equation}

The problem now is clear. To compute these averages we need to know the time-evolution of a macroscopic number of variables, usually interacting with each other. Here emerges the fundamental principle of the statistical mechanics: the system, when it is at equilibrium, will eventually reach every possible set of microscopical variables, namely \textit{configuration}\index{configuration}. If we could know which is the probability of each configuration\index{configuration} to appear, we could trade the integral over infinitely large periods with the weighted sum over all the possible configurations\index{configuration} of the system.

In the canonical ensemble\index{canonical ensemble}, the probability distribution of the configurations\index{configuration} $\{s_i\}$ is
\begin{equation}
P(\{s_i\}) = \dfrac{e^{-\beta \mathcal{H}(\{s_i\})}}{Z} \, , \labeq{prob_configuration}
\end{equation}
where $\beta$ is the inverse of the temperature $T$ of the heat bath\footnote{Which coincides with the temperature of the system in thermal equilibrium.} in units such that the \textit{Boltzman constant} $k_B=1$ and $Z$ is a normalization factor called \textit{Zustandssumme} or \textit{partition function}\index{partition function}\footnote{For convenience, we are assuming a discrete number of possible configurations\index{configuration} in the phase-space\index{phase space}. In general, the expression of the partition function\index{partition function} involves an integral instead of a sum.}
\begin{equation}
Z = \sum_{\{s_i\}} e^{-\beta \mathcal{H}(\{s_i\})} \, . \labeq{partition_function}
\end{equation}

Therefore, the computation of the averages is
\begin{equation}
\braket{\mathcal{O}} = \dfrac{ \sum_{\{s_i\}} \mathcal{O}(\{s_i\}) e^{-\beta \mathcal{H}(\{si\})}}{Z} \, . \labeq{average_o}
\end{equation}

It is worthy to note that we have fully characterized the system at equilibrium through the partition function\index{partition function} however, the very related quantity $F$, the \textit{free energy\index{free energy}}\footnote{Note that we use $\log$ as the symbol for the natural logarithm.}
\begin{equation}
F = - \dfrac{1}{\beta} \log Z(\{s_i\}) \, ,
\end{equation}
turns out to be much more practical because it is directly related to measurable quantities, fundamental in the thermodynamics of the system e.g. the \textit{energy}\index{energy} $U$ or the \textit{magnetization}\index{magnetization} $M$
\begin{equation}
U = - \left. \dfrac{\partial \left( \beta F\right)}{\partial T}\right|_H \quad , \quad M = - \left. \dfrac{\partial \left( \beta F\right)}{\partial H}\right|_T \labeq{thermodynamic_quantities} 
\end{equation}

\subsubsection{Disorder and self-averaging}
The systems appearing along this thesis present a particularity: the interaction between any pair of particles is random. This randomness\index{randomness}, extended in the whole system is what we call \textit{disorder}\index{disorder} and needs to be characterized by additional variables, taking into account the interaction between particles. We denote those variables $\{J\}$.

The disorder\index{disorder} variables $\{J\}$ can exhibit a dynamical evolution and the time-scale of its evolution will determine if we are treating with \textit{annealed} disorder or \textit{quenched} disorder.

On the one side, the annealed disorder\index{disorder!annealed} occurs when the time-scale of the dynamical evolution of $\{J\}$ is shorter than the observation time. Therefore, the interactions can be regarded as a sort of dynamic variables and we can average over them the same that we average over the configuration\index{configuration} of the system
\begin{equation}
\overline{\mathcal{O}_J} = \int \mathcal{O}(J) P(J) dJ \, , \labeq{average_disorder}
\end{equation}
where $P(J)$ is the probability distribution that follows the disorder variables $J$. Therefore, in this situation, the free energy\index{free energy} of the system is
\begin{equation}
F = - \dfrac{1}{\beta} \log \overline{Z_J} \, . \labeq{free_energy_annealed}
\end{equation}

The computations associated with the annealed disorder have no additional difficulty, we have just added a fashion hat to our quantities, but the treatment is essentially the same. 

On the other side, we have the quenched disorder\index{disorder!quenched}, in which the time-scale of the dynamical evolution of $\{J\}$ is much larger than the observation time. In this case, each system is different due to the disorder\index{disorder!quenched} 
\begin{equation}
F_J = - \dfrac{1}{\beta} \log Z_J \, .
\end{equation}

\textit{A priori}, there is no hope of universality\index{universality} in those systems and we are forced to study them individually. However, the \textit{self-averaging}\index{self-averaging} property emerges to rescue us. In the thermodynamic limit\index{thermodynamic limit}, for a system with $N$ degrees of freedom\index{degree of freedom}, we have
\begin{equation} 
\lim_{N \to \infty}\dfrac{F_J}{N} = f_\infty \, , \labeq{self_average}
\end{equation}
being $f_\infty$ the free-energy\index{free energy!density} density in the thermodynamic limit\index{thermodynamic limit}, which is independent of the disorder\index{disorder} variables $\{J\}$.

An argument supporting this property can be found in \cite{brout:59}. The reasoning is the following: any macroscopic system can be divided into a statistically large number $n$ of macroscopic systems. Due to the quenched disorder\index{disorder!quenched}, every subsystem will have a different free energy\index{free energy} $F_{J_j}$ with $j=(0,1,\dots,n)$. Can we relate those free energies\index{free energy} with the free energy\index{free energy} of the whole system? 

The free energy\index{free energy} is just the logarithm of the partition function\index{partition function} whose dependence of the interactions is codified in the Hamiltonian\index{Hamiltonian} $\mathcal{H}_J(\{s_i\})$. If we assume short-range interactions, which is a physically reasonable assumption, the total free energy\index{free energy} will be the sum of the free energies\index{free energy} of the subsystem plus the contribution to the free energy\index{free energy} of the interface between the subsystems. As long as $N \to \infty$, the interface between subsystems will be negligible against the volume of those subsystems and, therefore \refeq{self_average} holds. The implicit corollary is that computing the free energy\index{free energy} of a large enough system is equivalent to compute the sum of the free energy\index{free energy} for smaller systems.

If the distribution probability of the couplings\index{couplings} $J$ is not pathological, we expect
\begin{equation}
\overline{F_J^2} - \overline{F_J}^2 \propto \dfrac{1}{N} \, .
\end{equation}

The computation of the disorder\index{disorder!average} average of the logarithm 
\begin{equation}
F = \overline{F_J} = -\dfrac{1}{\beta} \overline{\log Z_J} \, ,
\end{equation}
is one of the biggest difficulties on studying statistical mechanics on disordered\index{disorder!systems} systems and we deal with it by using the so-called \textit{replica method}\index{replica!method/trick}.

It is worthy to note that, the crucial hypothesis for Brout's argument is that the contribution of the boundaries of those subsystems is negligible. There exist some situations in which this assumption is no longer valid (see for instance \cite{binder:86} for a more detailed explanation). For example, when a phase transition\index{phase transition} occurs, the boundary conditions\index{boundary conditions} become crucial and the system is no longer self-averaging\index{self-averaging}. 

It is well known that, for a spin glass below the critical\index{critical temperature} temperature, some quantities are non-self-averaging\index{self-averaging!non-}. This feature is closely related to the concept of \textit{dispersed metastate\index{metastate!dispersed}} (see \refch{metastate}).

\subsubsection{The replica method}
In order to compute $\overline{F_J}$ we use the so-called \textit{replica method}\index{replica!method/trick} or \textit{replica trick}, that was firstly introduced in the context of \gls{SG}s in \cite{edwards:75}.

The method is based on the expression
\begin{equation}
\log Z = \lim_{n\to 0} \dfrac{Z^n -1}{n} \, , \labeq{replica_expression}
\end{equation}
which is direct if we use the Taylor expansion of the right-side expression
\begin{equation}
\dfrac{Z^n -1}{n} = \dfrac{e^{n \log Z}-1}{n} = \dfrac{1 + n \log Z + O(n^2)-1}{n} = \log Z + O(n) \, . \labeq{replica_expression_explained}
\end{equation}

At this point, we consider $n$ identical systems $Z_J^{(a)}$ $a=(0,1,\dots,n)$ also called \textit{replicas}\index{replica} i.e. systems with the same realization of the disorder\index{disorder} $J$, and we define
\begin{equation}
Z_n \equiv \overline{Z_J^n} = \overline{\prod_{a=1}^n Z_J^{(a)}} \, ,
\end{equation}
and 
\begin{equation}
F_n = -\lim_{n\to 0} \dfrac{1}{n \beta}  \log Z_n \, .
\end{equation}

By using \refeq{replica_expression}, is easy to see now that $F_n = \overline{F_J}$
\begin{equation}
F_n = -\lim_{n\to 0} \dfrac{1}{n \beta}  \log \overline{Z_J^n} = -\dfrac{1}{\beta} \lim_{n \to 0} \dfrac{\log \left(1+n\overline{\log Z_J}+ O(n^2) \right)}{n} = - \dfrac{1}{\beta} \overline{\log Z_J} = \overline{F_J} \, . \labeq{replica_trick}
\end{equation}

The computation of $F_n$ is easier than the computation of $\overline{F_J}$ if we first compute $\log \overline{Z_J^n}$ with $n$ integer and then we take the limit $n \to 0$. This is, indeed, a very doubtful step in the mathematical sense. We consider $n$ to be an integer in order to define $Z_n$, then, we take the analytical extension of $Z_n$ for $n \in \mathbb{R}$ and finally we take the limit of $n\to 0$. In \refsubsec{theoretical_models} we will introduce a practical use for the replica method\index{replica!method/trick}.

In the cases in which the free energy\index{free energy} is an analytic function of the temperature\footnote{which usually happens in the high-temperature phase\index{phase!high-temperature/paramagnetic} of magnetic systems.} $\beta$, the replica method\index{replica!method/trick} is exact. Moreover, when other methods are available to compute the free-energy\index{free energy}, the results coincide with the predictions of the replica method\index{replica!method/trick}.

There exist also an alternative approximation to the replica method\index{replica!method/trick} which not requires the \textit{trick} of the duality nature of $n$ integer-real, see \cite{dotsenko:93,coolen:93,dotsenko:94}.

\subsection{Theoretical models} \labsubsec{theoretical_models}
Here, we discuss the main models that capture the \gls{SG} physics and that are relevant in the development of the field. The trade-game is clear, on the one hand, the model should be detailed enough to exhibit the main \gls{SG} phenomenology. We have discussed which are the basic ingredients for that: randomness\index{randomness} and frustration\index{frustration}. On the other hand, the model should be also simple enough to be analytically tractable. The \gls{EA}\index{Edwards-Anderson!model} model is the first we are going to introduce here, for historical reasons, but also due to its relevance in the actual context of \gls{SG} and the present thesis. We are also going to define the \gls{SK}\index{Sherrington-Kirkpatrick} model, which represents the mean-field\index{Mean-Field!model} approximation, which is analytically solvable, and which characterizes the \gls{SG}s at infinite dimension.

Nonetheless, the aim of this part is not to deeply review all the theoretical results in \gls{SG}s, that can be found in several places \cite{mezard:87,dotsenko:01,dedominicis:06}, but to give some historical context and present results affecting the very nature of the \gls{SG}s, an issue still debated nowadays at finite dimensions.

\subsubsection{Edwards-Anderson model}
This model was proposed by Edwards and Anderson\index{Edwards-Anderson!model} in \cite{edwards:75}. Now, the general degrees of freedom\index{degree of freedom} we defined in \refsubsec{tools_statistical_mechanics} are, indeed, spins which lie in a regular lattice\index{regular lattice} and that interact with each other through the couplings\index{couplings} $\{J\}$. The Hamiltonian\index{Hamiltonian} of this model is
\begin{equation}
\mathcal{H} = - \sum_{\braket{i,j}} J_{ij} \vec{s}_i \vec{s}_j - \sum_i \vec{h}_i \vec{s}_i \, , \labeq{ea_Hamiltonian_theory}
\end{equation}
where $\vec{s}_i$ is a unitary vector, $\vec{h}$ is an external magnetic field and the sum over $\braket{i,j}$ denotes the sum over the pairs of spins $s_i$, $s_j$ that are bounded by a coupling\index{couplings} $J_{i,j}$ and that actually depends on the concrete form of the considered lattice. Along this thesis we will focus on the case $\vec{h}_i = \vec{0}$, therefore, from now on we address the particular case of non external magnetic field for the sake of simplicity.

If the spin vector is 3-dimensional, the system is called \textit{Heisenberg\index{Heisenberg} spin glass}, if the vector is 2-dimensional it is called $XY$ \textit{spin glass} and, if the spin only can take values $s_i = \pm 1$ we say that it is an \textit{Ising\index{Ising} spin glass}. From now on, we will focus on the Ising\index{Ising} \gls{SG}, which is actually a reasonable assumption in many systems (see \refsubsec{source_randomness}) and is the particular case of our numerical simulations in the research developed throughout this thesis. 

Moreover, the most popular choice is to consider only nearest-neighbor interactions between the spins\footnote{The rationale of this approach is the short-ranged nature of the interactions between spins.}, with the quenched variables $J_{ij}$ following a Gaussian\index{Gaussian!distribution} probability distribution, however, the particular shape of the distribution is not very important and a very popular choice, apart from the Gaussian\index{Gaussian!distribution}, is the bimodal one ($J_{ij} = \pm J$) that we will use in the numerical simulations.

The \gls{EA}\index{Edwards-Anderson!model} model also brought a proposal for the order parameter\index{order parameter!Edwards-Anderson} controlling the phase transition\index{phase transition}: the \textit{overlap\index{overlap}}. The traditional order-parameters displayed in \refeq{no_magnetic_order} are not valid in \gls{SG} because, by definition, \gls{SG} exhibit no long-range order. Nonetheless, the frozen and disorder\index{disorder} nature of the glassy phase\index{phase!low-temperature/spin-glass} suggests a different order parameter\index{order parameter!Edwards-Anderson} based on time correlations on the same site.
\begin{equation}
q_{EA} = \dfrac{1}{N} \lim_{t \to \infty}  \sum_i \braket{s_i(t=0)s_i(t)}_t \, . \labeq{order_parameter_SG}
\end{equation}

The question that is answered by \refeq{order_parameter_SG} is, therefore, how similar is the configuration\index{configuration} of the system at a time $t$ compared to the configuration\index{configuration} at time $t=0$? This time average does not seem very useful, but fortunately, by the same reasoning made in \refsubsec{tools_statistical_mechanics} we can trade the time average by the weighted sum over all the possible configurations\index{configuration}
\begin{equation}
q_{EA} = \dfrac{1}{N} \sum_i \braket{s_i}^2 \, . \labeq{order_parameter_SG_2}
\end{equation}

We expect $q_{EA} = 0$ for $T > \Tc$ i.e. in the paramagnetic phase\index{phase!high-temperature/paramagnetic} $\braket{s_i}=0$. The expectation for $T \to 0$ is $q_{EA} \to 1$.

As a final remark, let us note that the very existence of a phase transition\index{phase transition} in the Ising\index{Ising} \gls{EA}\index{Edwards-Anderson!model} model was not completely accepted (even with the existence of an earlier consensus~\cite{kawashima:96,iniguez:96,iniguez:97,berg:98,janke:98,marinari:98}) until the beginning of the XXI century~\cite{palassini:99,ballesteros:00}.

\subsubsection{Mean Field: the Sherrington-Kirkpatrick model}

Even with the aim of simplicity in mind, the \gls{EA}\index{Edwards-Anderson!model} model is not simple to solve, nor its mean-field\index{Mean-Field!model} version, the \gls{SK}\index{Sherrington-Kirkpatrick} model:
\begin{equation}
\mathcal{H} = - \sum_{i<j} J_{ij} s_i s_j \, .
\end{equation}

The reader may have noticed one main difference between this Hamiltonian\index{Hamiltonian} and that appearing in \refeq{ea_Hamiltonian_theory}, in addition to the absence of an external field $\vec{h}$. Now, the sum is performed over every pair of sites with the restriction of $i<j$ to avoid double counting. In consequence, no useful concept of distance between spins exists, the interactions are infinite-ranged.

Moreover, for convenience, the couplings\index{couplings} $J_{ij}$ follow a Gaussian\index{Gaussian!distribution} distribution probability
\begin{equation}
P(J_{ij}) = \dfrac{1}{\sqrt{2\pi \sigma^2}} \exp \left[-\dfrac{(J_{ij}-\mu)^2}{2\sigma^2} \right]  \, ,
\end{equation}
where $\mu$ and $\sigma^2$ stand for the mean and the variance of the Gaussian\index{Gaussian!distribution} distribution and take the values
\begin{equation}
\mu = \overline{J_{ij}} = 0 \quad , \quad \sigma^2 = \overline{J_{ij}^2} = \dfrac{1}{N} \, .
\end{equation}

This particular choice of $\sigma^2$ makes the free-energy\index{free energy!density} density $f=F/N$ independent of the total number of spins $N$.

\subsubsection{The replica symmetric solution}

With this information, we are able to compute the key quantity, the free energy\index{free energy}, or equivalently, $Z_n$
\begin{align}
 Z_n & = \int_{\mydots} \int P(J) \prod_{a=1}^n \sum_{\{s^a\}} \exp \left( \beta \sum_{i<j}^N J_{ij} s^a_i s^a_j \right) \dd \{J\} = \nonumber \\ 
 & = \left(\dfrac{N}{2\pi}\right)^{N(N-1)/4} \sum_{\{s^a\}} \int_{\mydots} \int \exp \left(\beta \sum_{a=1}^n \sum_{i<j}^N J_{ij} s_i^a s_j^a - \dfrac{1}{2}N \sum_{i<j}^N J^2_{ij} \right) \dd \{J\} \, ,
\end{align} 
where the notation $\int_{\mydots} \int$ and $\dd \{J\} = \prod_{i<j} \dd J_{ij}$ denotes that we are dealing with multiple integrals, one for each pair $i<j$ and the sum over the configurations\index{configuration} now runs over all the possible configurations\index{configuration} for each replica\index{replica} with superindex $a$.

After some computations for the exchange in the sum order, we get the expression
\begin{equation}
Z_n = \sum_{\{s^a\}} \exp \left[ \dfrac{\beta^2 N n}{4} + \dfrac{\beta^2 N}{2} \sum_{a<b}^n \left( \dfrac{1}{N}\sum_i s_i^a s_i^b \right)^2 \right] \, .
\end{equation}

We can now step back by introducing an integral variable depending on the pair of replicas\index{replica} $a,b$ instead of the pair of sites $i,j$ with the help of the Hubbard-Stratonovich\index{Hubbard-Stratonovich} transformation
\begin{equation}
Z_n = \beta \sqrt{\dfrac{N}{2\pi}} \int_{\mydots}\int \sum_{\{s^a\}} \exp \left( \dfrac{\beta^2 N n}{4}-\dfrac{\beta^2 N}{2} \sum_{a<b}^n Q_{ab}^2 + \beta^2 \sum_{a<b}^n \sum_i^N Q_{ab}s_i^a s_i^b \right) \dd \{Q\} \, , \labeq{partition_function_mean_field_1}
\end{equation}
where now, $\int_{\mydots}\int$ and $d\{Q\} = \prod_{a<b} \dd Q_{ab}$ represent multiple integrals, one for each pair $a<b$.

After some basic algebra, we can write \refeq{partition_function_mean_field_1} as
\begin{equation}
Z_n = \int_{\mydots}\int e^{-\beta N n f_n(\{Q\})} \dd \{Q\} \, , \labeq{partition_function_mean_field_2}
\end{equation}
where we can see that $f_n(\{Q\})$ has a free-energy-like\index{free energy} structure
\begin{equation}
f_n(\{Q\})= -\dfrac{\beta}{4} + \dfrac{\beta}{2n}\sum_{a<b}^n Q_{ab}^2 - \dfrac{1}{\beta N n} \log \left[ \sum_{\{s^a\}} \exp \left( \beta^2 \sum_{i}^N \sum_{a<b}^n Q_{ab} s^a_i s^b_i \right) \right] \, ,
\end{equation}
with the effective Hamiltonian\index{Hamiltonian}
\begin{equation}
\mathcal{H}_{\mathrm{eff}} (\{Q\}) = -\beta \sum_i^N \sum_{a<b}^n Q_{ab}s_i^a s_i^b \, . \labeq{effective_Hamiltonian_mean_field}
\end{equation}

In the thermodynamic limit\index{thermodynamic limit} we can compute the integral of \refeq{partition_function_mean_field_2} through the saddle-point\index{saddle point} approximation. The minimization of $Z_n$ in \refeq{partition_function_mean_field_2} is given by the equations $\partial f_n(\{Q\})/ \partial Q_{ab}=0$ for all the pairs $a<b$
\begin{equation}
\dfrac{\partial f(\{Q\})}{\partial Q_{ab}} = \dfrac{\beta}{n} Q_{ab} - \dfrac{\beta}{nN} \dfrac{\sum_{\{s^a\}} \sum_i^N s_i^a s_i^b e^{-\beta \mathcal{H}_{\mathrm{eff}}(\{Q\})}}{\sum_{\{s^a\}} e^{-\beta \mathcal{H}_{\mathrm{eff}}(\{Q\})}} = 0 \, ,
\end{equation}
which leads to the solution
\begin{equation}
Q_{ab} = \dfrac{1}{N} \sum_i^N \braket{s_i^a s_i^b}_Q \, , \labeq{mean_field_solution_general}
\end{equation}
where $\braket{\cdot}_Q$ represents the usual average defined in \refeq{average_o} but with the effective Hamiltonian\index{Hamiltonian} $\mathcal{H}_{\mathrm{eff}}$.

Unfortunately, this is all that we can do with an arbitrary matrix $Q$. The first and most natural ansatz for the form of $Q$, as all the replicas\index{replica!equivalence} were supposed to be equivalent, is the \textit{replica symmetric} ansatz\index{replica!symmetric ansatz}. In that case, we have the symmetric form
\begin{equation}
Q_{ab} = (1-\delta_{ab})q \, , \labeq{replica_ansatz}
\end{equation}
where $\delta_{ab}$ is the Kronecker delta.

This particular form of $Q$ leads, with standard methods of Gaussian\index{Gaussian!integration} integration\footnote{see for example~\cite{dotsenko:01} for a detailed computation.}, to the saddle-point\index{saddle point} equation
\begin{equation}
q = \int_{-\infty}^\infty \dfrac{1}{\sqrt{2\pi}} e^{-z^2/2} \tanh^2 \left( \beta z \sqrt{q} \right) \dd z \, .
\end{equation}

This equation can be numerically solved and it is straightforward to prove that, for $\beta<1$ i.e. $T>\Tc=1$, the only solution is the trivial one $q=0$ corresponding to the paramagnetic phase\index{phase!high-temperature/paramagnetic}. On the contrary, for $T<\Tc$ not only $q \neq 0$, but also $\lim_{T \to 0} q(T) = 1$.

A deeper connection can be established by computing the disorder\index{disorder!average} average of the \gls{EA} order parameter\index{order parameter!Edwards-Anderson} introduced in \refeq{order_parameter_SG_2}
\begin{equation}
\overline{q_{EA}} = \dfrac{1}{N} \sum_i^N \overline{\braket{s_i}^2} = \dfrac{1}{N} \sum_i \overline{\left(\dfrac{\sum_{\{s\}} s_i \exp \left[-\beta \mathcal{H}_J(\{s\})\right] }{Z_J}\right)^2}
\end{equation}
Now, we multiply the numerator and the denominator by $Z_J^{n-2}$ in order to use the replica\index{replica!method/trick} trick
\begin{equation}
\overline{q_{EA}}  =  \dfrac{1}{N} \sum_i^N \overline{\left( \dfrac{\sum_{\{s^a\}}s^r_i \exp \left[ -\beta \mathcal{H}_J^a(\{s^a\}) \right]}{Z_J^n}\right)^2} \, ,
\end{equation}
where $s^r$ represents an arbitrary replica\index{replica} $r$. When the limit $n\to 0$ is taken, the denominator $Z_J^n$ tends to $1$ and we finally get
\begin{equation}
\overline{q_{EA}} = \dfrac{1}{N} \sum_i^N \overline{\braket{s_i}^2} = \dfrac{1}{N} \sum_i^N \lim_{n \to 0} \overline{\braket{s_i^\alpha s_i^\beta}} \, ,
\end{equation}
however, performing the disorder\index{disorder!average} average in the \gls{SK}\index{Sherrington-Kirkpatrick} formalism, as we have just done above, leads to
\begin{equation}
\overline{q_{EA}} = \dfrac{1}{N} \sum_i^N \lim_{n \to 0} \braket{s_i^\alpha s_i^\beta}_Q \, ,
\end{equation}
and finally, from \refeq{mean_field_solution_general} and \refeq{replica_ansatz} we have that $q = \overline{q_{EA}}$. $q$ is nothing but the order parameter\index{order parameter!Edwards-Anderson} in the \gls{EA}\index{Edwards-Anderson!model} model.

This solution is, unfortunately, wrong. The first sign of a deep error in the replica symmetric\index{replica!symmetric solution} solution was the computation of low-temperature entropy\index{entropy} that turned out to be negative $S(T=0) = -1/2\pi <0$. Moreover, a detailed analysis of the solution \cite{dealmeida:78} showed that it is unstable in the low-temperature phase\index{phase!low-temperature/spin-glass}.

\subsubsection{Parisi's solution: the Replica Symmetry Breaking}

\captionsetup{justification=centering}

The unsatisfactory previous results suggested that the replica symmetry \index{replica!symmetric ansatz}should be broken and some attempts can be found in \cite{bray:78} and \cite{blandin:78} who actually proposed the first step of the general iterative solution. That solution came from Parisi \cite{parisi:79,parisi:80,parisi:80b} and it is known as the \gls{RSB}\index{replica!symmetry breaking (RSB)} solution. The starting point is the replica symmetry\index{replica!symmetric matrix} matrix $Q_{ab}^{\text{RS}}$, see \reffig{rs_qab}
\begin{figure}[h]
\begin{equation*}
Q_{ab}^{\text{RS}} = \left(\begin{array}{cccccccc}
0 & & & & \multicolumn{4}{c}{\multirow{4}{*}{\Huge $q_0$}}\\
 & 0 & &  & \\
 &  & 0 &  & \\
 & &  & 0 & \\
\multicolumn{4}{c}{\multirow{4}{*}{\Huge $q_0$}} & 0 &  &  &  \\
 & & & &   & 0 &  &   \\
& & &  &  &  & 0 &   \\
 & & & &  &  &  & 0
\end{array}\right) 
\end{equation*}
\caption[\textbf{Replica symmetry ansatz for the matrix $Q_{ab}$.}]{\textbf{Replica symmetry ansatz\index{replica!symmetric ansatz} for the matrix $Q_{ab}$.}}
\labfig{rs_qab}
\end{figure}

From here the proposed matrix $Q_{ab}^{\text{RSB}}$ is constructed through successive iterations. The first step, called the one-step \gls{RSB}\index{replica!symmetry breaking (RSB)} consists of dividing the $n$ replicas\index{replica} into $n/m_1$ groups, where $n$ and $n/m_1$ are supposed to be integers at this point. The $n/m_1$ groups are $m_1 \times m_1$ squares placed in the diagonal of the matrix $Q_{ab}^{\text{1-step}}$. All the elements of the matrix where $a=b$ remain equal to $0$, the elements in the $m_1 \times m_1$ squares with $a \neq b$ are equal to $q_1$ and the rest of the elements are equal to $q_0$ . The compact form to represent the values of $Q_{ab}^{\text{1-step}}$ is
\begin{equation}
Q^{\text{1-step}}_{ab} =
\begin{cases}
0 & \text{if $a=b$}\\
q_0 & \text{if $\Ceil{a/m_1} \neq \Ceil{b/m_1}$} \\
q_1 & \text{if $\Ceil{a/m_1} = \Ceil{b/m_1}$}
\end{cases}
\, , \labeq{1step_RSB_compact} 
\end{equation}
being $\Ceil{x}$ the ceiling function. The schematic representation of \refeq{1step_RSB_compact} is in \reffig{1step_qab}.
\begin{figure}[h]
\begin{equation*}
Q_{ab}^{\text{1-step}} = \left(\begin{array}{cccc|cccc}
0 &  &\multicolumn{2}{r|}{\multirow{2}{*}{\Large $\ \ q_1$}} & \multicolumn{4}{c}{\multirow{4}{*}{\Huge $q_0$}}\\
 & 0 & &  & \\
\multicolumn{2}{c}{\multirow{2}{*}{\Large $q_1$}}  & 0 & & \\
 &  &  & 0 & \\
\hline
\multicolumn{4}{c|}{\multirow{4}{*}{\Huge $q_0$}} & 0 & & \multicolumn{2}{c}{\multirow{2}{*}{\Large $q_1$}}\\
 & & & &  & 0 &   &  \\
& & & & \multicolumn{2}{c}{\multirow{2}{*}{\Large $q_1$}}  & 0 &   \\
 & & & &  & &  & 0
\end{array}\right)
\end{equation*}
\caption[\textbf{First step of Replica Symmetry Breaking.}]{\textbf{First step of Replica Symmetry Breaking.\index{replica!symmetry breaking (RSB)}}}
\labfig{1step_qab}
\end{figure}

\begin{figure}[h]
\begin{equation*}
Q_{ab}^{\text{2-steps}} = \left(\begin{array}{cc|cc|cc|cc}
0 & q_2 &\multicolumn{2}{c|}{\multirow{2}{*}{\Large $\ \ q_1$}} & \multicolumn{4}{c}{\multirow{4}{*}{\Huge $q_0$}}\\
q_2 & 0 & &  & \\
\cline{1-4}
\multicolumn{2}{c|}{\multirow{2}{*}{\Large $q_1$}}  & 0 &q_2 & \\
 &  & q_2 & 0 & \\
\hline
\multicolumn{4}{c|}{\multirow{4}{*}{\Huge $q_0$}} & 0 &q_2 & \multicolumn{2}{c}{\multirow{2}{*}{\Large $q_1$}}\\
\multicolumn{4}{c|}{} &q_2  & 0 &   &  \\
\cline{5-8}
\multicolumn{4}{c|}{} & \multicolumn{2}{c|}{\multirow{2}{*}{\Large $q_1$}}  & 0 & q_2  \\
\multicolumn{4}{c|}{} &  & &q_2  & 0
\end{array}\right) 
\end{equation*}
\caption[\textbf{Second step of Replica Symmetry Breaking.}]{\textbf{Second step of Replica Symmetry Breaking.\index{replica!symmetry breaking (RSB)}}}
\labfig{2step_qab}
\end{figure}
\captionsetup{justification=raggedright}

We can repeat the computation of the free energy\index{free energy}\footnote{It is worthy to note that, contrary as usual, in the framework of the \gls{RSB}\index{replica!symmetry breaking (RSB)} formalism, the free-energy\index{free energy} should be \textit{maximized}. The formal reason is the number of components of the matrix $Q_{ab}$ becomes negative in $n \to 0$ limit, see \cite{mezard:87,dotsenko:01}.} with this matrix (see e.g. \cite{dotsenko:01}) and the related thermodynamic quantities. The zero-temperature entropy\index{entropy} is $S^{\text{1-step}}(T=0) \approx -0.01$ i.e. $|S^{\text{1-step}}(T=0)| < |S^{RS}(T=0)|$ and the instability of the \gls{SG} phase\index{phase!low-temperature/spin-glass} is also reduced\footnote{Actually, what is reduced is the most negative eigenvalue of the free-energy\index{free energy} Hessian matrix near the critical\index{critical temperature} temperature.}.

The \gls{RSB}\index{replica!symmetry breaking (RSB)} procedure can be generalized to an infinite number of steps. To obtain the two-steps \gls{RSB}\index{replica!symmetry breaking (RSB)}, we proceed in the same way as we did in the one-step \gls{RSB}\index{replica!symmetry breaking (RSB)} for each of the diagonal blocks of size $m_1 \times m_1$, now, dividing them in blocks of size $m_2 \times m_2$. An schematic view of the $Q_{ab}^{\text{2-steps}}$ is in \reffig{2step_qab}.

Successive steps of \gls{RSB}\index{replica!symmetry breaking (RSB)} lead to $S(T=0) \to 0$ and a less unstable solution in the low-temperature phase\index{phase!low-temperature/spin-glass}. It took a while, but finally, it was rigorously proved that the infinite-steps \gls{RSB}\index{replica!symmetry breaking (RSB)} produces the correct solution for the free energy\index{free energy} in the \gls{SK}\index{Sherrington-Kirkpatrick} model \cite{guerra:02,guerra:03,talagrand:06}.

The infinite-step solution, therefore, depends on an infinite number of parameters $q_i$. Each of those $q_i$ appear with a different weight in the \gls{pdf} of the overlap\index{overlap} $q$, that takes the form
\begin{align}
p(q) & = \dfrac{1}{n(n-1)} \sum_{a \neq b} \delta(Q^{\infty\text{-steps}}_{ab}-q) = \nonumber \\
& = \dfrac{n}{n(n-1)}\left[ (n-m_1)\delta(q-q_0) + (m_1-m_2)\delta(q-q_1) + \dots \right] \, . \labeq{pdf_q}
\end{align}
The $n \to 0$ limit here is a delicate procedure in which we move the $n\times n$ matrix $Q^{\text{RSB}}_{ab}=Q^{\infty\text{-steps}}_{ab}$ to a $0 \times 0$ matrix-space. Moreover, the construction of the matrix suggests the assimilation of $m_0 = n$ and, with the restriction of $n$ to be an integer, $m_\infty=1$. Obviously, $m_k > m_{k+1}$, so $n=m_0 > m_1 > \dots > m_{\infty} = 1$. The analytical extension of $n$ and the limit $n \to 0$ implies that there is no reason to still considering $m_k$ with $k=0,1,\dots$ to be integers and, therefore, $m_k \in [0,1]$. The direct implication is the reversing of the order of the coefficients $m_k$, the \gls{pdf} now is
\begin{equation}
p(q) = m_1 \delta(q-q_0) + (m_2-m_1)\delta(q-q_1) + \dots \, ,
\end{equation} 
and the \gls{SG} order parameter\index{order parameter} is no longer a discrete set of parameters but a function $q(x)$ with $x \in [0,1]$ defined as
\begin{equation}
q(x) = q_k \quad, \quad 0 \leq m_k < x < m_{k+1} \leq 1 \, .
\end{equation}

We stop here our brief recall of the \gls{RSB}\index{replica!symmetry breaking (RSB)} results in the \gls{SK}\index{Sherrington-Kirkpatrick} model, nonetheless, there exists a huge number of interesting results, for example in the rich physical interpretation (see e.g. \cite{parisi:83,rammal:86}) or numerical results that agree with the \gls{RSB}\index{replica!symmetry breaking (RSB)} predictions \cite{vannimenus:81,sommers:84,crisanti:02,aspelmeier:08}.

\subsection{Theoretical pictures in finite-dimension spin glasses} \labsubsec{theoretical_pictures}
The \gls{RSB}\index{replica!symmetry breaking (RSB)} computation gives us the solution to the mean-field\index{Mean-Field!model} model, but the behavior of the \gls{SG}s in the low-temperature phase\index{phase!low-temperature/spin-glass} at finite dimensions is still a widely debated issue. Here, we briefly review the differences between the diverse pictures explaining the equilibrium \gls{SG}-phase\index{phase!low-temperature/spin-glass} and we show the main predictions that will be crucial to elucidate their validity through experiments, analytical results, or in the case of this thesis, numerical simulations.

\subsubsection{The Droplet picture}
After Parisi's solution for the mean-field\index{Mean-Field!model} model, numerical studies of domain\index{magnetic domain!walls} walls in \gls{SG}s and their scaling properties \cite{mcmillan:84,mcmillan:85,bray:84,bray:87}, based on Migdal-Kadanoff\index{Midgal-Kadanoff!zzzzz@\Also{Wilson-Kadanoff}|gobbleone} renormalization\index{renormalization group} computations (that are exact for dimension $d=1$), were the seed for a completely different approach to explain the low-temperature phase\index{phase!low-temperature/spin-glass} in short-ranged Ising\index{Ising} \gls{SG}s. This picture, introduced along seminal works by Fisher, Huse, Bray and Moore \cite{fisher:86,bray:87,fisher:88} is known as the \textit{droplets}\index{droplet!picture} picture.

The droplet\index{droplet!picture} picture understands the \gls{SG} phase\index{phase!low-temperature/spin-glass} from its ground-state\index{ground-state}. The basic object, the \textit{droplet}\index{droplet}, consists of a compact domain\index{magnetic domain!compact} of linear size $L$ of coherently flipped spins with respect to the ground-state\index{ground-state}, which constitutes the lowest-energy\index{energy} excitation at this length-scale $L$. Actually, the droplets\index{droplet} are expected to have fractal boundaries with a surface area of order $L^{d_s}$, $d-1 \leq d_s < d$ \cite{fisher:86}.

The droplets\index{droplet} with zero energy\index{energy} occurs with a probability $P \propto L^{-\theta}$ being $\theta< (d-1)/2$ the so-called \textit{stiffness exponent\index{stiffness!exponent}} and the free-energy\index{free energy} cost of generating a droplet\index{droplet} of linear size $L$ is $F_L \sim \varUpsilon L^\theta$ where $\varUpsilon$ is the \textit{stiffness modulus\index{stiffness!modulus}}. The computation of $\theta$ have been performed numerically for $d=3$ resulting in $\theta=0.27$ \cite{boettcher:04,boettcher:05}, $\theta=0.26$ \cite{monthus:14}. For $d=2$ the exponent $\theta$ is negative ($\theta \sim -0.28$ \cite{boettcher:04}), thus, in the thermodynamic limit\index{thermodynamic limit} the free-energy\index{free energy} cost for generating a droplet\index{droplet} tends to zero and the \gls{SG} transition\index{phase transition} disappears.

The most relevant results of the droplet\index{droplet!picture} pictures are the following:
\begin{itemize}
\setlength\itemsep{0.3cm}

\item The spatial correlation decays with the exponent $\theta$ as 
\begin{equation}
C(r_{ij}) = \overline{\braket{s_i s_j}^2} - \overline{\braket{s_i}^2} \overline{\braket{s_j}^2} \propto r_{ij}^{-\theta} \, .
\end{equation}

\item As a direct consequence of the long-distance vanishing limit of the correlation\index{correlation function!four point} functions \cite{fisher:86} the overlap\index{overlap!distribution} distribution is trivial i.e. corresponds to a delta function $p(q) = \delta(q^2 - q^2_{EA})$. The many-states nature of the \gls{SG} phase\index{phase!low-temperature/spin-glass} displayed by the \gls{RSB}\index{replica!symmetry breaking (RSB)} is no longer valid in the droplet\index{droplet!picture} picture where only a pair of equilibrium states, related by spin-flip, appears\footnote{It is commonly said that, according to the droplet\index{droplet!picture} picture, the \gls{SG} is a ``ferromagnet in disguise``, that is, a system with a complicated spin configuration\index{configuration} for the ground-state\index{ground-state} due to the randomness\index{randomness!bond} of the couplings\index{couplings} but that can be mapped to a ferromagnet by performing gauge transformations, similar to the Mattis model\index{Mattis model} \cite{mattis:76}.}. 

\item Related to the dynamics, the aging\index{aging} in the droplet\index{droplet!picture} picture is explained through the growth of coherent domains\index{magnetic domain} of spins. Moreover, the coarsening\index{coarsening} domains\index{magnetic domain!compact} would be compact objects where the overlap takes one of the two possible values associated with the two pure states allowed $q = \pm q_{EA}$ \cite{fisher:88}.

\item The presence of an external magnetic field suppresses the transition\index{phase transition} to the \gls{SG} phase\index{phase!low-temperature/spin-glass}. The argument underlying this prediction is a generalization of the Imry-Ma \cite{imry:75} argument. The energy cost of reversing the spins inside a droplet\index{droplet} is, by an assumption of the droplets\index{droplet!picture} model, proportional to $L^\theta$, and the magnetization\index{magnetization} of the droplet\index{droplet} scales as $L^{d/2}$. By introducing the Zeeman\index{energy!Zeeman} energy, we can write the free-energy\index{free energy} cost for flipping the droplet\index{droplet} in the presence of a small external magnetic field $h$ as $L^\theta - hL^{d/2}$. Since $\theta < (d-1)/2 < d/2$, the \gls{SG} becomes unstable under the presence of any field $h$.

\item The \gls{SG} phase\index{phase!low-temperature/spin-glass} exhibits a chaotic behavior under small changes of the temperature \cite{banavar:87,bray:87}. This feature is a direct consequence of free-energy\index{free energy} scaling ansatz $F_L \propto L^\theta$. The \textit{naive} expectation for the free-energy\index{free energy} of the droplet\index{droplet} with surface $L^{d_s}$ would be $F_L \propto L^{d_s}$, and since $d_s \geq d - 1 > (d-1)/2 > \theta$, the difference between the \textit{naive} expectation and the scaling ansatz is the presence of large cancellations of the contribution to the free-energy\index{free energy} in different parts of the boundaries. This precarious equilibrium would be sensitive to small changes in the temperature due to changes in the sign of the free-energy\index{free energy} at large scales (see e.g. \cite{katzgraber:07} for further details). Thus, one would expect a complete reorganization of the spin equilibrium configurations\index{configuration} upon small changes of the temperature. This sensitivity of the system is known as \textit{temperature chaos}\index{temperature chaos}. In~\refch{Introduction_chaos} the reader may find a deeper discussion about this issue.
\end{itemize}

\subsubsection{The RSB picture}
The \gls{RSB}\index{replica!symmetry breaking (RSB)} solution for the mean-field\index{Mean-Field!model} model is expected to be correct in short-ranged models like the \gls{EA}\index{Edwards-Anderson!model} model for dimensions higher than the \textit{upper critical dimension}\index{critical dimension!upper} $d>d_u = 6$. However, its validity in lower dimensions (in particular we are interested in dimension $d=3$) is still a debated issue.

The \gls{RSB}\index{replica!symmetry breaking (RSB)} theory in short-ranged finite dimensions is obtained from perturbative computations from the original mean-field\index{Mean-Field!solution} solution but the physical behavior drawn is very similar to the mean-field\index{Mean-Field} predictions. The most remarkable results are:
\begin{itemize}
\setlength\itemsep{0.3cm}
\item In the \gls{SG} phase\index{phase!low-temperature/spin-glass}, the order parameter\index{order parameter} is a function $q(x): [0,1] \longrightarrow [-q_{EA},+q_{EA}]$. In particular, in the low-temperature phase\index{phase!low-temperature/spin-glass}, each pair of states will have an overlap\index{overlap} $q \in [-q_{EA},+q_{EA}]$ which follows the \gls{pdf} of \refeq{pdf_q} and that can be written as
\begin{equation}
p(q) = \dfrac{\dd x(q)}{\dd q} \, ,
\end{equation}
by the introduction of the inverse function of $q(x)$
\begin{equation}
x(q) = \int_0^q P(q') \dd q' \, .
\end{equation}
This non-trivial \gls{pdf} is the sign of one of the most distinctive features of \gls{RSB}\index{replica!symmetry breaking (RSB)}: in the low-temperature phase\index{phase!low-temperature/spin-glass} there exist infinitely many states.
\item The organization of those infinitely many states is studied through the \gls{pdf} of three pure states, see e.g. \cite{rammal:86,dotsenko:01}. The main result is that, for any arbitrary tern of states $\alpha$, $\beta$ and $\gamma$, the overlap\index{overlap} between them must fulfill the condition
\begin{equation}
q_{ab} = q_{bc} \leq q_{ac} \quad \forall \,\, (a,b,c) \in \mathrm{Sym}(\{\alpha,\beta,\gamma\}) \, \labeq{ultrametric}
\end{equation}
being Sym$(\{\alpha,\beta,\gamma\})$ the set of all permutations of the three states. Equivalently, 
\begin{equation}
q_{ab} \geq \min(q_{bc},q_{ac}) \quad \forall \,\, (a,b,c) \in \mathrm{Sym}(\{\alpha,\beta,\gamma\}) \, .
\end{equation}
This property defines a measure over the space of states and those spaces that present this particular metric are known as \textit{ultrametric} spaces. Therefore, in the space of \gls{SG} states, there exist no triangles with all three sides being different.

The usual way to visualize the ultrametricity\index{ultrametricity} in \gls{SG} is displayed in \reffig{ultrametric} from \cite{mydosh:93}. For each pair of states $\alpha$ and $\beta$, the overlap\index{overlap} $q_{\alpha \beta}$ is obtained by going back in the tree until reaching the first common level. The ultrametric property of \refeq{ultrametric} can be easily checked if one picks any three states (labeled with a number) in the referred figure.
\begin{figure}[h]
\includegraphics[width=0.5\textwidth]{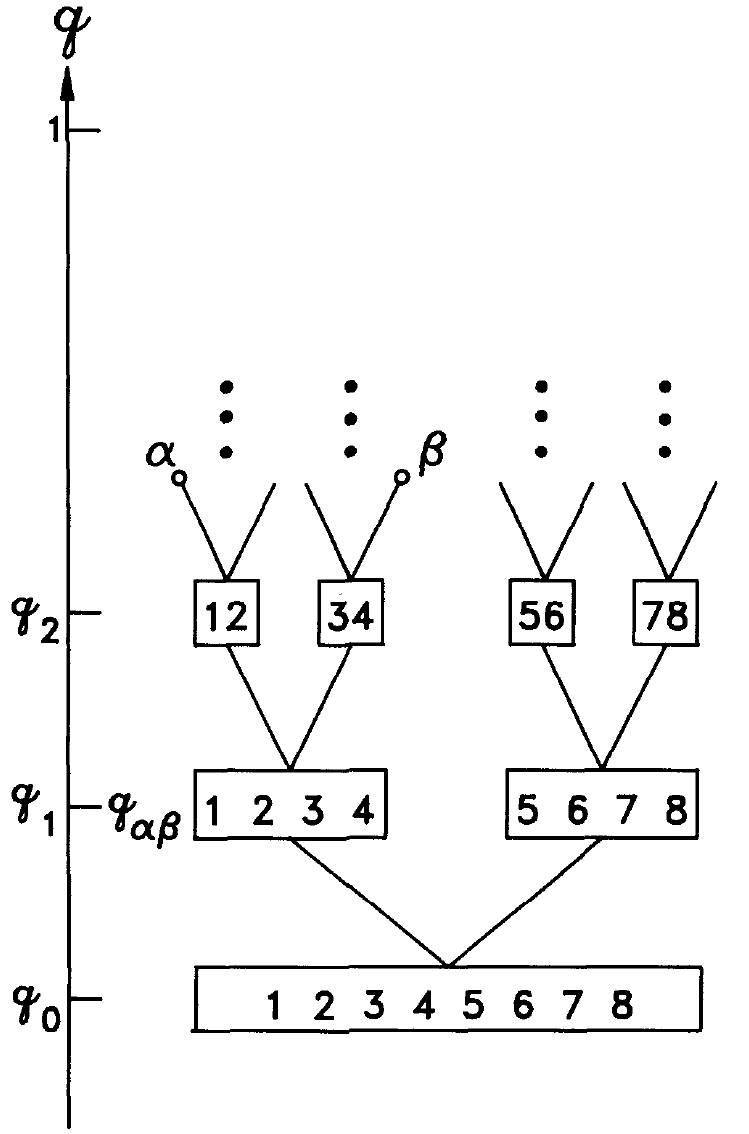}
\caption[\textbf{Ultrametric organization of Replica Symmetry Breaking states.}]{\textbf{Ultrametric organization of Replica Symmetry Breaking\index{replica!symmetry breaking (RSB)} states.} The tree representation of the Replica Symmetry Breaking states\index{replica!symmetry breaking (RSB)}. For any pair of states $\alpha$ and $\beta$ their corresponding overlap\index{overlap} is obtained by downing the tree until reaching the encounter point. Figure from \cite{mydosh:93}.}
\labfig{ultrametric}
\end{figure}

\item The ultrametricity\index{ultrametricity} is argued to be related to the origin of the temperature chaos\index{temperature chaos} phenomenon in the \gls{RSB}\index{replica!symmetry breaking (RSB)} picture, see for instance \cite{vincent:97}. The ultrametric hierarchical structure of states is temperature-dependent, that is, the free-energy\index{free energy!landscape} landscape changes with the temperature as sketched in \reffig{ultrametric_temperature}. 

In the thermodynamic limit\index{thermodynamic limit}, any small change of the temperature will relocate the state to a different local minimum, leading to a complete reorganization of its equilibrium configuration\index{configuration}. Furthermore, this explanation of the temperature chaos\index{temperature chaos} phenomenon would also explain the experiments of memory\index{memory effects} and rejuvenation\index{rejuvenation}, commonly associated with it \cite{picco:01,takayama:02,maiorano:05,jimenez:05} but not unanimously \cite{berthier:02}.

In this picture, the system at a temperature $T$ would explore the metastable\index{metastability} states\footnote{In the thermodynamic limit\index{thermodynamic limit}, the system would need infinite time to ``jump'' from one state to another.}. When the temperature is lowered by an amount $\delta T$, the system would move in the branch corresponding to its actual state and restart the aging\index{aging} process, this is the \textit{rejuvenation}\index{rejuvenation} effect. When the temperature is back to its previous value $T$, the system comes back to the initial state by moving in the same branch of the tree, this is the \textit{memory}\index{memory effects} effect.

\begin{figure}[h]
\includegraphics[width=0.7\textwidth]{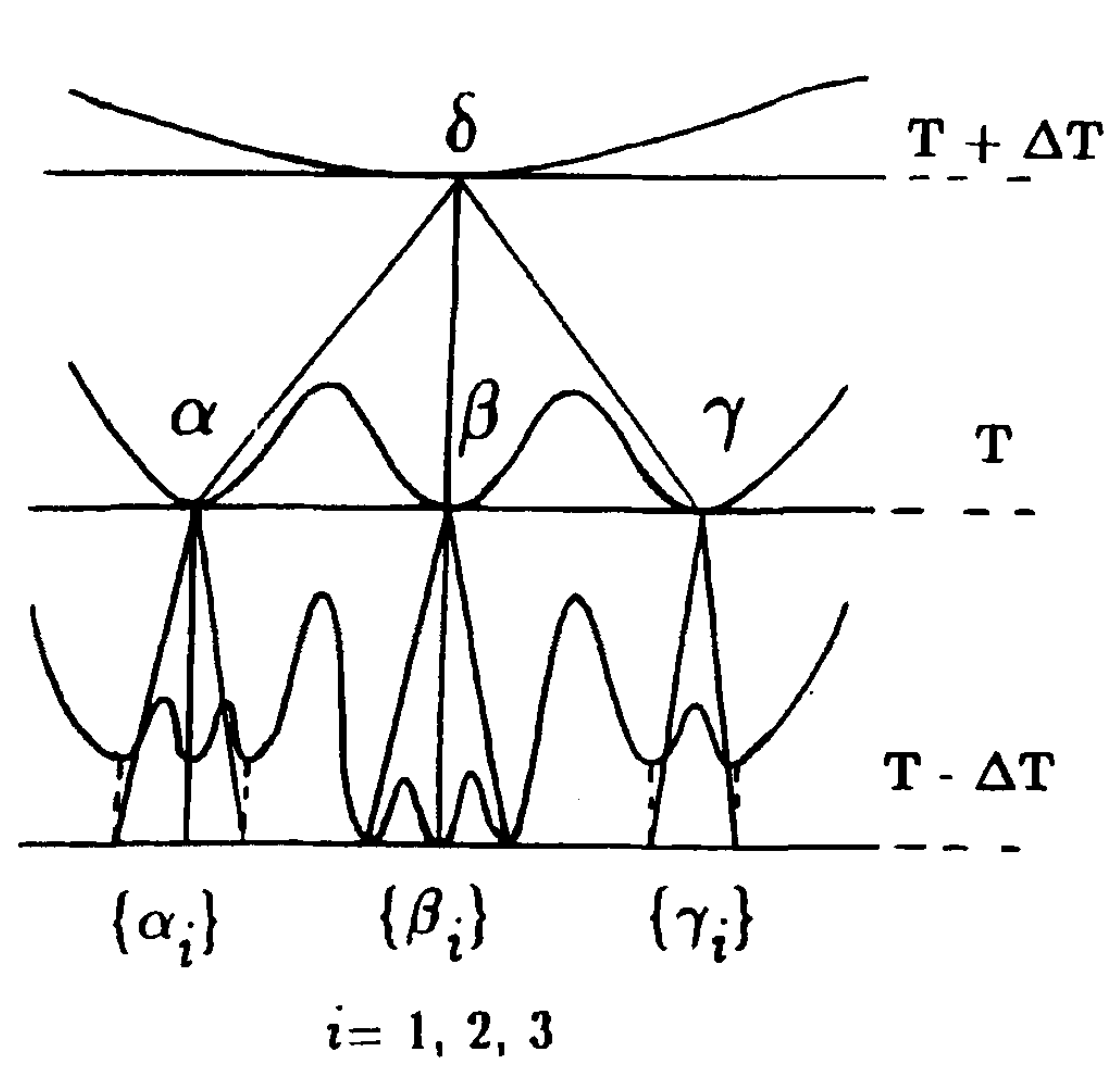}
\caption[\textbf{Sketch of ultrametric structure as a function of the temperature.}]{\textbf{Sketch of ultrametric structure as a function of the temperature.} The hierarchical structure of states as a function of temperature is commonly argued to be related to the temperature chaos\index{temperature chaos} phenomenon in the Replica Symmetry Breaking\index{replica!symmetry breaking (RSB)} picture. Figure from \cite{vincent:97}.}
\labfig{ultrametric_temperature}
\end{figure}

\item In the \gls{RSB}\index{replica!symmetry breaking (RSB)} picture, contrarily to the expected in the droplet\index{droplet!picture} picture, the \gls{SG} phase\index{phase!low-temperature/spin-glass} is not destroyed by a small magnetic field. The temperature-magnetic field plane is separated by the so-called \textit{de Almeida-Thouless} line \cite{dealmeida:78}. The part of the plane with a large magnetic field $h$ and a large temperature is paramagnetic-like while the opposite limit presents a \gls{SG} behavior.

\item Similarly to the droplet\index{droplet!picture} picture, the aging\index{aging} in the \gls{RSB}\index{replica!symmetry breaking (RSB)} picture is explained through the growth of coherent domains\index{magnetic domain} of spins. However, the predictions of both pictures split when trying to explain the nature of those domains\index{magnetic domain}. The \gls{RSB}\index{replica!symmetry breaking (RSB)} theory expects space-filling domains\index{magnetic domain!space-filling} i.e. the fractal dimension of the surface is $d_s=d$. 
\end{itemize}

\subsubsection{Problems with early interpretation: the concept of metastate}
However, the classic interpretation of \gls{RSB}\index{replica!symmetry breaking (RSB)} described above presents some issues. The properties of the theory were thought to be present in infinite systems but the procedure to obtain them was to average over the disorder\index{disorder!average} and, only after that, the infinite-size limit was taken. The problem in disordered\index{disorder!systems} systems is that, even if that limit exists for averaged quantities or distributions of quantities, it does not imply that an infinite system from which we obtain these quantities or distributions exists.

In order to solve this problem, mathematical approaches irrupted the physical debate through the concept of metastate\index{metastate}, firstly proposed in a general context of disordered\index{disorder!systems} systems by Aizenman and Wehr~\cite{aizenman:90} and later introduced in the specific context of \gls{SG}s to deal with this problem, by Newman and Stein~\cite{newman:92,newman:96,newman:98,newman:03}.

By using the metastate\index{metastate} formalism, two main pictures were introduced that rigorously solve the problem of taking the infinite-size limit: the metastate\index{metastate} interpretation of \gls{RSB}\index{replica!symmetry breaking (RSB)}, and the chaotic-pairs picture. The reader may find a detailed discussion in~\refch{metastate}.

\section{Numerical simulations in spin glasses} \labsec{numerical_spinglass}
The previous sections showed us that, in general, the main theoretical results were far apart from the main experimental results. One role of numerical simulations is to fill that gap. On the one side, experiments are restricted to off-equilibrium conditions, and access to microscopical configurations\index{configuration} is prohibited. On the other side, the theoretical works have been focused to understand the nature of the low-temperature phase\index{phase!low-temperature/spin-glass} in \gls{SG}. Furthermore, the analytical results are only exact in unrealistic models: droplets\index{droplet!picture}, exact in one dimensional \gls{SG}s, and \gls{RSB}\index{replica!symmetry breaking (RSB)}, exact in the \gls{SK}\index{Sherrington-Kirkpatrick} model and, with almost total consensus, in \gls{EA}\index{Edwards-Anderson!model} model for dimensions $d>6$.

Numerical simulations, mostly focused on Monte\index{Monte Carlo} Carlo simulations \cite{landau:05}, allow us to study off-equilibrium and equilibrium \gls{SG}s. Moreover, from numerical data we can access the microscopical configurations\index{configuration} and we have total control of the system. However, there are also obstacles in the path of numerical work. The equilibrium simulations are restricted to small system-sizes $L$ and temperatures $T$ not very far from the critical\index{critical temperature} temperature due to the sluggish dynamics exhibited in the low-temperature phase\index{phase!low-temperature/spin-glass}, thus, the suspect of finite-size and critical effects hovers over the results. In the off-equilibrium case, again due to the extremely slow dynamics of \gls{SG}s, the time-scale of the numerical work was traditionally very far from the time-scale of experiments.

Fortunately, this situation has improved significantly during the last years. The year-to-year increase of the computational power, the emergence of special-purpose\index{special-purpose computer} computer like Janus\index{Janus} \cite{janus:06,janus:09} and Janus\index{Janus} II \cite{janus:14}, and the implementation of algorithms like Parallel\index{parallel!tempering} Tempering have allowed to simulate larger systems with unprecedented precision and to achieve time-scales comparable with the experimental ones \cite{janus:08b,janus:09b,janus:18}.

This thesis aims to be a step forward in the conversion of the numerical simulations from an extremely useful tool for theoretical studies to a bridge between theory and experiments. Along this document we will present several original works with at least one of the following tasks on mind:
\begin{enumerate}
\item \textbf{Simulated systems must capture the observed phenomenology of real \gls{SG}.} Physics is an experimental science and, therefore, every model that we take into consideration to explain the \gls{SG}s physics must exhibit the same phenomenology observed in real systems. The state of the art for numerical simulations allows extrapolations to experimental scales (see~\refch{aging_rate}) and the comparison between numerical and experimental results. 
\item \textbf{Relate theoretical results with experiments.} A few years ago, the idea of the statics-dynamics equivalence\index{statics-dynamics equivalence} was proposed in numerical simulations \cite{janus:08b,janus:10,janus:17}. Numerical evidence is suggesting that off-equilibrium systems of coherence length\index{coherence length} $\xi$ could be regarded as a set of equilibrated systems of linear size $L \sim \xi$. This concept allows us to relate theoretical predictions in equilibrated \gls{SG}s with off-equilibrium measures that can be compared with experimental results (see \refch{aging_rate}).
\item \textbf{Discern between proposed theoretical pictures.} Although the new avenues open by the improvement of the computational power are exciting, we still can take advantage of the traditional purposes of the numerical works. We have discussed how different pictures provide different predictions in the \gls{SG} phase\index{phase!low-temperature/spin-glass} (\refsubsec{theoretical_pictures}). Throughout this thesis we will face those predictions and we will compare them with numerical results (see~\refch{metastate}, \refch{aging_rate}).
\item \textbf{Open new paths for experimental work.} The level of control that we have over simulated systems makes the numerical work an ideal field to find new phenomena that can be later addressed by experiments (see~\refch{mpemba} and \refch{out-eq_chaos}).
\item \textbf{Develop new tools for numerical simulations.} Of course, the numerical simulations are not perfect, and not only the development of the hardware is capable of improving their performance. The numerical research to find new methods have been fundamental, from a historical point of view. Here we also focus on improve the numerical simulations, for example, in the study of the Temperature Chaos\index{temperature chaos} (see~\refch{equilibrium_chaos}), but also by implementing well-established methods in our works (\refch{AP_statistics}, \refch{AP_technical_details_aging}, \refch{AP_PT}, \refch{AP_technical_details_out-eq_chaos}, \refch{AP_multispin_coding}).
\end{enumerate}

Throughout this thesis we will be focus on the numerical study of the 3D \gls{EA}\index{Edwards-Anderson!model} model by using Monte\index{Monte Carlo} Carlo methods, in the rest of this section, we will introduce the model and the observables computed with the goal to avoid repetitions in the subsequent chapters.

\subsection{3D Edwards-Anderson model} \labsubsec{3D_EA_model}
All the numerical simulations carried out in the original works of this thesis are performed in the three-dimensional Ising\index{Ising} \gls{EA}\index{Edwards-Anderson!model} model. In our simulations, the spins are disposed in a cubic lattice $\Lambda_L$  with \gls{PBC}\index{boundary conditions!periodic} where the vertices corresponds to the location of the Ising\index{Ising} spins $s_i = \pm 1$. The edges of the cubic lattice correspond to the quenched couplings\index{couplings} $J_{ij}$ and the energy\index{energy} of the system is defined by the Hamiltonian\index{Hamiltonian}
\begin{equation}
\mathcal{H}_{\{J\}} (\{s\})= - \sum_{\braket{i,j}} J_{ij}s_is_j \, . \labeq{EA_Hamiltonian}
\end{equation}
In our simulations, the couplings\index{couplings} $J_{ij}$ are independent and identically distributed random variables drawn from a bimodal distribution ($J_{ij}= \pm 1$ with a $50 \%$ probability). This model exhibits a spin-glass transition\index{phase transition} at temperature $\Tc = 1.1019(29)$ \cite{janus:13}.

Each realization of the couplings\index{couplings} $\{J\}$ is called a \textit{sample}\index{sample} and allows us to estimate the average over the disorder\index{disorder!average}. For each sample\index{sample}, we simulate statistically independent system copies, each of them evolving under the same couplings\index{couplings} $\{J\}$ but with different thermal noise. Each of these copies is called a \textit{replica}\index{replica}. The need of simulating different replicas\index{replica} will be exposed below.

The parameters of the simulation (number of samples\index{sample}, number of replicas\index{replica}, simulated temperatures, the size of the lattice, \dots) and the corresponding Monte\index{Monte Carlo} Carlo method used will be specified in the following chapters, depending on the simulation carried out. 

\subsection{Monte Carlo simulations} \labsubsec{Monte_Carlo}
As we have already anticipated, the studies carried out along this thesis are performed through Monte\index{Monte Carlo} Carlo simulations \cite{landau:05,amit:05}. Here, we briefly introduce this method for the reader unfamiliar with the Monte\index{Monte Carlo} Carlo methods. We will quickly explain the basics of Markov\index{Markov chain} chains, we will introduce the Metropolis\index{Metropolis-Hastings} algorithm and the Parallel\index{parallel!tempering} Tempering. Nonetheless, advanced applications of Markov\index{Markov chain} chains and, specifically Parallel\index{parallel!tempering} Tempering, related to the thermalization\index{thermalization} process will be described in~\refch{equilibrium_chaos}.
\subsubsection{Markov chains}
We are interested in the study of a very specific spin system in a lattice, as we have just discussed. The problem is that even for fairly small systems, the integrals involved in the computation of typical quantities have a very high dimensionality, which makes the numerical methods of integration very inefficient. We use, instead, a well-established method to obtain a sample\index{sample} of configurations\index{configuration} with the appropriate \gls{pdf}: a dynamic Monte\index{Monte Carlo} Carlo method.
 
To that purpose, we consider a random walk\index{random walk} in the configuration\index{configuration!space} space which hopefully allows us, for each temperature $T=1/\beta$, to \textit{move} from an arbitrary point in the configuration\index{configuration!space} space\footnote{In general, we have no \textit{a priori} information about how the typical configuration\index{configuration} should be at the desired temperature.} to the relevant configurations\index{configuration} at that temperature (i.e. the \textit{thermalization}\index{thermalization}) and to sample\index{sample} configurations\index{configuration} according to the Boltzmann-Gibbs\index{Boltzmann!-Gibbs distribution} distribution [see \refeq{prob_configuration}].

Besides, our random walk\index{random walk} should be Markovian\index{Markovian}\footnote{The next state depends only on the actual state and not on the history of the random walker\index{random walk}.} and will be represented by a transition matrix\index{transition matrix} $\pi$. The transition matrix\index{transition matrix} is 2-dimensional and the rows (equivalently the columns) correspond to every possible configuration\index{configuration} of the system. The elements $\pi_{XY}$ denote the probability of change from a configuration\index{configuration} $X$ to another configuration\index{configuration} $Y$ in the next step of the random walk\index{random walk}. Therefore $\pi_{XY} \geq 0$ $\forall X,Y$ and $\sum_{Y} \pi_{XY} = 1$.

As long as the system is Markovian\index{Markovian}, the probability for going from one configuration\index{configuration} $X_0$ to one configuration\index{configuration} $X_n$ in $n>1$ steps is just the sum of the probabilities of the system running over all the possible paths from $X_0$ to $X_n$ in $n$ steps. In those paths, due to the \textit{no-memory} characteristic of the Markovian\index{Markovian}\index{Markov chain} chains, the probability is just the product of the probabilities $\pi^{(n)}_{X_0X_n} = \sum_{ \{ X_1,X_2,\dots,X_{n-1} \} } \pi_{X_0X_1} \pi_{X_1X_2} \cdots \pi_{X_{n-1}X_n}$. 

In addition to those properties, which are inherent to all the Markov\index{Markov chain} chains, we require other properties that are fundamental for the thermalization\index{thermalization} process.

First, the so-called \textit{balance\index{balance condition}} condition
\begin{equation}
\dfrac{\exp\left(-\beta \mathcal{H}(Y) \right)}{Z} = \sum_{X} \pi_{XY} \dfrac{\exp\left((-\beta \mathcal{H}(X) \right)}{Z} \, . \labeq{balance_condition}
\end{equation}
This condition express the fact that, if we have a set of configurations\index{configuration} distributed according to the Boltzmann-Gibbs\index{Boltzmann!-Gibbs distribution} probability distribution at one step $t$, the set of configurations\index{configuration} after one step ($t+1$) of the random walk\index{random walk} for each element of the set will be also distributed according to the Boltzmann-Gibbs\index{Boltzmann!-Gibbs distribution} probability distribution.

Moreover, the \textit{irreducibility}\index{irreducibility}\index{irreducibility} condition assures that all the configurations\index{configuration} $Y$ are accessible from any configuration\index{configuration} $X$, mathematically this is expressed as 
\begin{equation}
\forall \, X,Y \, \exists \, n>0 \, : \, \sum_{ \{ X_1,X_2,\dots,X_{n-1} \} } \pi_{X_0X_1} \cdots \pi_{X_{n-1}X_n} > 0 \quad \mathrm{with} \quad X_n=Y \, . \labeq{irreducibility}
\end{equation}
There exists a more restricting condition, the \textit{aperiodicity}\index{aperiodicity} which needs the introduction of the concept of \textit{period}. A period $d_X$ (being $X$ a configuration\index{configuration}), is the greatest common divisor of the length of all the paths starting at configuration\index{configuration} $X$ and finishing at the same configuration\index{configuration} $X$. If that period is $d_X=1$ for all the states, we say that the Markov\index{Markov chain} chain is aperiodic\index{aperiodicity}.

The above properties allow us to introduce a specially useful theorem~\cite{sokal:97}: if one Markov\index{Markov chain} chain satisfies both, the aperiodic\index{aperiodicity} and the balance\index{balance condition} condition, then
\begin{equation}
\lim_{n\to \infty} \pi^{(n)}_{XY} = \dfrac{\exp\left( -\beta \mathcal{H}(Y)\right)}{Z} \, . \labeq{thermalization_condition}
\end{equation}
This theorem implies that the starting configuration\index{configuration} is immaterial, the random walk\index{random walk} will eventually sample the desired Boltzmann-Gibbs\index{Boltzmann!-Gibbs distribution} distribution. This theorem is not but a mathematical warranty of the fact that our system will thermalize.

\subsubsection{The Metropolis-Hastings algorithm}
In our numerical simulations, most of the time we are using the Metropolis-Hastings\index{Metropolis-Hastings} algorithm\footnote{Often called just Metropolis\index{Metropolis-Hastings}, for short.}, which is nothing but a dynamic Monte\index{Monte Carlo} Carlo method that fulfills the previously described conditions. This method is generally applicable to a multitude of contexts, but it is not our goal to provide general results on Monte\index{Monte Carlo} Carlo methods\footnote{To that purpose, the reader may consult \cite{sokal:97,amit:05,landau:05}.}. Hence, we focus here on our particular context and identify a configuration\index{configuration} $X$ with the set of all the spins in our lattice. 

There exist several possible ways to propose a change from the configuration\index{configuration} $X$ to the configuration\index{configuration} $Y$. One very common choice is to attempt the change of individual spins in the lattice in a sequential way. As far as we are focus on the Ising\index{Ising} model, for each spin there is only a possible change to perform: to flip the spin.

In the Metropolis-Hastings\index{Metropolis-Hastings} algorithm, for one temperature $T=1/\beta$, we proceed in the following way:
\begin{enumerate}
\item Compute the energy\index{energy} of the configuration\index{configuration} $X$, namely $\mathcal{H}(X)$
\item Flip the spin $i$ and compute the energy\index{energy} of the new configuration\index{configuration} $Y$: $\mathcal{H}(Y)$.
\item Draw a random number $r \in [0,1)$. If $r<\exp\left[ -\beta(\mathcal{H}(Y)-\mathcal{H}(X))\right]$ then we accept the change $X \to Y$, otherwise, we remain in the configuration\index{configuration} $X$.
\item Repeat the previous steps for all the spins $i$ in the lattice. We denote the realization of the Metropolis-Hastings\index{Metropolis-Hastings} algorithm to all the lattice a \textit{lattice sweep} or simply, a \textit{sweep}.
\end{enumerate}
This algorithm makes the system evolve whenever the energy\index{energy} is diminished, but also allows, with probability $e^{-\beta \Delta\mathcal{H}}$, local moves which increase the energy\index{energy} by an amount $\Delta \mathcal{H}$.

\subsubsection{Parallel Tempering}
The sluggish dynamics exhibited by the \gls{SG}s is a major obstacle in the classical dynamic Monte\index{Monte Carlo} Carlo methods, such as the Metropolis-Hastings\index{Metropolis-Hastings} algorithm, because the necessary time to thermalize, even small systems, would be prohibitive. Several algorithms try to palliate this problem, here we introduce the \gls{PT} algorithm \cite{hukushima:96,marinari:98b}.

The general idea of the \gls{PT} is to thermalize at the same time a set of $N$ identical copies which are at different temperatures $T_1 < T_2 < \cdots < T_N$ (or equivalently $\beta_N > \cdots > \beta_2 > \beta_1$). For those samples\index{sample} above the critical\index{critical temperature} temperature $T>\Tc$, the evolution of the system will be fast. On the contrary, for those systems that lie at temperatures $T<\Tc$, the configurations\index{configuration} will be almost frozen. The \gls{PT} algorithm consists of two alternating sets of steps.

First, each system copy independently undergoes standard Monte\index{Monte Carlo} Carlo dynamics (for example Metropolis\index{Metropolis-Hastings}) at its own temperature; one can use one or more Monte\index{Monte Carlo} Carlo steps each time. Second, pairs of spin configurations\index{configuration} attempt to exchange their temperatures by permuting the $N$ copies of configurations\index{configuration} in the temperature mesh. The exchange rule between two copies labeled as $\alpha$ and $\alpha'$ with configurations\index{configuration} $\{s^{(\alpha)}\}$ and $\{s^{(\alpha')}\}$ of the system follows the Metropolis\index{Metropolis-Hastings} scheme but we trade the accepting probability to
\begin{equation}
r < \exp \left\lbrace \left[\beta_{\alpha} - \beta_{\alpha'}\right] \left[\mathcal{H}(\{s^{(\alpha)}\}) - \mathcal{H}(\{s^{(\alpha')}\}) \right] \right\rbrace \, . \labeq{PT_exchange_rule}
\end{equation}

The goal of the \gls{PT} algorithm is to sample sets of $N$ configurations\index{configuration} (one for each copy of the system $\{s^{(\alpha)}\}_{\alpha=1}^N$), each one at its given temperature. Of course, those copies are not ordered in the temperature mesh with the $\alpha$ index because of the permutations introduced by the algorithm. In order to fully characterize the state of the system at a given time, we need to introduce $\pi(\alpha)$: the permutation of the $\alpha=1,2,\dots,N$ copies of systems in the temperature mesh. Now, the state of the \gls{PT} can be described by $X=\{ \pi, \{s^{(\alpha)}\}_{\alpha=1}^N\}$ and the stationary distribution of the \gls{PT} algorithm would be
\begin{equation}
P_{\mathrm{eq}}(X) = \dfrac{1}{N!}\prod_{\alpha=1}^N \dfrac{\exp \left[-\beta_{\alpha} \mathcal{H}(\{s^{\pi^{-1}(\alpha)}\})\right]}{Z_{\beta_{\alpha}}} \, , \labeq{PT_eq_distribution}
\end{equation}
being $Z_{\beta_{\alpha}}$ the partition function\index{partition function} at the temperature $\beta_{\alpha}$ and $\pi^{-1}(\alpha)$ the inverse permutation of $\pi$ that fulfills the condition $\pi^{-1}\left( \pi(\alpha)\right) = \alpha$ ; $\forall$ $\alpha$.

The rationale behind the \gls{PT} method is simple. Each system copy undergoes a random walk\index{random walk} in temperature space. When a system copy is at a low temperature, it only explores the nearby free-energy\index{free energy} local minima. When its temperature is high, however, free-energy\index{free energy!barrier} barriers disappear: the copy can freely wander in phase space\index{phase space}, and when it cools again it will typically fall in a different free-energy\index{free energy!valley} valley, with different local minima. For \gls{PT} to effectively thermalize, it is crucial that any copy of the system spends its time roughly evenly at every temperature: high temperatures are needed to ensure visiting all the phase space\index{phase space}; low temperatures are needed to visit its low free-energy\index{free energy} regions. 

In fact, \gls{PT} is currently used in a very large number of very different applications (for example in physics, biology, chemistry, engineering, statistics), and considerable efforts have been devoted to improving it from various communities. Various temperature-exchange rules have been developed and tested \cite{sugita:99,calvo:05,earl:05,brenner:07,bittner:08,malakis:13}. Furthermore, it has been suggested that a significant gain can be achieved by optimizing the choice of the $N$ temperatures \cite{katzgraber:06,sabo:08}. Further details on the \gls{PT} method can be found in~\refch{equilibrium_chaos} and \refsec{thermalizing_PT}.

\subsection{Observables}\labsubsec{observables_introduction}
Here, we present the observables that we will measure in most of the performed simulations. Nonetheless, in the subsequent chapters, we will present some observables that have to be announced in their context in order to be fully understood.

One more consideration needs to be done before defining the observables. The Hamiltonian\index{Hamiltonian} \refeq{EA_Hamiltonian} presents a gauge symmetry, specifically, the transformation 
\begin{equation}
s_i \longrightarrow \epsilon_i s_i \quad \quad J_{ij} \longrightarrow \epsilon_i \epsilon_j J_{ij} \, ,
\end{equation}
leaves it unchanged, being $\epsilon_i$ a random sign $\pm 1$ for each site of the lattice $i$. When several samples\index{sample} are simulated and averages over the disorder\index{disorder!average} are taken, this gauge symmetry is just the expression of redundant degrees of freedom\index{degree of freedom} of the system. If the measured observables are insensible to this symmetry, it would be possible to find pairs of $\{ \{J\}, \{s\}\}$ related by a gauge transformation and providing different \textit{weights} to the disorder\index{disorder!average} average i.e. totally equivalent configurations\index{configuration} from the energetic point of view, provide different results for some observables in different samples\index{sample}. Therefore, it is desirable to define gauge-invariant observables. The usual way to do that is by introducing \textit{replicas}\index{replica} (see \refsubsec{3D_EA_model}). 

The overlap\index{overlap!field} field between two replicas\index{replica} is defined as
\begin{equation}
q^{\sigma,\tau}_{\vec{x}}(t) = s_{\vec{x}}^\sigma(t) s_{\vec{x}}^\tau(t) \, , \labeq{def_overlap}
\end{equation}
where $\sigma$ and $\tau$ are labels to denote two different replicas\index{replica} and subscripts $\vec{x}$ represent the position in the lattice. The four-point spatial correlation\index{correlation function!four point} function is
\begin{equation}
C_4(T,\vec{r},t) = \overline{\braket{q^{\sigma,\tau}_{\vec{x}}(t)q^{\sigma,\tau}_{\vec{x}+\vec{r}}(t)}} \, , \labeq{def_C4}
\end{equation}
where $\overline{\left( \cdots \right)}$ is the disorder\index{disorder!average} average defined in \refeq{average_disorder} and $\braket{\cdots}$ the average over the thermal noise\footnote{This correlation\index{correlation function!four point} function will be measured in off-equilibrium systems along this thesis, therefore, the mean given by \refeq{average_o} that holds for equilibrium systems do not apply. We estimate the thermal noise by averaging over all the pairs of replicas\index{replica} $\sigma$ and $\tau$.}. In numerical simulations we can only estimate these means, since we only have finite number of samples\index{sample} (much smaller than the possible set of couplings\index{couplings} $\{J\}$) and a finite number of replicas\index{replica}. The long distance decay of $C_4(T,\vec{r},t)$ defines the coherence length\index{coherence length} $\xi(t)$, an observable of central importance as we will discuss below
\begin{equation}
C_4(T,\vec{r},t) \sim r^{-\vartheta} f(r/\xi(t)) \, . \labeq{long_distance_C4}
\end{equation}
The function $f(x)$ decreases faster than exponentially for large $x$, $f(x) \sim e^{-x^{\beta}}$ with $\beta \approx 1.7$ (see \cite{jimenez:05}). The exponent $\vartheta$ at the critical\index{critical temperature} temperature $T=\Tc$ is related to the anomalous dimension $\eta$ (see for example \cite{amit:05} for a definition), being $\vartheta(\Tc) = 1 + \eta$ with $\eta = -0.390(4)$ \cite{janus:13}. For $T<\Tc$, droplets\index{droplet!picture} picture predicts compact domains\index{magnetic domain!compact} and, therefore, $\vartheta=0$. On the contrary, \gls{RSB}\index{replica!symmetry breaking (RSB)} picture expects space-filling domains\index{magnetic domain!space-filling} where $C_4(T,\vec{r},t)$ vanishes at fixed $r/\xi$ as $t$ grows. The value of $\vartheta$ in this picture is given by the replicon\index{replicon}, a critical mode analogous to magnons in Heisenberg\index{Heisenberg} ferromagnets (see \cite{janus:10b} for a detailed discussion and \cite{dedominicis:06} for theoretical introduction of the replicon\index{replicon}). Previous numerical studies give us the value $\vartheta=0.38(2)$ \cite{janus:09} with a small dependence on the temperature $T$ that was vaguely attributed to the effect of the critical point. We will discuss widely about this exponent in~\refch{aging_rate}.

\subsubsection{Coherence length}\labsubsubsec{coherence_length}
The coherence length\index{coherence length} $\xi(t)$ is a fundamental observable in the dynamics of the \gls{SG}s because it rules the off-equilibrium phenomena a multitude of times as we will discuss along this thesis (see~\refch{aging_rate},~\refch{mpemba}, and \refch{out-eq_chaos}). Conceptually, the coherence length\index{coherence length} is a characteristic length-scale of the off-equilibrium systems that measure the linear size of domains\index{magnetic domain} of correlated spins. Although, the name ``coherence length\index{coherence length}''\footnote{This coherence length\index{coherence length} should not be confused with the so-called coherence length\index{coherence length} in optics.} is not universal and some authors denominate it \textit{dynamical correlation length} (see for example \cite{parisi:99c}) or, by abuse of language, simply \textit{correlation length} (see \cite{joh:99}).

The relation of the coherence length\index{coherence length} with the length of correlated spins deserves a real example which will (hopefully) help us to visualize the concept. Consider one three dimensional lattice of linear size $L=160$ in which we have simulated the \gls{EA}\index{Edwards-Anderson!model} model (see~\refsubsec{3D_EA_model}) with a Metropolis-Hastings\index{Metropolis-Hastings} algorithm for $t=2^{36}$ number of Monte\index{Monte Carlo} Carlo steps at temperature $T=0.7$. In the left panels of~\reffig{coherence_length_snapshot} two projections of configurations\index{configuration} at this time over the $xy$ plane are showed. The \textit{up} spins are plotted in yellow and the \textit{down} spins are plotted in blue. We can no see any pattern in those configurations\index{configuration} that appear to be random.

However, if we consider the overlap\index{overlap} between them (right panel of~\reffig{coherence_length_snapshot}) by using the same \textit{color code}, a pattern emerges. Islands of correlated spins appear in the plot. This, of course, is only a visual sketch of the concept but it encodes the idea that the four-point correlation\index{correlation function!four point} function (which is just a correlation\index{correlation function!four point} function of the overlaps\index{overlap}) has encrypted the correlation length. Yet, the quantitative obtaining of such observable have required a long time in both, numerical simulations and experiments.

The problem of finding characteristic lengths in off-equilibrium systems have been widely discussed. The integral estimators have been used since 1982 \cite{cooper:82}. Detailed numerical studies have concluded that a well-behaved estimator of $\xi(t)$, that is very convenient from the numerical point of view (see \cite{janus:09b}), should be computed through the integrals\footnote{The alert reader may notice that in this expression, the correlation\index{correlation function!four point} function $C_4(T,\vec{r},t)$ has changed its vectorial dependence of the distance $\vec{r}$ to a simple scalar dependence $r$. This rotational invariance is assumed and justified numerically with a careful study in \cite{janus:09b}.}
\begin{equation}
I_k(t) = \int_0^{\infty} r^k C_4(T,r,t) \dd r \, . \labeq{integral_estimator_xi}
\end{equation}
Let us identify the correlation\index{correlation function!four point} function $C_4(T,r,t)$ with its long range behavior displayed in \refeq{long_distance_C4}. In that case, taking $x=r/\xi$ would lead to
\begin{equation}
I_k(t) = \int_0^{\infty} \xi^{k-\vartheta} \left( r/\xi \right)^{k-\vartheta} f(r/\xi) \xi \dfrac{\dd r}{\xi} = \xi^{k+1-\vartheta} \int_0^{\infty} x^{k-\vartheta} f(x) \dd x \, .
\end{equation}

Therefore, the knowledge about the concrete form of the function $f(x)$ is no longer needed and one finds an estimator of $\xi$
\begin{equation}
\xi_{k,k+1}(t) = \dfrac{I_{k+1}(t)}{I_k(t)} \propto \xi(t) \, . \labeq{def_xi}
\end{equation}

However, we should not forget the previous assumption made concerning the correlation\index{correlation function!four point} function. The estimation of \refeq{def_xi} involves systematic errors and that result would be only valid in the $r \to \infty$ limit. The larger the value of $k$, the smaller is the deviation from the asymptotic behavior due to the term $r^k$, which rules the short distances. However,  increasing the exponent $k$ moves the peak of $r^k C_4(T,r,k)$ to larger values of $r$ where the relative error of the correlation\index{correlation function!four point} function is larger. A compromise solution of the value $k$ is needed, here, we use $k=1$. This decision is numerically justified in \cite{janus:09b} and used in several works \cite{janus:10b,janus:14b,fernandez:15,manssen:15,janus:17,fernandez:18b,fernandez:19}. Further details of computation in numerical simulations are provided in~\refsec{finite_size_effects}.

\begin{figure}[h]
\centering
\includegraphics[width=0.8\textwidth]{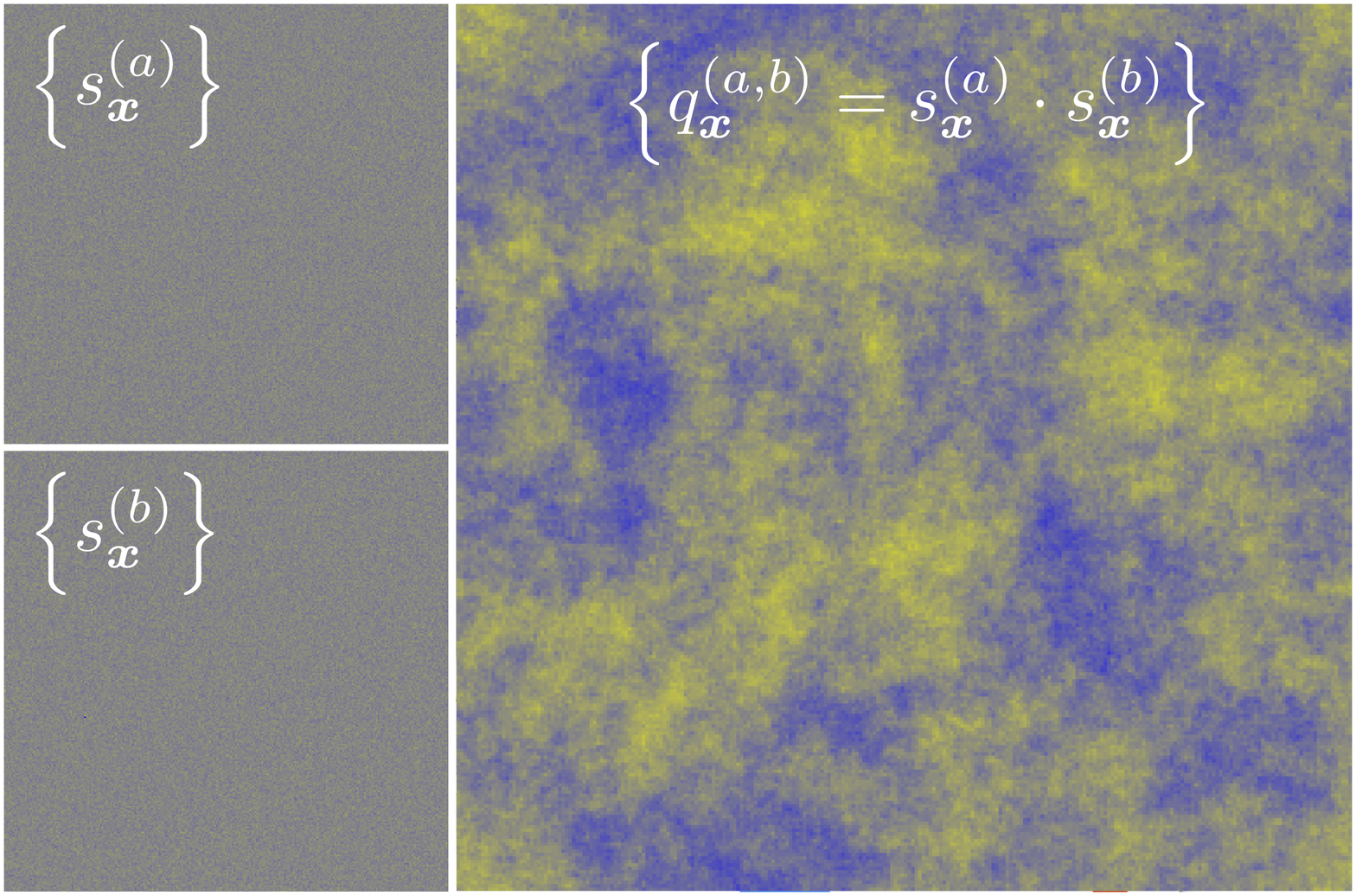}
\caption[\textbf{Spin-glass coherence length.}]{\textbf{Spin-glass coherence length\index{coherence length}.} \textbf{Top left:} A snapshot of a configuration\index{configuration} $\{s_{\boldsymbol{x}}^{(a)}\}$, which has evolved for $t=2^{36}$ Monte\index{Monte Carlo} Carlo steps at $T=0.7\approx 0.64T_\text{c}$.  We show the average magnetization\index{magnetization} on the $xy$ plane, averaging over $z$. \textbf{Bottom left:} Another configuration\index{configuration} $\{s_{\boldsymbol{x}}^{(b)}\}$ of the same sample\index{sample}, prepared in the same way as $\{s_{\boldsymbol{x}}^{(a)}\}$.  No visible ordering is present in either configuration\index{configuration} because the preferred pattern of the magnetic domains\index{magnetic domain} cannot be seen by eye ($s=1$ is plotted in yellow, and $-1$ in blue). \textbf{Right:} If one measures the overlap\index{overlap} between the two configurations\index{configuration}, and with the same color code used for the spins, the preferred pattern of the magnetic domains\index{magnetic domain}, of size $\xi$, becomes visible. Figure from~\cite{janus:19}.}
\labfig{coherence_length_snapshot}
\end{figure}

\addpart{Metastate}
\chapter{Metastate} \labch{metastate}
\setlength\epigraphwidth{.5\textwidth}

\epigraph{\textit{Si paso por Florida te recuerdo\\
Si paso por la Valle me es igual,\\
Que si estoy en Corrientes, que si estoy en Palermo\\
Por todo Buenos Aires conmigo siempre estás.}}{-- Julio Jaramillo, \textit{No me toquen ese vals} }


This chapter is dedicated to discussing the metastate\index{metastate}. We start by introducing the concepts of mixed and pure states in lattice systems, see \refsec{mixed_pure_states_metastate}, which would be needed in order to understand further discussions. Then, we describe the problem of taking the thermodynamic limit\index{thermodynamic limit} in disordered\index{disorder!systems} systems in \refsec{the_problem_metastate} and we introduce the proposed solution in \refsec{the_solution_metastate}. The different theoretical pictures described in \refsubsec{theoretical_pictures} provide different predictions for some observables in the metastate\index{metastate} formalism, we introduce those observables and discuss the different scenarios in \refsec{observables_predictions_metastate}. 

At this point, we present an original contribution to the metastate\index{metastate} problem developed during this thesis~\cite{billoire:17} by explaining the numerical setup \refsec{simulation_parameters_metastate} and by exploring the metastate\index{metastate} from the numerical point of view in \refsec{results_metastate}. Finally, we relate our numerical results to theory in~\refsec{relating_numerical_theory_metastate}.

\section{Mixed and pure states} \labsec{mixed_pure_states_metastate}

Consider a spin system in a lattice $\Lambda \subset \mathbb{Z}^d$ where the vertices correspond to the spins $s_i$ and the edges correspond to the couplings\index{couplings} between spins $J_{ij}$, leading to nearest-neighbor interactions defined through a Hamiltonian\index{Hamiltonian} $\mathcal{H}_{\mathcal{J}} = \sum_{\braket{i,j}} J_{ij} s_i s_j$. As far as we are dealing with general definitions, we are neither restricting the values of the spins, nor the values of the couplings\index{couplings}. 

One \textit{configuration}\index{configuration} of the system is determined by the value of the set of all the spins $\mathcal{S} = \{s_i\}$ as $i$ runs over all the lattice sites. In the same way, a concrete \textit{sample}\index{sample} is determined by the value of the set of all the couplings\index{couplings} $\mathcal{J} = \{J_{ij}\}$ as the pair $(i,j)$ runs over all the pairs of spins. 

The restriction of the lattice to finite size $L$ is simply done by considering a cubic lattice $\Lambda_L$ composed of $L^d$ spins. However, in the case of finite systems, an additional problem arises with the choice of the boundary conditions\index{boundary conditions}. It is possible to define several boundary conditions\index{boundary conditions} but we will consider a very common choice, the \gls{PBC}\index{boundary conditions!periodic}.

At a given temperature $T=1/\beta$, a Gibbs state $\Gamma_{L,\mathcal{J}}$ for a finite system $\Lambda_L$ is a probability distribution over the configurations\index{configuration} $\mathcal{S}$ where each configuration\index{configuration} has a probability to appear equal to 
\begin{equation}
\Gamma_{L,\mathcal{J}}(\mathcal{S}_{\Lambda_L}) = \dfrac{\exp\left(-\beta \mathcal{H}_{L,\mathcal{J}}(\mathcal{S})\right)}{Z_L} \, , \labeq{Gibbs_probability_state}
\end{equation}
being $\mathcal{H}_{L,\mathcal{J}}$ the Hamiltonian\index{Hamiltonian} of the system restricted to the lattice $\Lambda_L \subset \mathbb{Z}^d$ and $Z_L$ the partition function\index{partition function}\footnote{For the sake of simplicity, we tacitly assume that each spin can only take a finite set of possible values and, therefore, the set of configurations\index{configuration} $\mathcal{S}$ is countable and we can perform the sum. In the case of infinite-uncountable possible values for each spin, the trade between the sum and an integral is needed.} $Z_L=\sum_{\mathcal{S}} \exp \left( -\beta \mathcal{H}_{L,\mathcal{J}}\right)$.

When considering an infinite lattice, this definition of state is not useful anymore because the Hamiltonian\index{Hamiltonian} $\mathcal{H}_{\mathcal{J}}$ involves sums of infinite terms that do not converge. Nonetheless, there exist a well-established definition for state in infinite lattice: the \gls{DLR} states \cite{ruelle:04,sinai:14,friedli:17}. A probability distribution of states in an infinite-size lattice is a Gibbs state $\Gamma_{\mathcal{J}}$ if, for any finite subset $\Lambda_W \subset \mathbb{Z}^d$, two conditions are fulfilled:
\begin{enumerate}
\item The Hamiltonian\index{Hamiltonian} for a spin configuration\index{configuration} inside the subset $\Lambda_W$ conditioned to the values of the rest of spins in the infinite lattice $\mathbb{Z}^d \setminus \Lambda_W$, namely $\mathcal{H}_{W,\mathcal{J}} (\mathcal{S}_{\Lambda_W} \lvert \mathcal{S}_{\mathbb{Z}^d \setminus \Lambda_W})$, and the partition function\index{partition function} restricted to that subset $Z_{W}$ are finite for almost every $\mathcal{S}$.\\

In the case of the \gls{EA}\index{Edwards-Anderson!model} model the expression for $\mathcal{H}_{W,\mathcal{J}} (\mathcal{S}_{\Lambda_W} \lvert \mathcal{S}_{\mathbb{Z}^d \setminus \Lambda_W})$ is very easy due to the short-ranged nature of the Hamiltonian\index{Hamiltonian}. We have to take care only with the frontier $\partial \Lambda_W$ where the Hamiltonian\index{Hamiltonian} includes terms with spins out of $\Lambda_W$ and spins inside. Therefore, we have
\begin{equation}
\mathcal{H}^{\mathrm{EA}}_{W,\mathcal{J}} (\mathcal{S}_{\Lambda_W} \lvert \mathcal{S}_{\mathbb{Z}^d \setminus \Lambda_W}) = \sum_{J_{ij} \in \Lambda_W \setminus \partial \Lambda_W} J_{ij} s^{\mathrm{int}}_i s^{\mathrm{int}}_j + \sum_{J_{ij} \in \partial \Lambda_W} J_{ij} s^{\mathrm{int}}_i s^{\mathrm{out}}_j \, , \labeq{hamiltonian_conditioned}
\end{equation}
where the superindex \textit{int} stands for spins belonging to $\Lambda_W$ and the superindex \textit{out} stands for spins outside of $\Lambda_W$. The reader should notice that, if $\Lambda_W$ contains a finite number of spins, the Hamiltonian\index{Hamiltonian} of~\refeq{hamiltonian_conditioned} only involves finite sums.

\item The conditional probability $\Gamma_{W,\mathcal{J}}$ of a configuration\index{configuration} $\mathcal{S}$ in $\Lambda_W$ given the rest of the spins $\mathbb{Z}^d \setminus \Lambda_W$ is absolutely continuous\footnote{The reader may be confuse about the term ``absolutely continuous'' in this context. This continuity is defined over a measure of the random variables, the spins in our particular case. The reader should consult~\cite{sinai:14} for a much more detailed discussion with rigorous proofs.} and it is defined by the expression
\begin{equation}
\Gamma_{W,\mathcal{J}}(\mathcal{S}_{\Lambda_W} \lvert \mathcal{S}_{\mathbb{Z}^d \setminus \Lambda_W})  = \dfrac{\exp \left[-\beta \left( \mathcal{H}_{W,\mathcal{J}}(S_{\Lambda_W}) + \mathcal{H}_{W,\mathcal{J}}(S_{\Lambda_W} \lvert \mathcal{S}_{\mathbb{Z}^d \setminus \Lambda_W}) \right)  \right]}{Z_{W}} \, . \labeq{def_DLR}
\end{equation}
\end{enumerate}

From \refeq{def_DLR} one can conclude that, for any two spin configurations\index{configuration} $\mathcal{S}_{1,\Lambda_W}$ and $\mathcal{S}_{2,\Lambda_W}$ with the rest of the spins fixed $\mathcal{S}_{\mathbb{Z}^d \setminus \Lambda_W}$, the ratio of their conditional probabilities would be, simply
\begin{equation}
\dfrac{\Gamma_{W,\mathcal{J}}(\mathcal{S}_{1,\Lambda_W} \lvert \mathcal{S}_{\mathbb{Z}^d \setminus \Lambda_W})}{\Gamma_{W,\mathcal{J}}(\mathcal{S}_{2,\Lambda_W} \lvert \mathcal{S}_{\mathbb{Z}^d \setminus \Lambda_W})} = \exp \left[ -\beta \left( \mathcal{H}_{W,\mathcal{J}}(\mathcal{S}_{1,\Lambda_W})-\mathcal{H}_{W,\mathcal{J}}(\mathcal{S}_{2,\Lambda_W}) \right) \right] \, . \labeq{ratio_DLR}
\end{equation}

Does this definition hold for our particular case? It is easy to prove that the \gls{EA}\index{Edwards-Anderson!Hamiltonian} Hamiltonian\index{Hamiltonian} defined in \refeq{EA_Hamiltonian} is finite for any finite lattice $\Lambda_W$ and, therefore, the partition function\index{partition function}, which involves a finite number of summands since we are considering Ising\index{Ising} spins, is simply a finite sum of finite terms. The second condition is also fulfilled because $\mathcal{H}_{W,\mathcal{J}}(\mathcal{S}_{1,\Lambda_W})-\mathcal{H}_{W,\mathcal{J}}(\mathcal{S}_{2,\Lambda_W})$ is finite, since the lattice $\Lambda_W$ is finite by definition. The reader may find rigorous proof of the existence of the Gibbs states in the \gls{EA}\index{Edwards-Anderson!model} model in~\cite{ruelle:04}.

Actually, given a Hamiltonian\index{Hamiltonian} $\mathcal{H}_{\mathcal{J}}$ and given a temperature $T=1/\beta$, the previous definition of infinite state allows the existence of many different Gibbs states $\Gamma_\mathcal{J}$. The set of all of the possible Gibbs states is convex and compact, so there exist \textit{extremal} Gibbs states that are not convex combinations of any other Gibbs states. The extremal Gibbs states are called \textit{pure} states while the non-extremal Gibbs states are called \textit{mixed} states and can be decomposed uniquely into a convex combination of pure states in the following way\footnote{We assume, for the sake of simplicity, that the decomposition is discrete.}
\begin{equation}
\braket{\cdots}_{\Gamma_{\mathcal{J}}} = \sum_{\alpha} \omega_{\alpha,\Gamma_{\mathcal{J}}} \braket{\cdots}_{\alpha} \, , \labeq{decomposition_pure_states}
\end{equation}
being $\alpha$ a pure state and $\omega_{\alpha,\Gamma_{\mathcal{J}}}$ an appropriate weight for that pure state that fulfills the condition $\sum_\alpha \omega_{\alpha,\Gamma_{\mathcal{J}}} = 1$, with $\omega_{\alpha,\Gamma_{\mathcal{J}}} \geq 0$ $\forall \alpha$.

Moreover, the overlap\index{overlap} between two pure states labeled as $\alpha$ and $\beta$ can be defined in the lattice $\Lambda_W$ as
\begin{equation}
q_{\alpha \beta} = \dfrac{1}{W^d} \sum_{x \in \Lambda_W} \braket{s_x}_{\alpha}\braket{s_x}_{\beta} \, , \labeq{overlap_pure_states}
\end{equation}
being $s_x$ the spin at the position $x$. The \gls{pdf} of the overlap\index{overlap!distribution} can, therefore, be defined as
\begin{equation}
P_{\Gamma_{\mathcal{J}}}(q) = \sum_{\alpha,\beta}\omega_{\alpha,\Gamma_{\mathcal{J}}} \omega_{\beta,\Gamma_{\mathcal{J}}} \delta(q-q_{\alpha \beta}) \, . \labeq{pq_pure_states}
\end{equation}

\section{The problem: Chaotic Size Dependence} \labsec{the_problem_metastate}

In the previous section, we have properly defined the infinite-size states through the \gls{DLR} states, however, that definition is not physically relevant because experiments are always conducted in large but finite systems. It would be desirable from the physical point of view, to connect the \gls{DLR} states with a sequence of growing systems of linear size $L$ as $L \to \infty$.

This connection can be made for transitional-invariant Hamiltonians\index{Hamiltonian} like the ferromagnet one. However, in systems with quenched disorder\index{disorder!quenched}, such as the \gls{EA}\index{Edwards-Anderson!order parameter} model, that connection is still much of a mystery \cite{aizenman:90}. The problem of taking the $L \to \infty$ limit have been already introduced in \refsubsec{theoretical_pictures}: the \gls{CSD}. 

Let us consider a system of linear size $L$ in a lattice $\Lambda_L$ and fix an internal region of linear size $W$, $\Lambda_W \subset \Lambda_L$. If we take the limit $L \to \infty$, remaining constant the couplings\index{couplings} of the lattice $\Lambda_W$, the state $\Gamma_{W,\mathcal{J}}$ changes chaotically as $L$ grows and also the observables measured in $\Lambda_W$. 

This extreme sensitivity of the system to the addition of couplings\index{couplings} at the boundaries as long as $L$ tends to infinity is called \gls{CSD}. This problem remained apparently oblivious to the \gls{SG} literature and was pointed out for the first time by \cite{newman:92}.

\section{The solution: the Metastate} \labsec{the_solution_metastate}
The solution to that problem is the concept of Metastate\index{metastate!Aizenman-Wehr} which was introduced by Aizenman and Wehr in~\cite{aizenman:90} when studying first-order transitions\index{phase transition!first order} in general disordered\index{disorder!systems} systems. Two years later, Newman and Stein introduced this concept of metastate\index{metastate!Newman-Stein} to solve the \gls{CSD} problem in the particular case of \gls{SG}s~\cite{newman:92}. The concept of the metastate\index{metastate} \cite{aizenman:90,newman:92,newman:96,newman:98,newman:03} is just a generalization of the concept of Gibbs state that we have exposed in \refsec{mixed_pure_states_metastate}. A Gibbs state in a finite system can be regarded as a probability distribution of configurations\index{configuration} $\mathcal{S}$, each one with an associated probability given by \refeq{Gibbs_probability_state}. In the same way, the metastate\index{metastate}, which we denote as $\kappa_{\mathcal{J}}(\Gamma_\mathcal{J})$, is a probability distribution over the states $\Gamma_{\mathcal{J}}$. 

The reader may notice that the description of the metastate\index{metastate} concept considers infinite-size states $\Gamma_{\mathcal{J}}$, this poses the problem of the construction of that metastate\index{metastate} from finite-size systems. There exist two definition of metastates\index{metastate} in the literature that propose the solution to that problem: the \gls{AW} metastate\index{metastate!Aizenman-Wehr} \cite{aizenman:90} and the \gls{NS} metastate\index{metastate!Newman-Stein} \cite{newman:92}. Although, it has been argued that both proposals should be equivalent (see \cite{read:14} for a detailed discussion).

\subsection{Newman-Stein Metastate}
We first introduce the \gls{NS} metastate\index{metastate!Newman-Stein}, so-named in \cite{newman:96}. Consider a growing sequence of $n$ systems of size $L_0<L_1<L_2< \dots L_n$. For the smallest system of size $L_0$, we fix the couplings\index{couplings} $\mathcal{J}$ randomly in the lattice $\Lambda_{L_0}$ and we define a hypercube of linear size $W < L_0$, usually called \textit{window}, in the lattice $\Lambda_W \subset \Lambda_L$. 

In this system, we are able to define the Gibbs state $\Gamma_{L_0,\mathcal{J}}$ and to restrict this state to the window $W$, obtaining a probability distribution on the spins in $\Lambda_W$ consisting in $2^{W^d}$ real numbers. This probability distribution is a Gibbs state $\Gamma_{W,\mathcal{J}}$. If we consider now the space of finite states $\Gamma_{W,\mathcal{J}}$, each time that a specific state $\Gamma_{W,\mathcal{J}}$ appears, we can ``record'' it as a Dirac delta $\delta$ of integral $1/n$. Repeating this procedure for the $n$ systems would lead to an empirical probability distribution over the space of finite states $\Gamma_{W,\mathcal{J}}$.

Then, the limit $n \to \infty$ is taken and that empirical distribution may tend to a limit. If this limit exists for every window $W$ as $W \to \infty$, then, the resulting distribution of states $\Gamma_\mathcal{J}$ is the so-called \gls{NS} metastate\index{metastate!Newman-Stein}.

\subsection{Aizenman-Wehr Metastate} \labsubsec{aw_metastate}
We introduce now the \gls{AW} metastate\index{metastate!Aizenman-Wehr} that was indeed introduced earlier \cite{aizenman:90} in the context of first-order transitions\index{phase transition!first order} but was later related with this concept of metastate\index{metastate} as a limiting object for quenched-disordered\index{disorder!quenched} systems \cite{newman:92}.

The definition of the \gls{AW} metastate\index{metastate!Aizenman-Wehr} requires the introduction of three length scales $W \ll R \ll L$ which will be the linear size of three lattices $\Lambda_W \subset \Lambda_R \subset \Lambda_L$ as sketched in \reffig{sketch_AW_metastate}.

\captionsetup{justification=centering}
\begin{figure}
\includegraphics[width=0.6\textwidth]{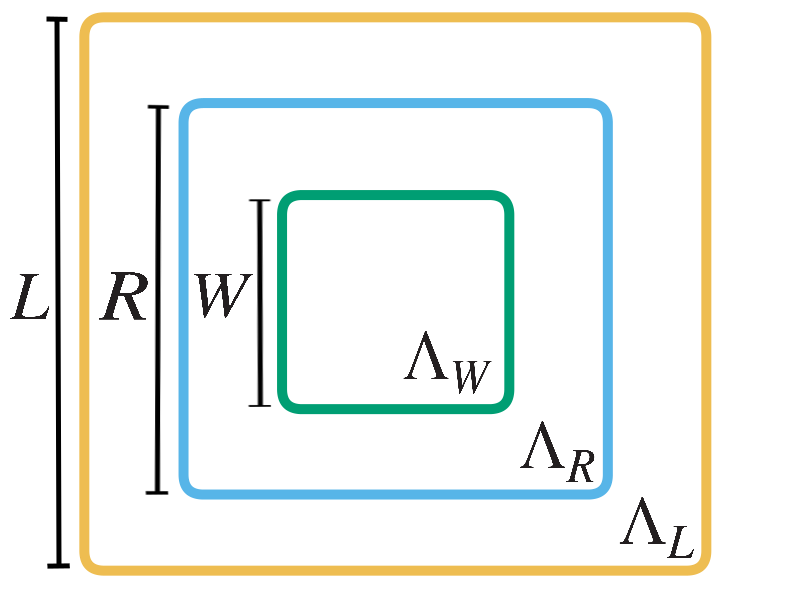}
\caption[\textbf{Sketch of the Aizenman-Wehr metastate construction.}]{\textbf{Sketch of the Aizenman-Wehr metastate\index{metastate!Aizenman-Wehr} construction.}}
\labfig{sketch_AW_metastate}
\end{figure}
\captionsetup{justification=raggedright}

Consider the lattice $\Lambda_L$ (with a given boundary conditions\index{boundary conditions}, usually \gls{PBC}\index{boundary conditions!periodic}). This lattice $\Lambda_L$ is divided into an inner region $\Lambda_R$ and an outer region $\Lambda_L \setminus \Lambda_R$. Now, consider the state $\Gamma_{L,\mathcal{J}}$ for a set of couplings\index{couplings} $\mathcal{J}$ in the lattice $\Lambda_L$. 

The first step in the construction of the \gls{AW} metastate\index{metastate!Aizenman-Wehr} is to take an average over the states $\Gamma_{L,\mathcal{J}}$. We keep constant all the couplings\index{couplings} that have both extremes inside $\Lambda_R$ while changing the rest of them, obtaining a set of states $\{\Gamma_{L,\mathcal{J}_1},\Gamma_{L,\mathcal{J}_2},\dots\}$. Then, we take the average of this set of states.

Finally, we take the limit of the three-length scales to infinity while respecting the relation $W \ll R \ll L$. The obtained distribution of states should not depend on the arbitrary choice of the fixed internal couplings\index{couplings} and will, hopefully, have a smooth limit: the metastate\index{metastate}.

The lattice $\Lambda_W$ plays a fundamental role in the \gls{AW} metastate\index{metastate!Aizenman-Wehr} as a \textit{measuring window}, we are restricting our attention to that hypercube when measuring the relevant quantities in order to avoid boundary effects that may appear as long as $R$ is finite.

\section{Theoretical pictures in the metastate formalism} \labsec{pictures_metastate}
As we have introduced in~\refsubsec{theoretical_pictures}, the mathematical concept of metastate\index{metastate} irrupted in the physics debate of the nature of a \gls{SG} at finite dimensions. The metastate\index{metastate} came to the \gls{SG} field to solve some mathematical ambiguities in the early formalism of \gls{RSB}\index{replica!symmetry breaking (RSB)}\index{replica!symmetry breaking (RSB)}.

The introduction of the concept of metastate\index{metastate}, although, brought also a deep debate about the validity of the \gls{RSB}\index{replica!symmetry breaking (RSB)}\index{replica!symmetry breaking (RSB)} metastate\index{metastate!RSB} picture. According to \cite{newman:03}, the existence of states composed of a mixture of infinitely many pure states would not be allowed. The Newman-Stein statement was not fully accepted in mathematical physics and it was debated in subsequent works. For example, see the work of Talagrand~\cite{talagrand:06}, who demonstrated that Parisi's formula reproduced exactly the free-energy\index{free energy} for the \gls{SK}\index{Sherrington-Kirkpatrick} model, and the work of Read, who questioned the arguments of Newman and Stein to rule out the \gls{RSB}\index{replica!symmetry breaking (RSB)}\index{replica!symmetry breaking (RSB)} picture (part IV of \cite{read:14}).

This debate leads to the proposal of an alternative picture: the \textit{chaotic pairs}\footnote{The term ``chaotic pair'' is intimately related with the droplet\index{droplet!picture} picture, but, as quoted by Newman and Stein \cite{newman:03} \textit{The term ‘chaotic pairs’ was chosen in reference to spin-symmetric boundary conditions\index{boundary conditions}, such as periodic. If one considers a fixed boundary condition metastate\index{metastate}, then it would be more appropriate to refer to this picture as ‘chaotic pure states’ because the Gibbs state in a typical large volume $\Lambda_L$ with fixed boundary conditions\index{boundary conditions} will be (approximately) a single pure state that varies chaotically with L.}}.

Here, we (very) briefly introduce both pictures that together with the classical Droplet\index{droplet!picture} picture stands as the mathematical consistent formalism to describe the \gls{SG} phase\index{phase!low-temperature/spin-glass}.

\subsection{The RSB metastate}
 
The adequacy of the \gls{RSB}\index{replica!symmetry breaking (RSB)}\index{replica!symmetry breaking (RSB)} formalism to the metastate\index{metastate!RSB} concept lead to the so-baptized as \textit{nonstandard \gls{RSB}\index{replica!symmetry breaking (RSB)}\index{replica!symmetry breaking (RSB)} picture} \cite{newman:03}. However, as quoted by \cite{read:14} we find this name unfortunate because this picture should be the standard in the \gls{RSB}\index{replica!symmetry breaking (RSB)}\index{replica!symmetry breaking (RSB)}. Thus, we adopt the terminology proposed by Read and we call this picture the \textit{\gls{RSB}\index{replica!symmetry breaking (RSB)}\index{replica!symmetry breaking (RSB)} metastate}\index{metastate!RSB}. 

In this picture, each of the states that form the metastate\index{metastate!RSB} is \gls{RSB}\index{replica!symmetry breaking (RSB)}\index{replica!symmetry breaking (RSB)}-like, that is, each state is composed of a hierarchical structure of pure states and presents all the properties exposed in~\refsubsec{theoretical_pictures}. Moreover, the metastate\index{metastate!dispersed} is \textit{disperse} i.e. the metastate\index{metastate!dispersed} is composed of infinitely many states. This dispersion has, indeed, deep consequences. For instance, the disperse metastate\index{metastate!dispersed} is responsible for the lack of the self-averaging\index{self-averaging!non-} property in certain quantities for the \gls{SG}s in the low-temperature phase\index{phase!low-temperature/spin-glass} (see, for example \cite{binder:86}, for the concept of non-self-average\index{self-averaging}). The reader may notice that, despite the deep meaning of these results, the study of a finite fixed-size system would not change the predictions of \gls{RSB}\index{replica!symmetry breaking (RSB)}\index{replica!symmetry breaking (RSB)} because, for that system, a single state is present and we have already stated that, in the \gls{RSB}\index{replica!symmetry breaking (RSB)}\index{replica!symmetry breaking (RSB)} metastate\index{metastate!RSB} picture, each state is \gls{RSB}\index{replica!symmetry breaking (RSB)}\index{replica!symmetry breaking (RSB)}-like.

\subsection{Chaotic pairs picture}
Chaotic pairs can be regarded as a generalization of the Droplet\index{droplet!picture} picture in the metastate\index{metastate!chaotic pairs} formalism, therefore, each of the states composing the metastate\index{metastate!chaotic pairs} would be trivial i.e. each state will be composed of a single pair of spin-flip related pure states. However, in this picture, the pair of pure states will change chaotically with $L$ due to the chaotic size dependence.

\section{Observables and predictions} \labsec{observables_predictions_metastate}
The formalism of the metastate\index{metastate} allows the computation of several quantities that would behave differently in the different theoretical pictures described above. Should we obtain these quantities with enough accuracy, it would be possible to discriminate (at least partially) between those competing pictures. This section is devoted to introducing the observables that we have measured in our numerical simulations and to recalling the predictions made for those observables.

\subsection{The observables} \labsubsec{observables_metastate}
The overlap\index{overlap} in the measure window $\Lambda_W$ is defined taking advantage of the metastate\index{metastate} setup. Consider one realization of the internal couplings\index{couplings} $i$ and a pair of external realizations of the outer couplings\index{couplings} $o$ and $o'$. One possible definition of the overlap\index{overlap} (that will be useful later) is
\begin{equation}
q^{i;o,o'} = \dfrac{1}{W^d} \sum_{x \in \Lambda_W} s_x^{\sigma;i,o}s_x^{\tau;i,o'}  = \dfrac{1}{W^d} \sum_{x \in \Lambda_W} q^{i,o,o'}_x \, , \labeq{metastate overlap}
\end{equation}
where $\sigma$ and $\tau$ denotes replicas\index{replica} of the system as in \refeq{def_overlap}.

At this point, we should introduce a new type of average that will be useful in the study of the metastate\index{metastate}. We have introduced the notion of metastate\index{metastate} $\kappa_{\mathcal{J}}(\Gamma_\mathcal{J})$, which is a distribution over states, and therefore, we have a new kind of average. In addition to the usual thermal average\footnote{In this chapter, we add the subscript $\Gamma_{\mathcal{J}}$ in order to avoid confusion with other averages.} $\braket{\cdots}_{\Gamma_{\mathcal{J}}}$ we can now average over the metastate\index{metastate!averaged state} $\left[ \cdots \right]_{\kappa}$. 

Those averages can be combined in several ways but one particular combination is of great interest here. The \gls{MAS} $\rho$ is defined via the average $\braket{\cdots}_{\rho} = \left[ \braket{\cdots}_{\Gamma_{\mathcal{J}}} \right]_{\kappa}$. 

The \gls{MAS} is a mixed Gibbs state that can be decomposed in terms of pure states\footnote{As stated in \cite{read:14}, the organization of the pure states in the \gls{MAS} is not ultrametric, on the contrary to each one of the states that compounds the metastate\index{metastate} according to the \gls{RSB}\index{replica!symmetry breaking (RSB)}\index{replica!symmetry breaking (RSB)} picture.}, as explained in \refsec {mixed_pure_states_metastate}.

The main objects in our numerical study will be the \gls{pdf} of the overlaps\index{overlap!distribution}
\begin{equation}
P(q) = \overline{\left[\braket{\delta(q-q^{i;o,o})}_{\Gamma_\mathcal{J}}\right]_{\kappa}}\, , \labeq{usual_pq}
\end{equation}
where we remark that the overlap\index{overlap} is defined with the same outer part $o$ for both replicas\index{replica} and $\overline{\left(\cdots\right)}$ denotes the disorder average\index{disorder!average}, as usual. This probability density function corresponds to the usual $P(q)$ in \gls{SG}s because we are forcing the outer couplings\index{couplings} to be the same, but we can also define a \index{metastate!averaged state}metastate-averaged $P(q)$ as
\begin{equation}
P_{\rho}(q) = \overline{\left[ \braket{\delta(q-q^{i;o,o'})}_{\Gamma_\mathcal{J}} \right]_{\kappa}} \, . \labeq{metastate_pq}
\end{equation}

The \gls{MAS} spin correlation\index{correlation function!four point} function is
\begin{equation}
C_{\rho}(x) = \overline{\left[ \braket{s^{\sigma;i,o}_0 s^{\sigma;i,o}_x}_{\Gamma_\mathcal{J}}\right]^2_{\kappa}} = \overline{\left[ \braket{q^{i,o,o'}_0 q^{i,o,o'}_x}_{\Gamma_\mathcal{J}}\right]_{\kappa}} \, .
\end{equation}
This correlation\index{correlation function!four point} function has been studied in other contexts \cite{dedominicis:98} (see also \cite{dedominicis:06,marinari:00}) and it has been found that $C_{\rho}(x) \sim \lvert x \rvert^{-(d-\zeta)}$ for large $\lvert x \rvert$. The exponent $\zeta$ has a central importance in the Read's discussion of \cite{read:14} as we discuss below.

Finally, from this correlation\index{correlation function!four point} function, we can define
\begin{equation}
\chi_{\rho} = \sum_{x \in \Lambda_W} C_{\rho} = \sum_{x \in \Lambda_W} \overline{\left[ \braket{s^{\sigma;i,o}_0 s^{\sigma;i,o}_x}_{\Gamma_\mathcal{J}}\right]^2_{\kappa}} =  \sum_{x \in \Lambda_W} \overline{\left[ \braket{q^{i,o,o'}_0 q^{i,o,o'}_x}_{\Gamma_\mathcal{J}}\right]_{\kappa}}\, .
\end{equation}
We can make some computations with this quantity, by decomposing it into pure states by using \refeq{decomposition_pure_states}.
\begin{equation}
\chi_{\rho} = \sum_{x \in \Lambda_W} \overline{\left[ \sum_{\alpha} \omega_{\alpha,\Gamma_{\mathcal{J}}} \braket{s^{\sigma;i,o}_0 s^{\sigma;i,o}_x}_{\alpha}\right]^2_{\kappa}}  \, . \labeq{susc_intermediate}
\end{equation}
It is easy to prove that, because the pure states exhibit the clustering property $\braket{s_0 s_x}_\alpha \sim \braket{s_0}_{\alpha}\braket{s_x}_{\alpha}$ as $\lvert x \rvert \to \infty$ and because as long as $W \to \infty$ the dominant terms in the sum are those with large $x$, the expression \refeq{susc_intermediate} can be written as:
\begin{equation}
\chi_{\rho} \approx \sum_{x \in \Lambda_W} \overline{\left[ \sum_{\alpha,\beta} \omega_{\alpha,\Gamma_{\mathcal{J}}} \omega_{\beta,\Gamma_{\mathcal{J}}} \braket{s^{\sigma;i,o}_0}_{\alpha} \braket{s^{\tau;i,o'}_0}_{\beta}  \braket{s^{\sigma;i,o}_x}_{\alpha} \braket{s^{\tau;i,o'}_x}_{\beta}\right]_{\kappa}} \, .
\end{equation}
Now, using the relation $\int_{-\infty}^{\infty} \delta(x-a) dx = 1$ $\forall a \in \mathbb{R}$ and the definitions of \refeq{overlap_pure_states} and \refeq{pq_pure_states} we can rewrite the previous expression 
\begin{equation}
\chi_{\rho} \approx W \int_{-\infty}^\infty q^2 P_{\rho}(q) \dd q \, .
\end{equation}

The quantity $\chi_{\rho}$ turns out to be the \gls{MAS} susceptibility\index{susceptibility}, and hence, it is related with the variance of the distribution function of the metastate\index{metastate}. Recalling the long term behavior of the correlation\index{correlation function!four point} function, the exponent $\zeta$ emerges again
\begin{equation}
\chi_{\rho} \sim W^\zeta \, . \labeq{susceptibility_scaling}
\end{equation}

\subsection{Theoretical predictions of the metastate} \labsubsec{theoretical_predictions_metastate}
As introduced in \refsubsec{theoretical_pictures} and in \refsec{pictures_metastate} we have three mathematical consistent pictures for the \gls{SG} phase\index{phase!low-temperature/spin-glass}. First, the droplet\index{droplet!picture} model whose metastate\index{metastate} is concentrated on a single trivial state\footnote{Recall that we name \textit{trivial state} to a mixture of two pure states related by global spin-flip symmetry.}. Second, the chaotic pictures \cite{newman:92,newman:96,newman:98,newman:03}, predicting a metastate\index{metastate!chaotic pairs} composed of infinitely many states i.e. disperse metastate\index{metastate!dispersed}, yet each state is trivial. Finally, the \gls{RSB}\index{replica!symmetry breaking (RSB)}\index{replica!symmetry breaking (RSB)} metastate\index{metastate!RSB} \cite{read:14}, which is disperse and every state drawn from it contains the Parisi hierarchical tree of pure states. There might exist alternatives to these three pictures but are much limited by recent rigorous results \cite{arguin:15}.

The Read's proposal \cite{read:14} to partially discriminate between them is to compute the exponent $\zeta$ because the logarithm of the number of pure states that one can discriminate in a window $\Lambda_W$ of linear size $W$ scales as $\sim W^{d-\zeta}$. Therefore, $\zeta<d$ implies an infinitely large number of states as $W \to \infty$, which in turn implies a disperse metastate\index{metastate!dispersed}.

\section{Numerical construction of the metastate} \labsec{simulation_parameters_metastate}
We simulate the \gls{EA}\index{Edwards-Anderson!order parameter} model introduced in \refsubsec{3D_EA_model} for the sizes $L=\{8,12,16,24\}$ using Monte\index{Monte Carlo} Carlo simulations consisting in Metropolis single spin-flip updates and \gls{PT} exchanges. Due to the \gls{PT} method, we have a temperature mesh of $n$ different values, however, in this work, we use only the lowest one $T=0.698 \approx 0.64\Tc$, well below the critical\index{critical temperature} temperature $\Tc = 1.102(3)$~\cite{janus:13}.

The thermalization\index{thermalization} protocol is explained with great detail in \refsec{time_scales_eq_chaos} and \refsec{variational_method_eq_chaos}, and for the largest systems, it required the use of multispin coding (see, \refch{AP_multispin_coding}). Moreover, deeper explanation of the simulation can be found in~\refsec{numerical_simulations_eq_chaos}.

Our computation is performed for $N_i = 10$ different realizations of the internal couplings\index{couplings} in the lattice $\Lambda_R$ (see \reffig{sketch_AW_metastate}) and, for each internal realization, we simulate a total of $N_o=128$ different outer-coupling\index{couplings} realizations. Therefore, we have a total of $N_\mathcal{J}=1280$ different samples\index{sample} and for each one, we simulate $\NRep=4$ different replicas\index{replica}.

We take $N_i \ll N_o$ because we expect all the internal disorder\index{disorder} to be ``typical''~\cite{read:14} when computing metastate\index{metastate!averaged state} averages at $R\gg 1$. However, we find sizable sample-to-sample fluctuations\index{sample-to-sample fluctuations} for the system sizes we consider. 

Numerically, the averages used in the computation of observables (see \refsubsec{observables_metastate}) are approached in the following way:
\begin{itemize}
\item Thermal averages over the Gibbs state, that we denote as $\braket{\cdots}_{\Gamma_{\mathcal{J}}}$ are estimated via Monte\index{Monte Carlo} Carlo thermal averages $\braket{\cdots}_{\text{MC}}$ in a fixed sample\index{sample}.
\item The metastate\index{metastate!averaged state} average $\left[ \cdots \right]_{\kappa}$ is estimated by averaging over the outer disorder\index{disorder!average} $\sum_o \sum_{o'} (\cdots)/N^2_o$.
\item The disorder\index{disorder!average} average $\overline{\left(\cdots\right)}$ is estimated by averaging over the internal disorder\index{disorder!average} $\sum_i (\cdots)/N_i$.
\end{itemize}
With these estimations, the observables would be
\begin{equation}
P(q) = \dfrac{\sum_i \sum_o \braket{\delta(q-q_{i;o,o})}_{\text{MC}}}{N_i N_o} \, ,
\end{equation}
\begin{equation}
P_{i,\rho}(q) = \dfrac{\sum_{o,o'} \braket{\delta(q-q_{i;o,o'})}_{\text{MC}}}{N^2_o} \quad , \quad P_{\rho}(q) = \dfrac{\sum_i P_{i,\rho}(q)}{N_i} \, ,
\end{equation}
\begin{equation}
C_{\rho}(x) = \dfrac{1}{N_i}\sum_i \dfrac{1}{N_o^2}\sum_{o,o'} \braket{s_0^{\sigma;i,o}s_x^{\sigma;i,o}s_0^{\tau;i,o'}s_x^{\tau;i,o'}}_{\text{MC}} \, .
\end{equation}
Finally, due to the number of replicas\index{replica} $\NRep=4$, it is worthy to note that, for each set of $\{i,o,o'\}$ we have $\NRep(\NRep-1)/2=6$ different estimations of the observables if $o=o'$ and $\NRep^2=16$ estimations otherwise.

\section{Exploring the numerical metastate} \labsec{results_metastate}
In this section, we will focus on the main results obtained from our work in the construction of the numerical metastate\index{metastate}. First of all, the exponent $\zeta$ is computed numerically in the metastate\index{metastate} setup. Moreover, one of the requisites for the \gls{AW} metastate\index{metastate!Aizenman-Wehr} is the relation of the three-length scales $W \ll R \ll L$. We investigate which are the ratios $W/R$ and $R/L$ that allow safe computations (given an accuracy level). 

Finally, we explore the scaling of the functions $P_{\rho}(q)$ and $P(q)$ as $W$ grows.

\begin{figure}
\centering
\includegraphics[width=0.85\textwidth]{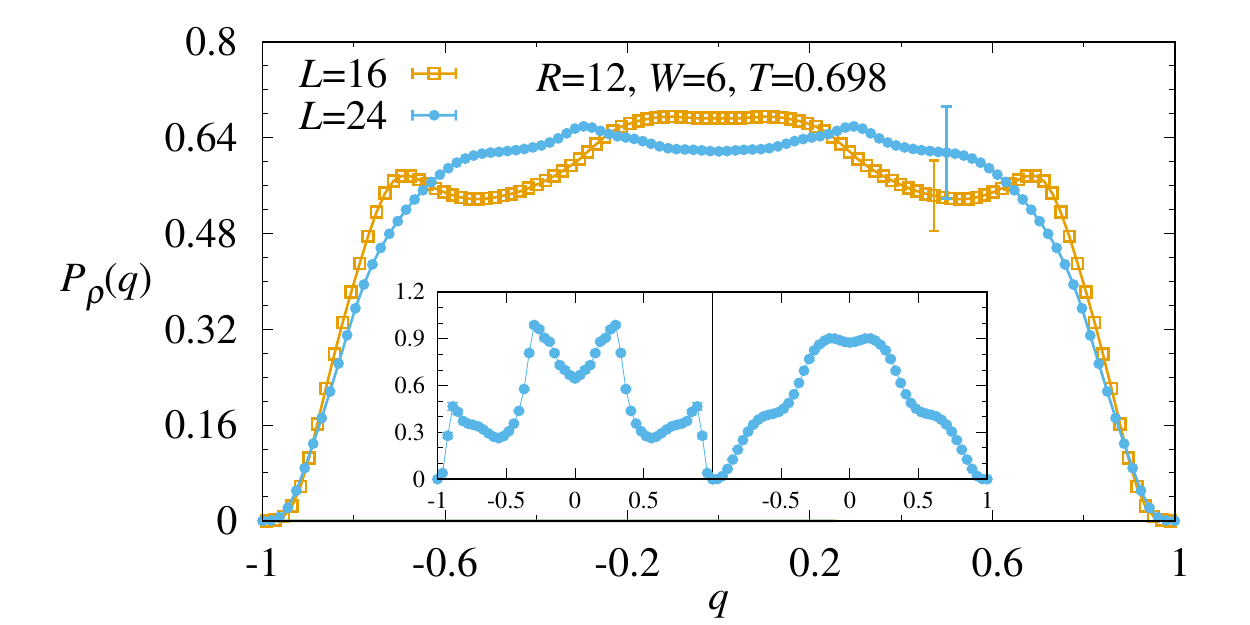}
\caption[\textbf{\boldmath $\mathbf{R/L}$ ratio in the metastate.}]{\textbf{\boldmath $\mathbf{R/L}$ ratio in the metastate\index{metastate}.} Main plot: The \gls{MAS} overlap\index{overlap!distribution} distribution $P_{\rho}(q)$ with $R=12$ at $T=0.698 \approx 0.64\Tc$ shows no statistically significant dependence on the lattice size $L$ (the error bars\index{error bars}, computed from fluctuations on the inner disorder\index{disorder}, are shown for only one value of $q$ for the sake of clarity). \textbf{Insets:} $P_{i,\rho}(q)$ for two specific configurations\index{configuration} of the inner disorder\index{disorder} (the error bars\index{error bars}, computed from fluctuations on the outer disorder\index{disorder}, are smaller than the data points).}
\labfig{RL_ratio}
\end{figure}

\subsection[The $R/L$ ratio]{The \boldmath $R/L$ ratio}
In the main plot of~\reffig{RL_ratio} we see that the \gls{MAS} $P_{\rho}(q)$ measured with $R=12$ and both $L=24$ and $L=16$ are statistically compatible, suggesting that the $R/L = 3/4$ is already a safe choice. As we have mentioned before, the error bars\index{error bars} are quite large due to an unexpected high dependence of $P_{i,\rho}(q)$ on the internal disorder\index{disorder} for the values of $W$ and $R$ that we simulate. The reader can check in the inset of~\reffig{RL_ratio} two examples of $P_{i,\rho}(q)$, where the error bars\index{error bars} are smaller than the data point and can appreciate the sample-to-sample\index{sample-to-sample fluctuations} fluctuation. 

An additional check can be performed by representing the susceptibility\index{susceptibility} $\chi_{\rho}$ against the size of the measuring window $W$ for different values of $L$ by keeping constant $R$, see~\reffig{RL_WR}. We see that, for $R/L \leq 0.75$ the data are statistically compatible, while data with $R/L = 6/7$ show significant deviations.

In view of the previous results, we establish the ratio $R/L=0.5$ as a safe choice for the following analysis.

\begin{figure}
\centering
\includegraphics[width=0.85\textwidth]{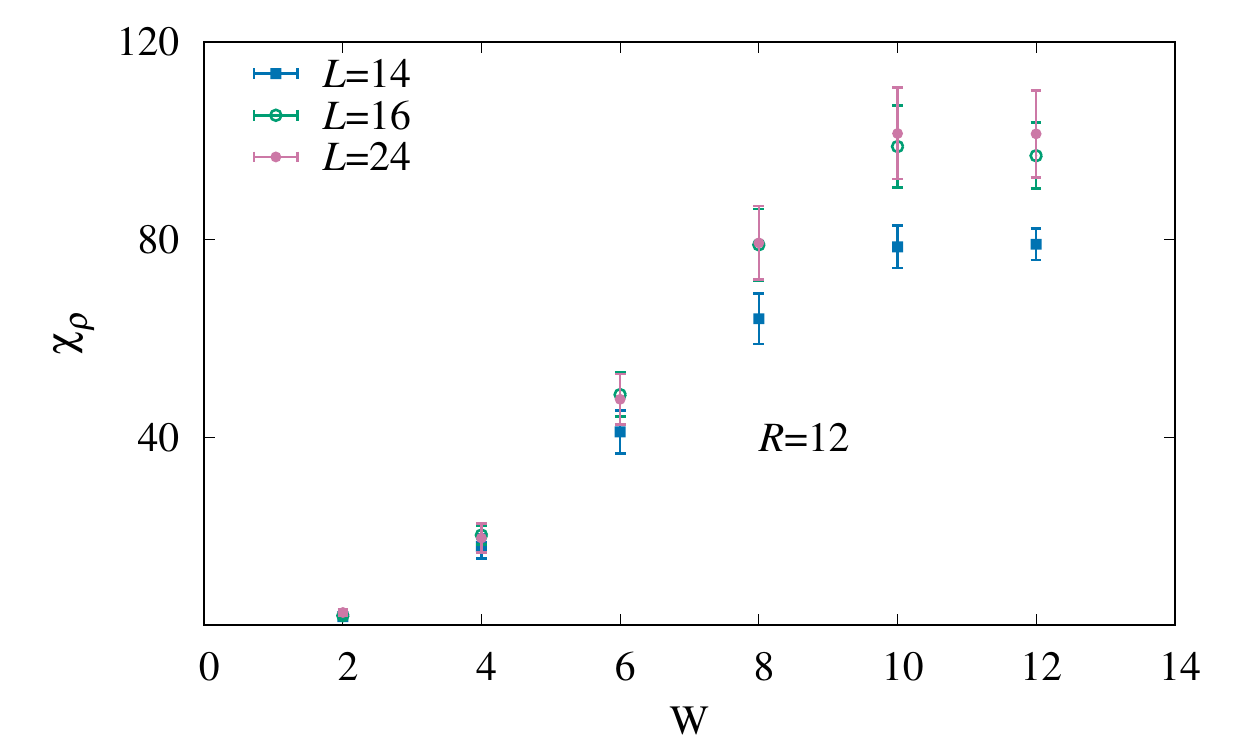}
\caption[\textbf{Susceptibility $\mathbf{\chi_{\rho}}$ for different length scales $\mathbf{\{W,R,L\}}$.}]{\textbf{Susceptibility\index{susceptibility} $\mathbf{\chi_{\rho}}$ for different length scales $\mathbf{\{W,R,L\}}$.} The susceptibility\index{susceptibility} $\chi_{\rho}$ is plotted against $W$ for a fixed $R=12$ in three different curves $L=14,16,24$. Deviation from the asymptotic behavior $R/L \ll 1$ are evident only for $R/L>0.75$.} 
\labfig{RL_WR}
\end{figure}

\subsection[The $W/R$ ratio]{The \boldmath $W/R$ ratio}
The susceptibility\index{susceptibility} scaling with $W$ in the $W/R \ll 1$ limit has been already expressed in~\refeq{susceptibility_scaling}. Therefore, the~\reffig{susceptibility_scaling} gives us relevant information about the validity range for $W/R$. We note that the expected power-law behavior in the $W\ll R$ limit actually extends up to $W/R \approx 0.75$, where corrections to the asymptotic power-law appear. Therefore, similarly to the $R/L$ case, the $W/R \approx 0.75$ stands as a safe choice.

\subsection{The exponent \boldmath $\zeta$}
In this section, we fix $R=L/2$ (which is in the safe side, given our bound $R< 3L/4$) and $W/R \approx 0.75$. We can now compute the exponent $\zeta$ by taking advantage of the relation~\refeq{susceptibility_scaling}. Fitting the data with $W/R \leq 0.75$ we found $\zeta = 2.3 \pm 0.3$. Moreover, we are concerned about the finite-size effects\index{finite-size effects} that our small-lattices simulations could suffer.

A finite-size scaling~\cite{cardy:12}\index{finite size scaling} is needed to study the size effects\index{finite-size effects}. Indeed, we expect for finite $R$ and $W$ a scaling behavior
\begin{equation}
\chi_\rho(W,R) = R^\zeta f(W/R)\, , \labeq{R-finite}
\end{equation}
which is compatible to~\refeq{susceptibility_scaling} for $f(x) \propto x^\zeta$ in the $x\to 0$ limit. \refeq{R-finite} is expected to be exact only in the limit of large $W$ and $R$~\cite{cardy:12}, hence one needs to check for size corrections. We do so with the quotients method\index{quotients method}~\cite{nightingale:76,ballesteros:96,amit:05} which produces effective $\zeta$ estimates at a well defined length scale. The size dependence can be assessed later on. Specifically, we take two sizes-pairs $(W_1,R_1)$, $(W_2,R_2)$ with the same value of $W/R$, which ensures the cancellation of scaling functions in the quotient
\begin{equation}
\frac{\chi_\rho(W_2=x R_2,R_2)}{\chi_\rho(W_1=x R_1,R_1)} = \frac{\left(W_2/x\right)^\zeta f(x) }{\left(W_1/x\right)^\zeta f(x)}=\left(\frac{W_2}{W_1}\right)^\zeta\, . \labeq{quotients}
\end{equation}

The resulting determination of $\zeta$, see \reftab{zeta_exponent_metastate}, is fully compatible with the computed result $\zeta=2.3 \pm 0.3$. Furthermore, no significant size-dependence emerges
from \reftab{zeta_exponent_metastate}. 

\begin{table}
\centering
\begin{tabular}{cclc}
 \toprule 
 \toprule
$W/R$ & $L/R$ & $(W_1,W_2)$ & $\zeta^\mathrm{eff}$\\\hline\hline
1/2 & 2 & (4,6) & 2.18(40)\\\hline
2/3 & 2 & (4,8) & 2.59(22)\\\hline
1   & 2& (8,12) & 2.37(26)\\
    &  &(6,8)  & 2.14(37)\\
    &  &(6,12) & 2.28(18)\\
\bottomrule
\end{tabular}
\caption[\textbf{Effective $\mathbf{\zeta}$ exponent.}]{\textbf{Effective $\mathbf{\zeta}$ exponent.} The effective $\zeta$ exponent depends on the two lengths $W_1$ and $W_2$ and
on the ratio $W_1/R_1=W_2/R_2$. }
\labtab{zeta_exponent_metastate}
\end{table}

Besides, in~\reffig{susceptibility_scaling}, the \gls{MAS} susceptibility\index{susceptibility} has been rescaled by using the previously defined scaling relations. A power-law behavior is exhibited for $W/R <0.75$ as expected and our $\zeta$ estimation interpolates the data nicely in that region.

\begin{figure}[h!]
\includegraphics[width=1.0\columnwidth]{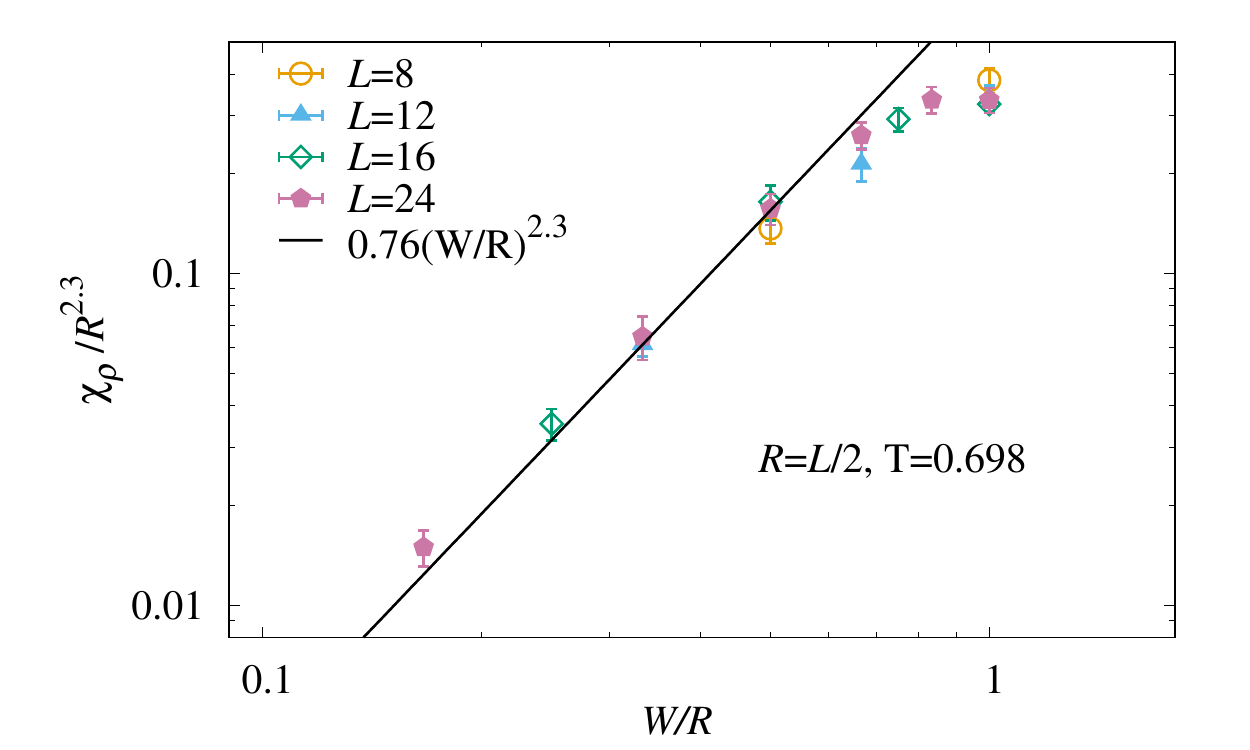}
\caption[\textbf{Scaling behavior of the \gls{MAS} susceptibility.}]{\textbf{Scaling behavior of the \gls{MAS} susceptibility\index{susceptibility}.} \gls{MAS} susceptibility\index{susceptibility} data measured with fixed $R/L=1/2$ at $T=0.698 \approx 0.64 \Tc$ as a function of the $W/R$ ratio.}
\labfig{susceptibility_scaling}
\end{figure}

\subsection{Size dependence of \boldmath $P(q)$ and \boldmath $P_{\rho}(q)$}
We show in~\reffig{pq_metastate}, for $L=24$ and $R/L=1/2$ the dependence of the functions $P_{\rho}(q)$ and $P(q)$ on $W$. The expectation for a dispersed metastate\index{metastate!dispersed}~\cite{read:14} is that both distributions are different in the thermodynamic limit\index{thermodynamic limit}. We found here that they are distinct objects even for moderate sizes of $W$.

\begin{figure}[h!]
\includegraphics[width=1.0\columnwidth]{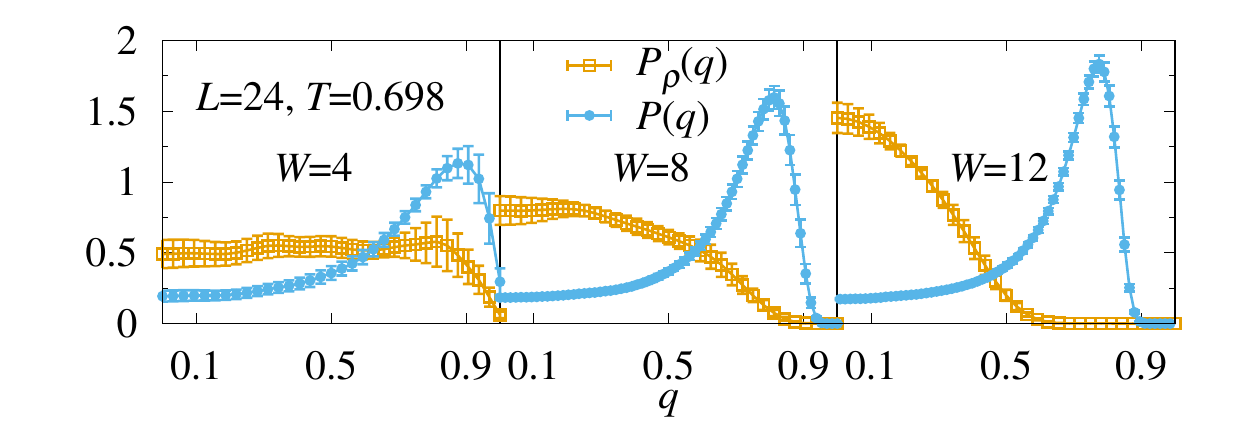}
\caption[\textbf{Size dependence of $\mathbf{P(q)}$ and $\mathbf{P_{\rho}(q)}.$}]{\textbf{Size dependence of $\mathbf{P(q)}$ and $\mathbf{P_{\rho}(q)}.$} Functions $P_{\rho}(q)$ and $P(q)$ for $L=24$, $R=L/2$ and $T=0.698 \approx 0.64 \Tc$. Different panels corresponds to different measuring window size $W=4,8,12$.}
\labfig{pq_metastate}
\end{figure}

\section{Relating numerical results and theory} \labsec{relating_numerical_theory_metastate}
To the best of our knowledge, this is the first numerical construction of the metastate\index{metastate} carried out in equilibrium simulations. Nonetheless, there exists one previous study in the non-equilibrium regime~\cite{manssen:15}, however, a deep relation between this \textit{dynamic metastate}\index{metastate!dynamic} and the \gls{AW} metastate\index{metastate!Aizenman-Wehr} is still to be explored. 

We have shown that the actual state of the art in the numerical \gls{SG}s allows the simulation of the \gls{AW} metastate\index{metastate!Aizenman-Wehr} in the \gls{EA}\index{Edwards-Anderson!order parameter} for $d=3$. We cannot extrapolate safely to the thermodynamic limit\index{thermodynamic limit} and the unexpected dependence of the \gls{MAS} on the inner disorder\index{disorder} is still important at the accessible system sizes. Nevertheless, we have found a convincing scaling law for the \gls{MAS} susceptibility\index{susceptibility} and we have estimated the exponent $\zeta(d=3)=2.3 \pm 0.3$, which strongly suggest $\zeta < d$, in addition, we have found no substantial size-dependence for this exponent.

\begin{figure}
\includegraphics[width=0.9\columnwidth]{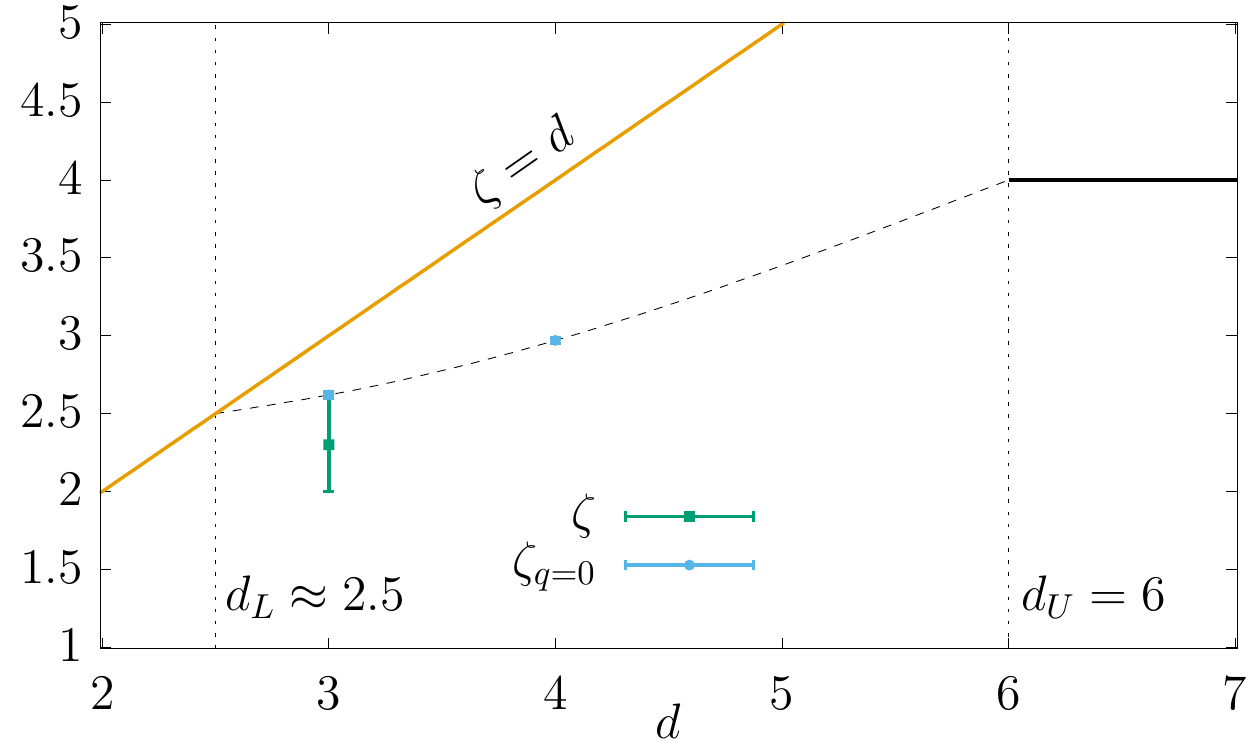}
\caption[\textbf{The exponent \boldmath $\zeta$ as a function of  $d$.}]{\textbf{The exponent \boldmath$\zeta$ as a function of $d$.} Different predictions of the exponent $\zeta$ for $d=3$ and $d=4$ are plotted. Above $d=6$, the mean-field\index{Mean-Field!solution} solution $\zeta=4$ is correct. The line $\zeta=d$ separates the disperse metastate\index{metastate!dispersed} for the trivial one.}
\labfig{zeta_exponent_metastate}
\end{figure}

In~\reffig{zeta_exponent_metastate} we have summarized our knowledge about the $\zeta$ exponent. Below the lower critical dimension\index{critical dimension!lower}\footnote{The lower critical dimension\index{critical dimension!lower} is the dimension below which there is no phase transition\index{phase transition} at finite temperature $T$.} $d_\mathrm{L}$ (at zero magnetic-field), the droplets\index{droplet!picture} picture is expected to be valid, and, therefore, the exponent $\zeta$ should be $\zeta \geq d$. In this work we have found $\zeta = 2.3 \pm 0.3$ (green-squared point in~\reffig{zeta_exponent_metastate}). Moreover, alternative estimations of the $\zeta$ exponent come from the decay of the four-spins spatial correlation\index{correlation function!four point} function at equilibrium.

In the equilibrium systems, the spins configurations\index{configuration} of different replicas\index{replica} follow a distribution function $P(q)$ (see~\refsubsec{theoretical_models}) and, therefore, the $C_4(T,r,\tw)$ has contributions of pair of replicas\index{replica} with all the possible values of $q$, weighted with the $P(q)$. However, the metastate\index{metastate} is in the $q=0$ by definition, and therefore, in order to compare both determinations, we have to restrict the computation of the correlation\index{correlation function!four point} functions to the zero overlap\index{overlap!zero sector} sector in the equilibrium results.

The four-point correlation\index{correlation function!four point} function conditioned to the $q=0$ sector is
\begin{equation}
C_4(T,r,\tw | q=0) \sim \lvert x \rvert^{-\vartheta} \, ,
\end{equation}
with $\vartheta=d-\zeta_{q=0}$, see~\refeq{long_distance_C4}. Previous studies found $\zeta_{q=0}(d=3) = 2.62 \pm 0.02$~\cite{janus:09b,janus:10} and $\zeta_{q=0}(d=4) = 2.62 \pm 0.02$~\cite{nicolao:14} (blue circles in~\reffig{zeta_exponent_metastate}). 

Finally, in $d\geq6$, where the mean-field\index{Mean-Field} computations are correct, we found $\zeta=4$~\cite{dedominicis:98,dedominicis:99}. A gentle extrapolation with the values of $\zeta_{q=0}(d=3,4)$ and the value of $z(d=6)=4$ (dashed line in~\reffig{zeta_exponent_metastate}) seems to meet, as expected, the yellow line corresponding to $d=\zeta$ around $d=2.5$, which is a general accepted estimation for the lower critical dimension\index{critical dimension!lower} from numerical and experimental results (see e.g.~\cite{franz:94,boettcher:04,boettcher:04b,boettcher:05,guchhait:14,maiorano:18}).

As we have previously discussed in~\refsubsec{theoretical_predictions_metastate}, the exponent $\zeta$ is related with the number of states that can be discriminated in a measuring window of size $W$, scaling that number with $\sim W^{d-\zeta}$~\cite{read:14}. This numerical estimation of $\zeta$ for $d=3$ supports the pictures of the metastate\index{metastate} with infinitely many states, namely \gls{RSB}\index{replica!symmetry breaking (RSB)} metastate\index{metastate!RSB} and chaotic pairs\index{metastate!chaotic pairs}.

\addpart{Off-equilibrium phenomena}
\chapter[Aging rate: exploring the growth of the coherence length]{Aging rate: exploring the growth \\ of the coherence length} \labch{aging_rate}
\setlength\epigraphwidth{.5\textwidth}

Experiments in \gls{SG}s are developed in out-of-equilibrium conditions most of the times\footnote{As we have mentioned before, recent experiments in thin-film geometry~\cite{guchhait:14,guchhait:15b} stands as honorable exceptions.}. Typically, the experimental setup consists of a system that it is rapidly cooled from $T_1>\Tc$ to $T_2<\Tc$, and its off-equilibrium evolution, which we have already termed as \textit{aging\index{aging}} (see~\refsubsec{aging_memory_rejuvenation}), is studied.

Under these non-equilibrium conditions, it was originally predicted in the context of the droplet\index{droplet!picture} theory (see~\refsubsec{theoretical_pictures}) that domains\index{magnetic domain} of correlated spins start to grow at the microscopic level~\cite{fisher:88}. Although with some differences in the nature of the domains\index{magnetic domain}, \gls{RSB} also expects a similar behavior. The linear size of those domains\index{magnetic domain} is known as \textit{coherence length\index{coherence length}}.

This coherence length\index{coherence length} have been measured in numerical simulations~\cite{huse:91,marinari:96,komori:99,komori:00} and also in experiments~\cite{joh:99,bernardi:01} long time ago. The initial expectation~\cite{fisher:88} for the growth of the coherence length\index{coherence length} with the time was $\xi \sim \left(\log \tw\right)^{1/\psi}$ and some numerical simulations found that ansatz to be compatible with their results (see, e.g.~\cite{kisker:96}). However, the mainstream usually accepts the alternative growth functional form described by $\xi \sim \tw^{1/z(T)}$ which better describes the results.

The aging rate\index{aging!rate} $z(T) = \dd \log \tw / \dd \log \xi$ was difficult to measure in traditional experiments based on the study of the shift of the peak in the relaxation\index{relaxation} rate $S(\tw)$~\cite{joh:99}. However, it can be now experimentally measured with excellent accuracy through the study of activation energies in \gls{SG}s with thin-film geometry~\cite{zhai:17}.

A strong discrepancy has been found between numerical and experimental measurements of $z(T)$. We solve that discrepancy in Ref.~\cite{janus:18} and this chapter is devoted to exposing the results of the cited reference that have been obtained in this thesis. In this chapter, we first motivate the work and describe the state of the art in~\refsec{why_study_aging} and~\refsec{how_can_study_aging}. Then, we describe the numerical simulation that we have performed in~\refsec{numerical_simulation_aging}. We describe the problem in~\refsec{controversy_aging} and finally we show the results in~\refsec{large_xi_limit}.

\section{Why should we care about off-equilibrium dynamics?} \labsec{why_study_aging}
We have stated several times that experiments and theory focus on different regimes, off-equilibrium, and equilibrium respectively. Moreover, we have also stated that, traditionally, simulations have been a powerful tool in theoretical research.

The increase of computational power has recently allowed us to promote the numerical simulations to a higher ``responsibility'' role in the development of the \gls{SG} field. Now, our simulations are in the border of the experimental regime~\cite{janus:08b,janus:09b,janus:17b,janus:18} and therefore, numerical work is in a privileged situation. On the one hand, it still has the classical advantages of the simulations: we are able to access microscopic configurations\index{configuration} that are difficult to access from the experiments, and we have total control of our system which is desirable in many senses (for example, our protocols are totally reproducible with no source of error). On the other hand, our numerical data can be now confronted with the experimental one through mild extrapolations (see e.g. ~\cite{janus:17b,janus:18,zhai-janus:20,zhai-janus:21}). This is extremely useful, not only because it allows us to test the numerical models but also because we can compute theoretical quantities, not accessible from experiments, in our experiment-compatible system at relevant time-scales.
 
Moreover, in the last years, the development of a statics-dynamics dictionary~\cite{barrat:01,janus:08b,janus:10,janus:10b,janus:17} has been a milestone in the development of the numerical research. According to the statics-dynamics equivalence\index{statics-dynamics equivalence}, the off-equilibrium properties of an (effective) infinite system that ages for a finite-time $\tw$ with a coherence length\index{coherence length} $\xi(\tw)$, are tightly bounded with the equilibrium properties of a finite-size system with linear length $L \sim \xi(\tw)$. The study of out-equilibrium systems may be helpful in order to extend the statics-dynamics dictionary and establish new relations.

\section{Coherence length, a fundamental quantity to study off-equilibrium dynamics}\labsec{how_can_study_aging}
The experimental study of out-equilibrium \gls{SG}s is usually focused on the characterization of magnetic responses when an external magnetic field is applied. As we have already discussed in~\refsec{experimental_spinglass}, both time scales, the aging\index{aging} before turn on (off) the magnetic field and the subsequent magnetic evolution of the system, turned out to be essential to describe the off-equilibrium phenomenon.

However, different aging\index{aging!rate} rates $z(T)$ for different temperatures make the coherence length\index{coherence length} $\xi \sim \tw^{1/z(T)}$ a much more convenient quantity to describe the ``aging\index{aging} state'' of the system. The meaning of the aging\index{aging!rate} rate $z(T)$ comes directly from the Arrhenius law. As we have mentioned many times along this thesis, the \gls{SG}s exhibit extremely slow dynamics. If we assume that the slow dynamics are due to the presence of many valleys in a rugged free-energy\index{free energy!landscape} landscape, it is natural to propose the Arrhenius law to characterize the typical time-scale that the system needs to overcome these free-energy barriers\index{free energy!barrier} and explore different valleys\index{free energy!valley}.
\begin{equation}
\tw = t(\xi) = \tau_0 \exp \left[ \beta \Delta(\xi) \right] \, , \labeq{arrhenius_law}
\end{equation}
being $\tau_0 \propto \beta_\mathrm{c}$ a microscopic time-scale and the exponent $\Delta(\xi)$ is the size of the energetic barriers\index{free energy!barrier} in units of $1/\beta=k_{\mathrm{B}}T$. Therefore, the aging\index{aging!rate} rate
\begin{equation}
z(T,\xi) = \dfrac{\dd \log \tw}{\dd \log \xi} = \dfrac{\dd \left[ \beta \Delta(\xi)\right]}{\dd \log \xi} \, , \labeq{def_aging_rate}
\end{equation}
give us the information of the evolution of the free-energy\index{free energy!barrier} barriers with $\log \xi$. Actually, as we will discuss in \refsec{large_xi_limit}, different hypothesis about the specific form of $\beta \Delta(\xi)$ will lead to very different behavior in the large-$\xi$ limit.

Moreover, as mentioned in the introduction of the chapter, experimental studies on thin-film geometry \gls{SG}s have achieved accurate measurements of this, once elusive, quantity~\cite{zhai:17}. Besides, it has been found that the experimental estimation of $\xi$ and the numerical one matches~\cite{janus:17b}.

The recent advances in the experimental determination of the coherence length\index{coherence length} $\xi$ have also brought a discrepancy in the estimation of the aging\index{aging!rate} rate $z(T)$ from numerical simulations and experiments~\cite{zhai:17}. To solve this discrepancy, it is fundamental to estimate $\xi$ with unprecedented accuracy. Two main factors have allowed us to compute the data needed to perform this research.

First, the dedicated hardware Janus\index{Janus} II~\cite{janus:14} has a central role in this work. The simulation of very large systems to very long times has been the result of thousands of computational hours with the largest special-purpose\index{special-purpose computer} machine focused on \gls{SG}s.

Second, our particular choice of the simulation parameters has turned out to be fortunate. The numerical effort is usually focused on increasing the number of samples\index{sample} $\NS$ as much as possible and simulating the minimum number of replicas\index{replica} needed to compute the observables, typically $\NRep=2$ or $\NRep=4$. However, we had in mind to study the \textit{temperature chaos}\index{temperature chaos} phenomenon (see~\refch{out-eq_chaos}), where the determination of the error of the observables of interest is greatly benefited by a maximization number of overlaps\index{overlap} $\Nov = \NRep (\NRep-1)/2$. Unexpectedly, this has led to a dramatic increase in precision. We analyze this reduction of the error in~\refsec{Nr_aging}.

This study is a clear demonstration of the importance of the high-precision results for the investigation of glassiness. Indeed, without the dramatic reduction of the error bars\index{error bars}, we would not be able to solve the discrepancy between numerical simulations and experiments.

\section{Numerical simulation} \labsec{numerical_simulation_aging}
In this work, we simulate in the FPGA-based\index{FPGA} [\gls{FPGA}] computer Janus\index{Janus} II an \gls{EA}\index{Edwards-Anderson!order parameter} model in three-dimensional spin glasses (\refsubsec{3D_EA_model}) for several temperatures $T$ in a lattice of linear size $L=160$, which is aimed to represent a system of infinite size. This assumption is sound, provided that $L\gg\xi$ (see~\refsec{finite_size_effects}). Note that this condition limits the maximum time at which we can safely ignore finite-size effects\index{finite-size effects}. The temperature remains constant throughout the whole simulation.

We shall perform direct quenches from configurations\index{configuration} of spins randomly initialized (which corresponds to infinite temperature) to the working temperature $T<\Tc$, where the system is left to relax for a time $\tw$. This relaxation\index{relaxation} corresponds with the (very slow) growth of glassy magnetic domains\index{magnetic domain} of size $\xi(\tw)$.

We compute a total of $\NS = 16$ different samples\index{sample}. For each sample\index{sample}, we shall consider $\NRep=256$ replicas\index{replica}. As we have already said, this simulation had the original aim to study the temperature chaos\index{temperature chaos} phenomenon under non-equilibrium conditions (see~\refch{out-eq_chaos}), however, the reader may notice that in that study we use a total number of replicas\index{replica} $\NRep=512$. Indeed, this study about the aging\index{aging!rate} rate was performed much earlier and we had at our disposal ``only'' $\NRep=256$.

The main observable of this study is the coherence length\index{coherence length} $\xi(T,\tw)$, estimated by $\xi_{12}(T,\tw)$, computed from integral estimators of the correlation\index{correlation function!four point} function $C_4(T,r,\tw)$. Both observables have been described with great detail in \refsubsec{observables_introduction}. The large number of replicas\index{replica} simulated has allowed us to follow the decay of $C_4(T,r,\tw)$ over six decades (see inset of~\reffig{growth_xi}). The $\xi$ estimation used in our work is plotted in~\reffig{growth_xi}.

\begin{figure}[t]
\includegraphics[width=0.8\linewidth]{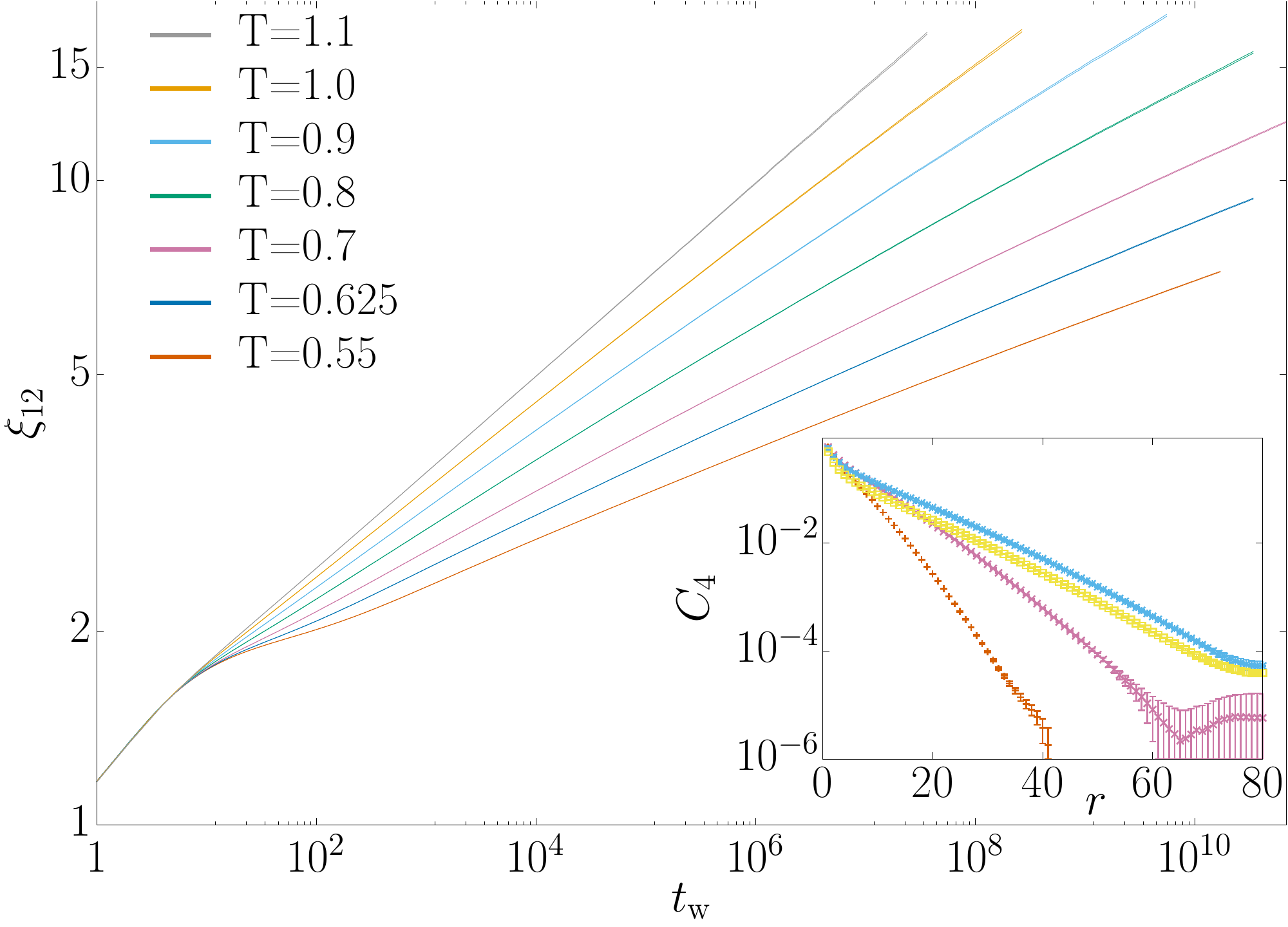}
\caption[\textbf{Growth of the coherence length \boldmath $\xi(T,\tw)$.}]{\textbf{Growth of the coherence length\index{coherence length} \boldmath $\xi(T,\tw)$.} Growth of the coherence length\index{coherence length} $\xi_{12}(T,\tw)$ with the waiting time $\tw$ after a quench to temperature $T$ in a log-log  scale [the critical\index{critical temperature} temperature is $\Tc=1.102(3)$]. Given the smallness of the statistical errors, instead of error bars\index{error bars} we have plotted two lines for each $T$, which enclose the error estimate.  At this scale, the curves seem linear for long times, indicating a power-law growth but, see~\reffig{a2}, there is actually a measurable curvature. \textbf{Inset:} Spatial four-point correlation\index{correlation function!four point} function of the overlap\index{overlap!field} field $C_4(T,r,\tw)$, plotted as a function of distance at the last simulated time for several temperatures. Note the six orders of magnitude in the vertical axis.}
\labfig{growth_xi}
\end{figure}

\section{The controversy of the aging rate}\labsec{controversy_aging}
The growth of the coherence length\index{coherence length} has been a debated issue in the \gls{SG} literature (see, e.g. \cite{fisher:88,huse:91,marinari:96,joh:99,bernardi:01,bouchaud:01,berthier:02}). However, despite the existence of different proposals, the simplest functional form that was able to fit the data was the power law
\begin{equation}
\xi(\tw,T) = A(T) \tw^{1/z(T)} \quad , \quad z(T) \approx z(\Tc)\dfrac{\Tc}{T} \, , \labeq{xi_powerlaw}
\end{equation}
being $z(T)=\dd \log \tw / \dd \log \xi$ the so-called aging\index{aging!rate} rate. Indeed, the experimental measurements concerns to the renormalized aging\index{aging!rate} rate
\begin{equation}
z_c = z(T) \dfrac{T}{\Tc} \, . \labeq{renormalized_aging_rate}
\end{equation}
Experiments performed in thin-film geometry systems~\cite{zhai:17} has measured $z_c \approx 9.62$, which is very far from the value predicted by numerical simulations $z_c = 6.86(16)$~\cite{janus:08b} and $z_c=6.80(15)$~\cite{lulli:16}. 

Those experiments are performed in \gls{CuMn} films with $20$ nm of thickness which translates to a distance of 38 lattice spacings (the typical Mn-Mn distance is 5.3 \r{A}). Therefore, we will need to extrapolate our results to $\xi_{12} \approx 38$ in order to confront the numerical simulations and the experiments.

\subsection[The growth of $\xi$ does not follow a power law]{The growth of \boldmath $\xi$ does not follow a power law} \labsubsec{no-powerlaw}
The increase of the precision of the data shows that the pure power law of~\refeq{xi_powerlaw} is not a faithful description anymore. Indeed, in order to discern if our data of~\reffig{growth_xi} presents a deviation from a power law, we propose a very naive ansatz
\begin{equation}
\log \tw(T,\xi_{12}) = a_0(T) + a_1(T)\log \xi_{12} + a_2(T) \log^2 \xi_{12}  \, , \labeq{naive_divergent}
\end{equation}
where $a_0(T)$, $a_1(T)$ and $a_2(T)$ are meaningless coefficients, only useful to interpolate our data. An absence of curvature [$a_2(T)=0$] would reduce~\refeq{naive_divergent} to~\refeq{xi_powerlaw}. On the contrary, $a_2(T)>0$ indicates a slowing down in the dynamics for increasing $\xi_{12}(T,\tw)$.

\reffig{a2} is telling us that $a_2 \geq 0$ and only vanishes for $T=\Tc$ with $z_c=6.69(6)$ which improves the accuracy of previous estimations. Therefore, the solution of the discrepancy of the results of $z_c$ seems to be the introduction of a (very mild) scale dependence in the \textit{effective} dynamical exponent which is defined as
\begin{equation}
z(T,\xi_{12}(\tw)) = \dfrac{\dd \log \tw}{\dd \log \xi_{12}} \, . \labeq{definition_aging_rate}
\end{equation}

The reader may think about the possibility that our deviation from the power-law behavior might be due to the existence of finite-size effects\index{finite-size effects}, however, two main reasons against this argument can be defended. First, the curvature decreases when increasing the temperature (see~\reffig{a2}) and one would expect the opposite behavior in presence of finite-size effects\index{finite-size effects}\footnote{Actually, finite-size effects\index{finite-size effects} would be controlled by $\xi/L$ which is smaller for the lower temperatures.}. Second, exhaustive checks have been done in order to establish our system size $L=160$ as a safe choice, see~\refsec{finite_size_effects}.

\begin{figure}[h!]
\includegraphics[width=0.8\linewidth]{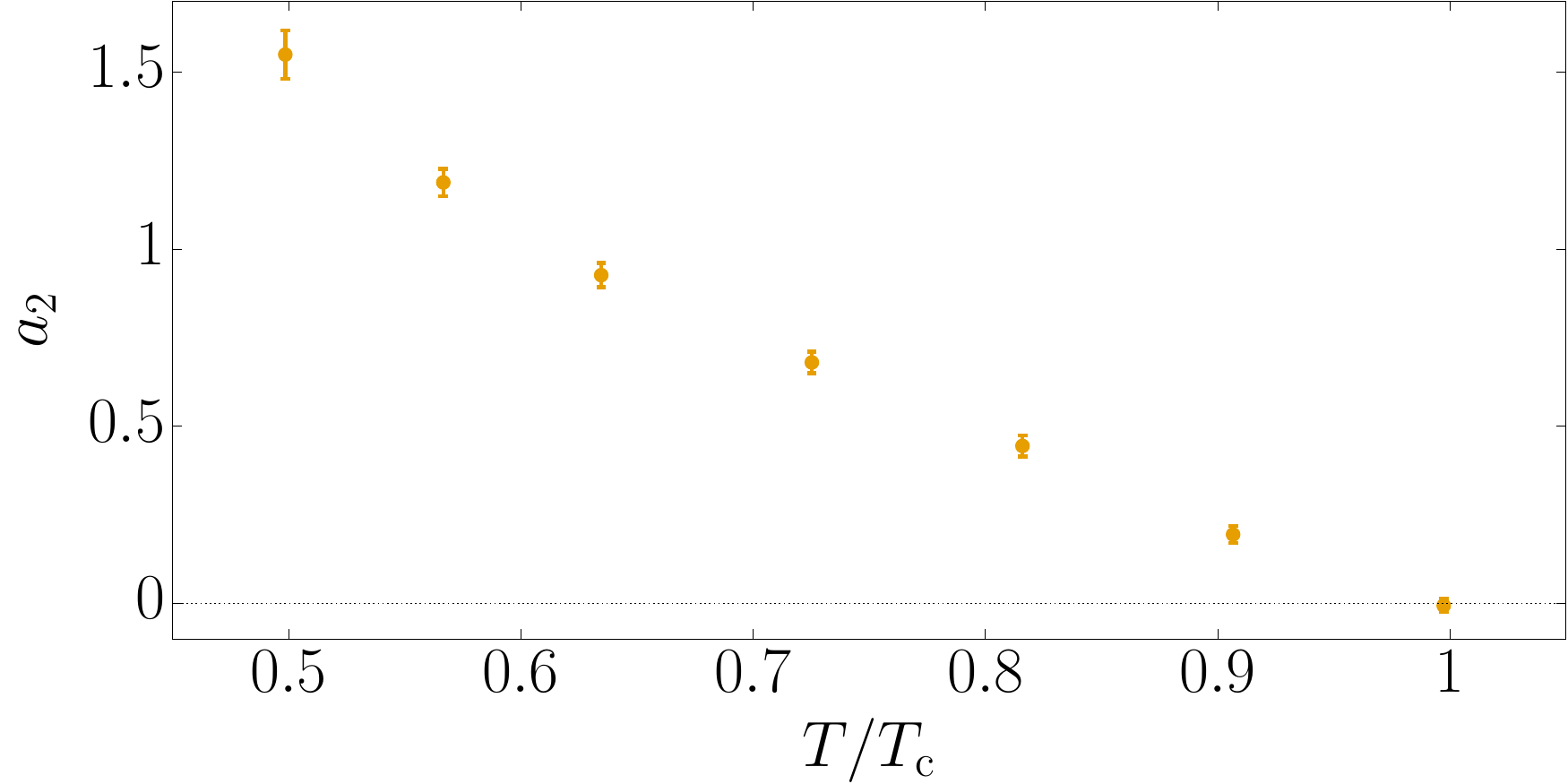}
\caption[\textbf{Deviation of \boldmath $\xi_{12}(\tw)$ from a simple power-law growth.}]{\textbf{Deviation of \boldmath $\xi_{12}(\tw)$ from a simple power-law growth.} We plot the quadratic parameter $a_2$ in a fit to~\refeq{naive_divergent}. This quantity is zero at the critical point, but  has a positive value at low temperatures, indicating that the growth  of $\xi_{12}$ slows down over the simulated time range.}
\labfig{a2}
\end{figure}

Of course, our naive ansatz of~\refeq{naive_divergent} is only useful for interpolations. If we want to explore the growth of $\xi_{12}(T,\tw)$ in the large-$\xi_{12}$ limit, we need some insight from theory.

\section[The large-$\xi$ limit]{The \boldmath large-$\xi$ limit} \labsec{large_xi_limit}
In this section, we explore the growth of $\xi$ in the large-$\xi$ limit. For that purpose, we need to propose extrapolations from our numerical data. We introduce and study two functional forms for the function $\log \tw$ that will allow us to extrapolate $z(T,\xi_{12}(\tw))$ in~\refsubsec{ansatzs_aging}. However, before extrapolating our data, we need to compute the exponent $\vartheta$ appearing in the long-distance decay of $C_4(T,r,\tw)$ [see \refeq{long_distance_C4}] because it is required by one of our ans\"atze. We recall the relation between $C_4(T,r,\tw)$ and $\vartheta$ for the reader's convenience and compute the exponent $\vartheta$ in \refsubsec{vartheta_aging}. Finally, we extrapolate our results and confront them with the experimental ones in~\refsubsec{extrapolations_aging}.
\subsection[Computing $\vartheta$]{Computing \boldmath $\vartheta$} \labsubsec{vartheta_aging}
The correlation\index{correlation function!four point} function $C_4(T,r,\tw)$ presents the following long-distance decay
\begin{equation}
C_4(T,r,\tw) = r^{-\vartheta} f(r/\xi(T,\tw)) \, , \labeq{recall_long_distance_C4}
\end{equation}
where $f(x)$ is an unknown function which vanishes faster than exponentially. As we have already introduced in~\refsubsec{observables_introduction}, the exponent $\vartheta$ at $\Tc$ is given by $\vartheta = 1 + \eta$ where $\eta=-0.390(4)$~\cite{janus:13} is the anomalous dimension. For $T<\Tc$ the two main pictures, namely droplets\index{droplet!picture} and \gls{RSB}\index{replica!symmetry breaking (RSB)}, have differing expectations: coarsening\index{coarsening} domains\index{magnetic domain!compact} with $\vartheta=0$ is the \index{droplet!picture}droplets' prediction while in the \gls{RSB}\index{replica!symmetry breaking (RSB)} theory, $\vartheta$ is given by the replicon\index{replicon} (see~\cite{janus:10b} for a detailed discussion). The best previous numerical study of $\vartheta$ found $\vartheta=0.38(2)$~\cite{janus:09b}.

If we recall the integral $I_k= \int_0^{\infty} r^k C_4(T,r,t) \dd r$ [see~\refeq{integral_estimator_xi}] it is easy to prove that $I_2(T,\xi_{12}) \propto \xi_{12}^{3-\vartheta}$ and, therefore, we can estimate the value of $\vartheta$ from the numerical derivative of $I_2$
\begin{equation}
\vartheta=3-\dfrac{\dd \log I_2}{\dd \log \xi_{12}} \, .
\end{equation}

Our estimations of $\vartheta$ show that, for $T=\Tc$, its value is compatible with $1+\eta$, as expected. However, for $T<\Tc$ we found a slow decrease as $\xi$ increases or $T$ decreases, i.e. $\vartheta \to \vartheta(T,\xi_{12})$.

This result is unsatisfactory because we expect $\vartheta$ to be constant (possibly 0) in the large-$\xi$ limit. For the sake of clarity, we will call that theoretically expected value $\vartheta_{\infty}$. The simplest explanation is that the values of $\vartheta(T,\xi_{12})$ are affected by critical effects of the fixed point at $T=\Tc$ with $\vartheta(\Tc,\xi_{12}) \approx 0.61$. For large $\xi$, we expect those critical effects to vanish and the value of $\vartheta(T,\xi_{12}\to \infty)$ should be dominated by the $T=0$ fixed point, i.e. $\vartheta(T,\xi_{12}\to \infty) = \vartheta_{\infty}$.

In analogy with the ferromagnetic phase\index{phase!ferromagnetic} of the $O(N)$ model, we study this crossover from $\vartheta(T=\Tc)$ to the unknown $\vartheta(T=0)$ in terms of the Josephson length $\ell_J$~\cite{josephson:66}. The Josephson length is expected to behave as $\ell_J \sim (\Tc - T)^{-\nu}$ with $\nu=2.56(4)$~\cite{janus:13} for temperatures $T \to \Tc$. The scaling corrections for $\ell_J(T)$ at lower temperatures are explained in~\refsec{Josephson_length}.

We can test the crossover hypothesis by considering the ratio of two different estimations of $\xi$: $\xi_{23}/\xi_{12}$. We plot this ratio against the scaling variable $x=\ell_J/\xi_{12}$. This ratio should be scale-invariant in the large-$\xi_{12}$ limit because different determinations of $\xi$ should grow with the same rate but with a different prefactor (see~\cite{janus:09b} and~\refsubsec{observables_introduction}).

\reffig{testing_crossover} shows us that the ratio grows towards the critical value (represented with a gray line) for the curves with the largest temperatures when $x$ is large, i.e. for a given curve, which is $T$-constant, when $\xi_{12}$ is small. Then, it relaxes to the value corresponding to the $T=0$ fixed point (small $x$). The lowest temperatures, namely $T=0.55$, $T=0.625$ and $T=0.7$, seem to be free of critical effects.

\begin{figure}[h!]
\includegraphics[width=0.8\linewidth]{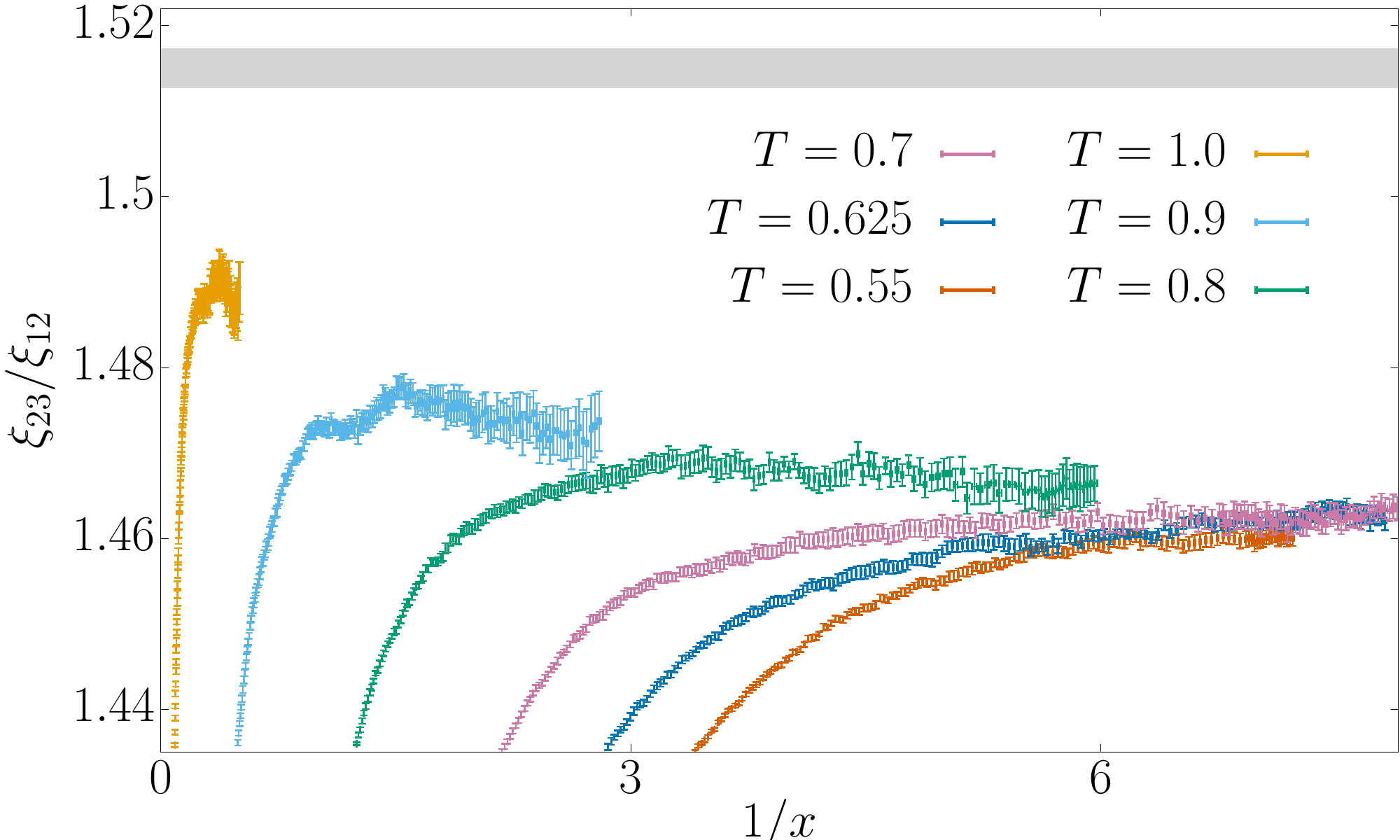}
\caption[\textbf{Testing the crossover hypothesis with \boldmath $\xi_{23}/\xi_{12}$.}]{\textbf{Testing the crossover hypothesis with \boldmath $\xi_{23}/\xi_{12}$.} We consider the ratio $\xi_{23}/\xi_{12}$ between two definitions of the coherence length\index{coherence length}, which should be constant in the large-$\xi_{12}$ (or $x\to0$) limit. For $T$ close to $\Tc$, this ratio initially grows, approaching the $T=\Tc$ value (represented by the thick gray line) and eventually relaxes towards the $T=0$ fixed point.}
\labfig{testing_crossover}
\end{figure}

This positive result encourages us to perform a similar analysis for $\vartheta(T,\xi_{12})$. If the hypothesis of the crossover is correct, we should observe a collapse of the $\vartheta(T,\xi_{12})$ values when we plot them in terms of the scaling variable $x=\ell_J/\xi_{12}$. 

The functional form of $\vartheta(T,\xi_{12})$ should be, in the \gls{RSB}\index{replica!symmetry breaking (RSB)} picture
\begin{equation}
\vartheta(x) = \vartheta_{\infty} + b_2 x^{2-\vartheta_{\infty}} + b_3 x^{3-\vartheta_{\infty}} + \cdots \, . \labeq{vartheta_RSB}
\end{equation}
The reader may find a derivation of this expression in~\refsec{Josephson_length}. On the contrary, for the droplets\index{droplet!picture} picture, it should be 
\begin{equation}
\vartheta(x)=C x^\zeta \, , \labeq{vartheta_droplet}
\end{equation}
where $\zeta\approx 0.24$~\cite{boettcher:04} is the stiffness coefficient\index{stiffness!coefficient}.

\reffig{vartheta_collapse} shows us a nice collapse of the $\vartheta(x)$ values. Moreover, we have to keep in mind that our goal is to extrapolate the values of the aging\index{aging!rate} rate to the experimental $\xi$ length-scale which roughly corresponds to $\xi_{\mathrm{films}}\approx 38$. 

For the \gls{RSB}\index{replica!symmetry breaking (RSB)} picture, a fit of the $\vartheta(x)$ values to~\refeq{vartheta_RSB} gives us the value $\vartheta_{\infty} \approx 0.30$, although we take $\vartheta^{\mathrm{upper}}=0.35$ as our upper bound for $\vartheta_{\infty}$. For the droplets\index{droplet!picture} picture, a fit to~\refeq{vartheta_droplet} can be performed only by considering\footnote{Similar results were found in~\cite{janus:10}} $\zeta \approx 0.15$. It is worthy to note that we find this exponent very sensitive to the fitting range. We extrapolate the droplet\index{droplet!picture} $\vartheta(x)$ to $\xi = \xi_{\mathrm{films}}=38$ and we obtain $\vartheta(\xi_{\mathrm{films}}=38) \approx 0.28$.

However, because our determination of $\xi$ through the estimator $\xi_{12}$ may differ from the experimental estimation of $\xi$ by a small constant factor, we consider also $\xi_{\mathrm{films}}=76$ and we obtain $\vartheta(\xi_{\mathrm{films}}=76) \approx 0.25$.

The same conclusions found in~\cite{janus:10,janus:10b} stands here: for the experimental relevant scales, the physics is well described by a non-coarsening\index{coarsening} picture with $0.25<\vartheta(\xi_{\mathrm{films}})<0.35$ depending on the theory we use to extrapolate the data and the exact value chosen for the experimental scale $\xi_{\mathrm{films}}$.

\begin{figure}[h!]
\includegraphics[width=0.8\linewidth]{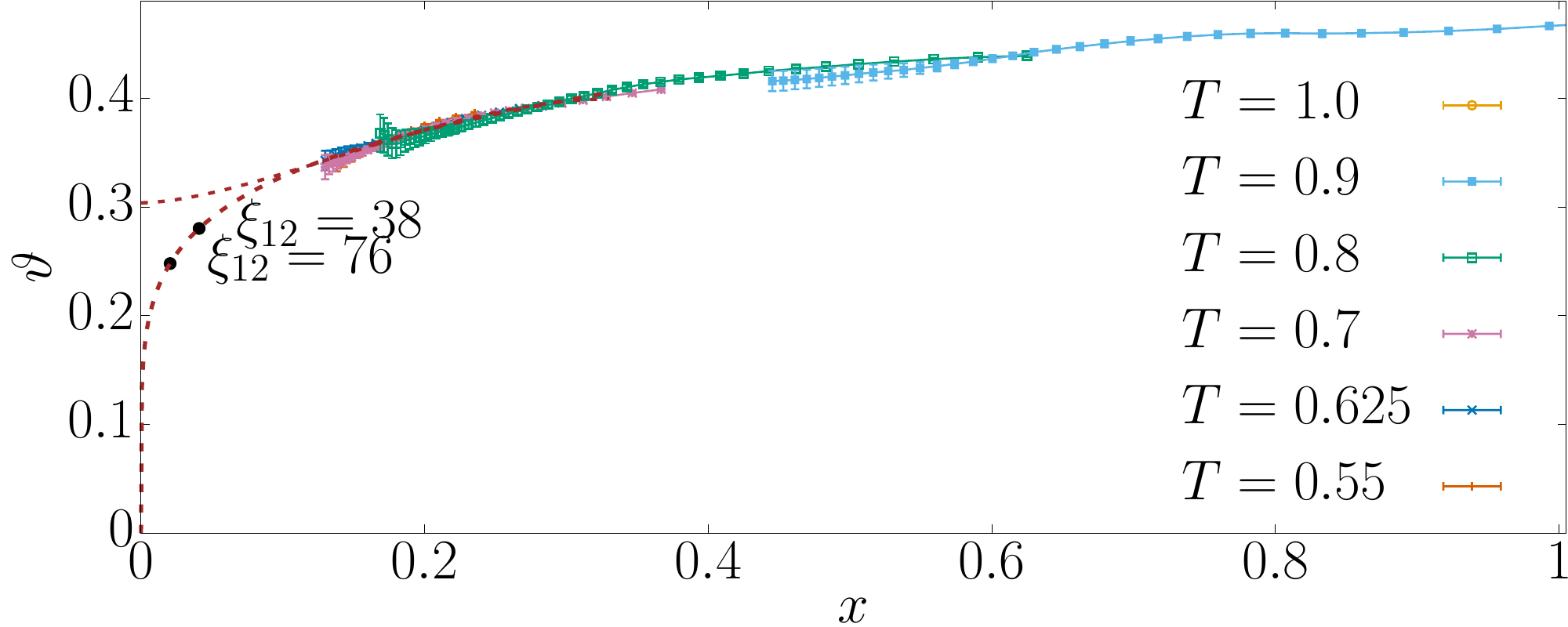}
\caption[\textbf{Crossover between the \boldmath $T=\Tc$ and the $T=0$ fixed points controlled by a Josephson length $\ell_\text{J}(T)$.}]{\textbf{Crossover between the \boldmath $T=\Tc$ and the $T=0$ fixed points controlled by a Josephson length $\ell_\text{J}(T)$.} We plot the evolution of the replicon\index{replicon} exponent $\vartheta$ for several temperatures against the relevant scaling variable $x=\ell_\text{J}(T)/\xi_{12}$. We show two possible extrapolations (dashed lines) to infinite $\xi_{12}$: one with finite $\vartheta$, as expected in the RSB picture, and one with $\vartheta=0$, as expected in the droplet\index{droplet!picture} picture. For the latter, we also show the extrapolated value for the experimental scale corresponding to experiments in \gls{CuMn} films~\cite{zhai:17}, which we estimate between $\xi_{12}=38$ and $\xi_{12}=76$.}
\labfig{vartheta_collapse}
\end{figure}

\subsection{The convergent and the divergent ansatz} \labsubsec{ansatzs_aging}
Now, we discuss the possible extrapolations of $z(T,\xi_{12})$ to the large-$\xi_{12}$ limit. The most natural assumption is to consider that $z(T,\xi_{12} \to \infty) = z_{\infty}(T)$ with a convergence $z(T,\xi_{12}) - z_{\infty}(T) \propto \xi_{12}^{-\omega}$. Taking into account the~\refeq{definition_aging_rate}, the expresion of $\tw$ should be
\begin{equation}
\log \tw = C_1(T) + z_{\infty}(T) \log \xi_{12} + C_2(T)\xi_{12}^{-\omega} \, , \labeq{convergent_ansatz}
\end{equation}
where $\omega$ is the exponent that controls finite-$\xi_{12}$ corrections. The value of $\omega$ for the critical\index{critical temperature} temperature $T_c$ is $\omega=1.12(10)$~\cite{janus:13,fernandez:15,lulli:16}. In the \gls{SG} phase\index{phase!low-temperature/spin-glass}, the leading behavior is given by $\omega=\vartheta$, see \cite{janus:10b} for a detailed discussion.

The effective exponent in the \textit{convergent} ansatz would be, therefore
\begin{equation}
z_{\mathrm{conv}}(T,\xi_{12}) = \dfrac{\dd \log \tw}{\dd \log \xi_{12}} = z_{\infty}(T) - \omega C_2(T) \xi_{12}^{-\omega} \, . 
\end{equation}

The fits to~\refeq{convergent_ansatz} have two main sources of error. First, the value of $\vartheta$ has associated some uncertainty. We choose $\vartheta=0.35$, as explained above. Second, we consider possible systematic effects due to the fitting range. We follow objective criteria to select a minimum $\xi_{12}^{\min}$ for the fitting range. Further details can be found in~\refsec{parameters_aging}.

An alternative approach was proposed in~\cite{bouchaud:01,berthier:02}. In those works, the authors proposed a crossover to activated dynamics. This approach is a refinement of the droplet\index{droplet!picture} proposal that we expose at the beginning of the chapter [$\xi \sim (\log \tw)^{1/\psi}$]. In this case, the authors propose
\begin{equation}
\tw = \tau_0 \xi^{z_c} \exp \left(\dfrac{\Upsilon(T) \xi^{\psi}}{k_{B}T} \right) \, , \labeq{original_divergent}
\end{equation}
being $\Upsilon(T) = \Upsilon_0(1-T/T_c)^{\psi\nu}$ and $k_{B}$ the Boltzmann\index{Boltzmann!constant} constant. Here, we express the original ansatz in logarithmic form with generic coefficients
\begin{equation}
\log \tw = D_1(T) + z_c \log \xi_{12} + D_2(T) \xi_{12}^{\psi} \, . \labeq{divergent_ansatz}
\end{equation}
The exponent $\psi$ has been used before in experiments~\cite{schins:93} and numerical simulations~\cite{rieger:93} with values $\psi \approx 1$. Moreover, the reader may noticed that $D_2(T) \sim (T_c-T)^{\psi\nu}$ [see~\refeq{original_divergent}]. This can be regarded as another way to present the crossover between the $T=T_c$ fixed point and the $T=0$ fixed point. We need $\xi_{12}(T_c-T)^{\nu} \gg 1$ in order to perceive deviations from the pure power-law with an aging\index{aging!rate} rate $z_c$ equal to the critical one.

The effective exponent in the \textit{divergent} ansatz would be, therfore
\begin{equation}
z_{\mathrm{div}}(T,\xi_{12})=\dfrac{\dd \log \tw}{\dd \log \xi_{12}} =z_c + D_2(T) \psi \xi_{12}^{\psi} \, .
\end{equation}

We find fair fits to our simulated data for both ans\"atze,~\refeq{convergent_ansatz} and~\refeq{divergent_ansatz}. Indeed, for~\refeq{divergent_ansatz} we find $\psi \approx 0.4$. The next step is clear, we need to test both proposals at the experimental length-scales.

\subsection{Extrapolation to experimental regime} \labsubsec{extrapolations_aging}
We perform extrapolations of the renormalized aging\index{aging!rate} rate $z_c(T,\xi) = z(T,\xi) T/T_c$ at the experimental length-scale $\xi_{\mathrm{films}}$. The value of $z(T,\xi)$ is computed for both $z_{\mathrm{conv}}(T,\xi_{\mathrm{films}})$ and $z_{\mathrm{div}}(T,\xi_{\mathrm{films}})$.

In a similar way as we did in the previous analysis of~\refsubsec{vartheta_aging}, we use safe estimations of $\xi_{\mathrm{films}}$ and we consider $\xi_{\mathrm{films}}=38$ and $\xi_{\mathrm{films}}=76$. All the relevant data from the fits can be found in~\refsec{parameters_aging}, however, we plot in~\reffig{zc} the relevant information.

The main plot shows the renormalized aging\index{aging!rate} rate $z_c(T,\xi_{\mathrm{films}})$ plotted against the reduced temperature $T/\Tc$. We see that the convergent ansatz of~\refeq{convergent_ansatz} is very successful in reproducing the experimental behavior for both $\xi_{12}=38$ and $\xi_{12}=76$.  Compatible results with experiments are found for a wide range of temperatures.

The inset shows the divergent ansatz of~\refeq{divergent_ansatz}. In this case this proposal for $z_c(T,\xi_{\mathrm{films}})$ is not able to reproduce the constant value found in experiments for $T/T_c < 0.8$.

\begin{figure}[h!]
\includegraphics[width=0.8\linewidth]{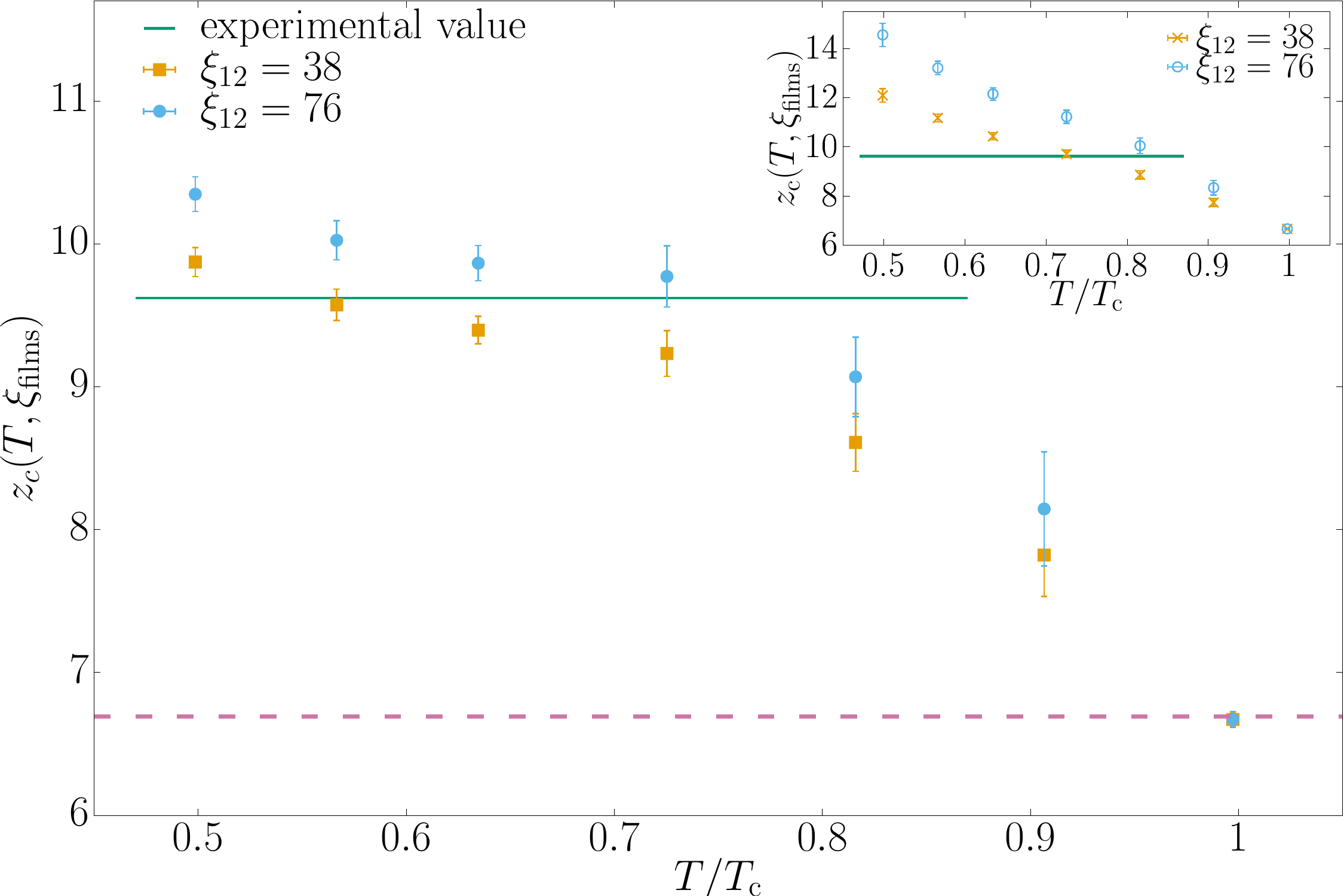}
\caption[\textbf{Numerical and experimental aging rate.}]{\textbf{Numerical and experimental aging rate.} Value of the experimental aging\index{aging!rate} rate for \gls{SG}s $z_c(T,\xi_{\mathrm{films}}) = z(T,\xi_\text{films}) T/\Tc$, extrapolated from our data for values of the coherence length\index{coherence length} corresponding to thin  \gls{CuMn} films. The main plot considers an ansatz~\refeq{convergent_ansatz} with a finite $z_{\infty}(T)T/\Tc$, which agrees very well with the experimental value of $z_c(T)\approx 9.62$~\cite{zhai:17}, indicated by the straight line, whose width represents the experimental temperature range. Notice that critical effects are only visible for $T>0.7$. \textbf{Inset:} Same plot but now considering a crossover to activated dynamics~\refeq{divergent_ansatz}, as in~\cite{bouchaud:01}. This is less successful at reproducing the roughly constant $z_c(T)$ observed in experiments.}
\labfig{zc}
\end{figure}

\chapter[The Mpemba effect]{The Mpemba effect} \labch{mpemba}

\epigraph{\textit{Aquí el que no carneguea, borreguea.}}{-- Acervo popular}

Consider two beakers of water that are identical to each other except for the fact that one is hotter than the other. If we put both of them in contact with a thermal reservoir (for example, a freezer) at some temperature lower than the freezing point of the water, under some circumstances, it can be observed that the, initially, hotter water freezes faster than the colder one. This phenomenon is known as the Mpemba\index{Mpemba effect} effect~\cite{mpemba:69}.

The history of this phenomenon is, indeed, very curious and constitutes one paradigmatic example of the importance of the scientific method in the development of science. Although the first written record comes from Aristotle, it would be probably a well-known fact for most of the people~\cite{aristotle-lee:89}. This effect was sporadically mentioned through the ages~\cite{bacon-burke:62,bacon:11,descartes:65} but it received little attention from the scientific community until the second half of the XX century.

In 1969 this phenomenon was brought back to the scientific debate by Erasto Mpemba, a young student in Tanzania, and Denis Osborne, a teacher of the University College Dar es Salaam, Tanzania~\cite{mpemba:69}. The same year, Dr. Kell reported the same phenomenon in an independent publication~\cite{kell:69}. 

Different arguments were given to explain this phenomenon~\cite{kell:69,deeson:71,firth:71,walker:77}, but there is no consensus neither in the explanations~\cite{osborne:79,freeman:79,wojciechowski:88} nor in the very existence of the effect.~\cite{burridge:16}. We will briefly discuss the situation later in \refsec{water_complicated_mpemba}.

This phenomenon is not specific to water and has been reported in other systems like nanotube resonators~\cite{greaney:11}, clathrate hydrates~\cite{ahn:16}, granular fluids~\cite{lasanta:17} and colloidal systems~\cite{kumar:20}. This chapter is devoted to discussing the Mpemba\index{Mpemba effect} effect in \gls{SG}s. Besides, the Mpemba\index{Mpemba effect} effect constitutes a great example to stress the importance of the coherence length\index{coherence length} as a fundamental quantity to describe the off-equilibrium phenomena in \gls{SG}s.

We begin with a brief historical introduction\footnote{The reader may consult also a fantastic historical review in~\cite{jeng:06}.} to the phenomenon in \refsec{historical_introduction_mpemba} and with some of the proposed explanations in~\refsec{water_complicated_mpemba}. Then, we explain the numerical simulation, performed in the Janus\index{Janus} II custom-built computer, that has allowed the study of the Mpemba\index{Mpemba effect} effect in \gls{SG} (see~\refsec{numerical_simulation_mpemba}). At this point, we are ready to discuss the results. We first identify in \refsec{identifying_mpemba} the Mpemba\index{Mpemba effect} effect in \gls{SG} by choosing the adequate quantity to represent the \textit{temperature} of our system. We found in \refsec{coherence_length_mpemba} that the quantity controlling the phenomenon is, indeed, the coherence length\index{coherence length} $\xi$. Finally, we study the Inverse Mpemba\index{Mpemba effect} effect in \refsec{inverse_mpemba}.

The results described in this chapter were published in~\cite{janus:19}.

\section{A historical introduction}\labsec{historical_introduction_mpemba}

The first record of the Mpemba\index{Mpemba effect} effect is attributed to Aristotle in his \textit{Metereologica} around 350 B.C.~\cite{aristotle-lee:89}. His discussion there, suggests that the phenomenon was a well-known fact and he used it as an example to illustrate his theory of \textit{antiperistasis}\footnote{The concept of antiperistasis refers to the reaction between two opposite \textit{forces}, when one increases, the other have to do it.}:

\textit{``If the water has been previously heated, this contributes to the rapidity with which it freezes [sic] for it cools more quickly (Thus so many people when they want to cool water quickly first stand it in the sun and the inhabitants of Pontus when they encamp on the ice to fish --they catch fish through a hole which they make in the ice-- pour hot water on their rods because it freezes quicker, using the ice like solder to fix their rods.) \dots ''}
\\[5pt]
\rightline{{ --- Aristotle, Metereologica Book I, Chapter XII}}

The Mpemba\index{Mpemba effect} effect probably remained in the popular heritage through the centuries, but actually, did not receive too much attention from academics. Yet, it was mentioned in some important texts, for example in the \textit{Opus Majus} of Roger Bacon~\cite{bacon-burke:62} or in the \textit{Novum Organum} of Francis Bacon~\cite{bacon:11}. 

Some years after Francis Bacon mentioned the phenomenon, Descartes wrote about it in \textit{Les Meteores}~\cite{descartes:65}. Indeed, he stressed the importance of the experiments, and actually, he proposed a specific experiment that is not exactly the standard Mpemba effect (which is the most commonly studied). 

Descartes proposed to fill a beaker (and he also specified that should have a long straight neck) with hot water that has been kept over the fire for a long time, then the water should be let to reach room temperature. He proposed to do the same with another beaker of water but now, without boiling it. Then, both beakers should be put in contact with the [sic] ``freezing cold air'' and one observes that the beaker which has held for a long time over the fire, freezes first.

He stated in a letter to Mersenne (1638) that he performed that experiment and he defended that there was nothing incorrect in his methods. 

However, the Mpemba\index{Mpemba effect} effect was relegated to oblivion in mainstream physics by the emergence of thermodynamics, which was supported by an unprecedented success, both in describing reality and in the creation of modern machines. Apparently, the Mpemba\index{Mpemba effect} effect contradicts the knowledge provided by thermodynamics. It has been suggested~\cite{kuhn:12} that the theoretical views of scientists may condition the experiments that they decide to perform, and this could be the answer to the small numbers of experiments researching this interesting effect before the XX century. Actually, this points also to one of the weaknesses of the human being, that affects the application of the scientific method: the confirmation bias.

In 1963, one young student of a secondary school in Tanzania, Erasto Mpemba, put this phenomenon under the scrutiny of the scientific community~\cite{mpemba:69}. In his school, he used to make ice-cream by boiling milk, mixing it with sugar, and by putting the mix into the freezer. One day, some students were doing ice-cream but the space in the freezer was scarce. Despite the warnings for not introducing the hot mix directly in the freezer because that could damage it, Mpemba decided to do it anyway in order to do not lose his space in it.

He observed that his mix had frozen before that of other boys who had followed the \textit{standard protocol} and had let the mix cool at the room's temperature before introducing it in the freezer. Persistent questions of Mpemba to different physics teachers about this fact led to the same answer ``That is impossible'', one of the teachers said that ``That is Mpemba's physics, not universal physics''.

Nonetheless, Mpemba did not surrender and he took the opportunity to ask this question to a university professor, Denis Osborne, that went to Mpemba's High School to give a talk. From the first time that Mpemba observed the phenomenon, he did great advances on building up a specific protocol to develop the experiment in a reproducible way and he asked a very concrete question:

\textit{``If you take two beakers with equal volumes of water, one at 35ºC and the other at 100ºC, and put them into a freezer, the one that started at 100ºC freezes first. Why?''}~\cite{mpemba:69}.

Fortunately, Osborne did not dismiss the Mpemba's claim although he confessed that he thought at the first moment that Mpemba was wrong. Indeed, he asks a technician to make the experiment. The result was the following~\cite{mpemba:69}:

\textit{``The technician reported that the water that started hot did indeed freeze first and added in a moment of unscientific enthusiasm: But we'll keep on repeating the experiment until we get the right result.''}

Of course, further tests led to the same results and they started to think about an explanation that we will briefly sketch in the next section. The same year of the publication of the article of Mpemba and Osborne, Dr. Kell in Canada reported the same experiment~\cite{kell:69}.

A curious fact is that this phenomenon, though very far from academic physics, was conserved in the popular heritage through the years. Indeed, Mpemba remembers that the ice-cream makers in his town were aware of this effect and they used it commonly to obtain their ice-cream faster~\cite{mpemba:69}.

Several explanations came up in the following years (see~\refsec{water_complicated_mpemba}) and the phenomenon was found to be not exclusive from water. Indeed, it was found in nanotube resonators~\cite{greaney:11}, clathrate hydrates~\cite{ahn:16}, granular fluids~\cite{lasanta:17}, colloidal systems~\cite{kumar:20} and, as we will expose in this chapter, in spin glasses~\cite{janus:19}.

\section{Water is too complex} \labsec{water_complicated_mpemba}
After the work of Mpemba and Osborne~\cite{mpemba:69}, several explanations were proposed, however, one of the main difficulties in the study of the Mpemba\index{Mpemba effect} effect is the high number of parameters that may affect the phenomenon. The shape of the beaker, the composition of the water, its temperature distribution \dots All of these parameters might play an important role in the explanation of the effect and should be taken into account.

The mass of the water may be one of these parameters that could (at least partially) explain the effect. The hotter system shall lose more mass due to evaporation processes than the colder one and it has been suggested~\cite{kell:69} that this would be the reason for the Mpemba\index{Mpemba effect} effect to occur. However, other experiments claimed that this mass loss would be insufficient to explain Mpemba\index{Mpemba effect} effect~\cite{osborne:79,freeman:79,wojciechowski:88}.

Another parameter that may play a role in the phenomenon is the temperature distribution of the water. Cold water is denser than hot water\footnote{Above 4ºC.} and, when preparing the hot beaker, if the heating process is not uniform that may induce convection currents in it. Due to those convection currents and the different densities of the water as a function of the temperature, the top part of the hot beaker would be at a lower temperature than the bottom part. This could favor the creation of a layer of ice on the top of the hot beaker before than in the colder one. Besides, the convection currents may work together with other factors, like the above-mentioned evaporation, to provoke the Mpemba\index{Mpemba effect} effect in the water. These convection currents would in turn be affected by other parameters like the shape of the beaker. Actually, experiments in which the hotter beaker was stirred in order to make the temperature gradient disappear, showed a sizable raise of the time of freezing~\cite{deeson:71}.

The above examples are only two of the variety of explanations proposed for the Mpemba\index{Mpemba effect} effect. However, the situation is far from clear and there exist experiments claiming the nonexistence of the phenomenon, see for example~\cite{burridge:16}.

It is clear that the complexity of water makes it too difficult to study the Mpemba\index{Mpemba effect} effect and it would be desirable to have a much better-controlled system to study the phenomenon. Here, we take advantage of the numerical simulations in \gls{SG}s to address the Mpemba\index{Mpemba effect} effect and studying it with total control of the system.

\section{Numerical simulation}\labsec{numerical_simulation_mpemba}
In this work, we use the same simulations performed for the study of the aging\index{aging!rate} rate (\refch{aging_rate}) and we add some simulations with temperature-varying\index{temperature-varying protocol} protocols. We briefly remind here the parameters of the simulation for the reader's convenience.

We simulate in the \gls{FPGA}-based\index{FPGA} computer Janus\index{Janus} II an \gls{EA}\index{Edwards-Anderson!model} model in three-dimensional spin glasses (\refsubsec{3D_EA_model}) for several temperatures $T$ in a lattice of linear size $L=160$, which is aimed to represent a system of infinite size. 

We shall perform two different protocols. The isothermal\index{isothermal} protocol consists of a direct quench from configurations\index{configuration} of spins randomly initialized (which corresponds to infinite temperature) to the working temperature $T$, where the system is left to relax for a time $\tw$. This relaxation\index{relaxation} corresponds to the (very slow) growth of glassy magnetic domains\index{magnetic domain} of size $\xi(\tw)$. To uniquely identify this protocol, it is enough to label it with its temperature $T$.

The temperature-varying\index{temperature-varying protocol} protocol begins in the same way that the isothermal\index{isothermal} one, by quenching the system from configurations\index{configuration} of randomly initialized spins to the working temperature $T_1$. When the system reaches a certain coherence length\index{coherence length} $\xi'(\tw)$ we change the temperature of the thermal reservoir to a temperature $T_2$. Hence, this protocol should be labeled with a pair of temperatures (the initial one $T_1$ and the final one $T_2$), and with the coherence length\index{coherence length} at which the temperature-change was produced $\xi'(\tw)$. We use the following notation $T_1,\xi' \to T_2$.

We compute a total of $\NS = 16$ different samples\index{sample}. For each sample\index{sample}, we shall consider $\NRep=256$ replicas\index{replica}. As said in~\refch{aging_rate}, this simulation had the original aim to study the temperature chaos\index{temperature chaos} phenomenon under non-equilibrium conditions (see~\refch{out-eq_chaos}), however, the reader may notice that in that study we use a total number of replicas\index{replica} $\NRep=512$. Indeed, this study about the Mpemba\index{Mpemba effect} effect was performed much earlier and we had at our disposal ``only'' $\NRep=256$.

The main observables of this study are the coherence length\index{coherence length} $\xi(T,\tw)$ and the energy\index{energy!density} density $e(t)$. The coherence length\index{coherence length} is estimated by $\xi_{12}(T,\tw)$, computed from integral estimators of the correlation function $C_4(T,r,\tw)$. These two observables have been described with great detail in \refsubsec{observables_introduction}. The energy\index{energy!density} density is defined as
\begin{equation}
e(t,{\mathcal{J}}) = \dfrac{1}{L^3} \braket{\mathcal{H}_{\mathcal{J}}(t)} \quad  \, \quad e(t) = \overline{e(t,\mathcal{J})} \, , \labeq{energy-density_definition}
\end{equation}
where $\overline{(\cdots)}$ is the exact mean over the disorder\index{disorder!average}. Although this estimation is perfectly correct, we decided to increase the accuracy of our estimations by using a control variate\index{control variate}~\cite{fernandez:09,ross:14} $\ecvj$ depending on the sample\index{sample} $\mathcal{J}$ and with an exact disorder\index{disorder!average} average $\mu_{\ecvj}$. 

The studied quantity would be now
\begin{equation}
\tilde{e}(t,\mathcal{J}) = e(t,{\mathcal{J}}) - \left[ \ecvj - \mu_{\ecvj} \right] \quad \, \quad \tilde{e}(t) = \overline{\tilde{e}(t,\mathcal{J})} \, . \labeq{energy-density_control_variate}
\end{equation}

This new quantity $\tilde{e}(t,\mathcal{J})$ has the same disorder\index{disorder!average} mean that the usual energy-density\index{energy!density} $e(t,\mathcal{J})$ but has a significantly lower variance. The reader may find further details of the implementation of the control variate\index{control variate} in \refsec{improving_statistics}.

\section{Identifying the Mpemba Effect}\labsec{identifying_mpemba}
The aim of identifying the Mpemba\index{Mpemba effect} effect in \gls{SG} is composed of two main tasks. Firstly, we have to identify what ``temperature'' means in an out-of-equilibrium \gls{SG}. Secondly, we have to establish a protocol to mimic the traditional protocol of the classic Mpemba\index{Mpemba effect} effect. 

The natural candidate to take the place of the temperature, which is telling us if a system is hotter than another, is the energy-density\index{energy!density} $e(t)$ [or equivalently, $\tilde{e}(t)$] because it is the observable conjugated with (the inverse of the) temperature. Furthermore, at equilibrium, where the temperature $T$ of the thermal reservoir corresponds to the temperature of the system by definition, there exists a monotonically increasing correspondence between the energy-density\index{energy!density} and the temperature.

The protocol followed in our numerical experiment strongly resembles the original Mpemba's protocol~\cite{mpemba:69}. We study the evolution of three different off-equilibrium systems. The first one is quenched from infinite temperature (random configuration\index{configuration}) to a temperature (of the thermal reservoir) $T_1=1.3$, which is above the critical\index{critical temperature} temperature $\Tc = 1.102(3)$~\cite{janus:13}. This system is labeled with the number $1$. We let it evolve until it reaches an energy\index{energy!density} $\tilde{e}_1(t=0)\approx -1.6428$ and we set this time as our starting point of the (numerical) experiment. 

The second system (labeled with the number $2$) is prepared in a similar way but the temperature of the reservoir is now $T_2=1.2$ and we let the system reach a much lower energy\index{energy!density} $\tilde{e}_2(t=0)\approx -1.6714$. 

In the last one, labeled with the number $3$, the temperature of the reservoir is again $T_3=1.2$ but the starting point is at even lower energy\index{energy!density} $\tilde{e}_3(t=0) \approx -1.6738$. 

At that point, we quenched the three systems to a temperature $T_{\mathrm{f}} =0.7 \approx 0.64\Tc$, we let them evolve and we record their energies. The results are shown in \reffig{first_mpemba} where we can appreciate the classical Mpemba\index{Mpemba effect} effect. The \textit{hot start} system (system 1) crosses the other two curves in a way indicating a faster cooling process.

\begin{figure}[t!]
\centering
\includegraphics[width=0.8\columnwidth]{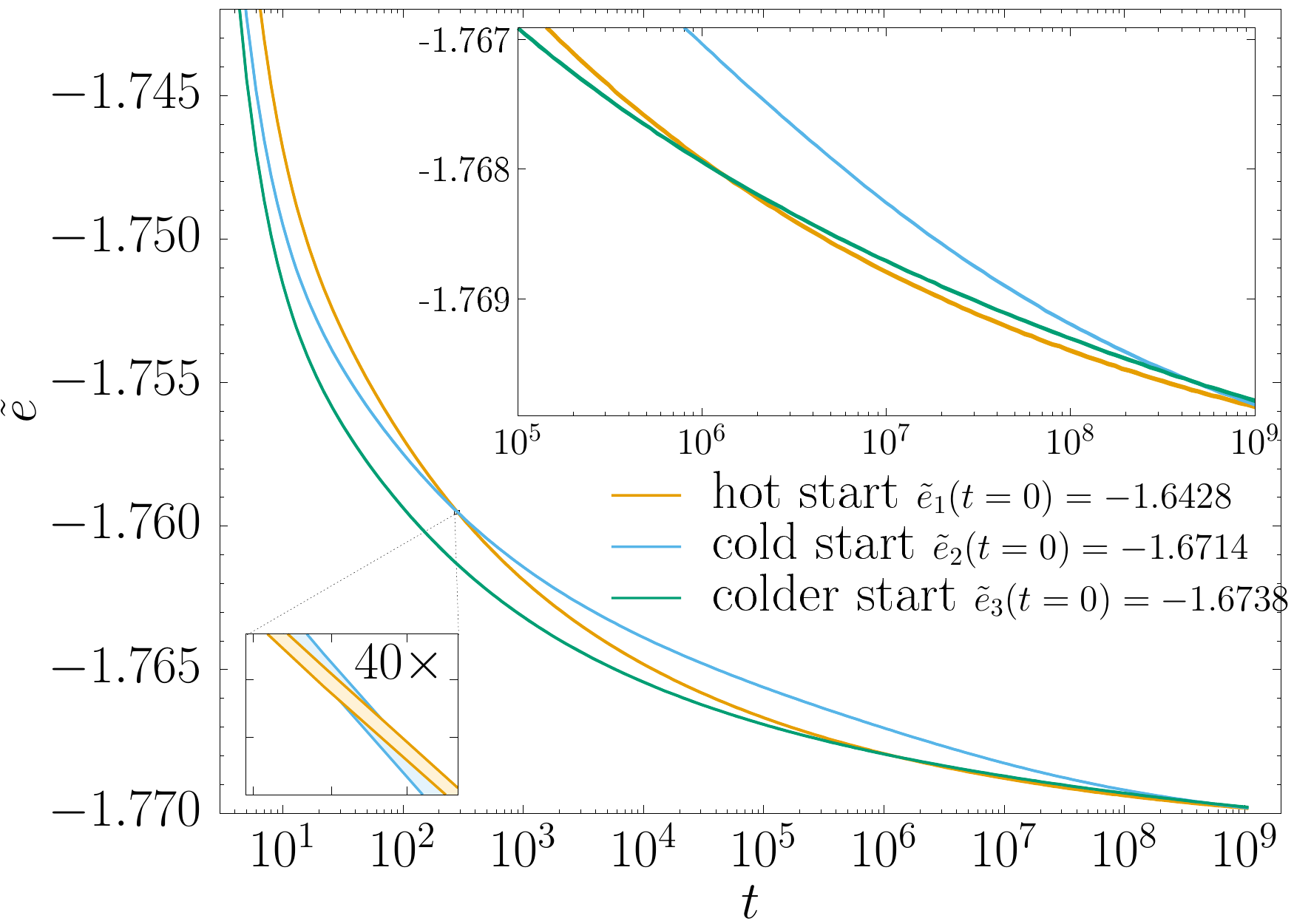}
\caption[\textbf{Classical Mpemba protocol}]{\textbf{Classical Mpemba protocol}. We show the time evolution of the energy\index{energy!density} of spin-glass systems initially prepared at a higher temperature ($T=1.3$, yellow line) or a lower temperature ($T=1.2$, blue and green lines), but always in the paramagnetic (high-temperature) phase\index{phase!high-temperature/paramagnetic} ($T_\text{c}\approx1.102$). In all three cases, the systems are initially left to evolve out of equilibrium until they reach the internal energies shown in the figure key. At $t=0$ all preparations are quenched, that is, put in contact with a thermal reservoir at temperature $T=0.7\approx0.64 T_\text{c}$.  As discussed in the text, the instantaneous energy-density\index{energy!density} is a measure of the (off-equilibrium) sample\index{sample} temperature. In agreement with the original Mpemba experiment~\cite{mpemba:69}, the system originally at the higher energy\index{energy!density} cools faster. \textbf{Bottom left inset:} Closeup of the first crossing between energy\index{energy!density} curves, showing the very small error bars\index{error bars}, equal to the thickness of the lines. \textbf{Top right inset:} Closeup of the second crossing between energy\index{energy!density} curves.}
\labfig{first_mpemba}
\end{figure}

The reader should notice that the crossing of $\tilde{e}_1(t)$ and $\tilde{e}_2(t)$ takes place at much longer times that the crossing of $\tilde{e}_1(t)$ and $\tilde{e}_3(t)$, even if the initial energies of $\tilde{e}_2(t)$ and $\tilde{e}_3(t)$ differ only by a $0.15\%$. We need a control parameter that helps us to quantitatively characterize the Mpemba\index{Mpemba effect} effect.

\section{Coherence length controls the Mpemba Effect in spin glasses}\labsec{coherence_length_mpemba}
The natural candidate, that characterizes the dynamical state of an off-equilibrium \gls{SG}, is the coherence length\index{coherence length} $\xi(t)$. Indeed, in terms of the coherence length\index{coherence length} $\xi(t)$ our three systems are very different. The \textit{hot start} system [$T_1=1.3$ and $\tilde{e}_1(t=0)\approx -1.6428$] has $\xi_1(t=0)=12$, the \textit{cold start} system [$T_2=1.2$ and $\tilde{e}_2(t=0)\approx -1.6714$] has $\xi_2(t=0)=5$ and the \textit{colder start} system [$T_3=1.2$ and $\tilde{e}_3(t)\approx -1.6738$] has $\xi_3(t=0)=8$.

This perspective is pointing us that the out-equilibrium \gls{SG}s for the study of the Mpemba\index{Mpemba effect} effect should not be labeled only with the temperature of the thermal reservoir, not even with the temperature of the thermal reservoir plus the energy density\index{energy!density} at some time $t$. We need the coherence length\index{coherence length} to fully characterize the state of the system and understand the Mpemba\index{Mpemba effect} effect in \gls{SG}s. Next, we test this hypothesis.

\subsection{A first test} \labsubsec{first_test_mpemba}
If our hypothesis is correct, crossing the critical\index{critical temperature} temperature should not matter in order to observe the Mpemba\index{Mpemba effect} effect. We only require that both starting points fulfill the next conditions: $\TA > \TB$ and $\xiA > \xiB$.

To test this hypothesis we focus on the low-temperature phase\index{phase!low-temperature/spin-glass}. We set the final temperature $T_\mathrm{f}=0.7$ and we simulate $4$ different systems, $3$ of them with the temperature-varying\index{temperature-varying protocol} protocol described in \refsec{numerical_simulation_mpemba} and the other with the isothermal\index{isothermal} protocol. We use the temperature-varying\index{temperature-varying protocol} notation also for the isothermal\index{isothermal} protocol in this case to stress that we set the time $t=0$ at a given coherence length\index{coherence length} $\xi$. The protocols are
\begin{itemize}
\item \textbf{Preparation A:} $T=0.7, \xi=6 \to T=0.7$ (this is the isothermal\index{isothermal} protocol).
\item \textbf{Preparation B:} $T=0.9, \xi=5 \to T=0.7$.
\item \textbf{Preparation C:} $T=0.9, \xi=8 \to T=0.7$.
\item \textbf{Preparation D:} $T=0.9, \xi=15 \to T=0.7$.
\end{itemize}

The time at which we quench the four systems to $T=0.7$ (in the case of the isothermal\index{isothermal} protocol simply corresponds to the time at which the system reaches $\xi=6$) will be $t=0$. Then, we plot $\tilde{e}(t)$ against time for the four preparations and we show the results in \reffig{test_mpemba}.

\begin{figure}[h!]
\centering
\includegraphics[width=0.8\columnwidth]{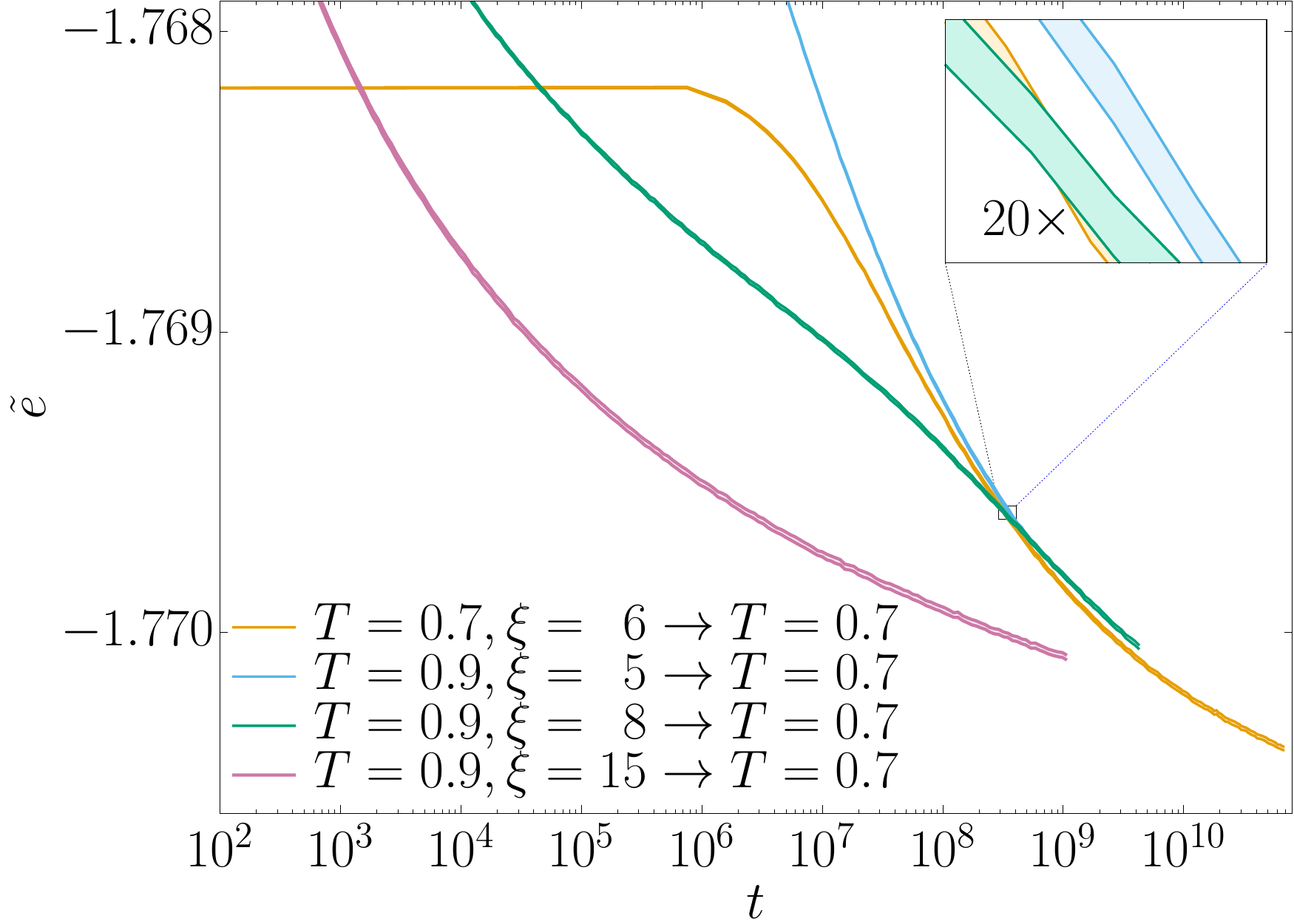}
\caption[\textbf{Mpemba effect in the spin-glass phase.}]{\textbf{Mpemba\index{Mpemba effect} effect in the spin-glass phase\index{phase!low-temperature/spin-glass}.} As in \reffig{first_mpemba}, but all four initial preparations are now carried out in the spin-glass phase\index{phase!low-temperature/spin-glass} ($T<T_\text{c}$). The preparation that cools faster is not the initially coldest one, but the one with the largest initial coherence length\index{coherence length}. \textbf{Inset:} A zoom of the \emph{second} crossing point between the curves for preparations $(T=0.9,\xi=8$; green curve) and $(T=0.7,\xi=6$; yellow curve). This second crossing is not the Mpemba\index{Mpemba effect} effect. Rather, the Mpemba\index{Mpemba effect} effect arises at the \emph{first} crossing at $t\approx 5\times 10^4$. The second crossing disappears if one plots parametrically $\tilde{e}(t)$ as a function of $\xi(t)$.}
\labfig{test_mpemba}
\end{figure}

Before initiating our discussion, we want to clarify the figure. The reader may wonder about the peculiar behavior of the curve belonging to the isothermal\index{isothermal} protocol. This peculiar behavior is just an optical artifact. Indeed, the times at which we record the configuration\index{configuration} of our system and measure the relevant quantities are exponential ($t=\Floor{2^{n/8}}$, with $n$ integer). Thus, for $T=0.7,\xi=6$ the measuring times are quite far from each other and the first point roughly corresponds to $t\approx 7 \cdot 10^5$. However, between two consecutive measurements, the change of the energy-density\index{energy!density} is very small and that explains the almost horizontal line.

We observe that preparations at initial temperature $T=0.9$ with coherence length\index{coherence length} $\xi>6$, namely preparation C and preparation D, cool faster than preparation A. Indeed, we observe the Mpemba\index{Mpemba effect} effect for preparation D at time $t\approx 10^3$, when it crosses the isothermal\index{isothermal} preparation. For preparation C we observe the same effect at time $t \approx 5 \cdot 10^4$. Nevertheless, for preparation B we observe no Mpemba\index{Mpemba effect} effect.

It is worthy to mention that a second crossing can be observed for higher times $t\approx 5\cdot 10^8$. This crossing does not correspond to the Mpemba\index{Mpemba effect} effect. Actually, we will observe that this crossing disappears under an appropriate representation.

\subsection[The $\tilde{e}-\xi$ phase-diagram]{The \boldmath $\tilde{e}-\xi$ phase-diagram}\labsubsec{e_xi_phase_diagram_mpemba}

Although we have identify the coherence length\index{coherence length} as the hidden parameter controlling the Mpemba\index{Mpemba effect} effect, we need to explore the relation between them to make our interpretation quantitative. Numerical and heuristic arguments \cite{marinari:96,parisi:97,janus:09b} suggest
\begin{equation}
\tilde{e}(t) = \tilde{e}_{\infty}(T) + \dfrac{e_1}{\xi^{d_\mathrm{L}}(t)} + \cdots \, , \labeq{energy_coherence_length_relation}
\end{equation}
where $d_{\mathrm{L}}\approx 2.5$~\cite{boettcher:05,maiorano:18} is the lower critical dimension\index{critical dimension!lower} at zero magnetic field, and the dots stand for scaling corrections, subdominant for large $\xi$. This relation makes sense only for the \gls{SG} phase\index{phase!low-temperature/spin-glass}~\cite{parisi:88}.

\subsubsection{The isothermal protocols}

\begin{figure}[t!]
\centering
\includegraphics[width=0.8\columnwidth]{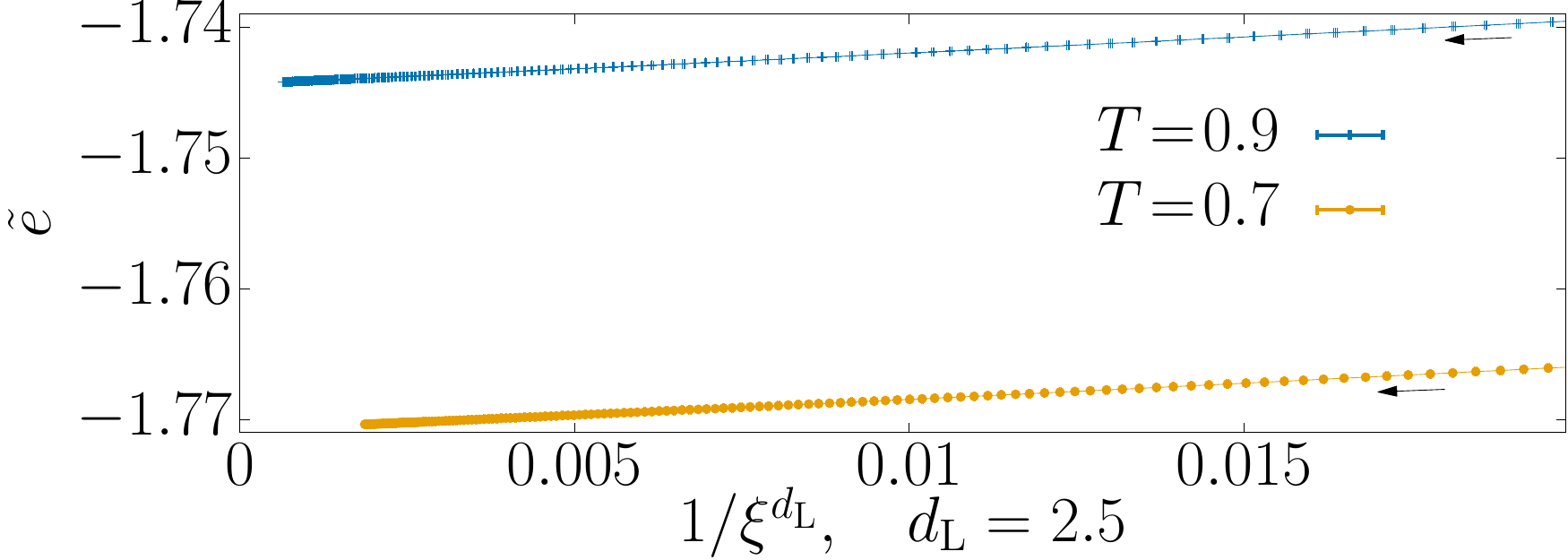}
\caption[\textbf{\boldmath Relationship between the energy density\index{energy!density} $\tilde{e}$ and the  $\xi$ for isothermal\index{isothermal} protocols.}]{\textbf{\boldmath Relationship between the energy density\index{energy!density} $\tilde{e}$ and the coherence length\index{coherence length} $\xi$ for isothermal\index{isothermal} protocols.} As suggested by~\refeq{energy_coherence_length_relation} for isothermal\index{isothermal} relaxations\index{relaxation} $\tilde{e}$ is an essentially linear function of $1/\xi^{2.5}$, (at least for the plotted range of  $\xi>4.8$). Furthermore, the dependence of the slope on temperature is marginal.}
\labfig{phase_diagram_isothermal}
\end{figure}

We test the relation defined by \refeq{energy_coherence_length_relation} in \reffig{phase_diagram_isothermal} by plotting the energy-density\index{energy!density} $\tilde{e}$ against $1/\xi^{d_{\mathrm{L}}}$, arrows indicate the direction of the points for increasing $t$. First, we observe that the isothermal\index{isothermal} protocols ($T=0.7$ and $T=0.9$) are (almost) straight lines in our representation. In addition, both isothermal\index{isothermal} protocols are (almost) parallel to each other. Of course, we need to make these observations quantitative. To that purpose we fit our data to
\begin{equation}
\tilde{e}(t) = \tilde{e}_{\infty}(T) + \dfrac{e_1}{\xi^{d_\mathrm{L}}(t)} + \dfrac{e_2}{\xi^{2d_\mathrm{L}}(t)} \, , \labeq{quadratic_fit_mpemba}
\end{equation}
which is just~\refeq{energy_coherence_length_relation} with a simple quadratic correction in $1/\xi^{d_\mathrm{L}}(t)$ that would be negligible for large $\xi$. 

Because this subdominant term (the quadratic one) becomes less important for the interesting limit (the large-$\xi$ limit), we decide to establish an objective criterion to select the fitting range. We perform the fit for the range $[\xi_{\min},\xi_{\max}]$ by setting $\xi_{\max}$ to the maximum $\xi$ simulated and by varying $\xi_{\min}$. We set $\xi_{\min}$ to be the lowest value of $\xi$ that stabilizes the values of $\tilde{e}_{\infty}$, $e_1$ and $e_2$ (within the error bars\index{error bars}), and, for the desired temperatures $T=0.7$ and $T=0.9$ we found $\xi_{\min}=6$. In addition, to describe the quality of the fit, we report the figure of merit $\chi^2$/d.o.f.\index{degree of freedom} The results can be consulted in \reftab{fit_results_mpemba}

\begin{table}[b!]
\begin{tabular}{ccccc}
\toprule
\toprule
$T$ & $\tilde{e}_{\infty}$ & $e_1$ & $e_2$ & $\chi^2/$d.o.f.\index{degree of freedom} \\
\hline
0.7 & $-1.7708070(7)$ & $0.2217(3)$ & $1.17(2)$ & $20.9(1)/119$ \\
0.9 & $-1.7443347(6)$ & $0.2251(3)$ & $1.08(2)$ & $11.0(4)/118$ \\
\bottomrule
\end{tabular}
\caption[\textbf{Mpemba parameters of the quadratic fit.}]{\textbf{Mpemba parameters of the quadratic fit.} We report the results of the fits to \refeq{quadratic_fit_mpemba}. For each fit, the figure of merit $\chi^2$/d.o.f.\index{degree of freedom} is also reported. Errors are computed by using the Jackknife\index{Jackknife} method.}
\labtab{fit_results_mpemba}
\end{table}

As we said, both curves are almost parallel ($e^{T=0.7}_1/e^{T=0.9}_1 \approx 0.9849$) and are also straight lines, because the effect of the curvature is around 1\% of the effect of the linear term $e_1/\xi^{d_{\mathrm{L}}}$ for $\xi \approx8$ that is a typical coherence length\index{coherence length} in our data.

\subsubsection{The temperature-varying protocols}

\begin{figure}[t!]
\centering
\includegraphics[width=0.8\columnwidth]{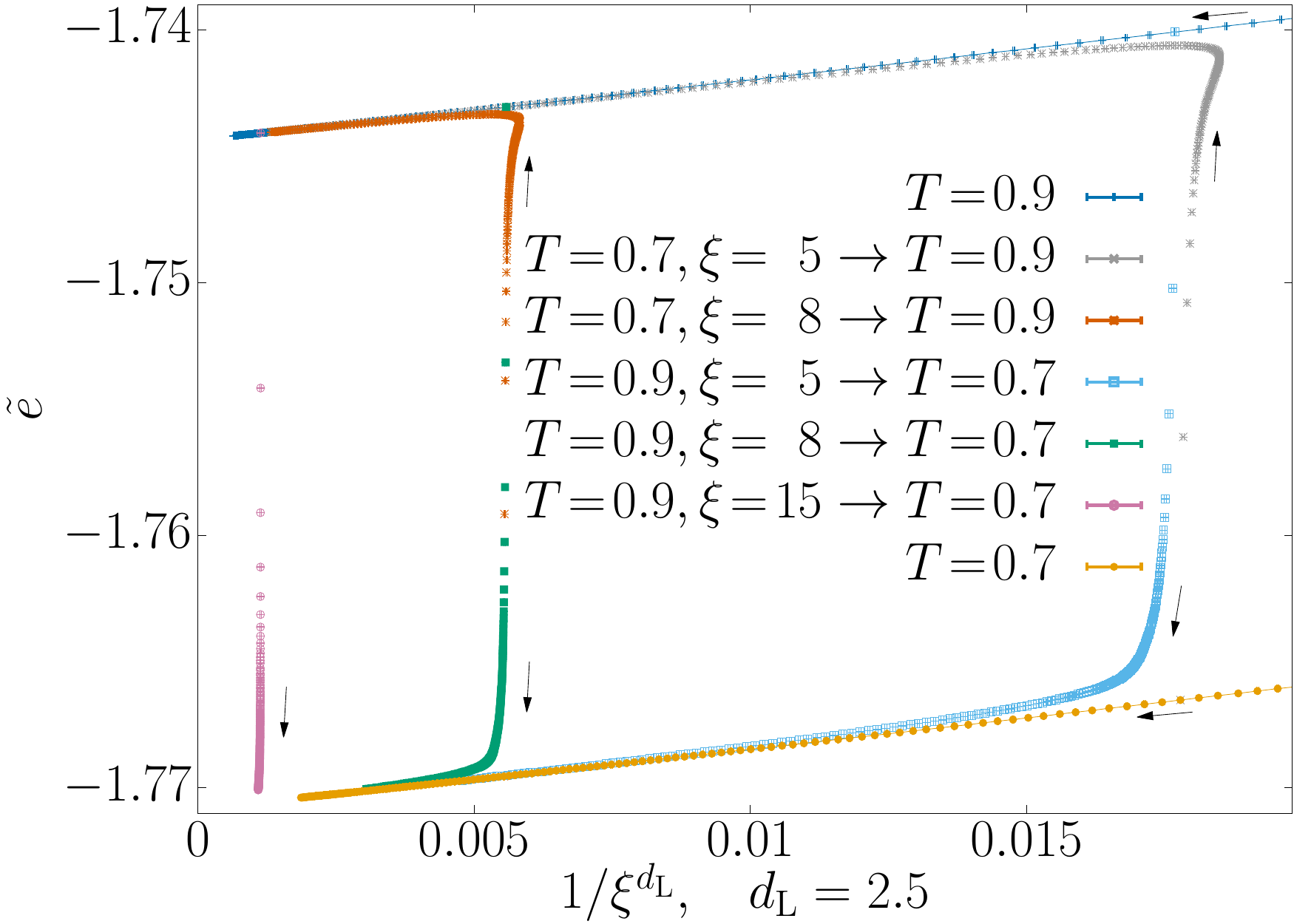}
\caption[\textbf{\boldmath Relationship between the energy density\index{energy!density} $\tilde{e}$ and the coherence length\index{coherence length} $\xi$ for temperature-varying\index{temperature-varying protocol} protocols.}]{\textbf{\boldmath Relationship between the energy\index{energy!density} density $\tilde{e}$ and the coherence length\index{coherence length} $\xi$ for temperature-varying\index{temperature-varying protocol} protocols.} Temperature-varying\index{temperature-varying protocol} protocols are seen to be essentially vertical moves between the straight lines corresponding to isothermal\index{isothermal} relaxations\index{relaxation} at the initial and final temperatures. These vertical moves are very quick initial transients, in which (in moves to higher temperatures only), $\xi$ slightly decreases and then increases again.}
\labfig{phase_diagram}
\end{figure}

We add the temperature-varying\index{temperature-varying protocol} protocols to the analysis, see \reffig{phase_diagram}. Those protocols where the temperature of the thermal reservoir decreases correspond to preparations B, C, and D in \reffig{test_mpemba}. We can see that the time-scales of the energy-density\index{energy!density} and the coherence length\index{coherence length} are totally decoupled. The energy-density\index{energy!density} $\tilde{e}$ is a \textit{fast variable} and, as a first approximation, when a quick temperature change takes place, $\tilde{e}$ instantaneously takes the value of the energy-density\index{energy!density} corresponding to its new thermal reservoir. However, the coherence length\index{coherence length} $\xi$ is a \textit{slow variable} that basically remains unchanged when a temperature change takes place. The combination of both effects is translated into almost vertical movements between isothermal\index{isothermal} protocols in \reffig{phase_diagram}.

In this representation, the crossing points in \reffig{test_mpemba} are not so evident. Now, the temperature-varying\index{temperature-varying protocol} protocols experiment a very fast decrease of the energy-density\index{energy!density}, while the isothermal\index{isothermal} protocols need longer times (equivalently, longer coherence length\index{coherence length}) to reach those values of the energy-density\index{energy!density} and, therefore, the temperature-varying\index{temperature-varying protocol} protocol ``cools'' faster. Of course, the previous analysis is a simplification, and measurable (still small) transient effects can be seen in \reffig{phase_diagram}, however, it provides a very simple explanation of the Mpemba\index{Mpemba effect} effect.

The curves corresponding to an increase in the temperature of the thermal reservoir are analyzed next.

\section{The Inverse Mpemba Effect} \labsec{inverse_mpemba}
\begin{figure}[b!]
\centering
\includegraphics[width=0.8\columnwidth]{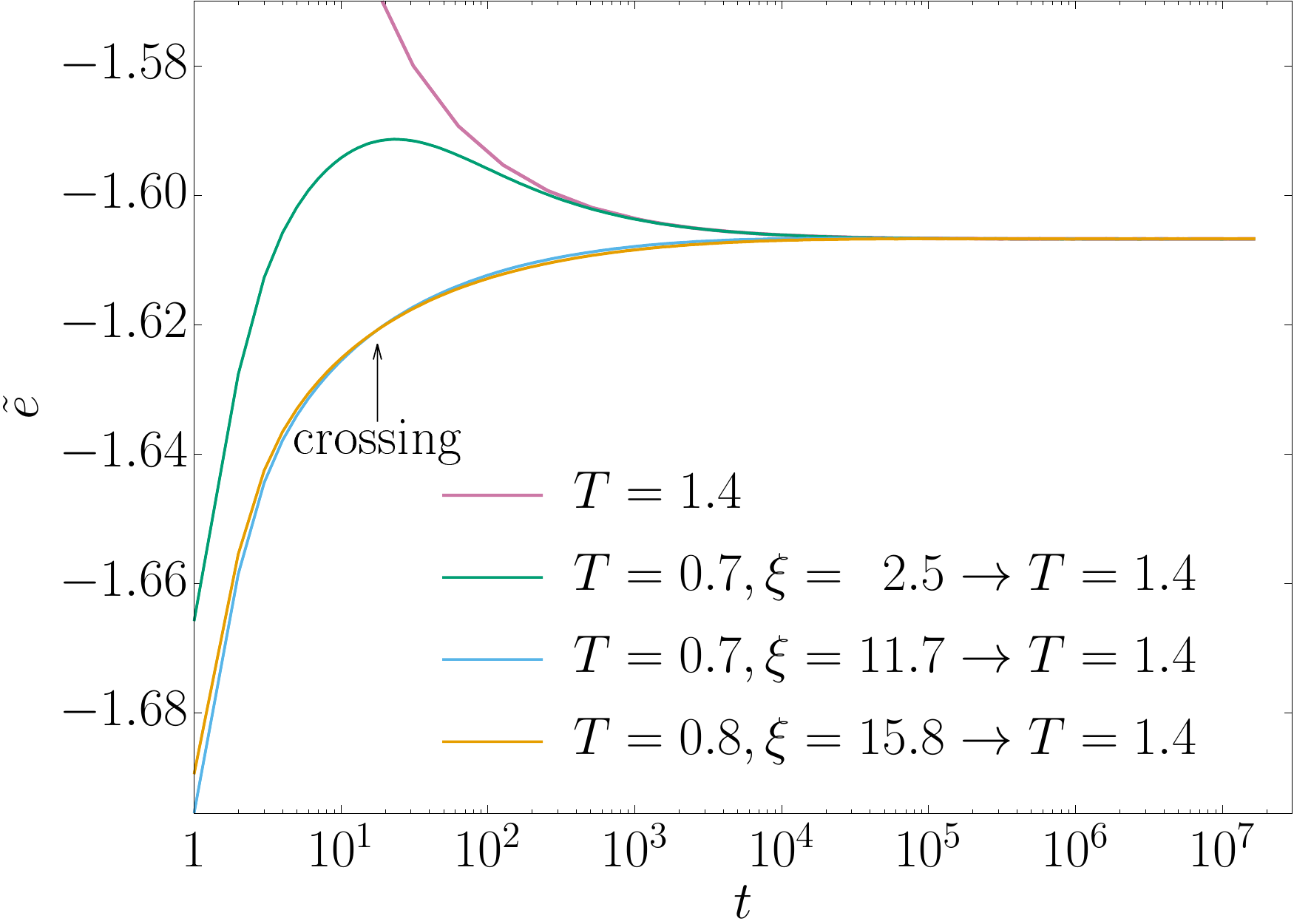}
\caption[\textbf{A tiny inverse Mpemba effect.}]{\textbf{A tiny inverse Mpemba\index{Mpemba effect} effect.} Time evolution of the energy\index{energy!density}, for the three different preparations (namely 1,2 and 3), compared with an isothermal\index{isothermal} protocol with $T=1.4$ (top curve). In the three preparations, the initial temperature is in the spin-glass phase\index{phase!low-temperature/spin-glass}, and the final temperature is $T=1.4>T_\mathrm{c}$. A very small Mpemba\index{Mpemba effect} effect is found at the time pointed by the arrow, only when warming up samples\index{sample} with similar starting energy\index{energy!density}.}
\labfig{inverse_me_ener}
\end{figure}

We focus now on the inverse Mpemba\index{Mpemba effect} effect protocol that was first suggested in~\cite{lu:17,lasanta:17}. Now, the final temperature of the thermal reservoir is chosen to be higher than the starting one. We see in \reffig{phase_diagram} that both curves corresponding to that protocol behave in a symmetrical way concerning to the classical protocol. In \reffig{phase_diagram} all the temperatures are below the critical one, and the natural question is, does the inverse Mpemba\index{Mpemba effect} effect survive for $T>\Tc$? The question is not trivial because \refeq{energy_coherence_length_relation} is not expected to hold for $T>\Tc$.

\begin{figure}[t!]
\centering
\includegraphics[width=0.8\columnwidth]{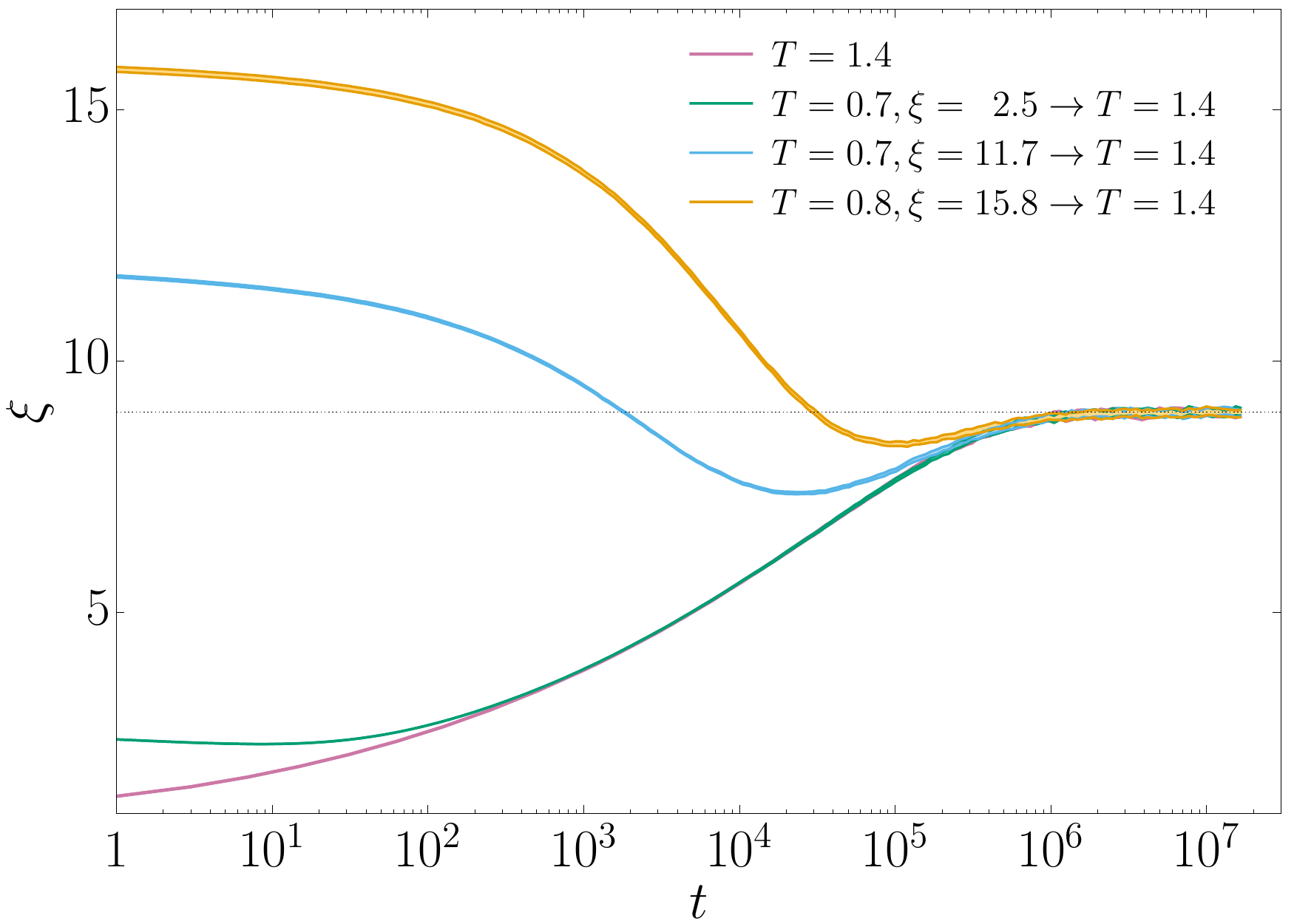}
\caption[\textbf{Coherence length: Undershooting and convergence to a master curve.}]{\textbf{coherence length\index{coherence length}: Undershooting and convergence to a master curve.} coherence length\index{coherence length}s $\xi$ of the experiments described in \reffig{inverse_me_ener}. The time evolution of $\xi$ tends to converge towards the curve corresponding to isothermal\index{isothermal} protocol with $T=1.4$ (bottom curve), giving rise to an undershoot of $\xi$ when its initial value is higher than the equilibrium $\xi$ at $T=1.4$.}
\labfig{inverse_me_xi}
\end{figure}

To answer this question, we use our temperature-varying\index{temperature-varying protocol} protocol, but this time the final temperature will be at $T>\Tc$. We propose three starting conditions:
\begin{itemize}
\item \textbf{Preparation 1:} $T=0.7, \xi=2.5 \to T=1.4$.
\item \textbf{Preparation 2:} $T=0.7, \xi=11.7 \to T=1.4$.
\item \textbf{Preparation 3:} $T=0.8, \xi=15.8 \to T=1.4$.
\end{itemize}
The reader should be aware that, although for $T<\Tc$ the coherence length\index{coherence length} grows without bonds\footnote{In an infinite system.}, this is not the case for $T>\Tc$. Specifically, for $T=1.4$ the asymptotic equilibrium value for the coherence length\index{coherence length} is $\xi=8.95(5)$. We also compare these temperature-varying\index{temperature-varying protocol} protocols with the isothermal\index{isothermal} protocol at $T=1.4$.

If we study the relaxation\index{relaxation} of the energy-density\index{energy!density} we can observe a small Mpemba\index{Mpemba effect} effect between protocols 2 and 3 for $t\approx 20$ (see \reffig{inverse_me_ener}). However, between protocols 1 and 3 or 1 and 2, the Mpemba\index{Mpemba effect} effect is clearly absent.

We can study also the relaxation\index{relaxation} of the coherence length\index{coherence length} $\xi$, see \reffig{inverse_me_xi}. In the paramagnetic phase\index{phase!high-temperature/paramagnetic}, the growth of the magnetic domains\index{magnetic domain} does not follow~\refeq{xi_powerlaw} as it becomes evident in the figure. In addition, we observe that all the protocols tend to the isothermal\index{isothermal} one very fast (for $t \leq 10^5$).

We can see here again that both time scales, the $\xi$ one and the $\tilde{e}$ one, are clearly decoupled. In \reffig{inverse_me_ener} and \reffig{inverse_me_xi} we can see that both quantities tend to their equilibrium values at very different time scales. Furthermore, if we focus on protocol 2 we can see that the undershoot present for the coherence length\index{coherence length} does not correspond to a similar behavior for the energy-density\index{energy!density}. Although for $T>\Tc$ the Mpemba\index{Mpemba effect} effect is strongly suppressed, this decoupling between both time scales seems to be necessary for the Mpemba\index{Mpemba effect} effect to take place.

\addpart{Temperature Chaos}
\chapter[Sweet introduction to Temperature Chaos]{Sweet introduction to \\ Temperature Chaos} \labch{Introduction_chaos}

\setlength\epigraphwidth{.5\textwidth}
\epigraph{\textit{Una racha de viento nos visitó \\
Pero nuestra veleta ni se inmutó.\\
La canción de que el viento se parara \\
Donde nunca pasa nada.}}{-- Extremoduro, \textit{Dulce introducción al caos}}

Before the second half of the XIX century, it was commonly accepted that the \textit{predictability} of a physical system was only constrained for technical reasons such as the limited knowledge of the position and speed of the particles. In the last part of the XIX century, however, Henri Poincaré, in his geometrical study of the stability of the Solar System, introduced the idea of the extreme sensibility of a system to small changes on its initial conditions.

That idea was relatively forgotten in the mainstream physics literature until $1963$ when Lorenz \cite{lorenz:63} realized that this extreme sensitivity was exhibited by a system of coupled differential equations. He was simulating a simplified model of convection rolls and he noticed that starting his numerical simulations from two slightly different initial conditions led to completely different results, even in relatively short times. This evidence about the impossibility of long-term predictions in certain systems was, indeed, very attractive for the physics community and, the interest in that research topic notoriously increased.

Although the concept of \textit{chaos} have considerably evolved through the years and it is well-defined in the mathematical context~\cite{hasselblatt:03}, it has been historically associated with the extreme sensitivity to small perturbations~\cite{strogatz:18}. \gls{SG}s borrow the term to describe the fragility of the glassy phase\index{phase!low-temperature/spin-glass} in response to perturbations.

The sensitivity of the \gls{SG} phase\index{phase!low-temperature/spin-glass} upon changes in the couplings\index{couplings}, namely \textit{disorder chaos}\index{disorder!chaos}~\cite{ney-nifle:97,ney-nifle:98,sasaki:05,katzgraber:07}, or in the external magnetic field~\cite{kondor:89,ritort:94,billoire:03}, have been widely studied and satisfactorily described. 

The temperature counterpart of this fragility is known as \gls{TC}\index{temperature chaos}, which means that the spin configurations\index{configuration} which are typical from the Boltzmann\index{Boltzmann!weight} weight at temperature $T_1$ are very atypical at temperature $T_2$ (no matter how close the two temperatures $T_1$ and $T_2$ are). This phenomenon has proved to be very elusive~\cite{bray:87b,banavar:87,kondor:89,kondor:93,ney-nifle:97,ney-nifle:98,billoire:00,mulet:01,billoire:02,krzakala:02,rizzo:03,sasaki:05,katzgraber:07,parisi:10,fernandez:13,billoire:14,wang:15,billoire:18,janus:21} and remains to be fully understood.

This chapter\footnote{The name of chapter is a small tribute to rock band Extremoduro and their song \textit{Dulce introducción al caos} (Sweet introduction to chaos).} pretends to be a very brief introduction to the \gls{TC}\index{temperature chaos} phenomenon by wandering through the main historical results in this field. The aim is to understand the starting points of the two following chapters (\refch{equilibrium_chaos} and \refch{out-eq_chaos}) that will be devoted to exposing the original results of this thesis in \gls{TC}\index{temperature chaos}.

\section{The origin of Temperature Chaos} \labsec{origin_tc}
The \gls{TC}\index{temperature chaos} phenomenon was originally predicted in finite-dimension \gls{SG}s by Bray, Moore, and Banavar\footnote{Although it was originally predicted in \gls{SG}s, other systems like polymers~\cite{sales:02,dasilveira:04} also exhibit it.}~\cite{bray:87,bray:87b,banavar:87} in the context of renormalization\index{renormalization group} studies and scaling arguments (the germ of the later-called \textit{droplets picture}\index{droplet!picture}, see~\refsubsec{theoretical_pictures}). For the sake of clarity, we briefly recall the main characteristics of the droplets\index{droplet!picture} to understand the emerging chaos feature in this theory (see for example \cite{katzgraber:07} for further details).

In the droplet\index{droplet!picture} picture, the low-temperature phase\index{phase!low-temperature/spin-glass} is understood in terms of excitation of the ground-state\index{ground-state}. The energy\index{energy} excitation occur through the so-called \textit{droplets}\index{droplet!picture}, which are compact domains\index{magnetic domain!compact} of spins with linear size $L$ that have been coherently flipped and whose boundaries are expected to be fractal, with a surface area of the order $L ^{d_s}$, $d-1 \leq d_s < d$. The free-energy\index{free energy} cost of generating such a droplet\index{droplet} is $F_L(T) \sim \Upsilon(T) L^{\theta}$ with $0<\theta<(d-1)/2$, being $\Upsilon$ and $\theta$ the stiffness modulus\index{stiffness!modulus} and the stiffness exponent\index{stiffness!exponent} respectively. The entropy\index{entropy} in the droplets\index{droplet!picture} picture scales with the size of the droplet\index{droplet} as $S = \sigma(T) L^{d_s/2}$, being $\sigma(T)$ the \textit{entropy\index{entropy!stiffness} stiffness}. 

The key to understanding the \gls{TC}\index{temperature chaos} in the droplets\index{droplet!picture} picture is to focus on the scaling behavior of the free energy\index{free energy}. One would \textit{naively} expect, for domains\index{magnetic domain!compact} of spin of volume $\sim L^{d_s}$, that the free-energy\index{free energy} would also scales as $\sim L^{d_s}$. However, as we have mentioned before, the free-energy\index{free energy} scales as $F_L(T) \sim L^{\theta}$ with $\theta < d_s$. This happens due to large cancellations of the contribution to the free\index{free energy} energy from different parts of the boundary of the droplet\index{droplet}, and this delicate equilibrium is the key to understand the \gls{TC}\index{temperature chaos} phenomenon. In this picture, \gls{TC}\index{temperature chaos} appears if the free energy\index{free energy} of a droplet\index{droplet} changes its sign upon a small change in the temperature. The length scale in which this happens is the so-called \textit{chaotic length} $\ell_c$.

The computation of $\ell_c$ is performed through thermodynamic arguments. The free\index{free energy} energy is $F(T) = U(T) - TS(T)$ being $U(T)$ the internal energy\index{energy} of the droplet\index{droplet} and $S(T)$ the entropy\index{entropy} of the same droplet\index{droplet}, both at temperature $T$. However, when the changes of temperature are small enough, the internal energy\index{energy} can be considered as an independent quantity with respect to the temperature, and therefore
\begin{equation}
F(T_2) = U(T_2) - T_2S(T_2) \approx U(T_1) - T_2S(T_2) = F(T_1) + T_1S(T_1) - T_2S(T_2) \, .
\end{equation}
Taking into account the scaling behavior of the free\index{free energy} energy and the entropy\index{entropy} in a droplet\index{droplet}, we can compute the length scale $\ell_c$ at which $F(T_2)$ inverts its sign
\begin{equation}
\ell_c = \left(\dfrac{\Upsilon(T_1)}{T_2 \sigma(T_2) - T_1 \sigma(T_1)}\right)^{1/\zeta} \quad \mathrm{being} \quad \zeta = d_s/2 - \theta \, . \labeq{def_chaotic_length}
\end{equation}
The usual approach\footnote{In~\cite{katzgraber:07} the authors avoid this simplification with small changes in the final result.} is to take $\sigma(T_2) \approx \sigma(T_1)$ when $\lvert T_2 - T_1 \rvert \ll 1$ and, therefore
\begin{equation}
\ell_c \sim \lvert T_2 - T_1 \rvert^{-1/\zeta} \, . \labeq{scaling_chaotic_length}
\end{equation}
The meaning of $\ell_c$ is clear: small changes in the temperature make domains\index{magnetic domain} of spins of length scales greater than $\ell_c$ to flip, leading to two separate regimes. On the one side, in the short-length regime, the chaos is absent or it is rather weak. On the other side, in the large-length regime, two equilibrium configurations\index{configuration} at temperatures $T_1$ and $T_2$ are completely uncorrelated, leading to a strong chaos phenomenon.

In this framework, a multitude of numerical work~\cite{ney-nifle:97,ney-nifle:98,aspelmeier:02,krzakala:04,sasaki:05,katzgraber:07,monthus:14} has been performed. Indeed, the scaling of the chaotic length showed in~\refeq{scaling_chaotic_length} was numerically found and the exponent $\zeta$ was computed. However, still this approach presents major problems. The equilibrium simulations performed at that time were limited to $L \sim 10$ which made the system to be in the $L \ll \ell_c$ regime, where the chaos is almost absent. Moreover, the scaling of~\refeq{scaling_chaotic_length} extends beyond the critical\index{critical temperature} temperature $\Tc$ where \gls{TC}\index{temperature chaos} should not occur. Thus, the numerical evidence supporting this picture is quite weak.

\section{Temperature Chaos in Mean Field}
In Mean-Field\index{Mean-Field!model} models\footnote{Those models that can be exactly solved through Mean-Field approximations.}, the \gls{TC}\index{temperature chaos} has proved to be particularly elusive. Specifically, the \gls{SK}\index{Sherrington-Kirkpatrick} model stoically resisted numerical attempts to characterize the \gls{TC}\index{temperature chaos} phenomenon~\cite{billoire:00,billoire:02}, later solved by~\cite{billoire:14} as we discuss below. The lack of numerical evidence of \gls{TC}\index{temperature chaos} and the publication of studies which, indeed, presented evidence against it~\cite{mulet:01,rizzo:01} led to the conclusion that \gls{TC}\index{temperature chaos} did not take place in the \gls{SK}\index{Sherrington-Kirkpatrick} model.

However, in 2003, a \textit{tour de force}~\cite{rizzo:03} showed that the \gls{SK}\index{Sherrington-Kirkpatrick} model presented an exceedingly small \gls{TC}\index{temperature chaos}, and it was necessary to compute up to the ninth order in a perturbative expansion in the replica\index{replica!symmetry breaking (RSB)} framework\footnote{Actually,~\cite{rizzo:01} found no \gls{TC}\index{temperature chaos} because the computations in this paper were performed ``only'' until the fifth order in the perturbation expansion.} to find it.

This study is based on the use of a large-deviation functional. The idea is that, under the \gls{TC}\index{temperature chaos} hypothesis, the overlap\index{overlap} between any pair of equilibrium states at temperatures $T_1$ and $T_2$ ($T_1 \neq T_2$) should be zero. Therefore, the shape of the probability distribution of overlaps\index{overlap!distribution} $q$ between equilibrium configurations\index{configuration} at different temperatures should tend to a Dirac's delta function peaked on $q=0$ as the size of the system grows. Moreover, the scaling of this probability distribution is given by the large-deviations formula
\begin{equation}
P(q) \sim \exp \left[-N\Delta F(q)\right] \, ,
\end{equation}
where $N$ is the size of the system and $\Delta F(q)$ takes account of the free-energy\index{free energy} cost of constraining two replicas\index{replica} to have a given mutual overlap\index{overlap} at equilibrium. This functional $\Delta F(q)$ is computed in~\cite{rizzo:03} through a perturbative approach. It is necessary to reach the ninth order to find a non-vanished term, hence, it was demonstrated that \gls{SK}\index{Sherrington-Kirkpatrick} model presents, though pathologically, the \gls{TC}\index{temperature chaos} phenomenon.

Nonetheless, the \gls{SK}\index{Sherrington-Kirkpatrick} model has not been the only Mean-Field\index{Mean-Field!model} model in which \gls{TC}\index{temperature chaos} has been studied. For example, it has been found that diluted Mean-Field\index{Mean-Field!model} \gls{SG}s present much stronger \gls{TC}\index{temperature chaos}~\cite{parisi:10}. Besides, in $p$-spin models, which are just a generalization of the \gls{SK}\index{Sherrington-Kirkpatrick} model in which interactions occur between $p \geq 3$ spins, different behaviors have been found. On the one hand, we have a recent mathematical proof of the absence of \gls{TC}\index{temperature chaos} in the homogeneous spherical $p$-spin model~\cite{subag:17}, in agreement with a previous claim based on physical arguments~\cite{kurchan:93}. On the other hand, \gls{TC}\index{temperature chaos} should be expected when one mixes several values of $p$~\cite{barrat:97}, as confirmed by a quite recent mathematical analysis~\cite{chen:14,panchenko:16,chen:17,arous:20}.

\section{Memory and rejuvenation}\labsec{memory_rejuvenation_introduction_chaos}
The reader may note that all the previous discussion about \gls{TC}\index{temperature chaos} assumes that it is an equilibrium phenomenon (since its very definition), however, most of the experimental work in spin-glasses is carried out under non-equilibrium conditions as we have already discussed in~\refsubsec{aging_memory_rejuvenation}\footnote{With the notable exception of
experiments in a thin-film geometry, see~\cite{guchhait:14}. In fact, the experimental study of \gls{TC}\index{temperature chaos} in thin films has been initiated~\cite{guchhait:15b}.}.

The spectacular rejuvenation\index{rejuvenation} and memory\index{memory effects} effects~\cite{jonason:98,lundgren:83,jonsson:00,hammann:00} (described in~\refsubsec{aging_memory_rejuvenation}) have been commonly related to the phenomenon of \gls{TC}\index{temperature chaos}~\cite{komori:00,berthier:02,picco:01,takayama:02,maiorano:05,jimenez:05}. Yet, the situation is far from clear.

The idea is that, due to the \gls{TC}\index{temperature chaos} phenomenon, the equilibrium configurations\index{configuration} at two slightly different temperatures $T_1$ and $T_2$ would be completely different, thus, the aging\index{aging} performed at temperature $T_1$ would not be useful at the temperature $T_2$, and the aging\index{aging} process restarts, leading to the so-called \textit{rejuvenation}\index{rejuvenation} phenomenon. The opposite behavior, the \textit{cumulative aging\index{aging!cumulative}}, means that the relaxation\index{relaxation} work carried out at temperature $T_1$ is still useful (partly useful at least) when the temperature is varied to $T_2$.

Indeed, some experiments~\cite{jonsson:02,bert:04} and most of the numerical work~\cite{komori:00,berthier:02,picco:01,takayama:02,maiorano:05,jimenez:05} trying to simulate temperature-varying\index{temperature-varying protocol} experimental protocols can be interpreted as cumulative aging\index{aging!cumulative}. At this point, opinions are split. Some authors find only cumulative aging\index{aging!cumulative} in their simulations~\cite{picco:01,takayama:02,maiorano:05}, while others find some traces of aging\index{aging} restart~\cite{komori:00,berthier:02}. However, this restarting of aging\index{aging} occurs on exceedingly short times~\cite{jimenez:05}. Perhaps more worryingly, it has been found numerically that a site-diluted ferromagnet (where no \gls{TC}\index{temperature chaos} is expected) behaves analogously to the spin glass~\cite{jimenez:05}.

If a strong connection between \gls{TC}\index{temperature chaos} and the experiments of memory\index{memory effects} and rejuvenation\index{rejuvenation} exists, some work is needed in order to establish it.

\section{Last steps}
This historical tour about \gls{TC}\index{temperature chaos} allows us to understand the general feeling in the field at the beginning of the 2010s. The \gls{TC}\index{temperature chaos} phenomenon seemed to be extremely weak, with gradual increasing effects that we were not able to perceive due to the difficulty of equilibrating large systems in the \gls{SG} phase\index{phase!low-temperature/spin-glass}. However, an alternative weak-\gls{TC}\index{temperature chaos} scenario~\cite{sales:02} could be compatible with the results. In this scenario, almost all the samples\index{sample} exhibit no \gls{TC}\index{temperature chaos} at all but a few of them suffer dramatic effects upon temperature changes. Actually, this scenario was not completely unknown and some numerical studies mentioned that situation~\cite{katzgraber:07}, but a quantitative study was lacking.

The main idea is that very few samples\index{sample} undergo \textit{chaotic events} i.e. at well-defined temperatures $T^*$, the samples\index{sample} suffer first-order\index{phase transition!first order} like transitions (rounded in finite-systems) such that the typical spin configurations\index{configuration} below and above $T^*$ differ. While the majority of samples\index{sample} do not have any chaotic event, some of them display one (or more) with a $T^*$ which seems to be located randomly within the \gls{SG} phase\index{phase!low-temperature/spin-glass}. Yet, the fraction of samples\index{sample} lacking chaotic events decreases upon increasing the system size. Indeed, one expects~\cite{rizzo:03,parisi:10} that the fraction
of samples\index{sample} lacking TC will decrease exponentially (in the system size).

It was in 2013 when the \gls{TC}\index{temperature chaos} was quantitatively studied as a rare-event-driven phenomenon~\cite{fernandez:13}. This numerical study was based on the study of large-deviations functional which was fundamental in order to deal with the wild sample-to-sample fluctuations\index{sample-to-sample fluctuations}. The mean problem in the numerical study of \gls{TC}\index{temperature chaos} in short-ranged \gls{SG} was, indeed, the statistical methods used to deal with the chaotic observables: the majority of non-chaotic samples\index{sample} killed any chaos signal when taking the disorder\index{disorder!average} average\footnote{As quoted by ``The Buggles'': \textit{Average Killed The Chaos Signal}.}.

Later, further works~\cite{billoire:14,martin-mayor:15,fernandez:16,billoire:18,janus:21} showed that the rare-event analysis was the appropriate protocol in order to study the \gls{TC}\index{temperature chaos} phenomenon. In the subsequent chapters (\refch{equilibrium_chaos} and \refch{out-eq_chaos}), we will develop two different rare-event analysis in 3D Ising\index{Ising} \gls{SG}s in order to study the \gls{TC}\index{temperature chaos} phenomenon.

\chapter{Dynamic variational study of Temperature Chaos} \labch{equilibrium_chaos}

The process of taking a \gls{SG} sample to equilibrium in numerical simulations requires the use of dynamic Monte\index{Monte Carlo} Carlo methods. Unfortunately, as we have already commented in \refsubsec{Monte_Carlo}, the sluggish dynamics exhibited by \gls{SG}s impedes the use of simple methods like the Metropolis-Hasting algorithm. One solution comes from the use of the \gls{PT} method, which equilibrates at once a set of $N$ copies of the system running at different temperatures.

However, the \gls{TC}\index{temperature chaos} phenomenon represents a major obstacle in the performance of \gls{PT}~\cite{fernandez:13}. In this chapter, we follow the ideas proposed in previous studies~\cite{fernandez:13,martin-mayor:15,fernandez:16} and we take advantage of that fact by quantitatively characterizing the \gls{TC}\index{temperature chaos} through a careful study of the process of thermalization\index{thermalization} of the system when using the \gls{PT} method. This work also extends the study of \gls{TC}\index{temperature chaos} performed in previous papers~\cite{fernandez:13,fernandez:16} by the development of a variational method\index{variational method}.

Moreover, we also focus on the very definition of \gls{TC}\index{temperature chaos} and we study it by comparing equilibrium configurations\index{configuration} at different temperatures. Both characterizations, namely dynamic and static, are found to correlate very well~\cite{fernandez:13,fernandez:16}. Here, we propose new observables to study the \textit{static} chaos and we found large correlations between the main observables of both characterizations, static and dynamic.

All the results exposed in this chapter came from the original work~\cite{billoire:18} which has been developed during this thesis.

\section{Numerical simulations} \labsec{numerical_simulations_eq_chaos}
In order to keep clean the rest of the chapter of technical details and to focus on the physical results, we explain here the simulations performed.

First of all, it is fundamental to mention that the data used here come from the study of the metastate\index{metastate} (see~\refch{metastate}) and, therefore, the structure of the couplings\index{couplings} is not conventional. We briefly recall here, for the reader's convenience, the particularities of this simulation.

The system, composed of $L^3$ spins, is divided into an inner region of $(L/2)^3$ spins and an outer region surrounding it. For each of the $10$ realizations of the inner disorder\index{disorder}, we have a set of $1280$ realizations of the outer disorder\index{disorder}. Hence, we have a total of $12800$ samples\index{sample} and for each one, we have simulated $\Nrep=4$ different replicas\index{replica}. 

A natural question is whether this particular setup's choice is affecting the results. One could imagine that those samples\index{sample} sharing the same inner disorder\index{disorder} would be strongly correlated and, hence, the statistics coming from only 10 different inner realizations could be not enough to deal with the sample-to-sample fluctuations\index{sample-to-sample fluctuations}. However, this choice is irrelevant for the studied observables in this work. The interested reader can find a detailed discussion in \refsec{selection_parameters}.

\begin{table}[t!]
\centering
\begin{tabular}{cccccccc}           
\toprule
\toprule
\multicolumn{7}{c}{MUSA-MSC} \\
\hline
$L$ & $L_{\text{int}}$ & $N_T$ & $T_{\mathrm{min}}$ & $T_{\max}$ & $N_\text{Met}$ ($ \times 10^6$) & $\text{ps/s}$  \\
\hline
24 & 12 & 24 & 0.698 & 1.538 & 500 & 104 \\ 
16 & 8 & 16 & 0.479 & 1.575 & 250 & 107  \\ 
16 & 8 & 13 & 0.698 & 1.575 & 250 & 119  \\ 
16 & 12 & 13 & 0.698 & 1.575 & 250 & 119 \\ 
14 & 12 & 13 & 0.698 & 1.575 & 500 & 120 \\
12 & 6 & 13 & 0.698 & 1.575 & 250 & 119  \\ 
8 & 4 & 13 & 0.698 & 1.575 & 250 & 126 \\
\bottomrule
\end{tabular}
\\[5mm]

\begin{tabular}{ccccccccc}  
\toprule
\toprule         
\multicolumn{8}{c}{MUSI-MSC} \\
\hline
$L$ & $L_{\text{int}}$ & $N_T$ & $N_\text{samp}$ & $N_\text{Met,min}$ & $N_\text{Met,mean}$ & $N_\text{Met,max}$ & $\text{ps/s}$ \\
\multicolumn{4}{c}{} & $\times 10^6$ & $\times 10^6$ & $\times 10^6$ & \\
\hline
24 & 12 & 24 & 2441 & 1000 & 4262 & 326000 & 57 \\ 
16 & 8 & 16  & 2898 & 500 & 5096 & 355500 & 304 \\ 
16 & 8 & 13 & 338 & 500 & 543 & 4000 & 306 \\ 
16 & 12 & 13 & 314  & 500 & 578 & 8000 & 306 \\ 
\bottomrule
\end{tabular}

\caption[\textbf{Parameters of the simulations MUSA-MSC and MUSI-MSC.}]{\textbf{Parameters of the simulations MUSA-MSC and MUSI-MSC.} $L$ is the lattice size; $L_\text{int}$ the size of the inner part of the lattice; $N_T$, $T_{\mathrm{min}}$ and $T_{\max}$ are the number of temperatures, the minimum and the maximum temperatures used in the \gls{PT} method; $N_\text{Met}$ is the number of Metropolis sweeps (at each temperature); $\text{ps/spin}$ is the average CPU time per spin-flip in MUSI-MSC, using an Intel Xeon CPU E5-2680 processors; $N_\text{samp}$ denotes the number of bad samples\index{sample} whose simulations had to be extended in order to thermalize and finally $N_\text{Met,min}$, $N_\text{Met,mean}$ and $N_\text{Met,max}$ are the minimum, mean and maximum number of Metropolis sweeps per temperature needed to reach thermalization\index{thermalization} (bad samples\index{sample}). The set of temperatures used is clearly the same in the MUSI-MSC and MUSA-MSC parts of this Table. The number of Metropolis sweeps between two consecutive \gls{PT} updates is always $N_\text{MpPT} = 10$.  For the MUSI-MSC simulation of $L=24$ we parallelized\index{parallel!computation}, using \emph{Pthreads}, by distributing the $N_T=24$ system copies among 12 CPU cores in the Intel Xeon CPU E5-2680.}
\labtab{parameters_simulation_MUSA_MUSI}
\end{table}

The samples\index{sample} have been equilibrated by using the \gls{PT} method with Metropolis updates between two consecutive \gls{PT} exchanges. We increase the performance of the Metropolis update via multispin coding and we apply two methods widely used in numerical simulations in statistical physics, namely the \gls{MUSA}\index{Multispin Coding!Multisample}~\cite{newman:99} and the \gls{MUSI}\index{Multispin Coding!Multisite}~\cite{fernandez:15}. The basic idea of these methods is the parallelization\index{parallel!computation} of operations that are, indeed, independent from each other by taken advantage of the streaming extensions of the current computer processors. Our simulations were carried out using either Intel Xeon E5-2680 or AMD Opteron Processor 6272. Further details can be found in \refch{AP_multispin_coding}.

The selection of the parameters of the simulation are in \reftab{parameters_simulation_MUSA_MUSI} and the reason for the choice of some of them is explained in~\refsec{selection_parameters}. Although all the simulations are included for completeness, some of them were only used in the metastate\index{metastate} study (see~\refch{metastate}). We focus here only in those simulations with $L_{\mathrm{int}} = L/2$. It is worthy to note that in most of this chapter, for the $L=16$ system, we are using the simulation with $N=16$ temperatures, barring the discussion on the impact of the minimum temperature of the \gls{PT} mesh in the \gls{TC}\index{temperature chaos} (see \refsec{correlation_dynamics_static}), where we will use the simulation with $N=13$.

\section{Monte Carlo, why have you forsaken me?}\labsec{Monte_Carlo_forsaken}
We have already sketched the main idea motivating this work. Traditional Monte\index{Monte Carlo} Carlo methods like the Metropolis-Hasting algorithm are not useful to study (at equilibrium) the low-temperature phase\index{phase!low-temperature/spin-glass} of a \gls{SG} because the presence of many free-energy\index{free energy!valley} local minima often causes the numerical simulation to get trapped and, as a consequence, the correct sampling of the phase space\index{phase space} gets severely harmed.

The \gls{PT} method solves this problem. The introduction of $N$ copies at different temperatures, and the possibility for each copy to exchange its temperature with another different copy, allows the copies to visit the high-temperature phase\index{phase!high-temperature/paramagnetic}, where it decorrelates very quickly from its previous state. When the copy \textit{comes back} to the low-temperature phase\index{phase!low-temperature/spin-glass}, it visits another different free-energy\index{free energy!valley} local minima and the performance of the thermalization\index{thermalization} process boosts.

The \gls{TC}\index{temperature chaos} phenomenon dramatically decreases that performance. Intuitively one can understand why the \gls{TC}\index{temperature chaos} represents a major obstacle in the \gls{PT} temperature flow~\cite{janus:10,fernandez:13,martin-mayor:15,fernandez:16}. Imagine we have two sets of configurations\index{configuration} (states) of two equilibrated systems at different temperatures $T_1<T_2$. If \gls{TC}\index{temperature chaos} occurs, then the typical configurations\index{configuration} will be very different for both temperatures. In the simpler scenario, we have a level crossing in which at a given temperature the free-energy\index{free energy} for both sets of configurations\index{configuration} are almost equal but their specific heats\index{specific heat} are rather different. Actually, this discrepancy in the value of the specific heat\index{specific heat} is very harmful to the \gls{PT} method (see, for example, \cite{rathore:05,katzgraber:06b,sabo:08,malakis:13}). As a consequence, to place a $T_1$\textit{-like} set of configurations\index{configuration} at $T_2$ is possible, but, due to the specific heat\index{specific heat} difference, it will have a hard traveling to higher temperatures (the inverse example is also true). In addition, equilibrium in \gls{PT} implies equilibrium at all the temperatures, hence, even a single \textit{dynamic chaotic event} as described above can ruin our expectations.

But these \textit{a priori} inconveniences are not a real reason to be sad. Actually, generally in Markov\index{Markov chain} chains, the exponential time $\tau_{\exp}$~\cite{sokal:97} tells us how long we have to simulate the system in order to reach the equilibrium. Those samples\index{sample} which suffer stronger \gls{TC}\index{temperature chaos} would also need longer times to thermalize and the study of the exponential time could help us to point those chaotic samples\index{sample}~\cite{fernandez:13,fernandez:16}. 

Unfortunately, $\tau_{\exp}$ has proven to be a very elusive quantity, and its accurate computation is not an easy task. It has been suggested that $\tau_{\exp}$ is best computed by studying the temperature-flow of the system copies of the \gls{PT} simulation~\cite{janus:10}. In the next section, we revisit the problem of the computation of $\tau_{\exp}$ and present a variational method\index{variational method} that can potentially save a large amount of time.

\section{Time scales in a Markov chain}\labsec{time_scales_eq_chaos}
We have stated in~\refsubsec{Monte_Carlo} that a well-behaved Monte\index{Monte Carlo} Carlo method, based on a Markov\index{Markov chain} chain, assures the system to reach the equilibrium (i.e. the configurations\index{configuration} of the system are distributed according to the Boltzmann-Gibbs\index{Boltzmann!-Gibbs distribution} distribution), no matter which particular initial condition we choose [recall~\refeq{thermalization_condition}]. However, knowing how long we need to simulate the system to reach that equilibrium is an important question.

Let us consider an equilibrated system, for a given quantity $f$ its \textit{autocorrelation function}\index{correlation function!time auto-} would be
\begin{equation}
C_f(t) = \mcav{f(t_1)f(t_2)} - \mcav{f(t_1)}^2 \, , \quad t=t_1 - t_2 \, , \labeq{unnormalized_autocorrelation_function}
\end{equation}
where $\mcav{\cdots}$ stands for the expectation value\footnote{It is worthy to note that $\mcav{\cdots}$ is different for the average $\braket{\cdots}$ defined in~\refeq{average_o}. It is true that, for an equilibrated system, the Markov\index{Markov chain} chain samples\index{sample} the configurational\index{configuration!space} space according to the Boltzmann-Gibbs\index{Boltzmann!-Gibbs distribution} distribution and, therefore, $\mcav{\cdots}$ can estimate $\braket{\cdots}$, but it is not true in general for a non-equilibrated system. We stress the difference between them by adopting different notations.} of the quantity $f$. The required equilibrium condition implies that $\mcav{f(t_1)} = \mcav{f(t_2)}$. 

The normalized autocorrelation function\index{correlation function!time auto-} $\hat{C}(f)$ is
\begin{equation}
\hat{C}_f(t) = \dfrac{C_f(t)}{C_f(0)} \, . \labeq{normalized_autocorrelation_function}
\end{equation}

This quantity decays exponentially for large $t$ as $\hat{C}_f(t) \sim \exp\left(-\abs{t}/\texpf\right)$. For each quantity we can extract the value of $\texpf$
\begin{equation}
\texpf = \limsup_{t \to \infty} \dfrac{t}{- \log \abs{\hat{C}_f(t)}} 
\end{equation}
and the supreme of $\texpf$ for every quantity $f$ 
\begin{equation}
\texp = \sup_f \texpf \, , \labeq{exponential_autocorrelation_time}
\end{equation}
is what we call \textit{exponential autocorrelation time}\index{autocorrelation time!exponential} which is the time scale ruling the transition of the system from an arbitrary configuration\index{configuration} to the equilibrium, that is, this is the quantity we are looking for.

Nonetheless, there are two main difficulties in the computation of $\texp$ from a practical point of view. The first one concerns the selection of the quantity $f$: we do not know which is the quantity $f$ that maximizes $\texpf$. The second one is a technical problem: the fit of a numeric autocorrelation function\index{correlation function!time auto-} to an arbitrary function from which we only know that its long-term behavior is an exponential decay is not an easy task. Procedures followed in previous works~\cite{janus:10} suggest considering the autocorrelation function\index{correlation function!time auto-} as a sum of two decaying exponentials from which we can extract the value of $\texp$. This procedure is very delicate because a large number of simulated samples\index{sample} requires an automatic way to obtain $\texp$ and this automatic protocol can fail, requiring human intervention and, most of the time, demanding to extend the simulations. The full protocol is carefully detailed in Appendix A of \cite{janus:10}.

It would be desirable to introduce a different time scale with better properties. At this point, we define the \textit{integrated autocorrelation time}\index{autocorrelation time!integrated} $\tintf$
\begin{equation}
\tintf = \dfrac{1}{2}+ \sum_{t=1}^\infty \hat{C}_f(t) \, . \labeq{integrated_autocorrelation_time}
\end{equation}

It is not hard to prove~\cite{sokal:97} that $\tintf$ controls the statistical errors in measuring the quantity $f$. Specifically, if an equilibrated system generates one configuration\index{configuration} at the Monte\index{Monte Carlo} Carlo time $t$, we need to wait until $t + 2\tintf$ to obtain a statistically independent configuration\index{configuration}, but only as far as the quantity $f$ is concerned.

The normalized autocorrelation function\index{correlation function!time auto-} can be expressed in terms of the eigenvalues $\lambda_n$ of the transition probability matrix\index{transition matrix} projected on the subspace orthogonal to its eigenvalue 1 ($1>|\lambda_1|\geq |\lambda_2|\geq\ldots$), see Ref.~\cite{sokal:97},
\begin{equation}
\hat C_f(t)=\sum_n A_{n,f} \lambda_n^{|t|}\,,\quad \sum_n A_{n,f}=1\,, \labeq{autocorrelation_function_decomposition}
\end{equation}
where the index $n$ runs from 1 to the size of the transition matrix\index{transition matrix}, in our case $N_T!2^{N_TL^D} - 1$.

The amplitudes $A_{n,f}$  depend on $f$, while the $\lambda_n$ are $f$-independent. We can plug~\refeq{autocorrelation_function_decomposition} into \refeq{integrated_autocorrelation_time} and, by computing the sum of a geometric series, we have
\begin{equation}
\tintf = \dfrac{1}{2} + \sum_n A_{n,f} \dfrac{\lambda_n}{1-\lambda_n} \, .
\end{equation}

Now, in practical applications the (leading) $A_{n,f}$'s and $\lambda_n$'s are real positive. Hence, $\lambda_n=\mathrm{e}^{-1/\tau_n}$ defines the characteristic time $\tau_n$. The exponential autocorrelation time\index{autocorrelation time!exponential} of the Markov\index{Markov chain} chain $\texp$ is just $\tau_1$, the largest of the $\tau_n$ (see \cite{sokal:97}). Now, for $\tau_n\gg 1$ we can perform a Taylor expansion and we obtain $\lambda_n/(1-\lambda_n)=\tau_n - \dfrac{1}{2} + \mathcal{O}(1/\tau_n)$. \refeq{normalized_autocorrelation_function} and \refeq{integrated_autocorrelation_time} become
\begin{equation}
\hat{C}_f(t) = \sum_n A_{n,f} e^{-\abs{t}/\tau_n} \>\> , \>\> \tintf = \sum_n A_{n,f} \tau_n \, . \labeq{autocorrelation_decomposition}
\end{equation}
The integrated autocorrelation\index{autocorrelation time!integrated} time $\tintf$ for the quantity $f$ is just an average of the decay modes of the correlation function, being the slower mode the exponential autocorrelation time\index{autocorrelation time!exponential} $\texp$ and the weights of that average the coefficients $A_{n,f}$. It is straightforward to prove that 
\begin{equation}
\tintf \leq \texp \, , \labeq{inequality_autocorrelation_time}
\end{equation}
and the equality is reached when $A_{1,f} = 1$.

$\tintf$ is the quantity we were looking for because, although it has the same disadvantage that $\texpf$ with respect to the chosen $f$, it is much simpler to compute. In addition, in this work we overcome the problem of the quantity $f$ by using a variational method\index{variational method} very similar to the Rayleigh-Ritz variational principle in Quantum Mechanics.

The last consideration has to be done before explaining our variational method\index{variational method}. In this work, although we have proposed a new thermalization\index{thermalization} protocol, we needed to compare our results with the reliable results provided by the computation of the exponential autocorrelation time\index{autocorrelation time!exponential}. The thermal protocol followed here is the same described in appendix A of~\cite{janus:10}.

\section{The variational method: dynamic Temperature Chaos} \labsec{variational_method_eq_chaos}
This sections aims to describe the variational method\index{variational method} used to compute our estimation of $\texp$. The idea is very simple, from~\refeq{autocorrelation_decomposition} and~\refeq{inequality_autocorrelation_time} we can deduce that a very good estimation of $\texp$ from $\tintf$ would imply to choose a quantity $f$ with $A_{1,f} \approx 1$ and $A_{n>1,f} \approx 0$. As we have already introduce above, it has been suggested that we should focus on the temperature flow along the \gls{PT} dynamics in order to compute the autocorrelation function\index{correlation function!time auto-}~\cite{janus:10,fernandez:13,martin-mayor:15}.

The computations of time autocorrelation functions\index{correlation function!time auto-} are usually performed with spin-configuration\index{configuration} functions $f$. In our protocol, we focus on the temperature random-walk\index{random walk!temperature} that each copy of the system performs over the temperature mesh in the \gls{PT} dynamics. At the first sight, it might be surprising that this temperature random-walk\index{random walk!temperature} can provide information about the thermalization\index{thermalization} of the system. The reader may find in \refsec{thermalizing_PT} a detailed discussion about this fact.

Let us consider, for a given sample\index{sample}, the $N$ system copies in the \gls{PT} dynamics. Our Markov\index{Markov chain} chain will be the temperature random-walk\index{random walk!temperature} through the index $i_t$ that indicates that, at time $t$, our system copy is at temperature $T_{i_t}$. At equilibrium, all the copies spend the same time in each temperature and, therefore, the probability for $i_t$ is just the uniform probability over the set $\{1,2,\dots,N \}$. In consequence, the expectation value of the quantity $f$ is just the arithmetic mean
\begin{equation}
\mcav{f} = \dfrac{1}{N} \sum_{i=0}^N f(i) \, . \labeq{expectation_value_f}
\end{equation} 
We shall consider, as well, functions that depends on pairs of system copies. For a given time $t$, these pairs will be described by two indices $i_t \neq j_t$. The equilibrium value for an arbitrary function of a pair of system copies is
\begin{equation}
\mcav{f_{\mathrm{pairs}}} = \dfrac{1}{N(N-1)}\sum_{i=0}^N \sum_{j\neq i}^N f_{\mathrm{pairs}}(i,j) \, .
\end{equation}
By looking at the definition of the time autocorrelation function\index{correlation function!time auto-} in~\refeq{unnormalized_autocorrelation_function} it is clear that, for computational purposes, it is convenient to define a function $f$ with $\mcav{f}=0$. Therefore, for each function $f$ we define
\begin{equation}
\tilde{f} = f - \mcav{f} \, . \labeq{f_expectation_0}
\end{equation}
We can define now our (not normalized) time autocorrelation function\index{correlation function!time auto-} as
\begin{equation}
C_f(t) = \dfrac{1}{N_s-t_0-t} \sum_{t'=t_0}^{N_s-t} \tilde{f}(i_{t'})\tilde{f}(i_{t'+t}) \, ,
\end{equation}
where $N_s$ is the total number of times we have \textit{measured} the state of the \gls{PT} indices $i_t$. Here, of course, we are considering the equilibrium autocorrelation function\index{correlation function!time auto-} and, therefore, $t_0 \gg \tintf$\footnote{The reader may find this statement a little bit contradictory since we are trying to estimate $\tintf$, however a self-consistent procedure is followed, similarly to that one explained in~\cite{janus:10}.}. Note that $C_f(t)$ is independent of the system copy as well as of the replica\index{replica} and, hence, we can improve our statistics by averaging over the $N \times \Nrep$ numerical estimations of $C_f(t)$. All the statements for $f$ depending on a single copy are straightforwardly transferable to functions depending of a pair of system copies. 

Once we have an estimation of $C_f(t)$ we can estimate the integrated autocorrelation\index{autocorrelation time!integrated} time
\begin{equation}
\tintf \approx \nmet \left( \dfrac{1}{2} + \sum_{t=0}^W \hat{C}_f(t) \right) \, , \labeq{window_integrated_autocorrelation_time}
\end{equation}
where $\nmet$ is the periodicity with which we record the time indices $i_t$ of our random walker\index{random walk} and $\hat{C}_f(t) = C_f(t)/C_f(0)$ is the normalized autocorrelation function. In our simulations $\nmet=25000$ Metropolis sweeps most of the times. The  original definition of $\tintf$ [see \refeq{integrated_autocorrelation_time}] involves an infinite sum, here we restrict the sum only to the first $W$ values, being $W$ a self-consistent window (see \cite{sokal:97}) that avoids the divergence of the variance of $\tintf$. We impose $\tintf < 10W$.

In order to compute $\tintf$ as closely as possible to the $\texp$ value, we consider three different parameters to optimize: the type of function $f$, the temperature $T^*$ at which $f$ is zero, and a Wilson-Kadanoff\index{Wilson-Kadanoff} renormalization\index{renormalization group} block length $\lblo$. We describe here the three parameters.

\begin{table}
\centering
\begin{tabular*}{0.6\columnwidth}{@{\extracolsep{\fill}}cc}
\toprule
\toprule
\textbf{Identifier} & \textbf{Function} \\
\toprule
$0$ & piecewise constant \\
$1$ & piecewise linear\\
$2$ & piecewise quadratic\\
$3$ & piecewise cubic\\
$|$ & OR in couples\\
$\&$ & AND in couples\\
$\wedge$ &  XOR in couples\\
$*$ &  Multiplication in couples \\
\bottomrule
\end{tabular*}
\caption[\textbf{Functions of the variational method.}]{\textbf{Functions of the variational method\index{variational method}.} Different choices of the function $f$ used in the variational method\index{variational method}.}
\labtab{functions_variational_method}
\end{table}

\begin{itemize}
\item \textbf{The type of function \boldmath $f$.} Similarly to the Rayleigh-Ritz variational principle in Quantum Mechanics we consider test-functions $f$, belonging to eight different classes (see~\reftab{functions_variational_method}). The four first functions depend only on a single system copy. Specifically, the function labeled as $0$ is just the complementary of the Heaviside function $1-\Theta(T^*)$. The function labeled as $1$ is a piecewise linear function that has already been used before in~\cite{janus:10}. As is quite evident by the description, the functions labeled as $2$ and $3$ are quadratic and cubic piecewise functions respectively. We will specify their specific functional form below. 

The last four functions depend on two system copies. For the functions labeled as $\lvert$, $\&$ and $\wedge$, each system copy of the pair has associated the value of the function labeled as $0$. Then, the value of the function is the corresponding binary operation of the pair of values obtained for each system copy. Finally, the $*$ function is just the multiplication of the piecewise linear function for each system copy (with the corresponding normalization).

\item \textbf{The temperature $\mathbf{T^*}$.} We require for the temperature $T^* \in \{T_1,T_2, \dots, T_{N/2}\}$ that $f(T^*) = 0$. The value of $T^*$ is our second variational parameter. In addition, the condition $\mcav{f_{T^*}}=0$ together with $f(T^*) = 0$ define our test-functions. Specifically, the linear piecewise function is
\begin{equation}
\begin{aligned}
T > T^*\,&: \quad &f_{T^*}(T) = a_+ (T-T^*) \, ,\\
T < T^*\,&: \quad &f_{T^*}(T) = a_- (T-T^*) \, .
\end{aligned}
\end{equation}
We require $a_+$ and $a_-$ to be positive and their ratio is fixed by the condition $\mcav{f_{T^*}}=0$. Indeed, we only need to fix the ratio, because the overall scale of the test function $f_{T^*}$ is irrelevant.

With an analogous procedure we can define the quadratic ($p=1$) and the cubic ($p=2$) functions
\begin{equation}
\begin{aligned}
T>T^* \, : \quad f_{T^*}(T) = a_+ (T-T^*)^p (2T_N - T^* -T) \, ,\\
T<T^* \, : \quad f_{T^*}(T) = a_- (T-T^*)^p (2T_1 - T^* -T) \, .\\
\end{aligned}
\end{equation}
Again, we choose $a_+>0$ and $a_->0$ and the ratio is fixed by $\mcav{f_{T^*}}=0$. We try all the possible values of $T^*$ in the lower half part of the set of temperatures in our \gls{PT} simulation.

\item \textbf{The renormalization\index{renormalization group} time-block \boldmath $\lblo$.} We can modify the value of the time autocorrelation function\index{correlation function!time auto-} by changing the function $f$ itself (as it has been introduced in the two previous parameters of the variational method\index{variational method}) but we can also modify it by changing the temporal series from which we compute that autocorrelation function\index{correlation function!time auto-}. We build Wilson-Kadanoff\index{Wilson-Kadanoff} blocks: the Monte\index{Monte Carlo} Carlo sequence $f_{T^*}(i_1),f_{T^*}(i_2),\dots,f_{T^*}(i_{N_s})$ is divided into blocks of $\lblo$ consecutive data (see e.g. \cite{amit:05}). For each block, we compute the average and we build a new sequence $f'_{T^*}(j_1),f'_{T^*}(j_2),\dots,f'_{T^*}(j_{N_s/\lblo})$ from which we compute the integrated autocorrelation time\index{autocorrelation time!integrated} just as we did for $\lblo=1$. Of course, after computing the integrated autocorrelation time\index{autocorrelation time!integrated}, we need to rescale it to recover the original time units. The idea of this parameter is to reduce the high-frequency fluctuations of the time autocorrelation function\index{correlation function!time auto-}.

However, the parameter $\lblo$ needs to be controlled, otherwise, it can become larger than $\tintfTlblo$ erasing all the information of the autocorrelation function\index{correlation function!time auto-} and giving us spurious results. If that happens, each block would be just the expectation value (well, a finite estimation) of the function $f$, and the autocorrelation function\index{correlation function!time auto-} would vanish for times $t\neq 0$. From~\refeq{window_integrated_autocorrelation_time} we deduce that, after converting $\tintf$ to the correct time units, we would have $\tintfTlblo = \lblo \nmet/2$, which diverges with $\lblo$. With the aim to control that spurious effect, we impose
\begin{equation}
\tintfTlblo > \dfrac{5}{2} \nmet \lblo \, . \labeq{tintfTlblo_condition}
\end{equation}
The values of $\lblo$ are taken from the list $\lblo =\{1,2,5,10,20,50,100,200,500,1000$
$,2000\}$.
\end{itemize}

Our estimation of the integrated autocorrelation time\index{autocorrelation time!integrated}, namely $\tintvar$, is just the highest value of all the $\tintfTlblo$. This estimation has great advantages from its predecessor (which corresponds with the linear piecewise function at the critical\index{critical temperature} temperature and $\lblo=1$). Firstly, our estimation is robust in the sense that it does not produce spurious values. Moreover, the process can be easily implemented in an automatic way which is a \textit{sine qua non} condition given the huge number of possible combinations of parameters.

The worse scenario would be to found that our effort has been in vain and the automatic process always chooses the piecewise linear function with $T^* = \Tc$ and $\lblo=1$. Fortunately, this is not the case, in~\reftab{frequency_functions_variational_method} we can see the numbers of times that our method chooses each function. Indeed, it is notorious that almost all of the time, the method chooses a single-copy function. The same happens with $T^*$, the chosen value is not always $\Tc$, actually, we called the chosen temperature the \textit{dynamic chaotic temperature} $T_d$ that will be useful in the following analysis. The effects of $\lblo$ or, more accurately, the effects of the discretization of $\lblo$ can also be observed in the results (for example \reffig{log_tau_I}). Specifically, the low density in the $\lblo$ mesh leads to small gaps in the determination of the autocorrelation time\index{autocorrelation time!integrated} $\tintvar$.

\begin{table}
\centering
\begin{tabular*}{0.8\columnwidth}{@{\extracolsep{\fill}}cccccccccc}
\toprule
\toprule
$L$ & $0$ & $1$ & $2$ & $3$ & $|$ & $\&$ & $\wedge$ & $*$ & Total  \\
\toprule
$16$ & $2032$ & $5320$ &  $3875$ & $1374$  & $4$ & $115$ & $74$ & $6$ & $12800$ \\
$24$ & $1556$ & $7196$ &  $3089$ & $820$  & $0$ & $127$ & $11$ & $1$ & $12800$ \\
\bottomrule
\end{tabular*}
\caption[\textbf{Frequency of choice of the variational method.}]{\textbf{Frequency of choice of the variational method\index{variational method}.} Number of times the variational method\index{variational method} has picked one of the eight choices among the functions $f$ described in the text. $L$ denotes the lattice size.}
\labtab{frequency_functions_variational_method}
\end{table}

\begin{figure}
\centering
\includegraphics[width=0.8\columnwidth]{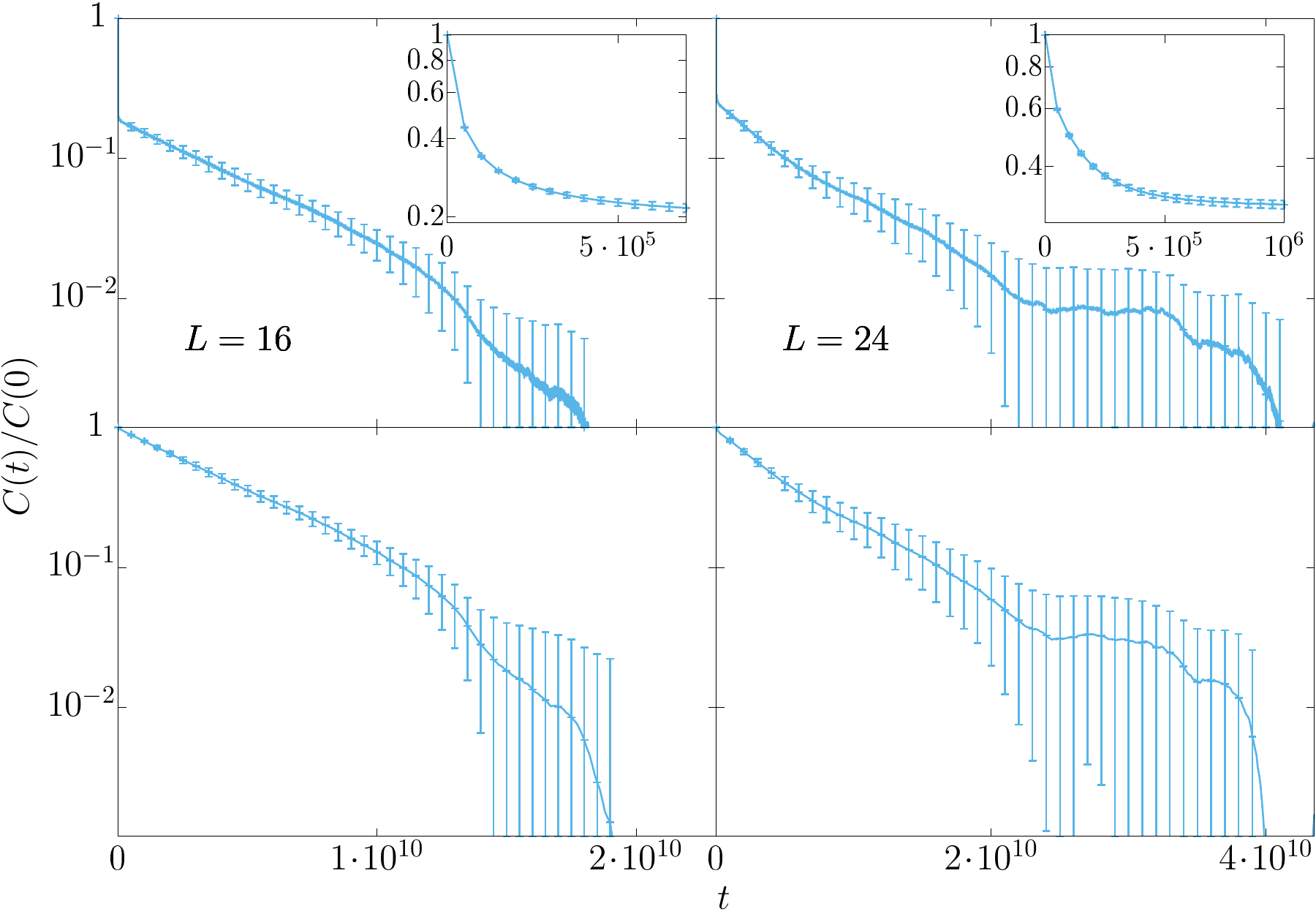}
\caption[\textbf{Improvement of the estimation of the time autocorrelation functions.}]{\textbf{Improvement of the estimation of the time autocorrelation functions.}\index{correlation function!time auto-} Auto-correlation function for the most chaotic sample\index{sample} for $L=16$ (left) and $L=24$ (right): (Top) Auto-correlation function computed using the method of \cite{janus:10} and (Bottom) using the variational method\index{variational method} presented here. Note that the improvement of the new method is notorious is we focus on the y-intercept (further details can be found in the text). \textbf{Inset:} Linear-log plot showing the small $t$ behavior of the autocorrelation function\index{correlation function!time auto-}.}
\labfig{example_autocorrelation_function}
\end{figure}

Another question has to be answered. It is true that our method chooses a variety of parameters, to the detriment of the classical choice but, is there a significant improvement in the estimation of $\texp$ or are all the estimations just small fluctuations of the previous estimation $\tint$? An example of the improvement obtained in the computation of the autocorrelation function\index{correlation function!time auto-} is shown in~\reffig{example_autocorrelation_function}. The main problem of the previous estimation becomes obvious from the figure: the value of $A_{1,f}$ [see~\refeq{autocorrelation_decomposition}] could be, indeed, fairly small $A_{1,f} \approx 0.1$. In the figure, this amplitude roughly corresponds to the abscissa of the initial point of the linear decreasing that we are able to see in the log-log scale (i.e. the beginning of the domination of the large time-scale corresponding to $\texp$). Our new estimations (bottom panels) are rather better.

This hand-waving argument can be made quantitative. Let us denominate $\tintold$ to the methodology of estimation of $\tint$ for previous works~\cite{janus:10} and $\tintvar$ to our variational-method\index{variational method} estimation. In~\reffig{histograma_taus_multiplot_g} we separate our samples\index{sample} in deciles according to its $\tintvar$ value so that the first decile corresponds to the $1280$ samples\index{sample} with smaller $\tintvar$. We have argued that the most chaotic samples\index{sample} will have larger $\texp$, so those deciles are our proposal for separating the \textit{most chaotic} and \textit{less chaotic} samples\index{sample}.

Then, we build up the histogram of the ratio $\tintold/\tintvar$ for the samples\index{sample} on a given decile. Top panels in~\reffig{histograma_taus_multiplot_g} show that the gain of considering $\tintvar$ is sizable but, if we focus on the most chaotic samples\index{sample} (i.e. the tenth decile, in the bottom panels) the benefits of our variational method\index{variational method} are more than evident with a significant fraction of the samples\index{sample} with $\tintold/\tintvar<0.1$.

\begin{figure}
\centering
\includegraphics[width=0.8\columnwidth]{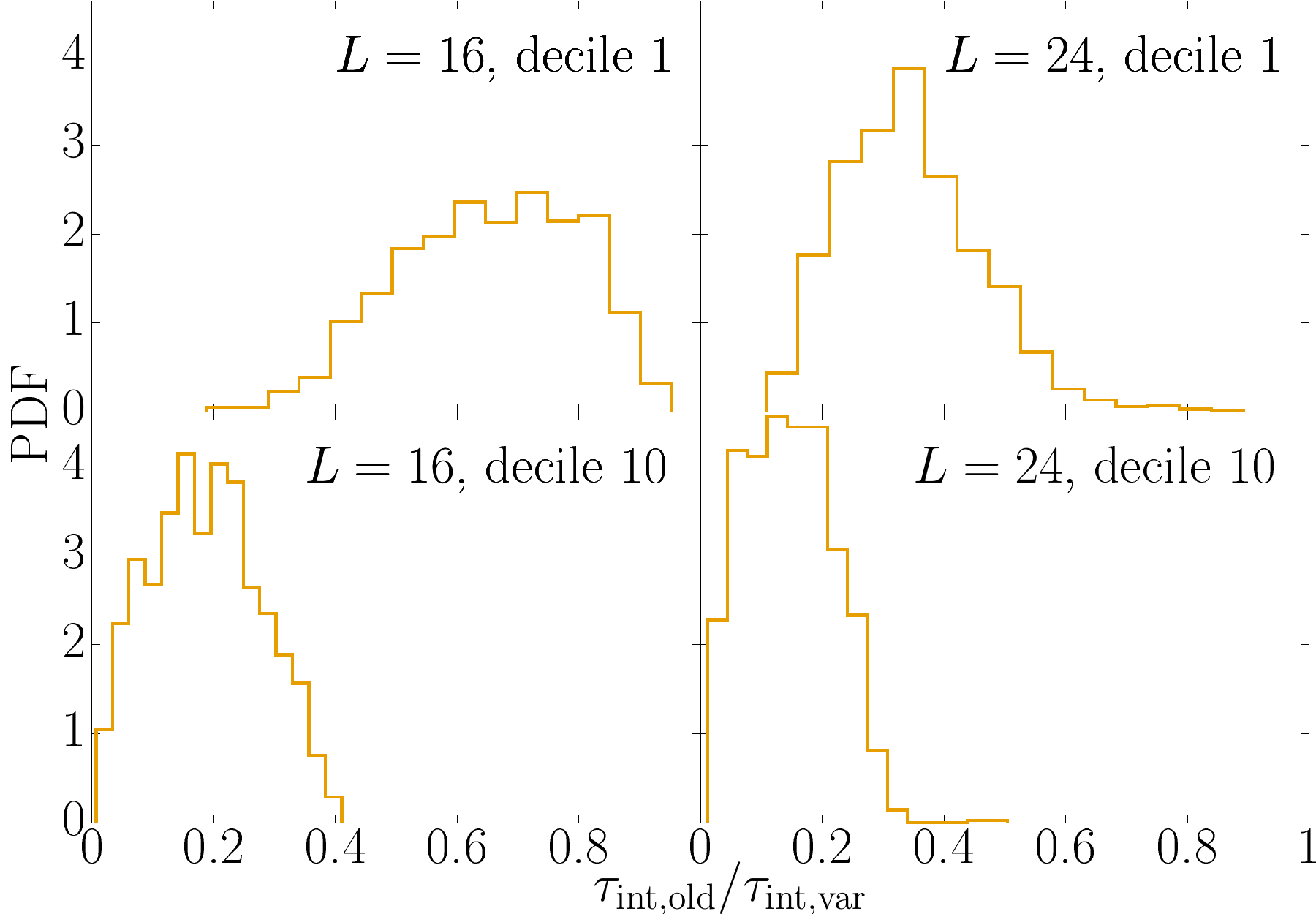}
\caption[\textbf{A quantitative argument for the variational method.}]{\textbf{A quantitative argument for the variational method\index{variational method}.} Conditional probability density function of the ratio $\tintold/\tintvar$, given that $\tintvar$ belongs to a given decile is plotted. We show the data for the first decile (left) and the tenth decile (right) for $L=16$ (top) and $L=24$ (bottom).}
\labfig{histograma_taus_multiplot_g}
\end{figure}

\section{Static Temperature Chaos}
At this point, we momentarily forget about previous disquisitions on \gls{TC}\index{temperature chaos} from the dynamical point of view and we come back to the classical static definition: \gls{TC}\index{temperature chaos} is the complete rearrangement of the equilibrium configurations\index{configuration} upon any change of temperature. This phenomenon has been traditionally studied~\cite{billoire:00,billoire:02,katzgraber:07} through the probability density function of the overlap\index{overlap!distribution} between the spin configurations\index{configuration} at temperatures $T_1$ and $T_2$,
\begin{equation}
q_{T_1,T_2} = \dfrac{1}{V} \sum_i s_i^{T_1} s_i^{T_2} \, . \labeq{overlap_2T}
\end{equation}
Due to the limitation of the maximum size that can be simulated, this quantity is strongly affected by finite-size effects\index{finite-size effects}. We focus therefore on other quantity, introduced in~\cite{ritort:94}: the \textit{chaotic parameter}
\begin{equation}
\xchaos = \dfrac{\braket{q^2_{T_1,T_2}}_J}{\sqrt{\braket{q^2_{T_1,T_1}}_J\braket{q^2_{T_2,T_2}}_J}} \, , \labeq{chaotic_parameter}
\end{equation}
where $\braket{\cdots}_J$ stands for the usual thermal average but we stress the sample\index{sample} dependency with the sub-index $J$. It has been proposed that the \gls{TC}\index{temperature chaos} phenomenon should be studied through a detailed analysis of the distribution of this sample-dependent chaotic parameter~\cite{fernandez:13,billoire:14}.

The reader should notice that $0 < \xchaos \lesssim 1$. The extreme values are clear; $\xchaos = 1$ means that both configurations\index{configuration} at temperatures $T_1$ and $T_2$ are indistinguishable i.e. absence of chaos. On the contrary, $\xchaos = 0$ means that both configurations\index{configuration} are completely different which would indicate strong chaos.

We select the most chaotic samples\index{sample} and the less chaotic ones accordingly to the estimation $\tintvar$ and we plot in~\reffig{X_TminT} their chaotic parameter as a function of temperature by keeping $T_1$ to the lower simulated temperature $T_{\min}$ and varying $T_2$. It is clear that qualitative different behaviors on the quantity $X_{T_{\min},T}^J$ are present in both sets. The less chaotic ones tend to decrease smoothly as $T_2$ increases while the more chaotic ones suffer sharp drops at well-defined temperatures, namely \textit{chaotic events}. In addition, it was empirically observed~\cite{fernandez:13} that chaotic events occurring at low temperatures are more harmful to the performance of \gls{PT}. With this information in mind, we are looking for a single number that could quantify the \textit{chaoticity} of a given sample\index{sample}. The introduced observable~\cite{fernandez:13} was the chaotic integral
\begin{equation}
I = \int_{T_{\min}}^{T_{\max}} \xchaosmin dT_2 \, . \labeq{chaotic_integral}
\end{equation}
This quantity will be smaller if the sample\index{sample} suffers a chaotic event that ``cuts'' the integral. Moreover, for the chaotic samples\index{sample} is usual that, once the chaotic event takes place at temperature $T^*$, the chaotic signal for temperatures $T>T^*$ is low but the fluctuations of the value are still present. To minimize this effect, we propose the parameter $I_2$ that reduces the integration range to the first half of the simulated temperatures.

Finally, looking at~\reffig{X_TminT}, we noticed that some samples\index{sample} presented strong decays but then, they maintain a relatively high value of the chaotic parameter for higher temperatures (for example, look at the purple curves in the top panels). To take into account that effect we define the quantity
\begin{equation}
K_i = 1-X_{T_{i},T_{i+1}} \, , \labeq{finite_differences}
\end{equation} 
which is essentially the finite difference of two consecutive points in the curve. After some trials based on heuristic arguments and after seeing a lot of $\xchaosmin$ vs $T$ curves, we define the quantity
\begin{equation}
I_X = aI_2 - b \min_i \left( -\log K_i^2\right) - c \sum_i \left(-\log K_i^2 \right) \, ,
\end{equation}
where the coefficients $a$, $b$ and $c$, that depend on the lattice size $L$, are obtained through a minimization of the correlation Pearson coefficient $r$ between $I_X$ and $\log(\tintvar)$ (as we will discuss in the following section).
\begin{figure}
\centering
\includegraphics[width=0.8\columnwidth]{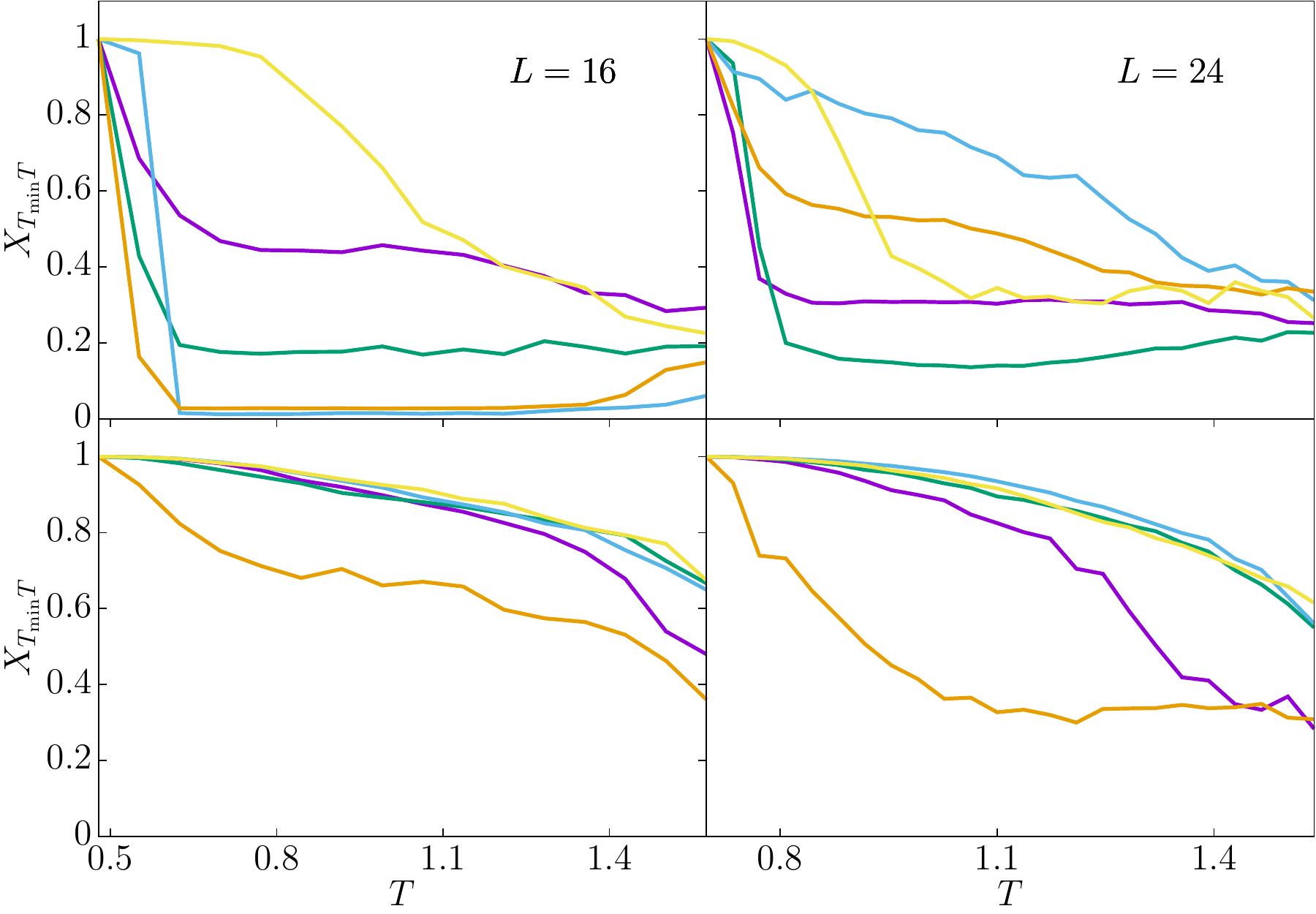}
\caption[\textbf{Temperature dependence of the chaotic parameter.}]{\textbf{Temperature dependence of the chaotic parameter.} Plot of $X_{T_{\min},T}^J$ versus $T$ for the five most chaotic samples\index{sample} (top) and the five less chaotic ones (bottom): $L=16$ case (left) and $L=24$ case (right).}
\labfig{X_TminT}
\end{figure}

\section{Correlation dynamics-statics} \labsec{correlation_dynamics_static}
We study here the correlation between the static characterization of chaos through the previously defined observables and the dynamic one, through the autocorrelation time\index{autocorrelation time!integrated} $\tintvar$ that we will call from now on, simply $\tint$. 

We also find it useful to address the failures in the process of finding quantities relating to both perspectives. Usually, space requirements in publications or the clarity of the message make the rules of choosing the appropriate results to show, and it is perfectly reasonable. However, this privileged format allows us to extend a little bit more and show the path of the research which often contains failures. In addition, addressing those quantities not related to chaos would be also helpful for future works.
\subsection{The failures}
Here, we address some \textit{a priori} reasonable quantities that turned out to be not related with \gls{TC}\index{temperature chaos}.

First, we recall the previously defined $T_d$ which is the temperature $T^*$ chosen by the variational method\index{variational method}. We compute also, from the static characterization, the temperature $T_s$ at which the \textit{bigger} chaotic event occurs\footnote{Note that, for the less chaotic samples\index{sample} with a smooth decay of the function $\xchaosmin$ against $T_2$, this chaotic event can be fairly small.} i.e. the temperature for which $\xchaosmin$ presents the maximum (negative) slope.

The correlation between both quantities is almost absent as can be seen in~\reffig{Td_Ts}. We can see an over-density, however, the points out of the principal density are too dispersed. For $L=16$ (top) the number of points within the lines is $8017$ ($62.63 \%$ of the total) while for $L=24$ (bottom) the number of points within the lines is $7539$ ($58.90 \%$ of the total). If we compute the correlation coefficients we obtain~\reftab{coef_correla_TdTs}. The errors in the determination of the correlation coefficients are computed from the Bootstrap\index{Bootstrap} method (see \refsec{estimating_errorbars}).

\begin{figure}[h!]
\centering
\includegraphics[width=0.6\columnwidth]{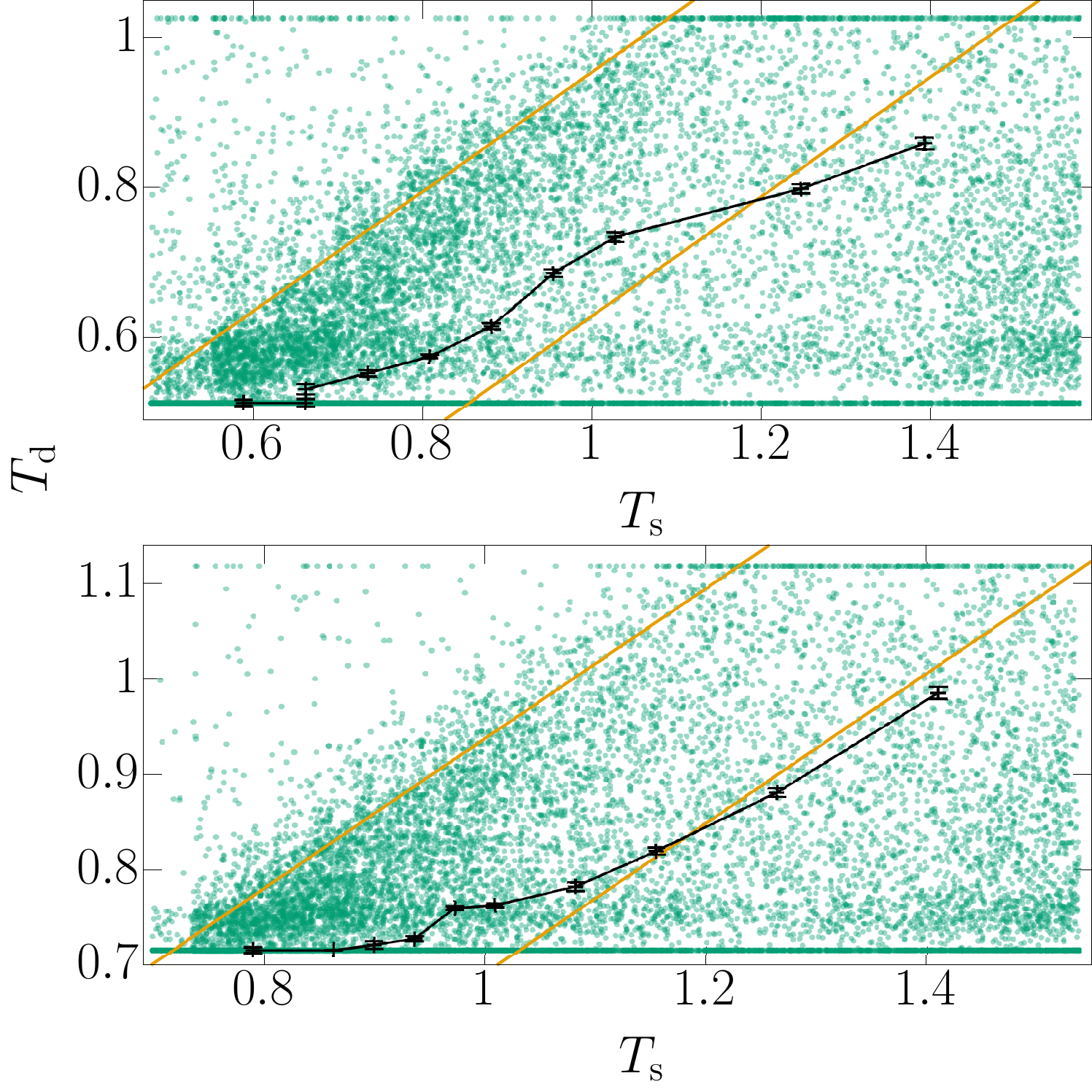}
\caption[\textbf{\boldmath Scatter plot of $T_d$ versus $T_s$.}]{\textbf{Scatter plot of $\mathbf{T_d}$ versus $\mathbf{T_s}$.} We present the $L=16$-data (top) and the $L=24$-ones (bottom). Points are calculated with a special procedure. First, samples\index{sample} are classified on deciles according to $\log(\tau_{\mathrm{int}})$. The points coordinates were obtained by computing the median $T_d$ and the median $T_s$ within each decile (errors from Bootstrap\index{Bootstrap}). The golden parallel lines enclose the area of over-density that presents a higher correlation for later recount.}
\labfig{Td_Ts} 
\end{figure}

\begin{table}
\centering
\begin{tabular*}{0.5\columnwidth}{@{\extracolsep{\fill}} ccccc}
\toprule
\toprule
&$L$  & & $r$ & \\
\toprule
&$16$ & & $0.348 \pm 0.008$ &\\
&$24$ & & $ 0.342 \pm 0.007$ & \\
\bottomrule
\end{tabular*}
\caption[\textbf{\boldmath $T_d$ vs $T_s$.}]{\textbf{$\mathbf{T_d}$ vs $\mathbf{T_s}$.} Correlation coefficients of the scatter plot of $T_d$ against $T_s$ for the simulated two lattice sizes.}
\labtab{coef_correla_TdTs}
\end{table}

We can try to relate $T_s$ to other dynamic estimation of chaos, for example, the integrated autocorrelation time\index{autocorrelation time!integrated} $\tint$. Unfortunately, although slightly better than our previous attempt, we observe a weak correlation between both estimators, $\tint$ and $T_s$, (see \reffig{log_tau_Ts}) and we can check it quantitatively through \reftab{coef_correla_Ts}.

\begin{figure}[h!]
\centering
\includegraphics[width=0.6\columnwidth]{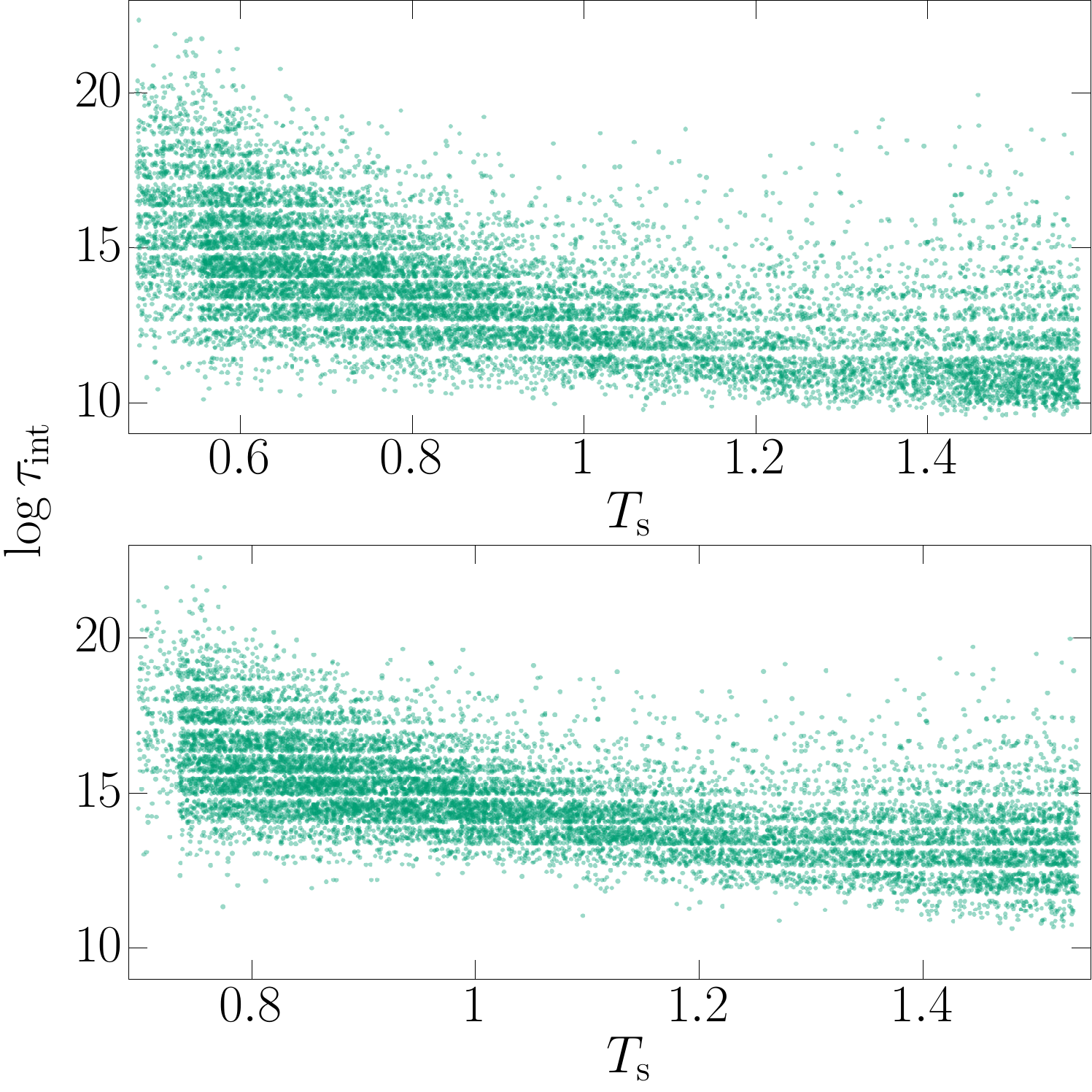}
\caption[\textbf{\boldmath Scatter plot of $\log(\tint)$ against $T_s$.}]{\textbf{Scatter plot of $\mathbf{\log(\tint)}$ against $\mathbf{T_s}$.} We show $L = 16$ (top) and $L=24$ (bottom).}
\labfig{log_tau_Ts} 
\end{figure}

\begin{table}
\centering
\begin{tabular*}{0.5\columnwidth}{@{\extracolsep{\fill}}ccccc}
\toprule
\toprule
&$L$   && $r$ & \\
\toprule
&$16$  && $-0.621 \pm 0.006$ &\\
&$24$  && $ -0.621 \pm 0.006$ &\\
\bottomrule
\end{tabular*}
\caption[\textbf{Correlation coefficients for the scatter plot of $\mathbf{\log(\tau_{\mathrm{int}})}$ versus $\mathbf{T_s}$ for the two simulated lattice sizes.}]{\textbf{Correlation coefficients for the scatter plot of $\mathbf{\log(\tau_{\mathrm{int}})}$ versus $\mathbf{T_s}$ for the two simulated lattice sizes.}}
\labtab{coef_correla_Ts}
\end{table}

\subsection{The success}
Here, we present our most successful attempts to relate both dynamic and static characterization of chaos. In \reffig{log_tau_I}, we confront the most representative
estimator for the dynamical chaos, namely the largest integrated autocorrelation time\index{autocorrelation time!integrated} $\tint$ found in our variational\index{variational method} study, with the static chaotic integrals $I$, $I_2$ and $I_X$. We can observe how spurious values of the original parameter $I$ (i.e. large values of $I$ associated with large $\tau_\mathrm{int}$) are displaced towards lower values when we use the improved parameters $I_2$ and $I_X$.

\begin{figure}[h!]
\centering
\includegraphics[width=0.8\columnwidth]{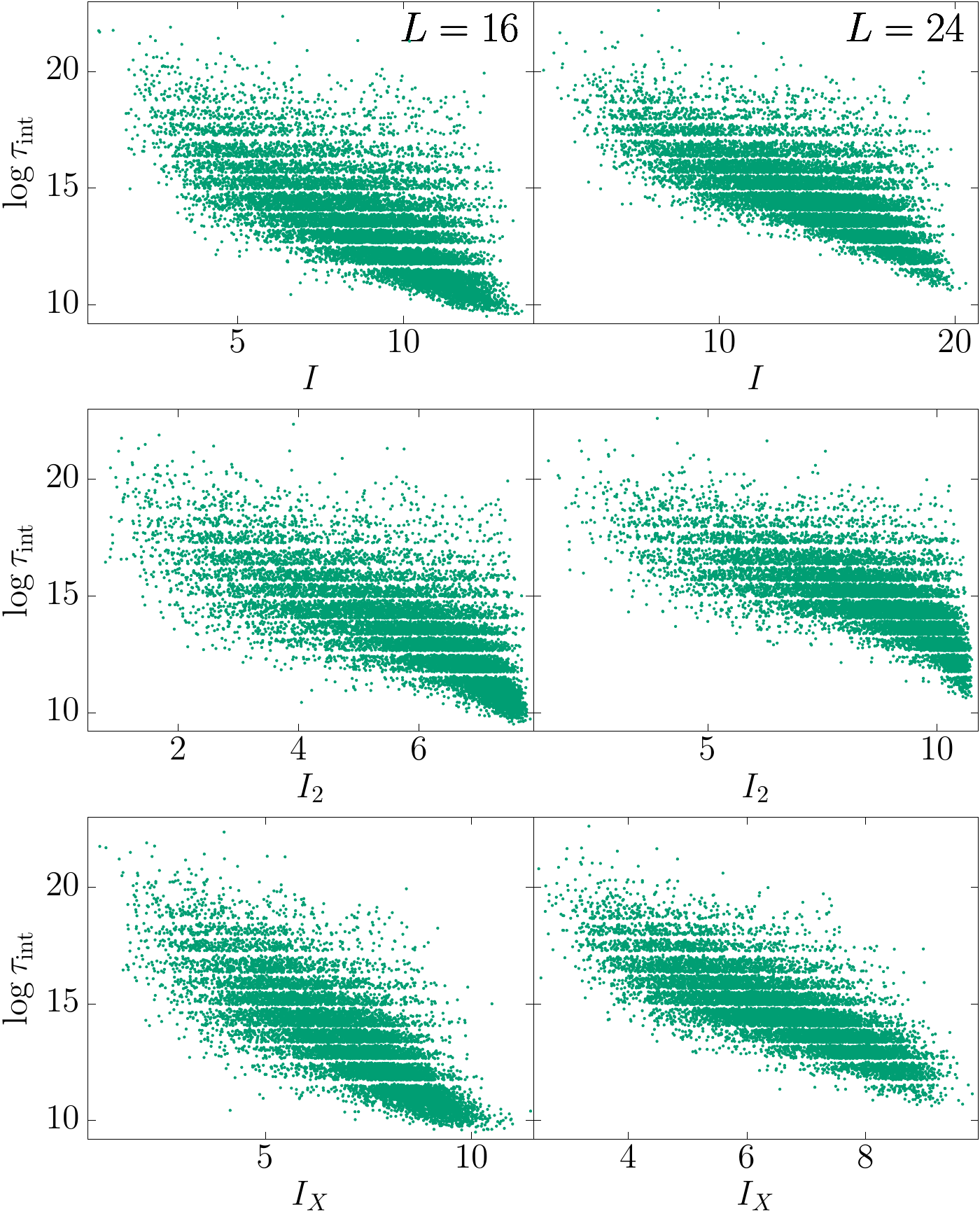}
\caption[\textbf{Scatter plot of $\mathbf{\log(\tintvar)}$ versus integrated chaotic parameters.}]{\textbf{Scatter plot of $\mathbf{\log(\tintvar)}$ versus integrated chaotic parameters.} We present data for two lattice sizes and for the three definitions of the integrated chaotic parameter defined in the text ($I, I_2$ and $I_X$). The pattern of depleted horizontal bands is due to our choice of a few $l_\mathrm{blo}$.}
\labfig{log_tau_I}
\end{figure}

The value of the correlation coefficients is reported in~\reftab{coef_correla} (as before, the errors are computed by using a Bootstrap\index{Bootstrap} method, see \refsec{estimating_errorbars}). We observe a strong anti-correlation in  $I_X$, which improves over the previous indicator of correlation $I$. \cite{fernandez:13} The improvement is less clear for $I_2$.

\begin{table}
\centering
\begin{tabular*}{0.6\columnwidth}{@{\extracolsep{\fill}}ccccc}
\toprule
\toprule
&$L$ & Integral & $r$  \\
\toprule
&$16$ & $I$   & $-0.714 \pm 0.005$ \\
&$16$ & $I_2$ & $-0.751 \pm 0.005$ \\
&$16$ & $I_X$ & $-0.795 \pm 0.004$ \\
\toprule
&$24$ & $I$   & $-0.725 \pm 0.005$ \\
&$24$ & $I_2$ & $-0.746 \pm 0.005$ \\
&$24$ & $I_X$ & $-0.786 \pm 0.004$ \\
\bottomrule
\end{tabular*}
\caption[\textbf{Correlation coefficients for $\mathbf{\log(\tint)}$ versus the integrated chaotic parameters.}]{\textbf{Correlation coefficients for $\mathbf{\log(\tint)}$ versus the integrated chaotic parameters.} Correlation coefficients are shown for each two lattice sizes and for the three definitions of the parameter ($I, I_2$ and $I_X$).}
\labtab{coef_correla}
\end{table}

\section{Finite size scaling} \labsec{scaling_eq_chaos}
\index{finite size scaling}
This section is devoted to study the size-scaling behavior of the dynamic characterization of \gls{TC}\index{temperature chaos}. It has been observed \cite{fernandez:13} that chaotic events are less common in small systems. This suggests a large $L$ limit for the chaotic behavior that we investigate here.

An implicit assumption of our study is that the scaling behavior of $\tint$ is mostly decided by the value $\Tmin$. Other details, such as the number of temperatures in the \gls{PT} mesh, are expected to play a minor role (if kept in a reasonable range). Our choice of simulated parameters (see, \reftab{parameters_simulation_MUSA_MUSI}) does not allow us to check the impact of the number of temperatures in the \gls{PT} mesh, but we can justify that $\Tmin$ has a deep impact on the determination of $\tint$.

\subsection{Temperature chaos depends on \boldmath $\Tmin$}
In order to study how the range of temperatures in the \gls{PT} affects the dynamics, we have confronted both simulations for $L=16$, one with $N=13$ and $\Tmin=0.698$, the other one with $N=16$ and $\Tmin=0.479$. We need to increase the number of temperatures $N$ in the mesh to keep the interval between adjacent temperatures fixed.

Since the simulation with $N = 16$ reaches a lower minimum temperature than the simulation with $N = 13$, we expect to find chaos events (i.e a jam in the \gls{PT} temperature flow) that the simulation with $N=13$ cannot ``see''. In~\reffig{cociente_taus} we show a scatter plot of $\log(\tau_{\mathrm{int},16}/\tau_{\mathrm{int},13})$ versus $T_d$ for the 12800 samples\index{sample} ($\tau_{\mathrm{int},16}$ and $\tau_{\mathrm{int},13}$ are the autocorrelation times for $N=16$ and $N=13$ respectively). 

\begin{figure}[h!]
\centering
\includegraphics[width=0.7\columnwidth]{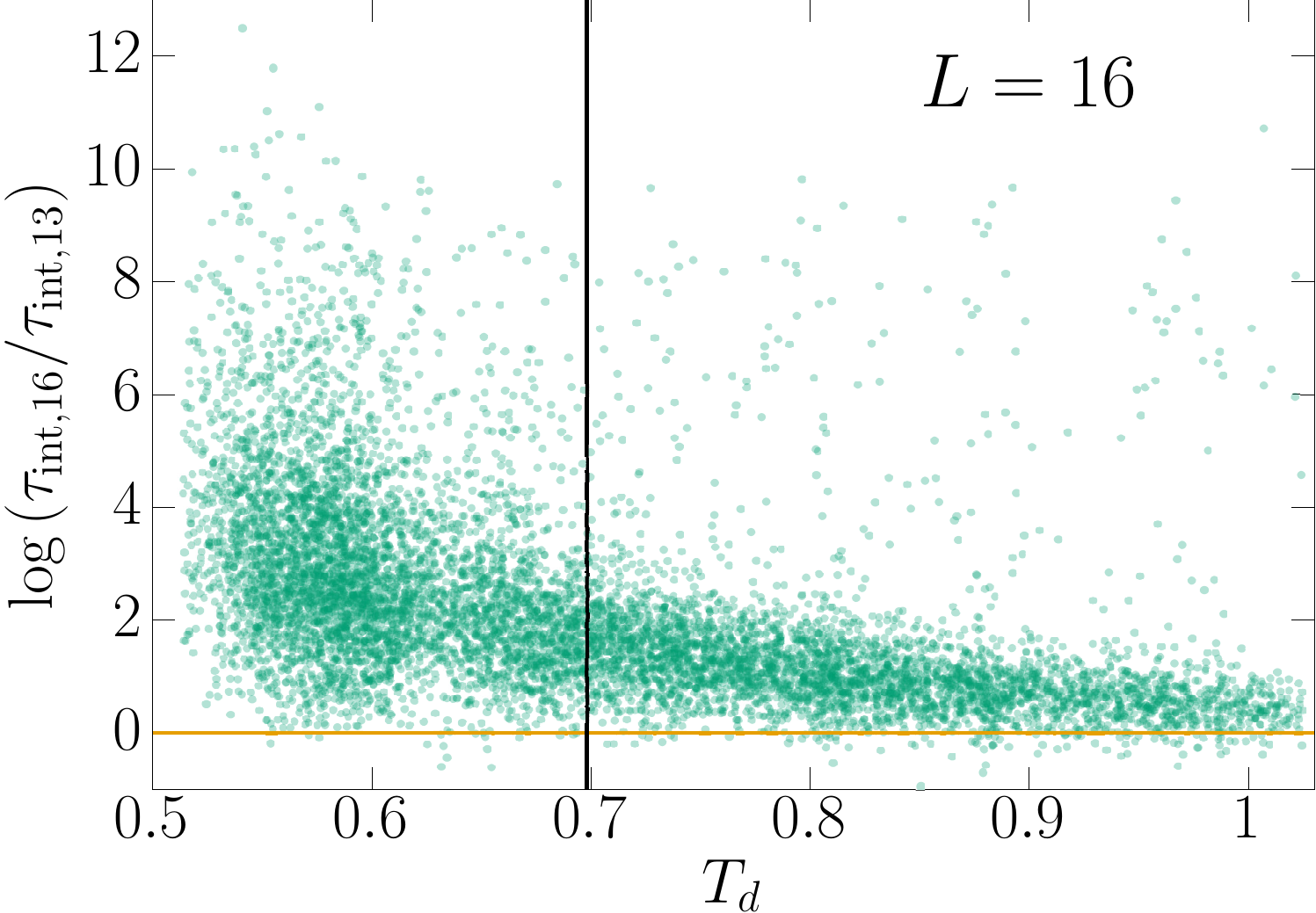}
\caption[\textbf{Scatter plot of $\mathbf{\log(\tintd/\tintt)}$ versus $\mathbf{T_d}$.}]{\textbf{Scatter plot of $\mathbf{\log(\tintd/\tintt)}$ versus $\mathbf{T_d}$.} The lattice size is $L=16$, $\tau_{\mathrm{int},16}$ is the relaxation\index{relaxation} time for $N=16$ ($T_\mathrm{min}=0.479$), $\tau_{\mathrm{int},13}$ is the relaxation\index{relaxation} time for $N=13$ ($T_\mathrm{min}=0.698$), $T_d$ is the temperature of chaos from a dynamical point of view (defined in the variational method\index{variational method}) of the simulation with $N = 16$. Both simulations have the same number of disorder\index{disorder} samples\index{sample}. The vertical black line represents the minimum temperature simulated in the $N= 13$ simulation. (We added a small Gaussian\index{Gaussian!noise} white noise to $T_d$, which is a discrete variable, to avoid the cluttering of data in vertical lines). }
\labfig{cociente_taus}
\end{figure}

For $T_d>0.698$ the ratio takes values of order one for most samples\index{sample}, while for $T_d < 0.698$ there is a huge number of samples\index{sample} with $\tau_{\mathrm{int},16} \gg \tau_{\mathrm{int},13}$,
i.e. there are a lot of samples\index{sample} with a chaotic behavior in a temperature-range  below $T_\mathrm{min}=0.698$.

The same idea can be analyzed from a different point of view. Imagine that we have studied with great care a given sample\index{sample} down to some temperature $T_\mathrm{min}$. Can we say something about possible chaotic effects at lower temperatures? The question is answered negatively in \reffig{no-prediction-from-small-Tmin}: the probability that a sample\index{sample} has a large $\tau_{\mathrm{int}}$ for the simulation with a lower $T_\mathrm{min}$ is not correlated to the value of $\tau_{\mathrm{int}}$ for the first simulation.

\begin{figure}[h!]
\centering
\includegraphics[width=0.7\columnwidth]{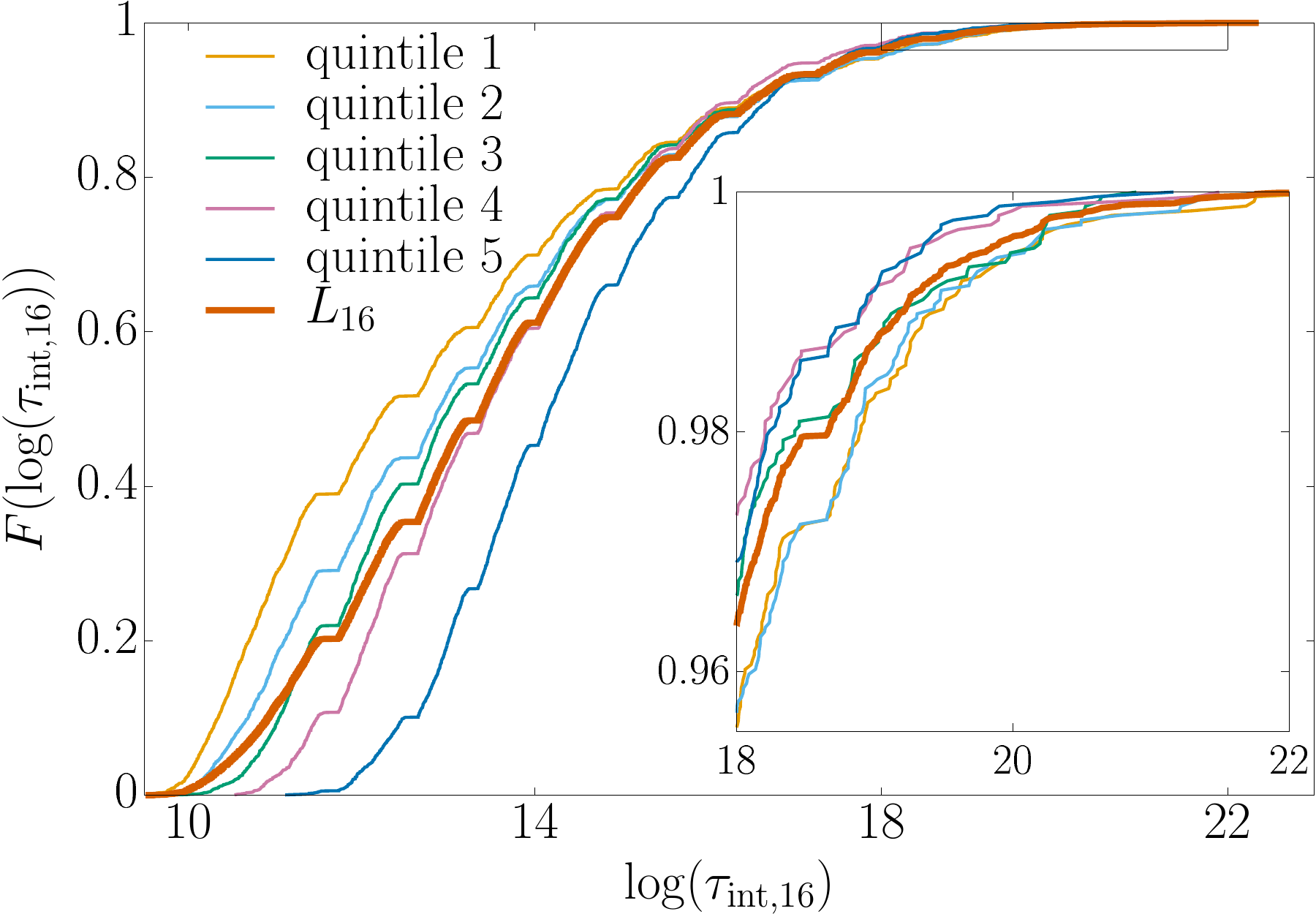}
\caption[\textbf{Conditional probability distribution of $\mathbf{\tint}$.}]{\textbf{Conditional probability distribution of $\mathbf{\tint}$.} The empirical probability distribution as a function of $\tint$ for the $N=16$ simulation, conditional to the $\tint$ obtained from $N=13$ simulation belonging to a given quintile. The non-conditional probability distribution function is also shown ($L_{16}$ curve).  \textbf{Inset.} Blowup of the top right part of the main figure. For the hard samples\index{sample}, the simulation with $T_\mathrm{min}=0.698$ conveys little or no information on the difficulty of the $T_\mathrm{min}=0.479$ simulation.}
\labfig{no-prediction-from-small-Tmin}
\end{figure}

\subsection{The scaling}
We have discussed that $\Tmin$ has a great impact on the $\tint$ value, therefore, we need to fix the same $\Tmin$ for all the simulations in order to establish fair comparisons. We study the \gls{PT} dynamics for $L=8,12,16,24$ and $32$ with $\Tmin\approx 0.7$. An important advantage of $T_\mathrm{min}\approx 0.7$ is that \gls{TC}\index{temperature chaos} has been already characterized at such temperatures, in the equilibrium setting~\cite{fernandez:13}. Lowering $T_\mathrm{min}$ would increase chaos effects, which would have been good in principle, but it would have been also extremely difficult to reach thermal equilibrium. Instead, increasing $T_\mathrm{min}$ to approach the critical point would make the results irrelevant, because samples\index{sample} displaying \gls{TC}\index{temperature chaos} would be too scarce (besides, we want to study the \gls{SG} phase\index{phase!low-temperature/spin-glass}, rather than critical effects).

The $L=32$ data are from Ref.~\cite{janus:10} and have been obtained with the dedicated Janus\index{Janus} computer~\cite{janus:09}. The Janus\index{Janus} simulation used heat bath dynamics, rather than Metropolis, and the \gls{PT} there had $N_T=34$ and $T_\mathrm{min}=0.703$. In order to be sure that heat bath autocorrelation times are consistent with Metropolis times (as we would expect) we simulated with Janus\index{Janus} ten randomly selected samples\index{sample} with both algorithms, finding that $\tau_\mathrm{Metropolis}\approx \tau_\mathrm{heat-bath}/3$.

We show in~\reffig{all_L_prob_tau} the cumulative distribution function of $\tau = \tint$, $F(\tau)$. It can be seen qualitatively from the figure that the maximum slope of $F$ decreases with $L$ for the small systems, and it stabilizes between $L=24$ and $L=32$; indeed these two distributions can be approximately superposed by a simple translation. This is reminiscent of a critical slowing-down~\cite{zinn-justin:05}
\begin{equation}\label{eq:zPT-def}
\tau\sim L^{z^\mathrm{PT}(T_\mathrm{min})}\,.
\end{equation}

It is not obvious \emph{a priori} that such a simple scaling should hold in the \gls{SG} phase\index{phase!low-temperature/spin-glass}. As a working, simplifying hypothesis we assume that the exponent $z^\mathrm{PT}$ only depends on the value of the lowest temperature in the \gls{PT} grid, $T_\mathrm{min}$ (and not on the number of temperatures).

The reader may warn that in~\reffig{all_L_prob_tau} the distribution functions are not drawn for small values of $F(\tau)$ in the $L=8$ and $L=16$ cases. The reason is that we could not find with our variational method\index{variational method} a $\tint$ that fulfills the condition of~\refeq{tintfTlblo_condition} and therefore we can not provide a safe computation of $\tau$. As long as we are concerned about the top part of the curve, we ignore this problem that does not appear in the simulation of $L=16$ with $N=16$, which is the simulation used in the rest of the study.

\begin{figure}
\centering
\includegraphics[width=0.7\columnwidth]{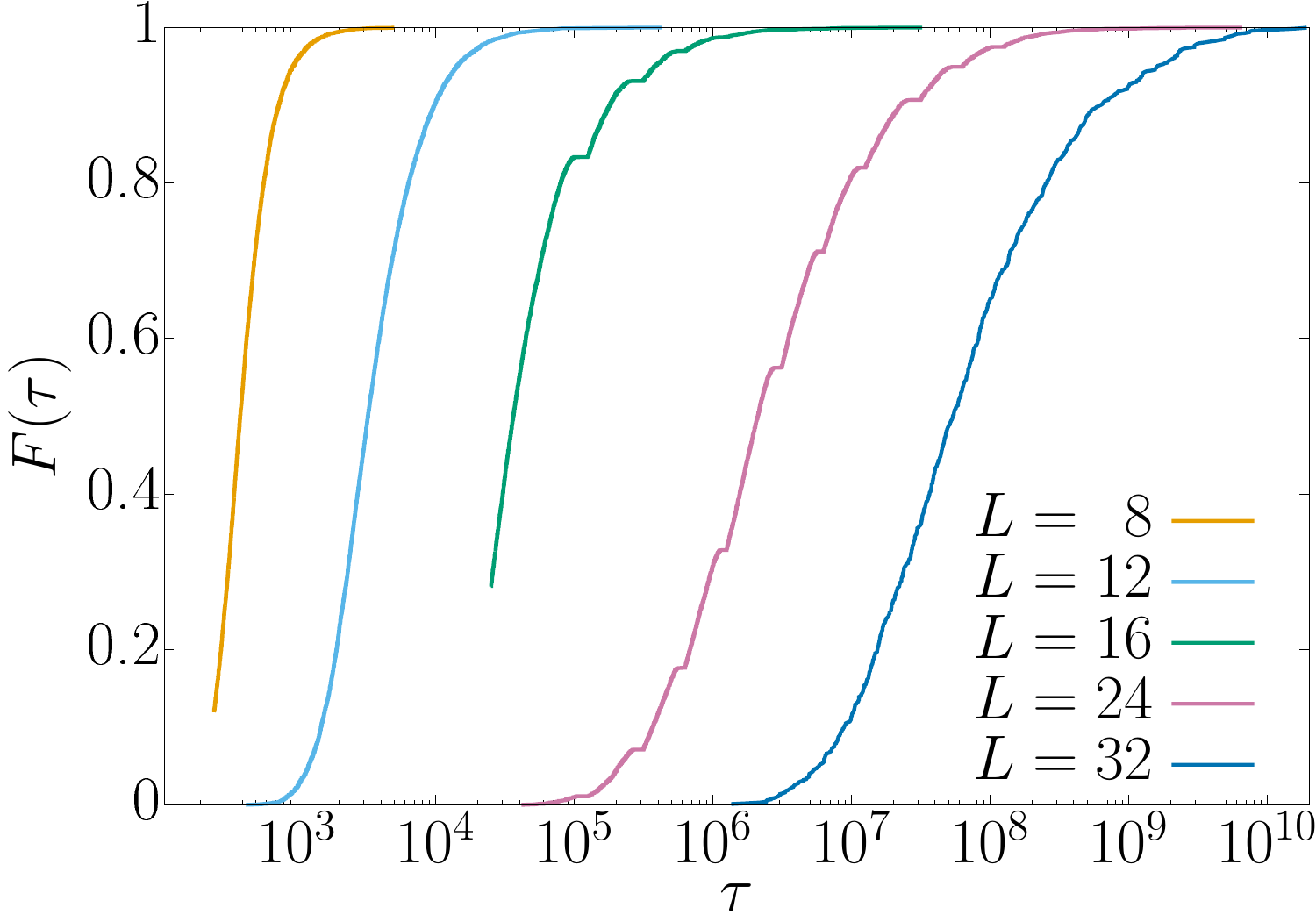}
\caption[\textbf{Empirical probability distribution of $\mathbf{\tau}$ for $\mathbf{L=8,12,16,24}$ and 32.}]{\textbf{Empirical probability distribution of $\mathbf{\tau}$ for $\mathbf{L=8,12,16,24}$ and 32.} For $L=8$ and $L=16$ some of the samples\index{sample} have $\tau$ smaller than our minimal resolution (if $\tau<n_{\mathrm{Met}}$ we cannot compute it safely). We show only the part of the distribution function that can be safely computed.}
\labfig{all_L_prob_tau}
\end{figure}

It is clear from~\reffig{all_L_prob_tau} that a simple translation would not attains the collapse of all the curves. Therefore, we study the exponent $\zpt$ for different parts of the distribution function, i.e. for different \textit{percentiles}. We use the different simulated sizes to compute an effective $\zpt$ exponent for each pair of lattices ($L_1,L_2$) by means of the definition
\begin{equation}
z^\mathrm{PT}(L_1,L_2,p)=\frac{\log(\tint(L_1,p)/\tint(L_2,p))}{\log(L_1/L_2)}\,,
\end{equation}
where $\tau(L_i,p)$ is determined by the implicit equation $F(\tau(L_i,p))=p/100$ where $p=1,\ldots,100$ is the so called percentile rank (i.e. $\tau(L_i,p)$ is the $p$-th percentile of the distribution for the size $L_i$).  

We have computed $z^\mathrm{PT}$ for three pairs of lattice sizes, (12,24), (16,24), and (24,32) and in~\reffig{z_percentile} we show the results as a function of the rank. As expected, we see that the scaling does not hold for the whole curve in general. However, we observe that for higher ranks, i.e. the most chaotic samples\index{sample}, the scaling hold for all the pairs of lattices. In addition, for the largest pair of lattices (24,32), the exponent $\zpt$ is independent of the percentile, within the statistical errors. This result is consistent with the previously drawn picture in~\cite{fernandez:13}: for small system sizes, it is almost impossible to find samples\index{sample} displaying strong \gls{TC}\index{temperature chaos} and one needs to go to larger lattices, like $L=32$ to find chaotic samples\index{sample} with a significant probability.

An interesting coincidence with the results of non-equilibrium simulations~\cite{janus:08b,janus:09b,fernandez:15,janus:17} could have a deep meaning. Indeed in non-equilibrium conditions one finds that the \gls{SG} coherence length\index{coherence length} $\xi$, in a lattice of size $L\gg\xi$, at temperature $T=0.7$ grows with the simulation time $t_\mathrm{w}$ as~\cite{janus:09b}
\begin{equation}
\xi(t_\mathrm{w})\propto t_w^{1/z(T)}\,,\quad z(T=0.7)=11.64(15)\,, \labeq{zeta_janus}
\end{equation}
where $z(T)$ is the so-called dynamic critical exponent\index{dynamic critical exponent}\index{dynamic critical exponent!zzzzz@\Also{aging rate}|gobbleone}, that turns out to be strongly temperature-dependent in the \gls{SG} phase\index{phase!low-temperature/spin-glass} $z(T)\propto T_\mathrm{c}/T$, see \refch{aging_rate}.

Our results for the lattice pair $(24,32)$ suggest that
\begin{equation}
z(T=0.7)\approx z^\mathrm{PT}(T_\mathrm{min}=0.7)\,. \labeq{z-coincide}
\end{equation}

\begin{figure}[h!]
\centering
\includegraphics[width=0.7\columnwidth]{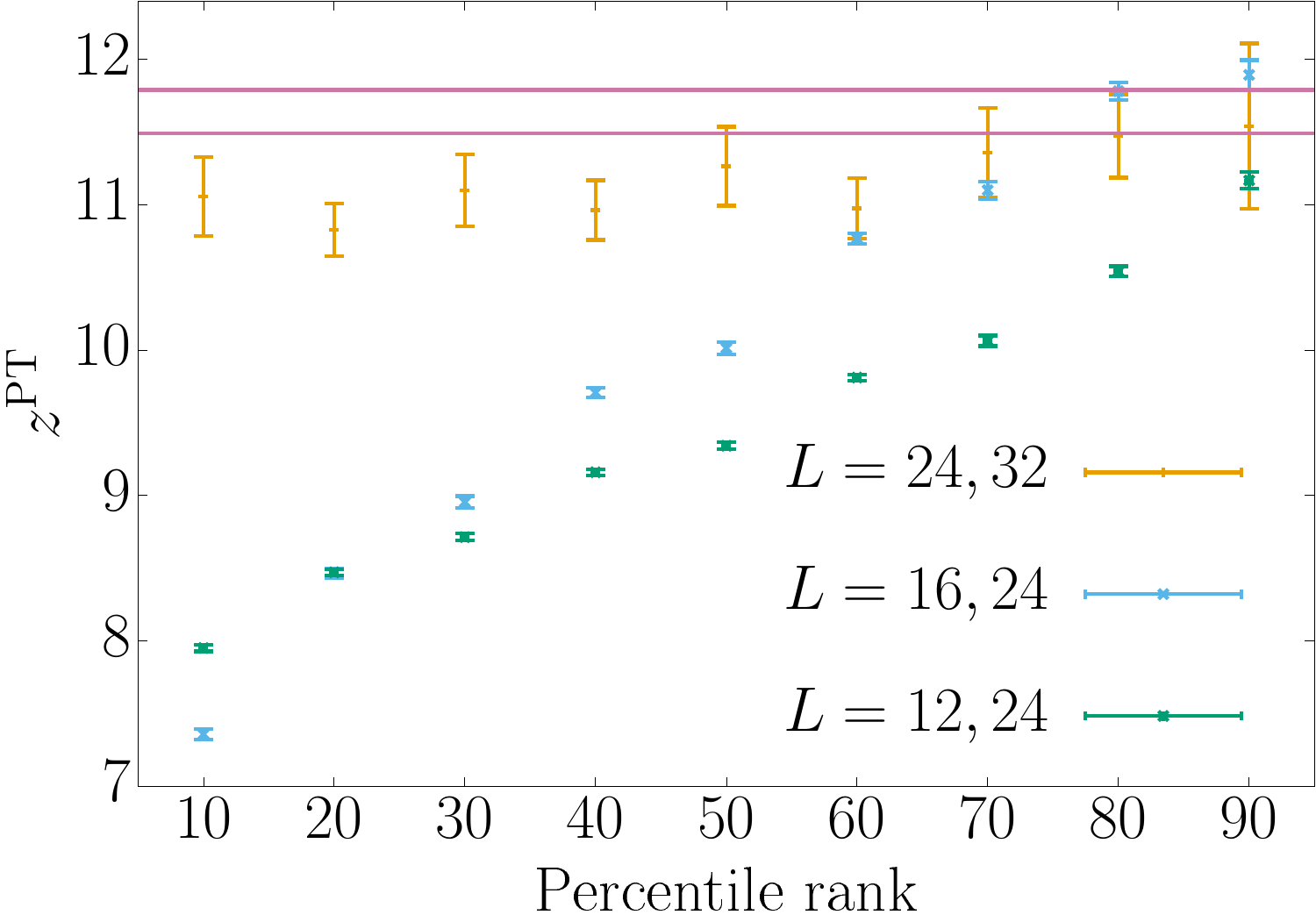}
\caption[\textbf{The effective exponent $\mathbf{\zpt}$.}]{\textbf{The effective exponente $\mathbf{\zpt}$.} The effective exponent $z^\mathrm{PT}(L_1,L_2,p)$ for three different pairs of lattice sizes (12,24), (16,24) and (24,32) as a function of the percentile rank $p$.  The two horizontal lines are the bounds for the off-equilibrium value $z=11.64(15)$ [see \refeq{zeta_janus}]. The numerical values of $z^\mathrm{PT}$ for the largest pair are compatible with the off-equilibrium value.}
\labfig{z_percentile}
\end{figure}

The surprising result of~\refeq{z-coincide} suggests the rescaling of the whole probability distribution as a test. This is done in figure \reffig{all_L_prob_tau_reescaled} (main) that shows $F(\tau)$ as a function of $y = \tau/L^z$. 

As expected, the data for $L=24$ and $L=32$ present a nice collapse. The curve corresponding to $L=16$ collapses into them for percentile ranks higher than $80$ only and the curve corresponding to $L=12$ collapses for percentile ranks higher than $90$. This is pointing to us that the chaotic behavior of the large $L$ limit is harder to find as we go to smaller $L$.

In figure~\ref{fig:all_L_prob_tau_reescaled} (inset), we show a log-log plot of $1\!-\!F(\tau)$ as a function of $\tau/L^z$, that emphasizes the large $\tau$ tail of the distribution. The fit presented shows that the probability density function of $\tau$ behaves, asymptotically for large $y$, like a fat tailed distribution:
\begin{equation}
\rho(y\equiv \tau/L^z) \sim y^{-1-a_1}\,,\quad a_1\approx 1.38 \,.
\end{equation}

The distribution seems to reach its asymptotic form for $L\geq 24$. Perhaps unsurprisingly, the thermodynamic (i.e. equilibrium) effective potential that characterizes temperature chaos turns out to be also asymptotic for $L\geq 24$~\cite{fernandez:13}.

\begin{figure}[h!]
\centering
\includegraphics[width=0.8\columnwidth]{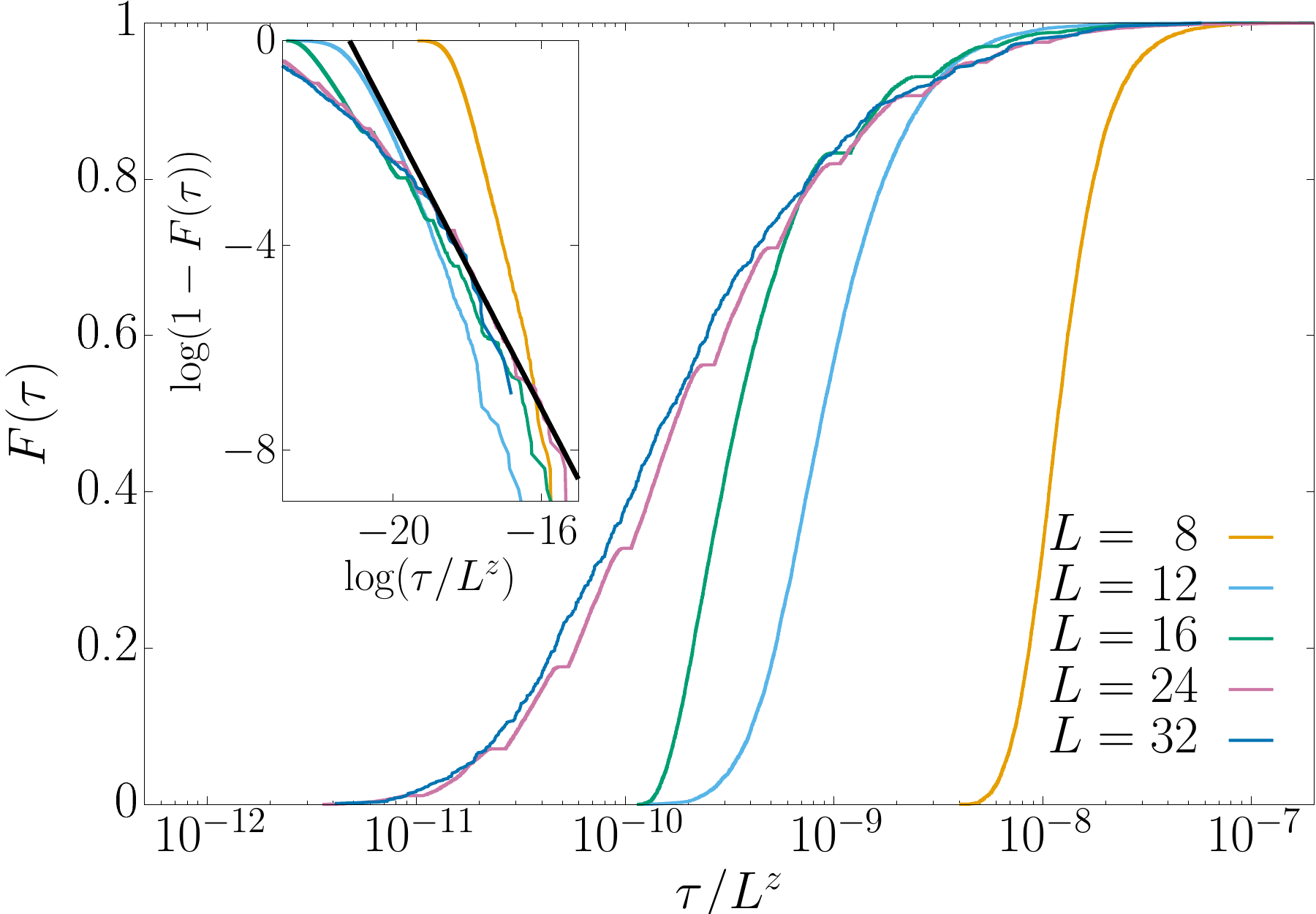}
\caption[\textbf{Probability distribution function of the rescaled variable \boldmath $y = \tau/L^{z}$}.]{\textbf{Probability distribution function of the rescaled variable \boldmath $y = \tau/L^{z}$}. $z$ is the dynamic exponent corresponding to $T_\mathrm{min}=0.7$, namely $z(T=0.7)=11.64(15)$. \textbf{(Inset)} Plot of $\log(1-F(\tau))$ versus $\log\left(\tau/L^{z}\right)$; the straight black line is a fit to the form $a_0 - a_1 \log(\tau/L^z)$ yielding $a_0 = -29.33$ and $a_1=-1.38$.}
\labfig{all_L_prob_tau_reescaled}
\end{figure}

\chapter[Temperature Chaos phenomenon in off-equilibrium Spin Glasses]{Temperature Chaos phenomenon in \\ off-equilibrium spin glasses} \labch{out-eq_chaos}

\gls{TC} phenomenon is described in terms of equilibrium configurations\index{configuration}, and therefore, all the previous work studying it is focused on equilibrated spin glasses. However, experiments are almost always carried out in non-equilibrium conditions. Moreover, the concept of \gls{TC}\index{temperature chaos} is not totally alien to the non-equilibrium regime in \gls{SG}. Actually, it has been related to the memory\index{memory effects} and rejuvenation\index{rejuvenation} effects~\cite{komori:00,berthier:02,picco:01,takayama:02,maiorano:05,jimenez:05}.

However, at this point, the relation of \gls{TC}\index{temperature chaos} with memory\index{memory effects} and rejuvenation\index{rejuvenation} effects is far from clear and no quantitative description of \gls{TC}\index{temperature chaos} under off-equilibrium conditions has been provided. In this chapter, we present a numerical work in off-equilibrium conditions that try to be the first step to fill that gap. All the simulations have been performed in the dedicated computer Janus\index{Janus} II~\cite{janus:14} and the high-accuracy achieved could not be possible without its computational power.

In the course of the chapter, we will explain why traditional approaches to study \gls{TC}\index{temperature chaos} do not work and we need, again, a rare-event analysis in order to fully understand the phenomenon. Moreover, the statics-dynamics equivalence\index{statics-dynamics equivalence}~\cite{barrat:01,janus:08b,janus:10b,janus:17} shows us the path to quantitatively study an equilibrium phenomenon in a non-equilibrium system.

The results of this work show us how, again, the coherence length\index{coherence length} $\xi$ rules the off-equilibrium phenomena in \gls{SG}s. In particular, a crossover behavior between a weak chaos regime and a strong chaos regime is found when $\xi$ grows. The characteristic length scale where this occurs, $\xi^*$, is related to its equilibrium counterpart, the chaotic length $\ell_c$ defined in~\refeq{def_chaotic_length}.

In \refsec{numerical_simulations_tc}, we give all the information about the performed numerical simulations. We explore the first \textit{naive} attempt to characterize \gls{TC}\index{temperature chaos} in off-equilibrium dynamics in~\refsec{average_killed_chaos_signal}. The computed observables to perform our rare-event analysis are introduced in~\refsec{observables_tc}. The rare-event analysis can be found in~\refsec{char} and its results in~\refsec{results}. In~\refsec{scaling_fixed_r} we explore the scaling behavior of our chaos-related quantities and, in \refsec{T-changes}, we focus on temperature changing protocols in order to make contact with the cumulative-aging\index{aging!cumulative} controversy (described in \refsec{memory_rejuvenation_introduction_chaos}) and lay the groundwork to future numerical works trying to relate simulations and experiments.

All the results provided in this chapter were originally published (in a reduced form) in~\cite{janus:21}.

\section{Numerical parameters of the simulation} \labsec{numerical_simulations_tc}
In this work, we simulate the \gls{EA}\index{Edwards-Anderson!model} model in three-dimensional spin glasses (\refsubsec{3D_EA_model}) for several temperatures $T$ in a lattice of linear size $L=160$, which is aimed to represent a system of infinite size. This assumption is sound, provided that $L\gg\xi$ (see~\reftab{xi_max}). Note that this condition limits the maximum time at which we can safely ignore finite-size effects\index{finite-size effects}. The temperature remains constant through the whole simulation, with the only exception of the runs reported and discussed in~\refsec{T-changes}.

We shall perform direct quenches from configurations\index{configuration} of spins randomly initialized (which corresponds to infinite temperature) to the working temperature $T<\Tc$, where the system is left to relax for a time $\tw$. This relaxation\index{relaxation} corresponds with the (very slow) growth of glassy magnetic domains\index{magnetic domain} of size $\xi(\tw)$.

We compute a total of $\NS = 16$ different samples\index{sample}. For each sample\index{sample} we shall consider $\NRep=512$ \emph{replicas}\index{replica} \footnote{It has been noted in~\cite{janus:18} (see also \refsec{Nr_aging}) that, for global observables (see~\refsubsec{observables-globales}), it is advantageous to have $\NRep\gg \NS$. However, working with $\NRep\gg \NS$ is not only a matter of numerical convenience for us. In fact, the local observables in~\refsubsec{observables-locales} are well defined only in the limit of $\NRep\to\infty$.}.

The simulation has been performed in the dedicated \gls{FPGA}-based\index{FPGA} computer Janus\index{Janus} II~\cite{janus:14} by using Metropolis dynamics. 

\begin{table}
\begin{center}
\begin{tabular}{l c r c c c r}
\toprule
\toprule
$T$ &  \hspace{1cm} & $\tw$ (MCs) & \hspace{1cm} & $\log_2(\tw)$ & \hspace{1cm} & $\xi_{\max}(\tw)$ \\
\toprule
$0.625$ & \hspace{1cm} & 42669909513 & \hspace{1cm} & 35.3 & \hspace{1cm} & 9.52(1) \\
\hline
$0.7$ & \hspace{1cm} & 48592007999 & \hspace{1cm} & 35.5 & \hspace{1cm} & 12.02(2) \\
\hline
$0.8$ & \hspace{1cm} & 34359738368 & \hspace{1cm} & 35\phantom{.5} & \hspace{1cm} & 15.84(5) \\
\hline 
$0.9$ & \hspace{1cm} & 17179869184 & \hspace{1cm} & 34\phantom{.5} & \hspace{1cm} & 20.34(6)  \\
\hline
$1.0$ & \hspace{1cm} & 4294967296 & \hspace{1cm} & 32\phantom{.5} & \hspace{1cm} & 24.4(1)\phantom{0} \\
\bottomrule
\end{tabular}
\caption[\textbf{Parameters of the simulations.}]{\textbf{Parameters of the simulations.} Maximum value of $\xi(\tw)$ simulated for each temperature. The central columns show the $\tw$ corresponding value for the computed $\xi(\tw)$.}
\labtab{xi_max}
\end{center}

\end{table}

\section{Taking spatial averages kills the chaotic signal} \labsec{average_killed_chaos_signal}

The first naive attempt to study \gls{TC}\index{temperature chaos} phenomenon in out-equilibrium \gls{SG}s consisted of studying global quantities affecting the whole system. However, in close analogy with equilibrium studies~\cite{fernandez:13}, we find that \gls{TC}\index{temperature chaos} is extremely weak when the full system is considered on average, see \reffig{xi_T1T2}, although the effect increases when the coherence length\index{coherence length} $\xi(\tw)$ grows (just as \gls{TC}\index{temperature chaos} becomes more visible in equilibrium when the system size increases).

In view of the above negative result, we have followed Ref.~\cite{fernandez:13} and performed a rare-event analysis that provides a satisfactory quantification of the \gls{TC}\index{temperature chaos} phenomenon. The
rationale for this approach is the statics-dynamics equivalence\index{statics-dynamics equivalence}~\cite{barrat:01,janus:08b,janus:10b,janus:17}: we expect to learn about the non-equilibrium dynamics of a spin glass (of infinite size), with a finite coherence length\index{coherence length} $\xi(\tw)$, by studying small samples\index{sample} of size $L\sim\xi(\tw)$ which can be equilibrated. 

\begin{figure}[t!] 
\centering 
\includegraphics[width=0.8\textwidth]{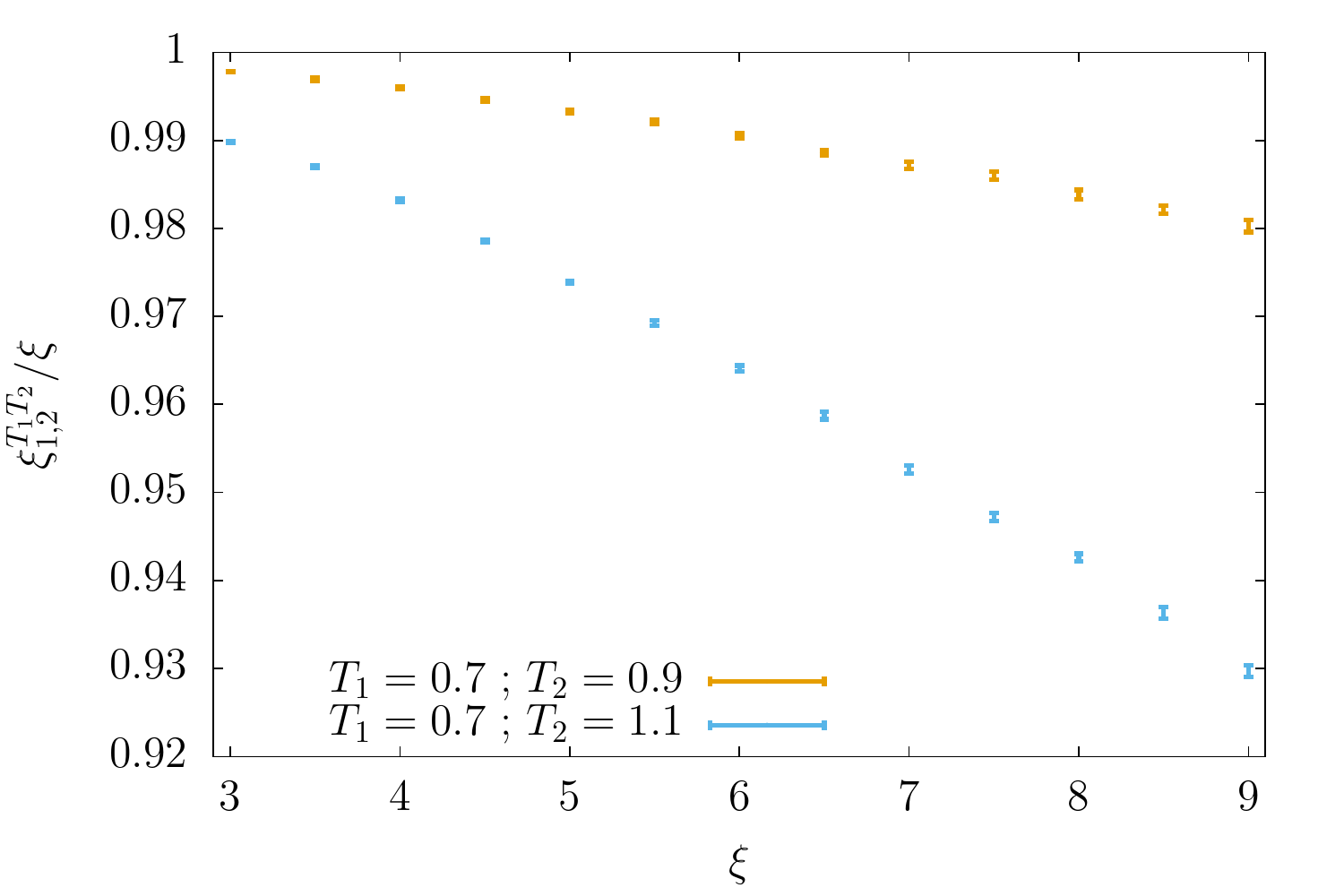}
\caption[\textbf{Non-equilibrium \gls{TC}\index{temperature chaos} is weak when averaging over the whole system.}]{\textbf{Non-equilibrium \gls{TC}\index{temperature chaos} is weak when averaging over the whole system.} We compare typical spin configurations\index{configuration} at temperature $T_1$ and time $t_{\mathrm{w},1}$ with configurations\index{configuration} at $T_2$ and time $t_{\mathrm{w},2}$. The comparison is carried through a global estimator of the coherence length\index{coherence length} of their overlap\index{overlap} $\xi^{T_1T_2}_{1,2}$, see \refeq{def_correlation_length_2T} (physically, $\xi^{T_1T_2}_{1,2}$ is the maximum length scale at which configurations\index{configuration} at temperatures $T_1$ and $T_2$ still look similar). The two times $t_{\mathrm{w},1}$ and $t_{\mathrm{w},2}$ are chosen in such a way that the configurations\index{configuration} at both temperatures have glassy-domains\index{magnetic domain} of the same size, namely $\xi_{1,2}(t_{\mathrm{w},1},T_1)=\xi_{1,2}(t_{\mathrm{w},2},T_2)=\xi$. The figure shows the ratio $\xi^{T_1T_2}_{1,2}/\xi$ as a function of $\xi$ for two pairs of temperatures $(T_1,T_2)$, recall that $\Tc\approx 1.1$, see~\refsubsec{3D_EA_model}. Under the hypothesis of fully developed \gls{TC}\index{temperature chaos}, we would expect $\xi^{T_1T_2}_{1,2}$ to be negligible as compared to $\xi$. Instead, our data show only a small decrease of $\xi^{T_1T_2}_{1,2}/\xi$ upon growing $\xi$ (the larger the difference $T_2-T_1$ the more pronounced the decrease).}
 \labfig{xi_T1T2}
\end{figure}

In our case, we shall be considering spatial regions (spheres) of linear size $\sim \xi(\tw)$, chosen randomly within a very large spin glass. Just as found with the small samples\index{sample} in equilibrium~\cite{fernandez:13}, we expect that a small fraction of our spheres will display strong \gls{TC}\index{temperature chaos}. The important question will be how this rare-event phenomenon evolves as $\xi(\tw)$ grows. In fact, we expect that our description of non-equilibrium \gls{TC}\index{temperature chaos} will allow us to perform sensible extrapolations to values of $\xi$ of experimental interest (for comparison, a typical experimental value is $\xi\sim 100$ lattice spacings, while in our simulations $\xi\sim 10$ lattice spacings).

\section{Observables} \labsec{observables_tc}
In this section, we briefly introduce the quantities that will help us to characterize \gls{TC}\index{temperature chaos} in off-equilibrium systems. The global observables (\refsubsec{observables-globales}) will be fundamental in order to characterize the relevant length (equivalently time) scales of the system. On the contrary, local observables (\refsubsec{observables-locales}) will be necessary in order to perform a rare-event analysis of the \gls{TC}\index{temperature chaos}.

\subsection{Global observables}\labsubsec{observables-globales}
The out-equilibrium time evolution is usually characterized by the growth of the coherence length\index{coherence length} $\xi(\tw)$ at temperature $T$, see~\refsubsec{observables_introduction}. In order to compute it, two basic observables are needed: the overlap\index{overlap!field} field and the four-point\index{correlation function!four point} spatial correlation function [see~\refeq{def_overlap} and~\refeq{def_C4}]. We repeat here the definitions for the reader's convenience
\begin{equation}
q^{\sigma,\tau}(x,\tw) = s_{x}^\sigma(\tw) s^\tau_{x}(\tw) \, ,
\end{equation}
\begin{equation} 
C_4(T,r,\tw) = \overline{\langle q^{\sigma,\tau}(x,\tw) q^{\sigma,\tau}(x+r,\tw)\rangle_T} \, . \labeq{def_corr_func}
\end{equation} 
In the previous definitions $ s_{x}^\sigma(\tw)$ is the spin of the replica\index{replica} $\sigma$ in the lattice position $x$ at time $\tw$, $\langle\dots\rangle_T$ is the average over thermal noise at temperature $T$, and $\overline{(\cdots)}$ is the average over the disorder\index{disorder!average}. Of course, the two replica\index{replica} indices $\sigma$ and $\tau$ should be different.

The correlation function in~\refeq{def_corr_func} can be extended for a pair of temperatures $T_1$ and $T_2$ in the following
way
\begin{equation}
C_4^{T_1T_2}(T_1,T_2,t_{\mathrm{w}1},t_{\mathrm{w}2},r) = \overline{\langle q^{\sigma(T_1),\tau(T_2)} (x,t_{\mathrm{w}1},t_{\mathrm{w}2}) q^{\sigma(T_1),\tau(T_2)}(x+r,t_{\mathrm{w}1},t_{\mathrm{w}2})\rangle_{T}} \, , \labeq{def_corr_func_2T}
\end{equation}
where now the thermal averages are taken at temperature $T_1$ for the replica\index{replica} $\sigma$, and at temperature $T_2$ for the replica\index{replica} $\tau$. From the four-point\index{correlation function!four point} correlation function we can compute the coherence length\index{coherence length} as have been explained in~\refsubsec{observables_introduction} and \refsec{finite_size_effects}.

Of course, the coherence length\index{coherence length} can be straightforwardly extended to a pair of temperatures $T_1$ and $T_2$ by using  $C_4^{T_1T_2}$ instead of $C_4$:
\begin{equation} 
I^{T_1T_2}_k( t_{\mathrm{w}1},t_{\mathrm{w}2}) = \int_{0}^{\infty}r^k\,C^{T_1T_2}_4(r,t_{\mathrm{w}1},t_{\mathrm{w}2})\,\mathrm{d} r \, , \labeq{def_integral_2T}
\end{equation}
and
\begin{equation}
\xi^{T_1T_2}_{k,k+1}(t_{\mathrm{w}1},t_{\mathrm{w}2}) = \dfrac{I^{T_1T_2}_{k+1}(t_{\mathrm{w}1},t_{\mathrm{w}2})}{I^{T_1T_2}_k(t_{\mathrm{w}1},t_{\mathrm{w}2})} \, . \labeq{def_correlation_length_2T}
\end{equation}
As a rule, we shall fix the two times $t_{\mathrm{w}1}$ and $t_{\mathrm{w}2}$ through the condition\footnote{Because our $\tw$ are in a discrete grid, we solve~\refeq{el_reloj_doble} for the \emph{global} overlaps\index{overlap} defined in~\refsubsec{observables-globales} through a (bi)linear interpolation.}:
\begin{equation}
\xi(t_{\mathrm{w}1},T_1)=\xi(t_{\mathrm{w}2},T_2)=\xi\,, \labeq{el_reloj_doble}
\end{equation}
that ensures that we are comparing spin-configurations\index{configuration} which are ordered on the same length scale.

\subsection{Local observables}\labsubsec{observables-locales}

In order to explore the heterogeneity of the system, we construct here local observables that will allow us to generalize the rare-event analysis in~\cite{fernandez:13}. Specifically, we shall be studying the properties of spherical regions.

We start by choosing $N_{\mathrm{sph}}=8000$ centers for the spheres, on each sample\index{sample}. The sphere centers are chosen randomly, with uniform probability, on the dual lattice~\footnote{The dual lattice of a cubic lattice with \gls{PBC}\index{boundary conditions!periodic} is another cubic lattice of the same size, and with \gls{PBC}\index{boundary conditions!periodic} as well. The nodes of the dual lattice are the centers of the elementary cells of the original lattice.}. The radius of the spheres is varied, but their centers are held fixed. Let $B_{s,r}$ be the $s$-th ball of radius $r$. 

Similarly as in the previous chapter (\refch{equilibrium_chaos}), our basic observable will be the overlap\index{overlap} between temperatures $T_1$ and $T_2$
\begin{equation}
q_{T_1,T_2}^{s,r}(\xi) = \dfrac{1}{N_r} \sum_{x\in B_{s,r}} s_{x}^{\sigma,T_1}(t_{\mathrm{w}1}) s_{x}^{\tau,T_2}(t_{\mathrm{w}2}) \>\> , \labeq{sphere_overlap}
\end{equation}
where $N_r$ is the number of spins within the ball of radius $r$, and the two times $t_{\mathrm{w}1}$ and $t_{\mathrm{w}2}$ are chosen according to~\refeq{el_reloj_doble}\footnote{Because the local overlaps\index{overlap} in~\refeq{sphere_overlap} have much larger fluctuations than the global overlaps\index{overlap} in~\refsubsec{observables-globales}, in this case we solve~\refeq{el_reloj_doble} in a cruder way. We just select the value of $t_{\mathrm{w}1}$ that yields the $\xi(t_{\mathrm{w}1},T_1)$ nearest to our target $\xi$ value. The same procedure is followed with $t_{\mathrm{w}2}$.}.
Next, again as in the previous chapter, we introduce the so called \textit{chaotic parameter}~\cite{ritort:94,ney-nifle:97,fernandez:13,billoire:14} which now is restricted to the balls $B_{s,r}$
\begin{equation} 
X^{s,r}_{T_1,T_2}(\xi) = \dfrac{\langle [q_{T_1,T_2}^{s,r}(\xi)]^2\rangle_T}{\sqrt{\langle[q_{T_1,T_1}^{s,r}(\xi)]^2\rangle_T \,\langle[q_{T_2,T_2}^{s,r}(\xi)]^2\rangle_T}} \, . \labeq{def_chaotic_parameter}
\end{equation} 
The extreme values of the chaotic parameter, just in close analogy with the equilibrium case, have a very clear interpretation: $X^{s,r}_{T_1,T_2}=1$ corresponds to a situation in which spin configurations\index{configuration} in the ball $B_{s,r}$, at temperatures $T_1$ and $T_2$, are completely indistinguishable (absence of chaos) while $X^{s,r}_{T_1,T_2}=0$ corresponds to completely different configurations\index{configuration} (which means strong \gls{TC}\index{temperature chaos}). A representative example of our results is shown in \reffig{hetereogeneity_chaos}.

\begin{figure}[h!] 
\centering 
\includegraphics[width=0.48\textwidth]{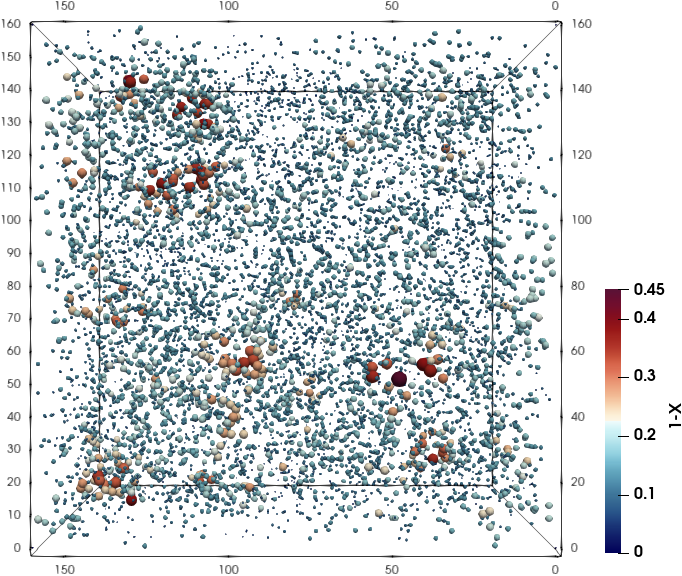}
\caption[\textbf{Dynamic temperature chaos is spatially heterogeneous.}]{\textbf{Dynamic temperature chaos is spatially heterogeneous.} The 8000 randomly chosen spheres in a sample\index{sample} of size $L=160$ are depicted with a color code depending on $1-X$ [$X$ is the chaotic parameter, \refeq{def_chaotic_parameter}, as computed for spheres of radius $r=12$, $\xi=12$ and temperatures $T_1=0.7$ and $T_2=1.0$]. For visualization purposes, spheres are represented with a radius $12(1-X)$, so that only fully chaotic spheres (i.e., $X=0$) have their real size.}
\labfig{hetereogeneity_chaos}
\end{figure}

We shall focus our attention on the distribution function
\begin{equation}
F(X,T_1,T_2,\xi,r)=\text{Probability}[X^{s,r}_{T_1,T_2}(\xi)<X] \, ,\labeq{F-def}
\end{equation}
and its inverse $X(F,T_1,T_2,\xi,r)$.

The alert reader will note that the computation of $X^{s,r}_{T_1,T_2}(\xi)$ requires the \emph{exact} computation of thermal expectation values, which would require an infinite number of replicas\index{replica}.  Fortunately, we have been able of extrapolating $X(F,T_1,T_2,\xi)$ to the limit $\NRep\to\infty$. Details about this extrapolation are provided in~\refsubsec{extrapolation}.

A final note: as explained in~\refsec{cambio_de_r}, we have found it advantageous to trade the sphere radius $r$ by $N_r^{1/3}$. Of course, $N_r$ equals $4\pi r^3/3$ for large $r$, but $N_r^{1/3}$ provides smoother interpolations at small $r$.

\section{The temperature-chaos distribution functions} \labsec{char}
\begin{figure}[t!]
\centering
\includegraphics[width=\textwidth]{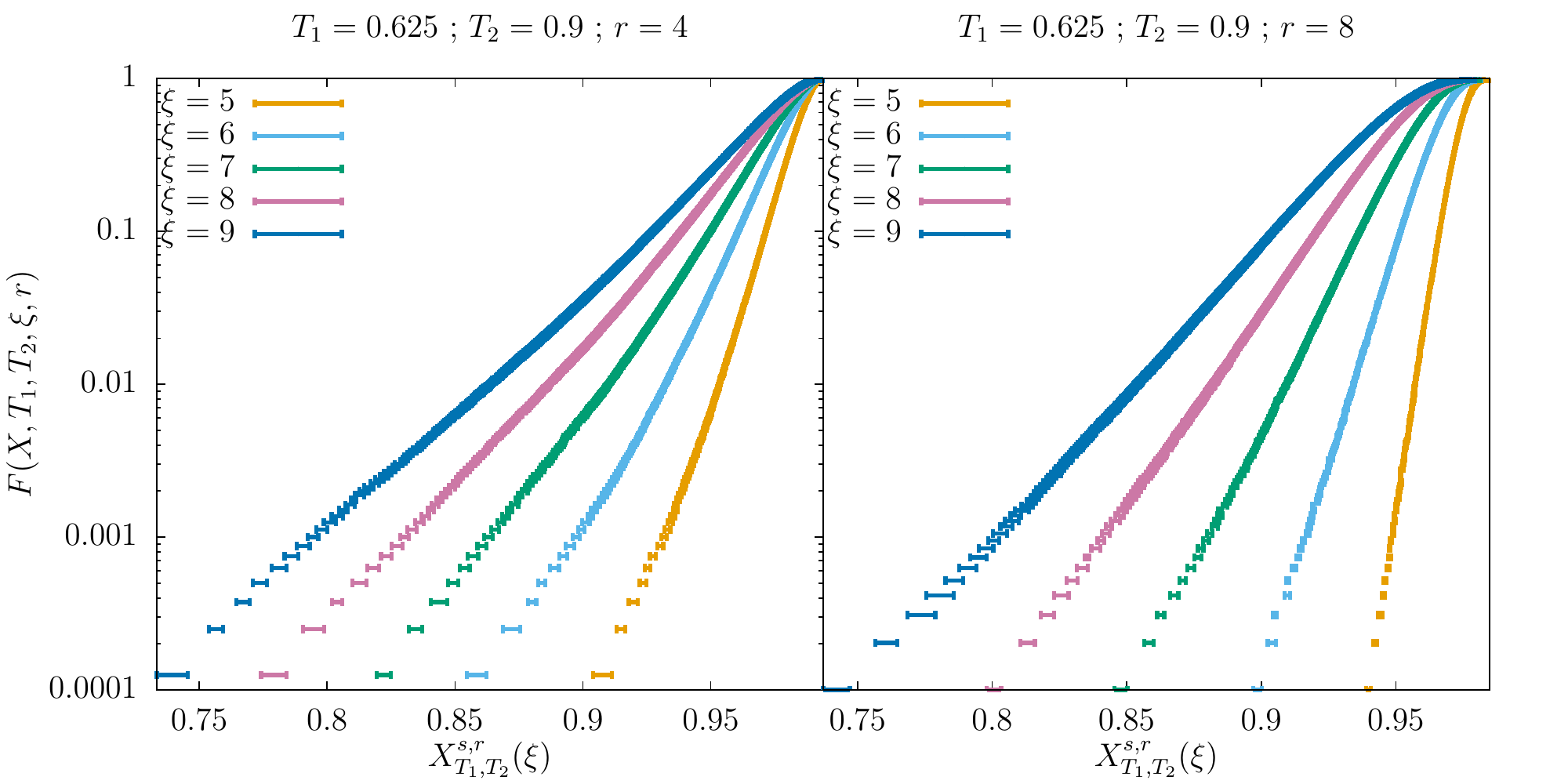}
\caption[\textbf{Temperature chaos increases with \boldmath $\xi(\tw)$.}]{\textbf{Temperature chaos increases with \boldmath $\xi(\tw)$.} The figure shows the distribution function $F(X,T_1,T_2,\xi,r)$, see~\refeq{F-def}, for $T_1=0.625$ and $T_2=0.9$, for spheres of radius $r=4$ (left) and $r=8$ (right), as computed for various values of $\xi$. The distributions have been extrapolated to $\NRep=\infty$. Error bars\index{error bars} are horizontal, because we have actually extrapolated the inverse function $X(F,T_1,T_2,\xi,r)$ (the extrapolation is not always safe for $F>0.3$, see~\refsubsec{extrapolation}; the same caveat applies to all the distribution functions shown in~\refsec{char} and~\refsec{results}). Most of the spheres have a chaotic parameter very close to $X=1$.}
\labfig{distribution_function_xi}
\end{figure}

\begin{figure}[t!]
\centering
\includegraphics[width=\textwidth]{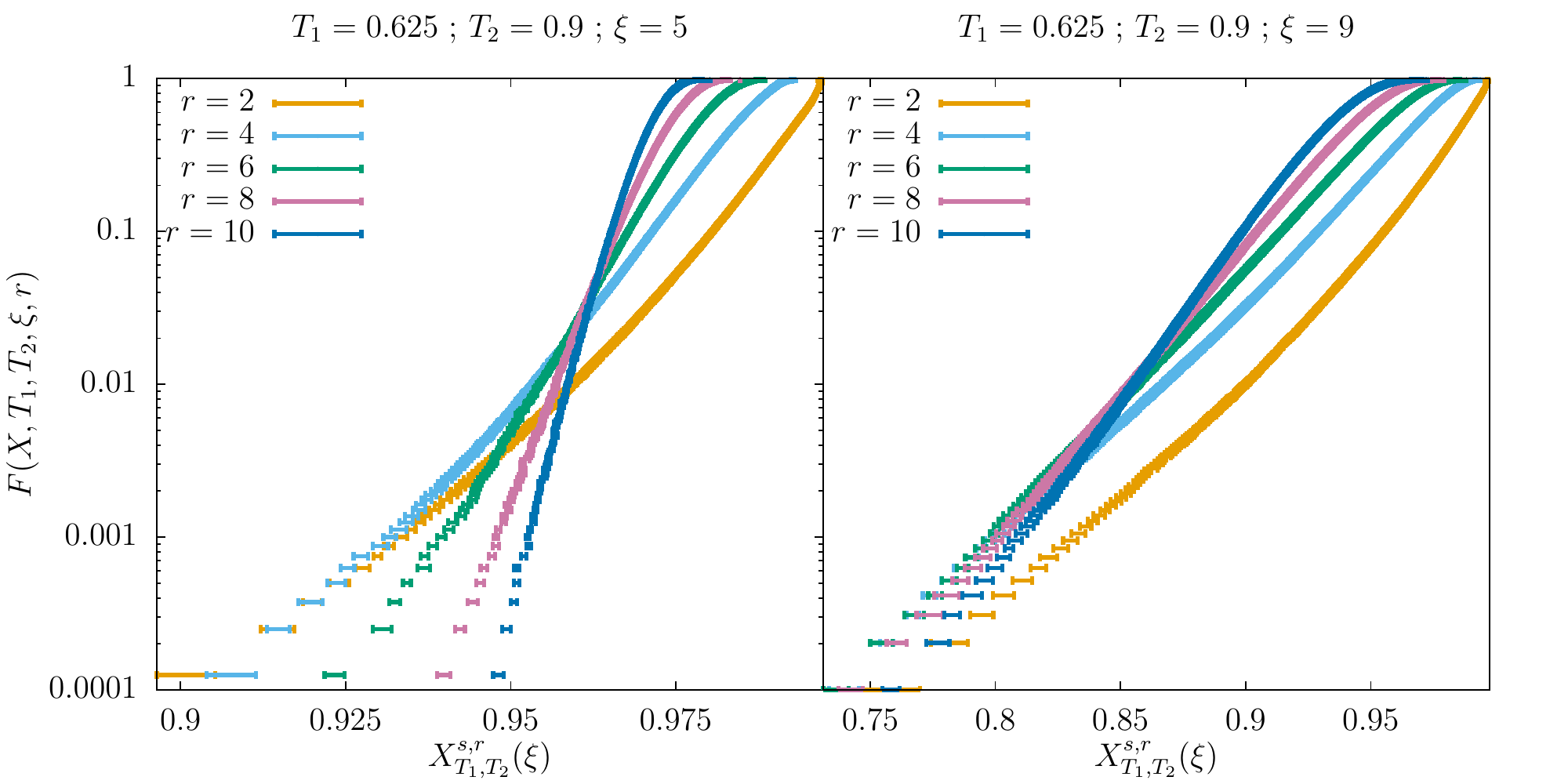}
\caption[\textbf{Dependency of Temperature Chaos on the size of the observation region.}]{\textbf{Dependency of Temperature Chaos on the size of the observation region.} The figure shows the distribution function $F(X,T_1,T_2,\xi,r)$ for $T_1=0.625$ and $T_2=0.9$, for coherence length\index{coherence length} $\xi=5$ (left) and $\xi=9$ (right), as computed for spheres of various radius $r$. If we focus on some low probability ($F=0.01$, for instance), we find that there is an optimal size for the observation of chaos (in the sense of a smallest chaotic parameter $X$).}
 \labfig{distribution_function_r}
\end{figure}

\begin{figure}[h!]
\centering
\includegraphics[width=\textwidth]{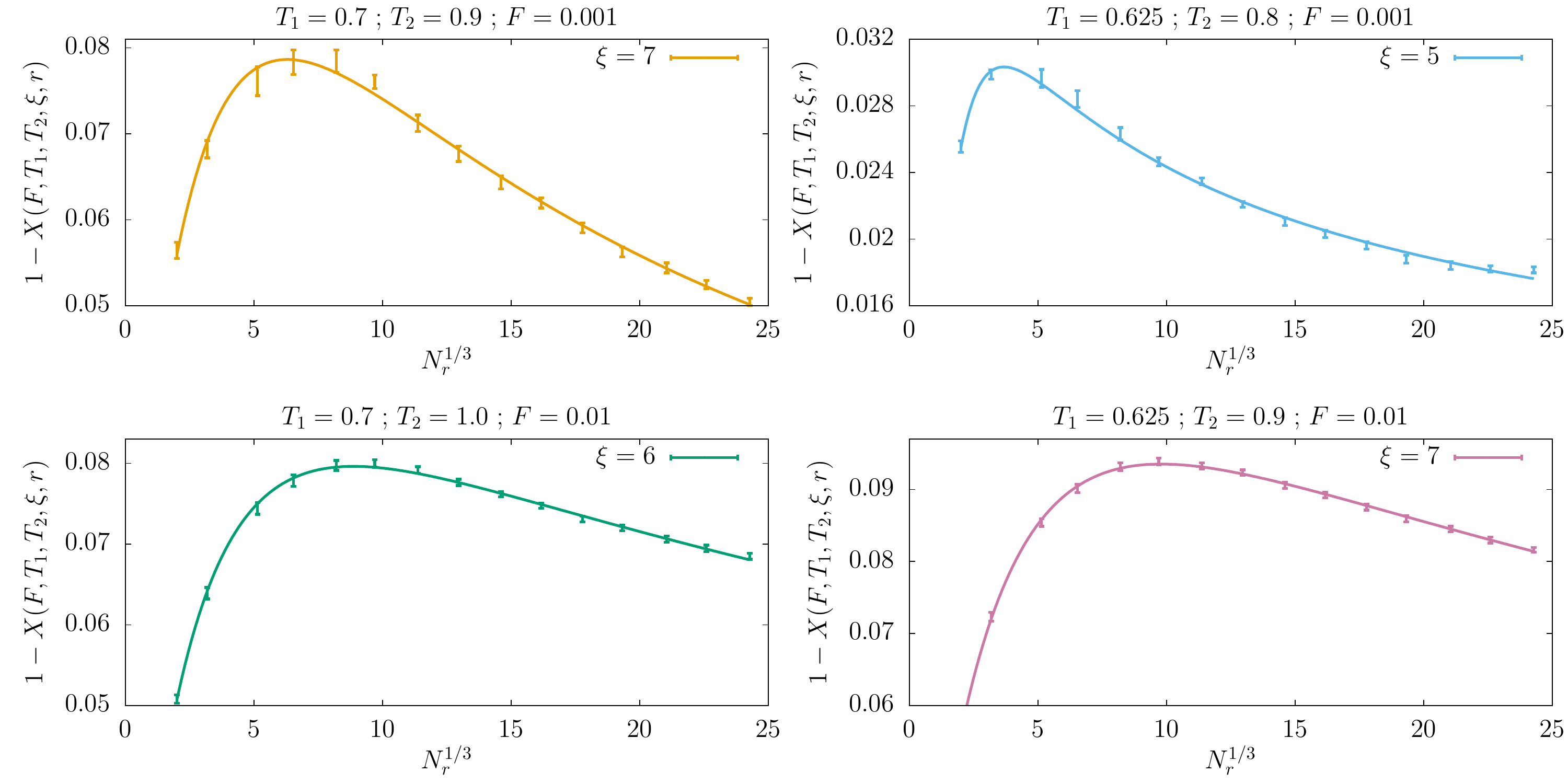}
\caption[\textbf{The complementary chaotic parameter $\mathbf{1-X(F,T_1,T_2,\xi,r)}$ for fixed $F$ as a function of $\mathbf{r}$.}]{\textbf{The complementary chaotic parameter $\mathbf{1-X(F,T_1,T_2,\xi,r)}$ for fixed $F$ as a function of $\mathbf{r}$.} The difference $1-X(F,T_1,T_2,\xi,r)$ [recall that $X(F,T_1,T_2,\xi,r)$ is the inverse of the distribution function, see~\refeq{F-def}] as a function of the cubic root of the number of points in the spheres $N_r^{1/3}$, as computed for different values of $F$, $T_1$, $T_2$ and $\xi$.  Our rationale for choosing $N_r^{1/3}$ as independent variable, rather than the radius of the spheres $r$, is explained in~\refsec{cambio_de_r}. In this representation, the size of the spheres which are optimal for the observation of chaos (for given parameters $F$, $T_1$, $T_2$ and $\xi$) appears as the maximum of these curves. Continuous lines are fits to~\refeq{functional_form}.}
\labfig{Xvsr_examples}
\end{figure}

Some examples of the distribution functions $F(X,T_1,T_2,\xi,r)$ can be found in~\reffig{distribution_function_xi} and~\reffig{distribution_function_r}, for typical (fixed) values of $T_1$ and $T_2$. Although most spheres are clearly non-chaotic ($X>0.9$), the situation is far more interesting for low probabilities (say $F=0.01$). For the sake of simplicity, consider first spheres of a fixed size (\reffig{distribution_function_xi}). For small $F$, we find that $X$ decreases significantly (and monotonically) upon growing $\xi$. The situation is more complex if we consider spheres of different sizes, for given $F$ and $\xi$.  As \reffig{distribution_function_r} shows, when the size of the spheres grows the chaotic parameter is non-monotonic.

The situation clarifies when we fix both the probability $F$ and the coherence length\index{coherence length} $\xi$, see~\reffig{Xvsr_examples}. Rather than the chaotic parameter, let us consider the difference $1-X$ (which grows when \gls{TC}\index{temperature chaos} becomes stronger). We find that $1-X$ peaks for one size of the spheres which indicates the optimal length scale for the study of \gls{TC}\index{temperature chaos} (however, see~\reffig{Xvsr_examples}, this peak is asymmetric and becomes broader when $\xi$ increases). Our main analysis in~\refsec{results} will correspond to the scaling with $\xi$ of these peaks.

Let us remark that, at least close to a maximum, any smooth curve is characterized by the position, height and width of the peak. In order to meaningfully compute these three parameters from our data (see e.g.~\reffig{Xvsr_examples}), we fit $1-X$ to
\begin{equation}
f(z) = \dfrac{az^b}{1+cz^d} \quad ,\quad z=N_r^{1/3}\,, \labeq{functional_form}
\end{equation}
($a$, $b$, $c$, and $d$ are the parameters of the fit). We extract the position, width and height from the fitted function $f(z)$. In order to compute errors in (say) the peak position $N_{r,\max}^{1/3}$ we use a Jackknife\index{Jackknife} method (see \refsec{estimating_errorbars} for further details): we perform a separated fit for each Jackknife\index{Jackknife} block, extract $N_{r,\max}^{1/3}$ from the fit for that block, and compute errors from the block fluctuations. Of course, Jackknife\index{Jackknife} blocks are formed from our $\NS=16$ samples\index{sample}. Let us stress that~\refeq{functional_form} is meant to be only a convenient way of characterizing the peak, without any deep meaning attached to it.

However, the reader may question whether or not the local peak description (i.e. position, height, and width) is sensible for the full curve. We provide some positive evidence in this respect in~\refsubsec{taylor}.

\subsection{Global versus local description of the peaks}\labsubsec{taylor} 
\begin{figure}[b!]
\centering
\includegraphics[width=0.8\textwidth]{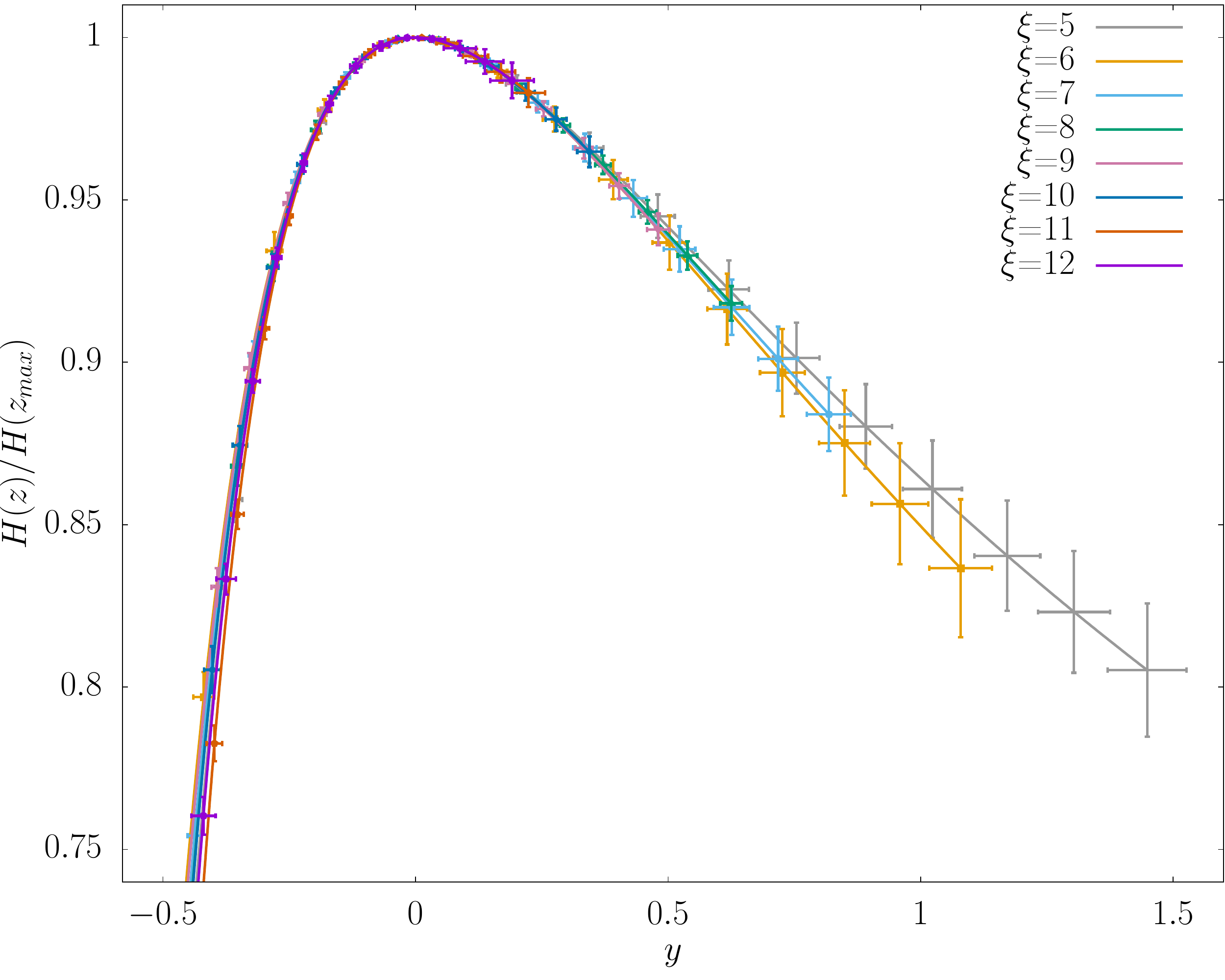}\\

\includegraphics[width=0.8\textwidth]{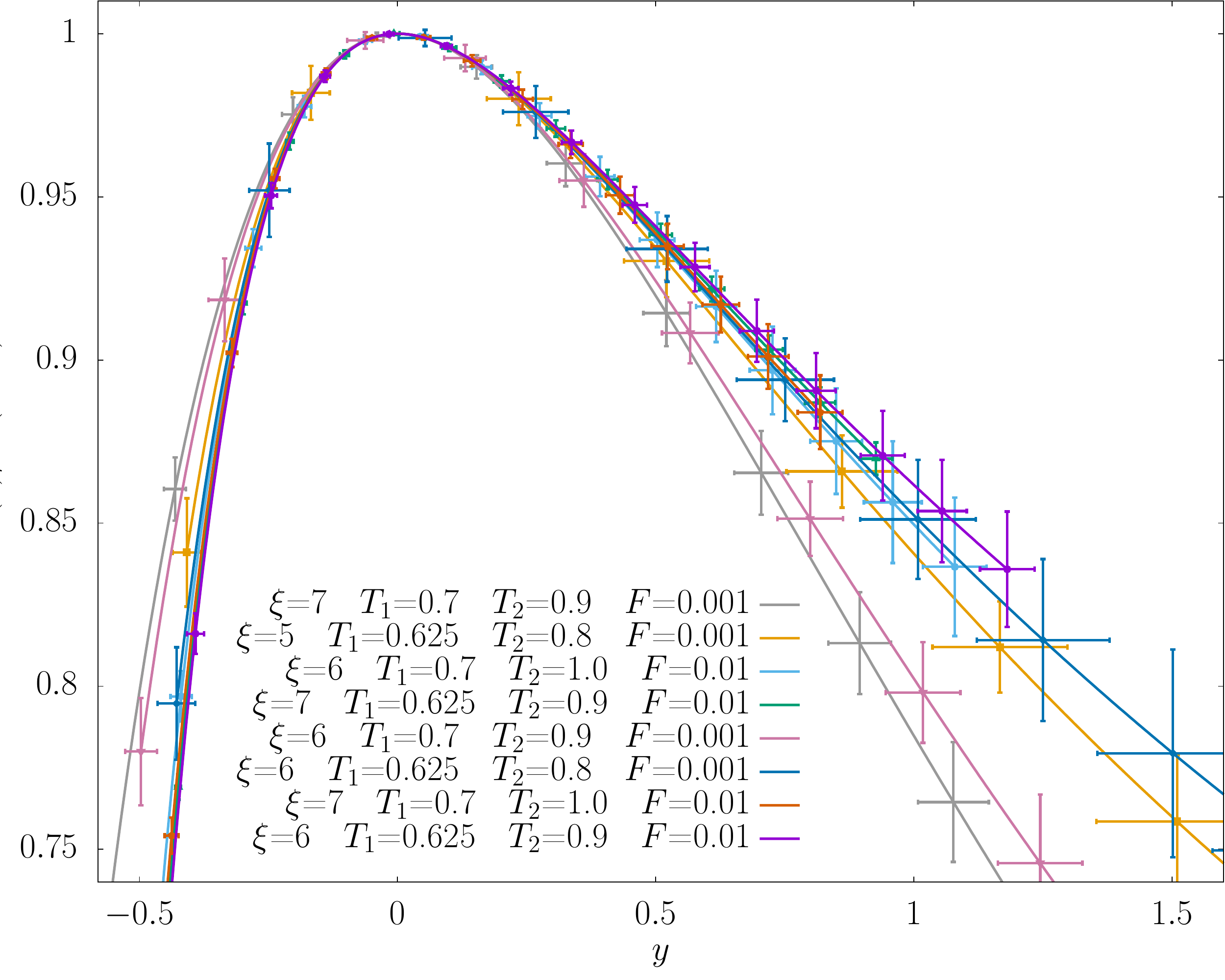}
\caption[\textbf{Universality in $1-X$ extends beyond the trivial Taylor's Universality.}]{\textbf{Universality in $1-X$ extends beyond the trivial Taylor's Universality.} The upper part shows $1-X$ in units of its peak value, for the temperatures $T_1=0.7$, $T_2=1.0$ and $F=0.01$. Taylor's theorem implies that, using the independent variable $y$ [see~\refeq{Taylor-universality}], the different curves should coincide close to $y=0$. However, we see that the coincidence holds beyond the quadratic approximation (as evinced by the strong asymmetry of the master curve). The lower panel shows the same set of temperatures $T_1$ and $T_2$ and probabilities $F$ shown in~\reffig{Xvsr_examples} (we have added data for several coherence length\index{coherence length}). Mixing different values of $F$, $T_1$ and $T_2$ leads to significant discrepancies for large values of $|y|$. Nevertheless, the collapse of the curves is still present in the range $y \in (-0.3,0.5)$ where the asymmetry is not negligible.}
\labfig{taylor}
\end{figure}

Consider any smooth, positive function $H(z)$, with a local maximum at $z=z_{\max}$. Close to this peak, Taylor's theorem implies some (trivial) Universality
\begin{equation}
\frac{H(z)}{H(z_{\max})}=1-\frac{1}{2} y^2+{\cal O}(y^3)\,,\text{ where } \,
y=\sqrt{\frac{|H''(z_{\max})|}{H(z_{\max})}}(z-z_{\max})\,. \labeq{Taylor-universality}
\end{equation}
Note that, in the language of the previous paragraph, the peak position is $z_{\text{max}}$, its heigth is $H(z_{\max})$ and its (inverse) width is $\sqrt{|H''(z_{\max})|/H(z_{\max})}$. Of course, in principle, there is no reason for~\refeq{Taylor-universality} to be accurate away from the peak. However, \refeq{Taylor-universality} suggests yet another representation for our $1-X$ curves, see~\reffig{taylor}. We note that, in this new representation, the $1-X$ curves are invariant under changes of coherence length\index{coherence length} $\xi$ (\reffig{taylor} upper panel). However, when considering changes in the temperatures $T_1$ and $T_2$ and the probability $F$, the curves mildly differ away from the peak (see~\reffig{taylor} lower panel). This (approximate) independence in $(T_1,T_2,F,\xi)$ is a fortunate fact because the complexity of the problem gets reduced to the study of the scaling with $\xi$ of the three peak parameters while keeping constant $(T_1,T_2,F)$.

\section{The off-equilibrium characterization of Temperature Chaos}\labsec{results}
In this section we present the scaling of the peak position (\refsubsec{peaks-position}), the peak height (\refsubsec{peaks-height}) and the peak (inverse) width (\refsubsec{peaks-width}) with the coherence length\index{coherence length} $\xi(\tw)$.

Due to the difficulty of characterizing peaks which exhibit weak \gls{TC}\index{temperature chaos}, in the following analysis we exclude the data corresponding to the pair of temperatures ($T_1=0.625,T_2=0.7$) at the probability level $F=0.01$ (see~\refsec{peak_characterization} for further details).

\subsection{The peak position}\labsubsec{peaks-position}
\begin{figure}[b!]
  \centering
  \includegraphics[width=\textwidth]{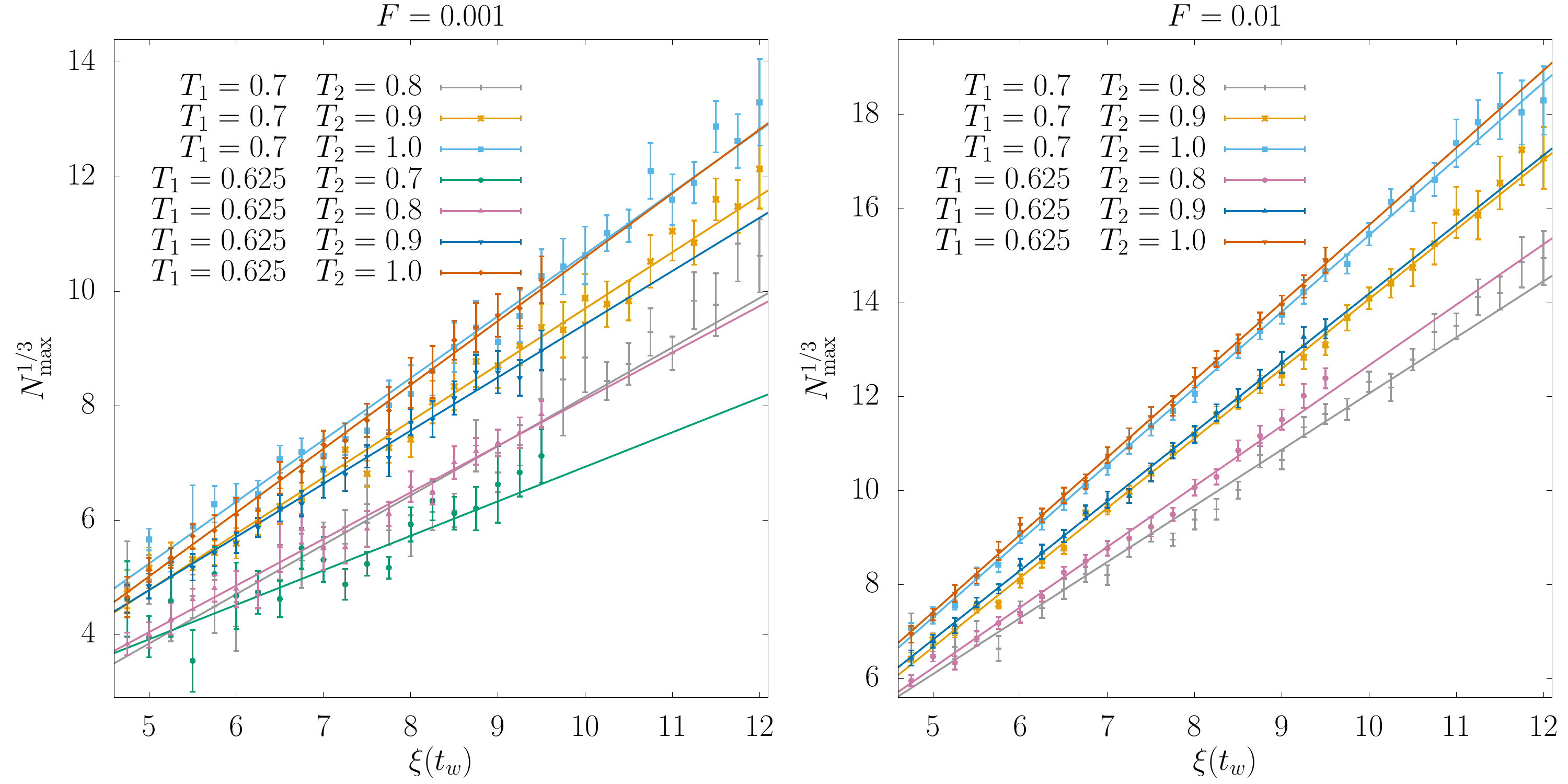}
  \caption[\textbf{Peak position increases with $\xi(\tw)$.}]{\textbf{Peak position increases with $\xi(\tw)$.} Position of the peak against $\xi(\tw)$ is plotted for all the simulated pair of temperatures $T_1$ and $T_2$ at different probability levels, $F=0.001$ (left panel) and $F=0.01$ (right panel).}
 \labfig{Nr_xi}
\end{figure}

Let us recall that the peak position indicates the most convenient length-scale for studying \gls{TC}\index{temperature chaos} (for a given coherence length\index{coherence length} $\xi$, probability $F$ and temperatures $T_1$ and $T_2$). Dimensional analysis suggests the linear fit as the natural ansatz to study the scaling of the peak position $N_{r,\max}^{1/3}$ with the coherence length\index{coherence length} $\xi(\tw)$ (indeed, both quantities are lengths):
\begin{equation}
N_{r,\max}^{1/3} = a \> \xi(\tw) + b \, \, . \labeq{Nr_xi}
\end{equation}
\reffig{Nr_xi} and~\reftab{parametros_Nmax} show the fits to~\refeq{Nr_xi}. In all cases, values of the parameter $b$ are compatible with $0$ (at the two-$\sigma$ level). In addition, the proportional parameter $a$ shows a monotone increasing behavior with $T_2-T_1$ and with the probability $F$. Hence, our naive expectation $N_{r,\max}^{1/3} \propto \xi(\tw)$ is confirmed.

\begin{table}[t!]
\begin{center}
\begin{tabular}{c c c c c c c c c c c}
\toprule
\toprule
$F$ & \hspace{1cm} & $T_1$ &  \hspace{1cm} & $T_2$ & \hspace{1cm} & $a$ & \hspace{1cm} &$b$ & \hspace{1cm} & $\chi^2/\mathrm{d.o.f.}$\index{degree of freedom} \\
\hline \hline
0.001 & \hspace{1cm} & 0.625 &  \hspace{1cm} & 0.7 & \hspace{1cm} & 0.60(12) & \hspace{1cm} & 0.9(9) & \hspace{1cm} & 22.12/19 \\
\hline 
0.001 & \hspace{1cm} & 0.625 &  \hspace{1cm} & 0.8 & \hspace{1cm} & 0.81(7) & \hspace{1cm} & 0.0(5) & \hspace{1cm} & 11.52/19 \\
\hline 
0.001 & \hspace{1cm} & 0.625 &  \hspace{1cm} & 0.9 & \hspace{1cm} & 0.93(10) & \hspace{1cm} & 0.1(6) & \hspace{1cm} & 5.35/19 \\
\hline 
0.001 & \hspace{1cm} & 0.625 &  \hspace{1cm} & 1.0 & \hspace{1cm} & 1.13(13) & \hspace{1cm} & -0.6(8) & \hspace{1cm} & 3.99/19 \\
\hline 
\hline
0.001 & \hspace{1cm} & 0.7 &  \hspace{1cm} & 0.8 & \hspace{1cm} & 0.86(8) & \hspace{1cm} & -0.5(6) & \hspace{1cm} & 43.40/28 \\
\hline 
0.001 & \hspace{1cm} & 0.7 &  \hspace{1cm} & 0.9 & \hspace{1cm} & 0.98(8) & \hspace{1cm} & -0.1(6) & \hspace{1cm} & 14.90/28 \\
\hline 
0.001 & \hspace{1cm} & 0.7 &  \hspace{1cm} & 1.0 & \hspace{1cm} & 1.08(7) & \hspace{1cm} & -0.2(6) & \hspace{1cm} & 22.32/28 \\
\hline 
\hline \hline
0.01 & \hspace{1cm} & 0.625 &  \hspace{1cm} & 0.8 & \hspace{1cm} & 1.29(5) & \hspace{1cm} & -0.2(3) & \hspace{1cm} & 22.30/19 \\
\hline 
0.01 & \hspace{1cm} & 0.625 &  \hspace{1cm} & 0.9 & \hspace{1cm} & 1.47(6) & \hspace{1cm} & -0.5(4) & \hspace{1cm} & 7.32/19 \\
\hline 
0.01 & \hspace{1cm} & 0.625 &  \hspace{1cm} & 1.0 & \hspace{1cm} & 1.65(6) & \hspace{1cm} & -0.8(4) & \hspace{1cm} & 4.83/19 \\
\hline 
\hline
0.01 & \hspace{1cm} & 0.7 &  \hspace{1cm} & 0.8 & \hspace{1cm} & 1.19(6) & \hspace{1cm} & 0.1(4) & \hspace{1cm} & 53.23/28 \\
\hline 
0.01 & \hspace{1cm} & 0.7 &  \hspace{1cm} & 0.9 & \hspace{1cm} & 1.48(9) & \hspace{1cm} & -0.7(6) & \hspace{1cm} & 17.19/28 \\
\hline 
0.01 & \hspace{1cm} & 0.7 &  \hspace{1cm} & 1.0 & \hspace{1cm} & 1.63(9) & \hspace{1cm} & -0.8(6) & \hspace{1cm} & 10.81/28 \\
\bottomrule
\end{tabular}
\caption[\textbf{Peak position characterization.}]{\textbf{Peak position characterization.} Parameters obtained in the fits of our data for $N_{r,\max}^{1/3}$ to \refeq{Nr_xi}. For each fit, we also report the figure of merit $\chi^2/\mathrm{d.o.f.}$\index{degree of freedom}}
\labtab{parametros_Nmax}
\end{center}
\end{table}

\subsection{The peak height}\labsubsec{peaks-height}

\begin{figure}[!h]
  \centering
  \includegraphics[width=1\textwidth]{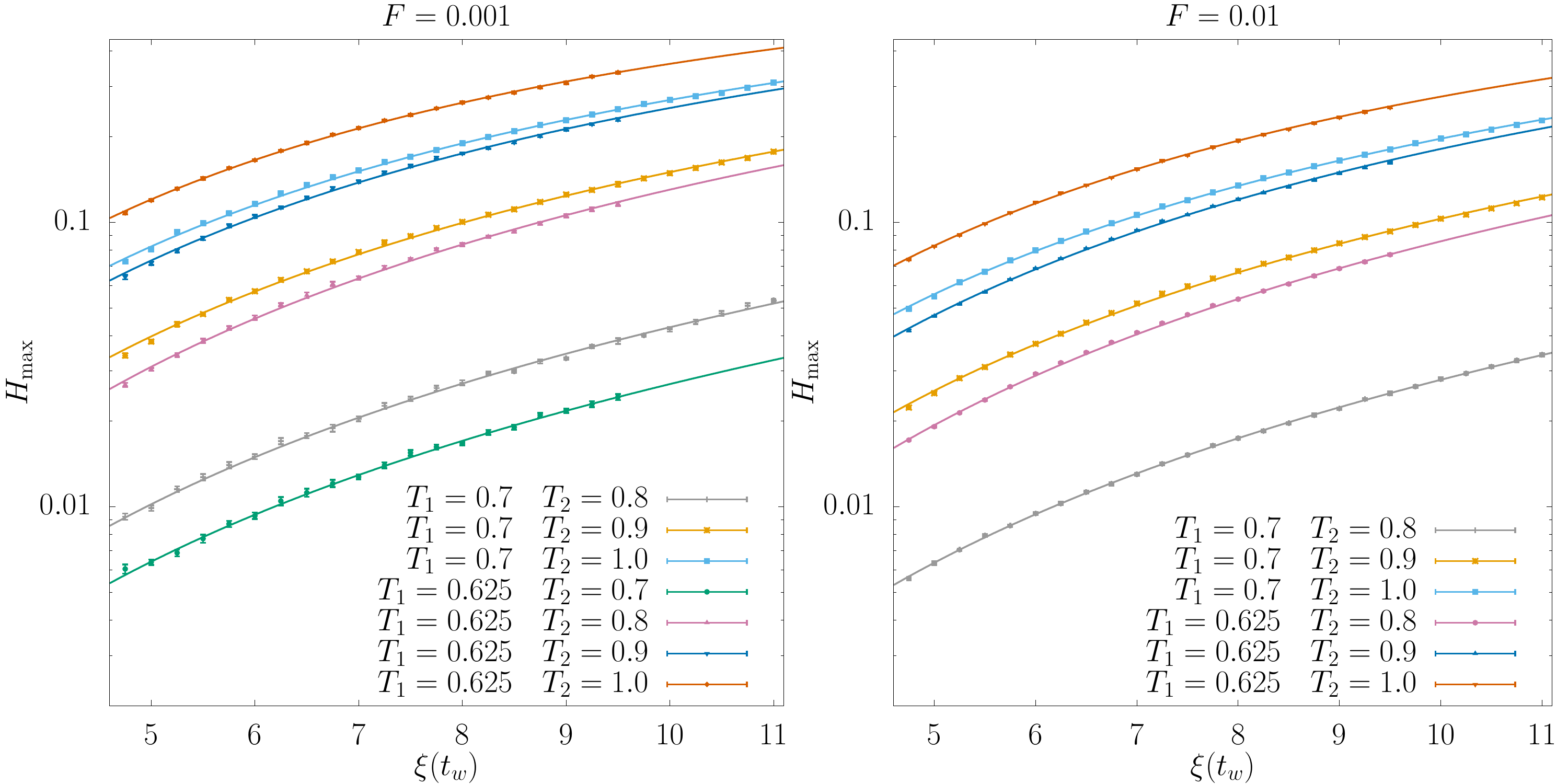}
  \caption[\textbf{Peak height increases with $\xi(\tw)$.}]{\textbf{Peak height increases with $\xi(\tw)$.} Height of the peak against $\xi(\tw)$ is plotted for all the simulated pair of temperatures $T_1$ and $T_2$ at different probability levels, $F=0.001$ (left panel) and $F=0.01$ (right panel). Curves display a monotone trend with the difference of temperatures $T_2-T_1$.}
 \labfig{fmax_xi}
\end{figure}

\begin{figure}[!h]
	\centering
	\includegraphics[width=0.7\textwidth]{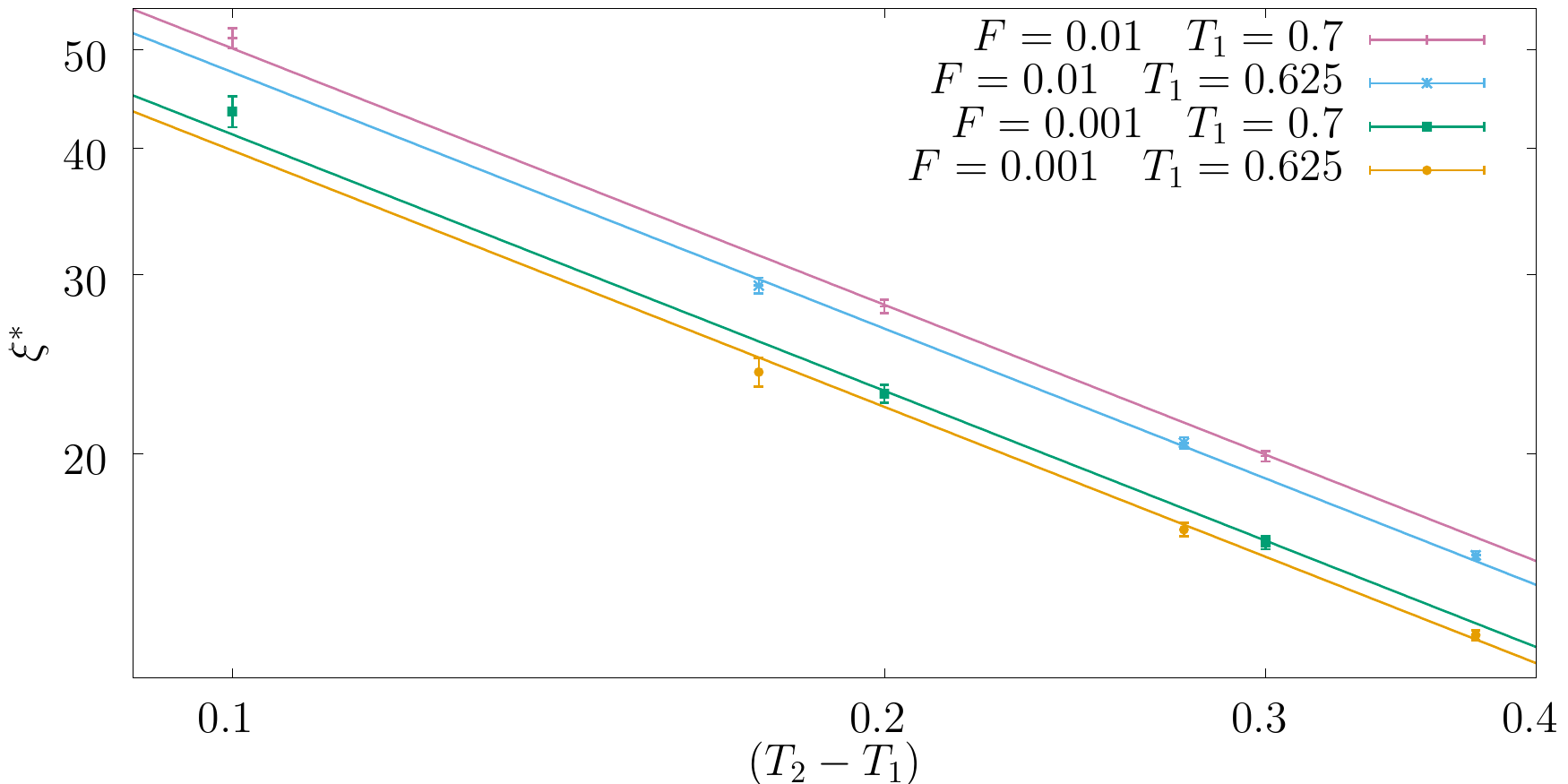}
	\caption[\textbf{The exponent $\mathbf{\zeta_{\text{NE}}}$ turns out to be independent of $\mathbf{F}$ and $\mathbf{T_1}$.}]{\textbf{The exponent $\mathbf{\zeta_{\text{NE}}}$ turns out to be independent of $\mathbf{F}$ and $\mathbf{T_1}$.} The characteristic length $\xi^*$ is plotted against the temperature difference $T_2-T_1$ in a log-log scale. Each curve is uniquely identified by the probability level $F$ and the smallest temperature of each pair $T_1$. Fits to~\eqref{eq:def_zeta}, enforcing a common exponent, are shown with continuous lines and result in a chaotic exponent $\zeta_\text{NE}=1.19(2)$.}
 \labfig{zeta_exponent}
\end{figure}

The peak height $H_{\max} \equiv H(z_{\max})$ is an indication of the strength of \gls{TC}\index{temperature chaos} (for a given coherence length\index{coherence length} $\xi$, probability $F$ and temperatures $T_1$ and $T_2$). In order to study the scaling with $\xi$, we have considered the following ansatz:
\begin{equation}
  H_{\max}(\xi) = \dfrac{\varepsilon(\xi)}{1+ \varepsilon(\xi)} \, , \text{ with } \varepsilon(\xi)=(\xi/\xi^*)^\alpha\,. \labeq{fmax_xi}
\end{equation}
The fit parameters are the characteristic length scale $\xi^*$ and the exponent $\alpha$. The rationale behind~\refeq{fmax_xi} is that, although in cases of extremely weak chaos $1-X$ may grow with $\xi$ as a power law, $1-X$ should eventually approach its upper bound $1-X=1$ (when chaos becomes strong).
Nevertheless, a consistency check necessary to give some physical meaning to~\refeq{fmax_xi}, is that exponent $\alpha$ should not depend neither on temperatures $T_1$ and $T_2$ nor on the chosen probability $F$.

We find fair fits to~\refeq{fmax_xi}, see~\reffig{fmax_xi} and~\reftab{parametros_fmax}. Fortunately, in all cases exponent $\alpha$ turns out to be very close to $\alpha \approx 2.1$ (actually, all the $\alpha$ obtained in the fits turn out to be compatible with $2.1$ at the two-$\sigma$ level). Under these conditions, we can interpret $\xi^*$ as a characteristic length indicating the crossover from weak to strong \gls{TC}\index{temperature chaos}, at the probability level indicated by $F$ (the relatively large value of exponent $\alpha$ indicates that this crossover is sharp). The trends for the crossover-length $\xi^*(F,T_1,T_2)$ are very clear: it grows upon increasing $F$ or upon decreasing $T_2-T_1$. At this point, we can try to be more quantitative. 

Indeed, because $\xi^*$ indicates the crossover between weak and strong chaos, it must be the non-equilibrium analogue of the equilibrium chaotic length $\ell_\text{c}(T_1,T_2)$~\cite{fisher:86,bray:87b} (see~\refsec{origin_tc}). Now, the equilibrium $\ell_\text{c}(T_1,T_2)$ has been found to scale for the 3D Ising\index{Ising} spin glass as 
\begin{equation}
\ell_\text{c}(T_1,T_2) \propto (T_2-T_1)^{-1/\zeta} \, \, , \labeq{def_zeta_equilibrium}
\end{equation}
with $\zeta \approx 1.07$~\cite{katzgraber:07} and $\zeta \approx 1.07(5)$~\cite{fernandez:13}. These considerations suggest the following ansatz for the non-equilibrium crossover length
\begin{equation}
\xi^*(T_1,T_2,F) = B(F) \, (T_2-T_1)^{-1/\zeta_{\text{NE}}} \, \, , \labeq{def_zeta}
\end{equation}
where $B(F)$ is an amplitude.

We have tested~\refeq{def_zeta} by computing a joint fit for four $(T_1, F)$ pairs as functions of $T_2-T_1$, allowing each curve to have its own amplitude but enforcing a common $\zeta_\text{NE}$ (see~\reffig{zeta_exponent}). The resulting $\chi^2/\text{d.o.f.} = 7.55/7$\index{degree of freedom} validates our ansatz, with an exponent $\zeta_\text{NE}=1.19(2)$ fairly close  to the equilibrium result $\zeta=1.07(5)$~\cite{fernandez:13}. This agreement strongly supports our physical interpretation of the crossover length. We, furthermore, find that $B$ is only weakly dependent on $T_1$.
Nevertheless, the reader should be warned that it has been suggested~\cite{fernandez:13} that the equilibrium exponent $\zeta$ may be different in the weak- and strong-chaos regimes.

\begin{table}[t!]
\begin{center}
\begin{tabular}{c c c c c c c c c c c}
\toprule
\toprule
$F$ & \hspace{1cm} & $T_1$ &  \hspace{1cm} & $T_2$ & \hspace{1cm} & $\xi^*$ & \hspace{1cm} &$\alpha$ & \hspace{1cm} & $\chi^2/\mathrm{d.o.f.}$\index{degree of freedom} \\
\hline \hline
0.001 & \hspace{1cm} & 0.625 &  \hspace{1cm} & 0.7 & \hspace{1cm} & 55(4) & \hspace{1cm} & 2.10(7) & \hspace{1cm} & 14.10/19 \\
\hline 
0.001 & \hspace{1cm} & 0.625 &  \hspace{1cm} & 0.8 & \hspace{1cm} & 23.5(7) & \hspace{1cm} & 2.22(5) & \hspace{1cm} & 38.00/19\\
\hline 
0.001 & \hspace{1cm} & 0.625 &  \hspace{1cm} & 0.9 & \hspace{1cm} & 16.8(3) & \hspace{1cm} & 2.09(4) & \hspace{1cm} & 28.88/19 \\
\hline 
0.001 & \hspace{1cm} & 0.625 &  \hspace{1cm} & 1.0 & \hspace{1cm} & 13.24(15) & \hspace{1cm} & 2.04(3) & \hspace{1cm} & 8.77/19 \\
\hline 
\hline
0.001 & \hspace{1cm} & 0.7 &  \hspace{1cm} & 0.8 & \hspace{1cm} & 43.5(15) & \hspace{1cm} & 2.12(5) & \hspace{1cm} & 41.05/28 \\
\hline 
0.001 & \hspace{1cm} & 0.7 &  \hspace{1cm} & 0.9 & \hspace{1cm} & 22.9(5) & \hspace{1cm} & 2.09(4) & \hspace{1cm} & 33.32/28 \\
\hline 
0.001 & \hspace{1cm} & 0.7 &  \hspace{1cm} & 1.0 & \hspace{1cm} & 16.3(2) & \hspace{1cm} & 2.04(4) & \hspace{1cm} & 22.32/28 \\
\hline 
\hline \hline
0.01 & \hspace{1cm} & 0.625 &  \hspace{1cm} & 0.8 & \hspace{1cm} & 28.4(4) & \hspace{1cm} & 2.26(2) & \hspace{1cm} & 49.15/19 \\
\hline 
0.01 & \hspace{1cm} & 0.625 &  \hspace{1cm} & 0.9 & \hspace{1cm} & 20.1(2) & \hspace{1cm} & 2.16(2) & \hspace{1cm} & 48.07/19\\
\hline 
0.01 & \hspace{1cm} & 0.625 &  \hspace{1cm} & 1.0 & \hspace{1cm} & 15.87(16) & \hspace{1cm} & 2.08(2) & \hspace{1cm} & 23.93/19 \\
\hline 
\hline
0.01 & \hspace{1cm} & 0.7 &  \hspace{1cm} & 0.8 & \hspace{1cm} & 51.4(12) & \hspace{1cm} & 2.17(3) & \hspace{1cm} & 8.06/28 \\
\hline 
0.01 & \hspace{1cm} & 0.7 &  \hspace{1cm} & 0.9 & \hspace{1cm} & 27.7(4) & \hspace{1cm} & 2.13(2) & \hspace{1cm} & 65.66/28 \\
\hline 
0.01 & \hspace{1cm} & 0.7 &  \hspace{1cm} & 1.0 & \hspace{1cm} & 19.9(2) & \hspace{1cm} & 2.05(2) & \hspace{1cm} & 31.78/28 \\
\bottomrule
\end{tabular}
\caption[\textbf{Peak height characterization.}]{\textbf{Peak height characterization.} Parameters obtained in the fits of our data for $H_{\max}$ to~\refeq{fmax_xi}. For each fit, we also report the figure of merit $\chi^2/\mathrm{d.o.f.}$\index{degree of freedom}}
\labtab{parametros_fmax}
\end{center}
\end{table}

\begin{table}[h!]
\begin{center}
\begin{tabular}{c c c c c c c c c}
\toprule
\toprule
$F$ & \hspace{0.25cm} & $T_1$ &  \hspace{0.25cm} & $B(F)$ & \hspace{0.25cm} & $\zeta_{\text{NE}}$ & \hspace{0.25cm} & $\chi^2/\mathrm{d.o.f.}$\index{degree of freedom} \\
\hline \hline
0.001 & & 0.625 & & 5.77(11) & & 1.19(2) & & 2.14/2\\
\hline
0.01 & & 0.625 & & 6.94(14) & & 1.19(2) & & 1.57/2\\
\hline
0.001 & & 0.7 & & 5.99(13) & & 1.19(2) & & 2.46/2\\
\hline
0.01 & & 0.7 & & 7.28(16) & & 1.19(2) & & 1.38/2\\
\hline
\hline
0.001 & & 0.625, 0.7 & & 5.85(11) & & 1.19(2) & & 11.54/5\\\hline
0.01 & & 0.625, 0.7 & & 7.08(14) & & 1.19(2) & & 21.19/5\\

\bottomrule
\end{tabular}
\caption[\textbf{The chaotic exponent $\zeta$.}]{\textbf{The chaotic exponent $\zeta$.} The smallest temperature $T_1$ is fixed in each fit in the upper part of the table. The two last rows in the table correspond to the fit including all the available pairs of temperatures (i.e. in these fits we mix data with $T_1=0.625$ and $T_1=0.7$). Points with $(T_1=0.625,T_2=1.0)$ for both $F=0.001$ and $F=0.01$ are not considered in these fits.}
\labtab{zeta}
\end{center}
\end{table}

\subsection{The (inverse) peak width}\labsubsec{peaks-width}

\begin{figure}[!h]
  \centering
  \includegraphics[width=1\textwidth]{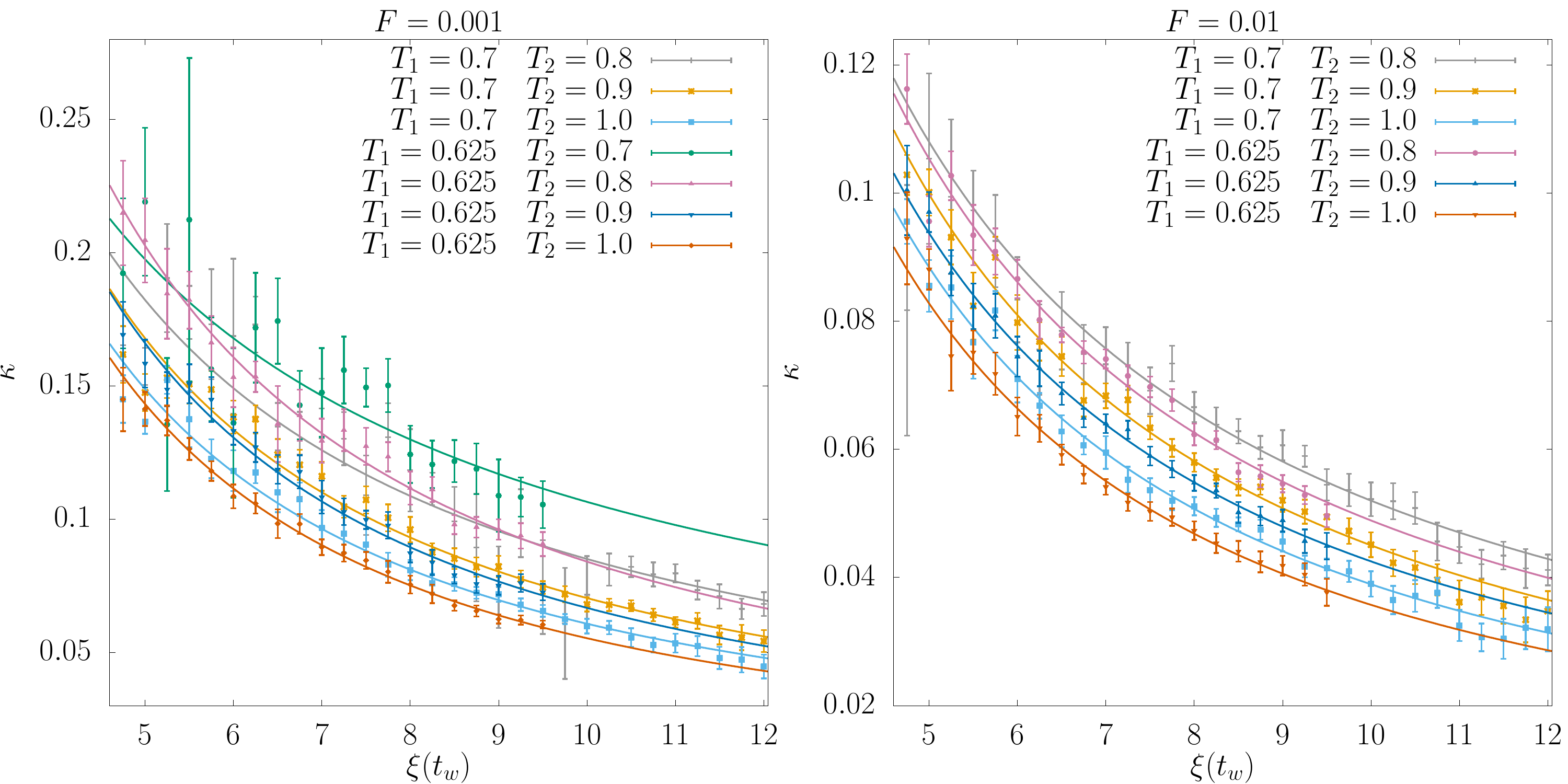}
  \caption[\textbf{Curvature $\kappa$ decays as a power law when increasing $\xi$.}]{\textbf{Curvature $\kappa$ decays as a power law when increasing $\xi$.} The inverse peak width $\kappa(\xi)$ is plotted against the coherence length\index{coherence length} $\xi(\tw)$ for all the simulated pairs of temperatures $T_1$ and $T_2$ at different probability levels, $F=0.001$ (left panel) and $F=0.01$ (right panel). Similar decaying exponent $\beta$, actually compatible at the two-$\sigma$ level (see~\reftab{parametros_width}), is displayed for all the pairs of temperatures in both probability levels.}
 \labfig{width_xi}
\end{figure}

The peak width provides the answer to the following question: how critical is it to select the right length-scale to study \gls{TC}\index{temperature chaos}? Obviously, if the peak width becomes larger than its position (see~\refsubsec{peaks-position}), this choice is no longer critical.

We study the inverse peak width (i.e. the curvature)
\begin{equation}
\kappa(\xi)=\sqrt{\dfrac{|H''(z_{\max})|}{H(z_{\max})}} \, , \labeq{def_curvature}
\end{equation}
and propose a power law decaying with $\xi(\tw)$ characterized by the ansatz
\begin{equation}
\kappa(\xi) = A(F) \, \xi^{-\beta} \, , \labeq{width_xi}
\end{equation}
where $A(F)$ is an amplitude while $\beta$ is the power law exponent. Results are shown in~\reffig{width_xi} and~\reftab{parametros_width}.

The value of $A(F)$ turns out to be compatible for all the pairs of temperatures $(T_1,T_2)$ at fixed probability $F$. Furthermore, at the current precision of the data, exponent $\beta$ does not exhibit any significant dependency on the temperature pair ($T_1,T_2$) or the probability $F$.

Let us now recall the linear relation between the peak position and the coherence length\index{coherence length}, see~\refeq{Nr_xi}. Consider the ratio between the position of the maximum and its width, $N_{r,\max}\kappa(\xi) \sim \xi^{1-\beta}$. The~\reftab{parametros_width} mildly suggests that $\beta$ is slightly greater than 1, which implies that the ratio goes to zero (very slowly) in the limit of large $\xi$. The parameter $\beta$ would have, indeed, a critical meaning for the large $\xi$ limit. If greater than one, as mildly suggested by the results, the chaotic behavior would be present at any scale $\xi_c$. On the contrary, if further works find $\beta < 1$, in the $\xi \to \infty$ limit, the chaos would only be visible at the coherence-length scale.

\begin{table}[h!]
\begin{center}
\begin{tabular}{c c c c c c c c c c c}
\toprule
\toprule
$F$ & \hspace{1cm} & $T_1$ &  \hspace{1cm} & $T_2$ & \hspace{1cm} & $A$ & \hspace{1cm} &$\beta$ & \hspace{1cm} & $\chi^2/\mathrm{d.o.f.}$\index{degree of freedom} \\
\hline \hline
0.001 & \hspace{1cm} & 0.625 &  \hspace{1cm} & 0.7 & \hspace{1cm} & 0.8(3) & \hspace{1cm} & 0.9(2) & \hspace{1cm} & 18.72/19 \\
\hline 
0.001 & \hspace{1cm} & 0.625 &  \hspace{1cm} & 0.8 & \hspace{1cm} & 1.6(4) & \hspace{1cm} & 1.27(14) & \hspace{1cm} & 8.07/19 \\
\hline 
0.001 & \hspace{1cm} & 0.625 &  \hspace{1cm} & 0.9 & \hspace{1cm} & 1.4(3) & \hspace{1cm} & 1.32(12) & \hspace{1cm} & 10.05/19 \\
\hline 
0.001 & \hspace{1cm} & 0.625 &  \hspace{1cm} & 1.0 & \hspace{1cm} & 1.3(2) & \hspace{1cm} & 1.37(9) & \hspace{1cm} & 5.60/19 \\
\hline 
\hline
0.001 & \hspace{1cm} & 0.7 &  \hspace{1cm} & 0.8 & \hspace{1cm} & 1.1(3) & \hspace{1cm} & 1.10(12) & \hspace{1cm} & 35.26/28 \\
\hline 
0.001 & \hspace{1cm} & 0.7 &  \hspace{1cm} & 0.9 & \hspace{1cm} & 1.26(16) & \hspace{1cm} & 1.25(7) & \hspace{1cm} & 25.90/28 \\
\hline 
0.001 & \hspace{1cm} & 0.7 &  \hspace{1cm} & 1.0 & \hspace{1cm} & 1.19(17) & \hspace{1cm} & 1.29(7) & \hspace{1cm} & 23.01/28 \\
\hline 
\hline \hline
0.01 & \hspace{1cm} & 0.625 &  \hspace{1cm} & 0.8 & \hspace{1cm} & 0.63(9) & \hspace{1cm} & 1.11(7) & \hspace{1cm} & 20.44/19 \\
\hline 
0.01 & \hspace{1cm} & 0.625 &  \hspace{1cm} & 0.9 & \hspace{1cm} & 0.59(10) & \hspace{1cm} & 1.14(8) & \hspace{1cm} & 6.08/19 \\
\hline 
0.01 & \hspace{1cm} & 0.625 &  \hspace{1cm} & 1.0 & \hspace{1cm} & 0.58(15) & \hspace{1cm} & 1.21(12) & \hspace{1cm} & 9.05/19 \\
\hline 
\hline
0.01 & \hspace{1cm} & 0.7 &  \hspace{1cm} & 0.8 & \hspace{1cm} & 0.59(11) & \hspace{1cm} & 1.05(11) & \hspace{1cm} & 21.26/28 \\
\hline 
0.01 & \hspace{1cm} & 0.7 &  \hspace{1cm} & 0.9 & \hspace{1cm} & 0.63(8) & \hspace{1cm} & 1.15(7) & \hspace{1cm} & 18.46/28 \\
\hline 
0.01 & \hspace{1cm} & 0.7 &  \hspace{1cm} & 1.0 & \hspace{1cm} & 0.59(12) & \hspace{1cm} & 1.18(9) & \hspace{1cm} & 17.93/28 \\
\bottomrule
\end{tabular}
\caption[\textbf{Peak width characterization.}]{\textbf{Peak width characterization.} Parameters obtained in the fits of our data for $\kappa(\xi)$ to~\refeq{width_xi}. For each fit, we also report the figure of merit $\chi^2/\mathrm{d.o.f.}$\index{degree of freedom}}
\labtab{parametros_width}
\end{center}
\end{table}

\subsection{On the relation between experimental and numerical results} \labsubsec{relation_experiment_numerical_simulations}
This characterization of \gls{TC}\index{temperature chaos} in the off-equilibrium dynamics of a large \gls{SG} paves the way to a major interplay between numerical simulations and experiments.

Although we have considered in this work fairly small values of the chaotic system fraction $F$, a simple extrapolation, linear in $\log F$, predicts $\xi^* \approx 60$ for $F=0.1$ at $T_1=0.7$ and $T_2=0.8$ (our closest pair of temperatures in~\reftab{parametros_fmax}). A spin-glass coherence length\index{coherence length} well above $60 a_0$ is experimentally reachable
nowadays~\cite{zhai:19,zhai:20b,zhai-janus:20,zhai-janus:21} ($a_0$ is the typical spacing between spins), which makes our dynamic \gls{TC}\index{temperature chaos} significant.

Indeed the \gls{TC}\index{temperature chaos}-closely related experimental study~\cite{zhai:20b} reported a value for exponent\footnote{The authors propose several schemes and different computations of the exponent are provided.} $\zeta_{\text{NE}}$ in fairly good agreement with our result of $\zeta_\text{NE}=1.19(2)$ in~\reffig{zeta_exponent}.

A deeper relation between the \gls{TC}\index{temperature chaos} phenomenon in experiments and in numerical simulations would be desirable. Actually, simple temperature-varying protocols\index{temperature-varying protocol} (in which temperature sharply drops from $T_2$ to $T_1$, see, e.g.~\cite{zhai:20b}) seems more accessible to a first analysis than memory\index{memory effects} and rejuvenation\index{rejuvenation} experiments~\cite{jonason:98,lundgren:83,jonsson:00,hammann:00}.

The rupture point between both approaches is the difference in the measured observables. An important problem is that the correlation functions that are studied theoretically are not easily probed experimentally. Instead, experimentalists privilege the magnetization\index{magnetization!density} density (which is a spatial average over the whole sample\index{sample}). The study of the magnetization\index{magnetization!density} density from a numerical point of view, on the other side, is clearly bounded by the computational power available nowadays since its global nature makes the chaos signal almost disappear for our achievable coherence lengths\index{coherence length}. Therefore an important theoretical goal is to predict the behavior of the non-equilibrium time-dependent magnetization\index{magnetization} upon a temperature drop.

\section{Scaling at fixed \boldmath $r$}\labsec{scaling_fixed_r}
\begin{figure}[!h]
  \centering
  \includegraphics[width=0.49\textwidth]{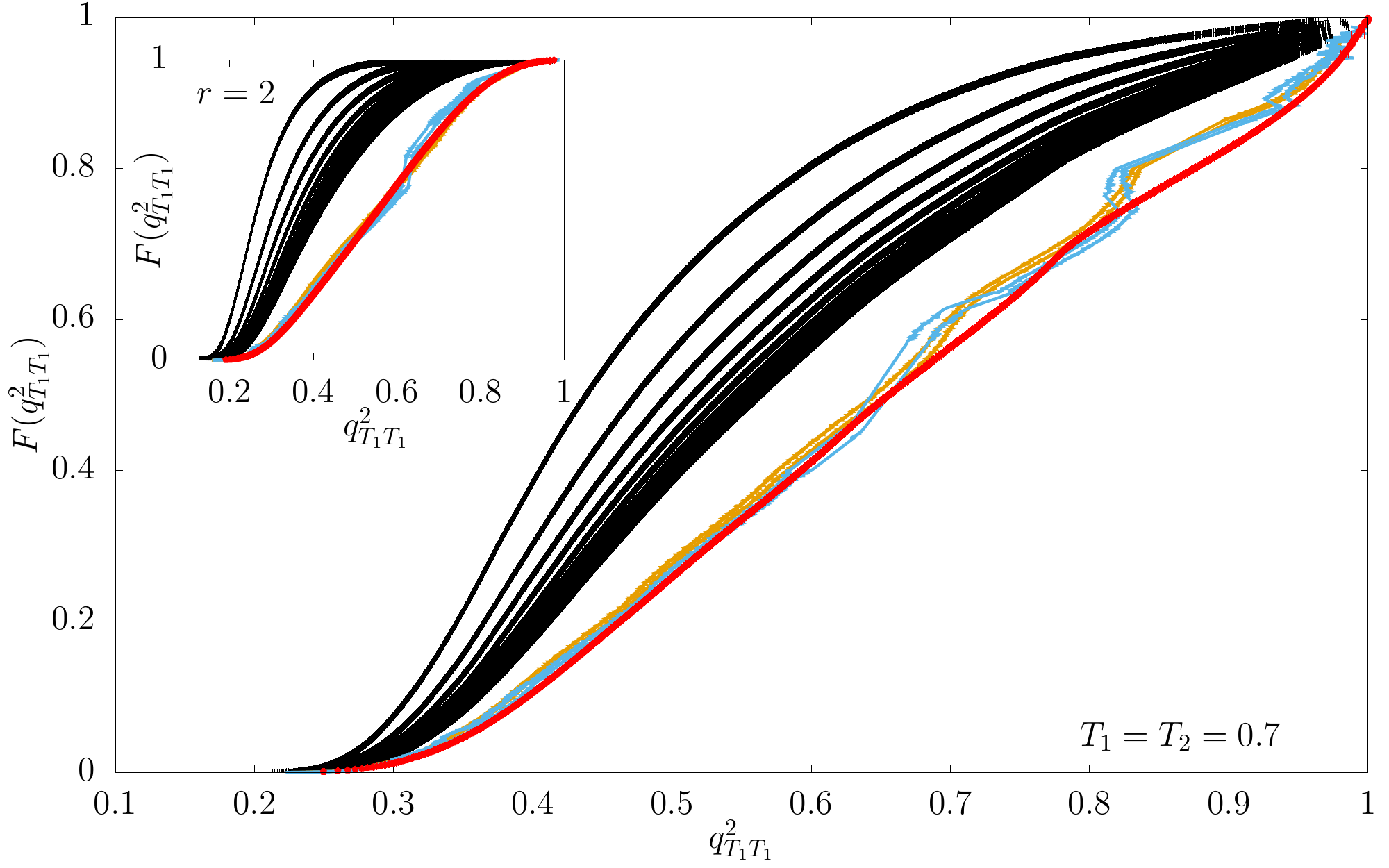}
  \includegraphics[width=0.49\textwidth]{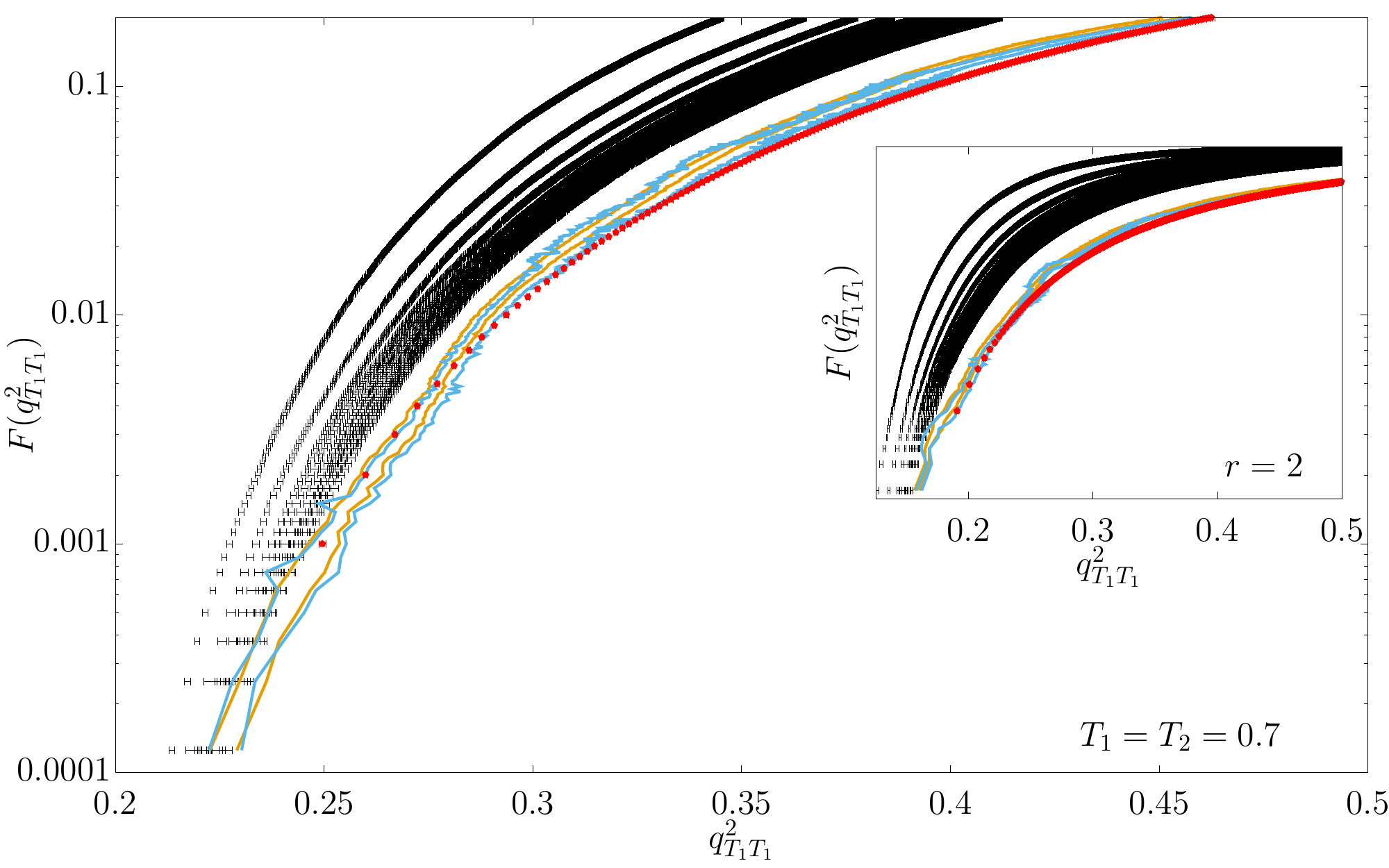}	
  \includegraphics[width=0.49\textwidth]{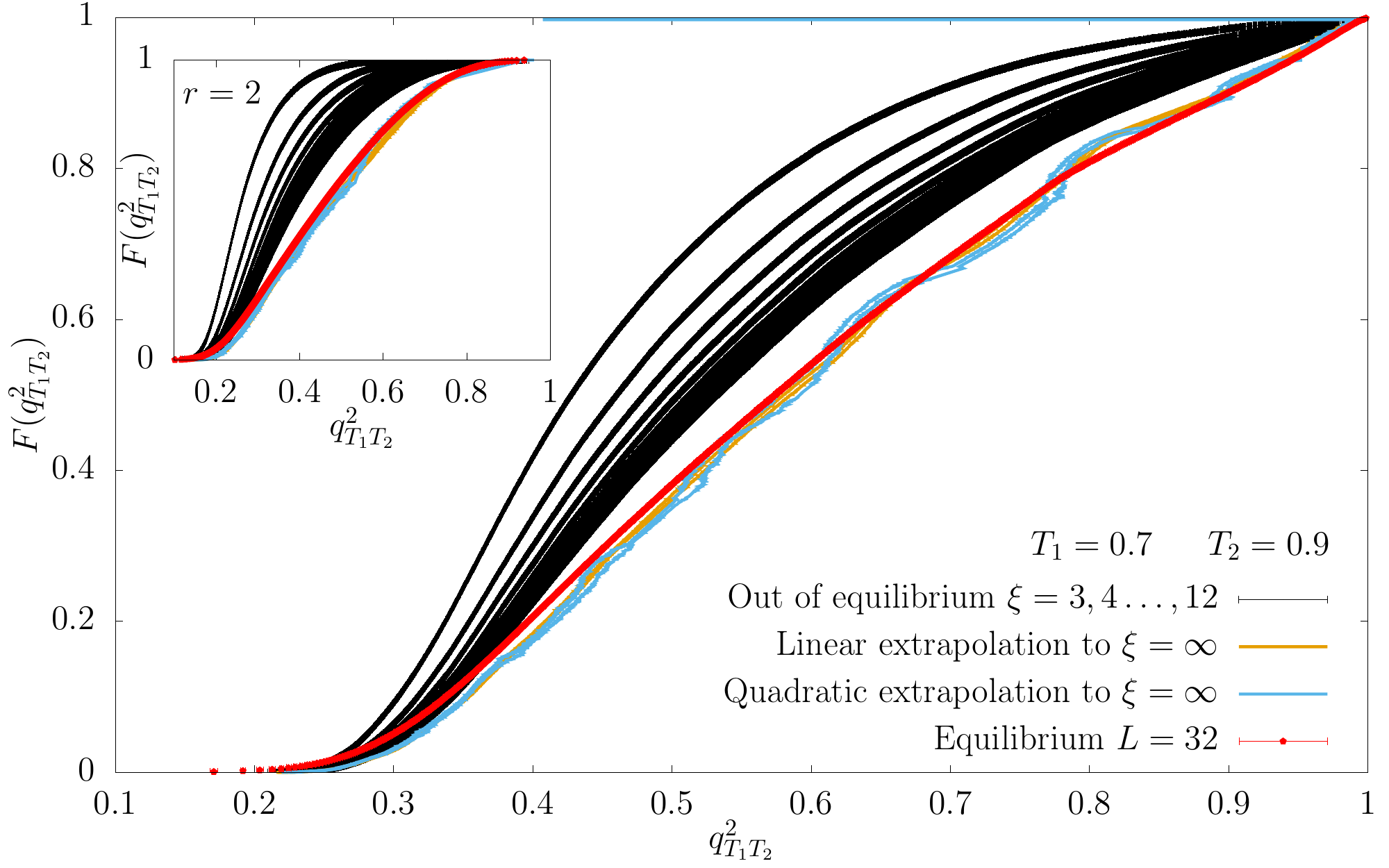}
  \includegraphics[width=0.49\textwidth]{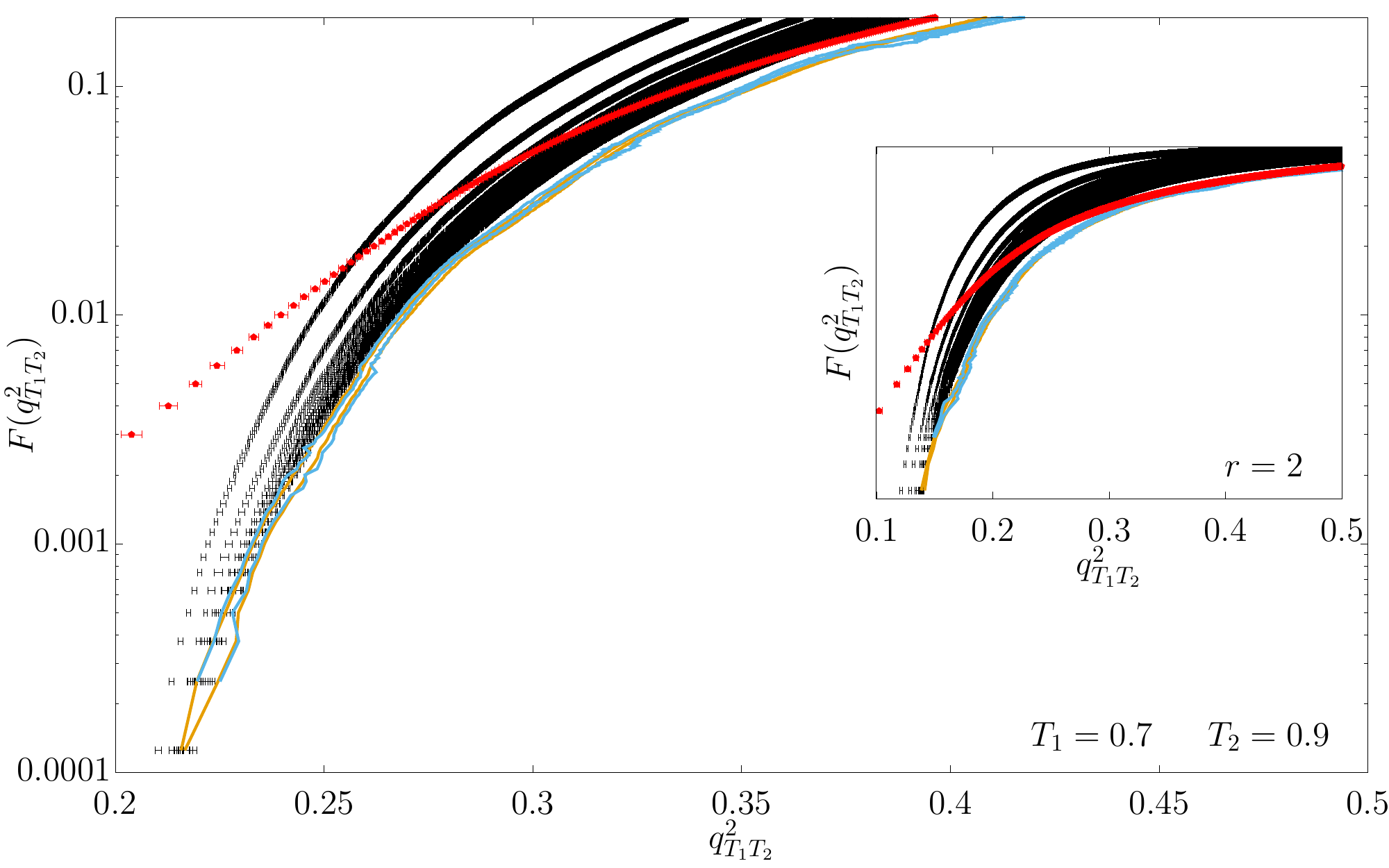}	
  \caption[\textbf{Extrapolation of $\mathbf{F(q^2)}$ to the $\mathbf{\xi \to \infty}$ limit.}]{\textbf{Extrapolation of $F(q^2_{T_1T_1})$ (up) and  $F(q^2_{T_1T_2})$ (down) to the $\xi \to \infty$ limit, both in linear scale (left) and logarithmic scale (right).} \textbf{Main plots:} We extrapolate the distribution functions for spheres of radius $r=1$ to the $\xi \to \infty$ limit. Both ans\"atze, the one in~\refeq{extrapolacion_estatica_lineal}, golden curves, and the one in~\refeq{extrapolacion_estatica_cuadratica}, blue curves, produce equivalent extrapolations. For the sake of clarity, we plot two curves for each extrapolation (with the same color) which corresponds to the computed central value plus (minus) the standard deviation. We compare the results with the equilibrium data from an \gls{EA}\index{Edwards-Anderson!model} model in a cubic lattice of linear size $L=32$ (red curve, see main text for further details). The equilibrium distribution and the extrapolated one are compatible for the $q^2_{T_1T_1}$ curves. Instead, $q^2_{T_1T_2}$ is compatible for percentiles of order one but not for probabilities smaller than $F=0.1$. \textbf{Insets:} As in the main plots, for spheres of radius $r=2$.}
\labfig{extrapola_estatica_q2}
\end{figure}

Until this point, the analysis of the chaotic parameter (its characterization through the size of the sphere and the dependence of the peak with the coherence length\index{coherence length}) have been inspired by the representation of the data given by the~\reffig{distribution_function_r} and the idea of an optimal scale to observe \gls{TC}\index{temperature chaos}. However, a different approach can be performed by regarding the representation of the data in~\reffig{distribution_function_xi}.

Fixing the two temperatures $T_1$ and $T_2$ and the size of the sphere, the distribution function $F(X,T_1,T_2,\xi,r)$ exhibit a monotonic behavior with the coherence length\index{coherence length} $\xi$ (see~\reffig{distribution_function_xi}). Specifically, regarding at the lowest possible size of the sphere $r=1$, we found there was no convergence (apparently) to a limit curve when increasing the coherence length\index{coherence length} $\xi$.

The absence of a limit curve led us to consider slow convergence as a hypothesis. In this section, we propose algebraic extrapolations to explain the convergence to the $\xi \to \infty$ limit. Moreover, we compare the extrapolations with the equilibrium data obtained from an \gls{EA}\index{Edwards-Anderson!model} Ising\index{Ising} model simulated with a \gls{PT} algorithm in a cubic lattice of linear size $L=32$ (see~\cite{janus:10,janus:10b}). We expect that, for the smallest sphere radius at least, $L=32$ will be representative of the thermodynamic limit\index{thermodynamic limit}.

Let us fix the probability level $F$ and the radius $r$ of the spheres, we propose two different extrapolations to the $\xi(\tw) \to \infty$ limit
\begin{equation}
\Omega(T_1,T_2,\xi,r,F) = \Omega(T_1,T_2,\xi= \infty,r,F) +  \Phi(F) \left(\dfrac{N_r^{1/3}}{\xi}\right)^{\delta(F)} \, , \labeq{extrapolacion_estatica_lineal}
\end{equation}
\begin{equation}
\Omega(T_1,T_2,\xi,r,F) = \Omega(T_1,T_2,\xi= \infty,r,F) + \Psi_1 (F) \left(\dfrac{N_r^{1/3}}{\xi}\right)^{\epsilon(F)} + \Psi_2(F) \left(\dfrac{N_r^{1/3}}{\xi}\right)^{2\epsilon(F)} \, , \labeq{extrapolacion_estatica_cuadratica}
\end{equation}
where $\Omega(T_1,T_2,\xi,r)$ is the extrapolated quantity $\Omega \in \lbrace q^2_{T_1T_1},q^2_{T_2T_2},q^2_{T_1T_2},X_{T_1T_2} \rbrace$ at the fixed $F$ probability level, $\Omega(T_1,T_2,\xi= \infty,r)$ is the value of that quantity in the $\xi \to \infty$ limit, $\Phi$, $\Psi_1$ and $\Psi_2$ are the coefficients of the fit, and $\delta$ and $\epsilon$ are the exponents of the fits.

From now on, we focus on the $r=1$ and $r=2$ cases. The extrapolations for $q^2_{T_1T_1}$ and $q^2_{T_2T_2}$ show the agreement between the extrapolations to the $\xi \to \infty$ limit and the equilibrium data, see~\reffig{extrapola_estatica_q2} (up panels), for the whole range. Besides, the quadratic and the linear extrapolations are statistically equivalent.

The extrapolation for $q^2_{T_1T_2}$ keeps the agreement between the extrapolations to the $\xi \to \infty$ limit and the equilibrium data for high values of the distribution function, but differs for short values of $F(q^2_{T_1T_2})$, see~\reffig{extrapola_estatica_q2} (down panels). Again, linear and quadratic extrapolation show no difference within the statistical error.

For the chaotic parameter $X_{T_1T_2}$ we find no convergence at all at the current precision of the data and the simulated scales of the coherence length\index{coherence length}.

We also explore the $r>2$ case, however, the greater the sphere size $r$, the bigger the simulated coherence length\index{coherence length} $\xi$ needed for a reliable fit to~\refeq{extrapolacion_estatica_lineal} and~\refeq{extrapolacion_estatica_cuadratica}. Extreme events stop providing reliable fits when increasing the sphere size $r$.

Indeed, this approach provides reasonable results in the extrapolation to the $\xi \to \infty$ limit for $q^2$ distribution functions when compared with the equilibrium data. However, the convergence is slow and the fits are difficult at our current precision.

Finally, we examine our results in search of scale invariance. Due to the compatibility of the linear and quadratic extrapolations, we will focus on the former for the sake of simplicity, see~\refeq{extrapolacion_estatica_lineal}. We study the $r=1$ and $r=2$ cases which more likely hold the limit $r/\xi \ll 1$. 

If the scale invariance is present in our results, we would expect the exponent $\delta$ to be independent of the radius $r$. We find a similar behavior for the exponent in the extrapolations for spheres of radius $r=1$ and $r=2$ (\reffig{factor_exponente} main plots), nevertheless the results are not compatible for all the percentiles $F(q^2_{T_1T_1})$ and $F(q^2_{T_1T_2})$.

We also expect the ratio of the coefficients $\Phi(r=2)/\Phi(r=1)$ to be constant if the scale invariance holds. We find small changes in the ratio between the amplitudes $\Phi(F,r)$, see~\reffig{factor_exponente} insets. The ratio remains around $1$ for all the percentiles, however, the results are not compatible for the whole range with $1$.

Our results mildly suggest the existence of scale invariance for the limit $r/\xi \ll 1$, in our cases represented by the simulations with sphere radius $r=1$ and $r=2$. However, the difficulty of the extrapolations and the noise in the amplitude and the exponent, prevent us from making robust statements.

\begin{figure}[h!]
  \centering	
  \includegraphics[width=1.0\textwidth]{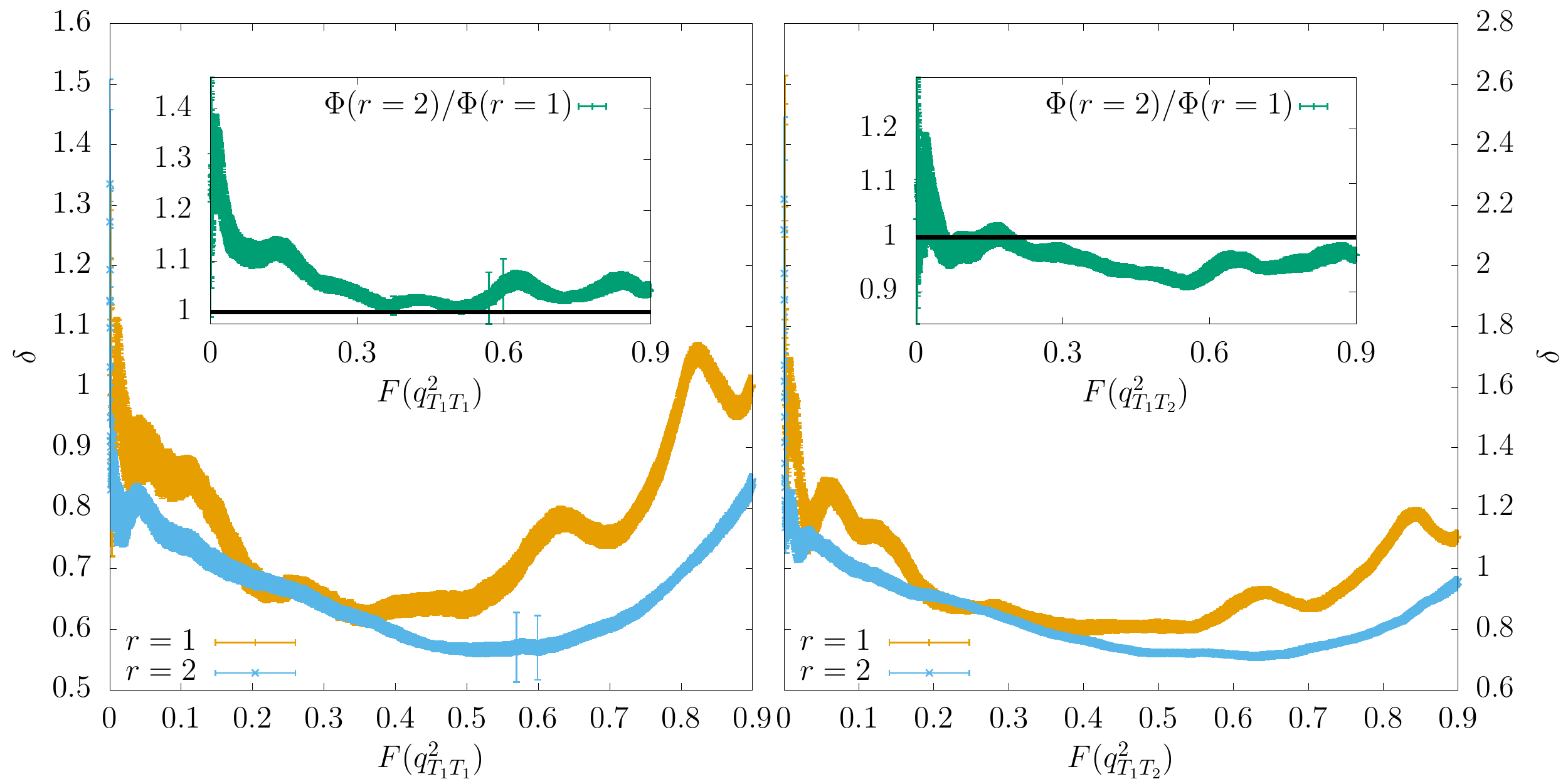}
  \caption[\textbf{Scale invariance test.}]{\textbf{Scale invariance test for parameters in~\refeq{extrapolacion_estatica_lineal}.} Indeed, scale invariance demands that both the exponent $\delta(F)$ and the amplitude $\Phi(F)$ should be independent of the sphere radius. The figure compares both quantities for the radius $r=1$ and $r=2$. \textbf{Main panels:} exponent $\delta(F)$ as computed for spheres of radius $r=1$ (golden curves) and $r=2$ (blue curves), as a function of $F(q^2_{T_1T_1})$ (left panel) or $F(q^2_{T_1T_2})$ (right panel). \textbf{Insets: }Ratio of amplitudes $\Phi(F,r=2)/\Phi(F,r=1)$ for the same fits to~\refeq{extrapolacion_estatica_lineal} reported in the main panels. The horizontal black line is set at $1$ as a reference.}
\labfig{factor_exponente}
\end{figure}

\section{Temperature changing protocols}\labsec{T-changes}

In this section, we explore the impact of temperature-varying\index{temperature-varying protocol} simulations in the \gls{TC}\index{temperature chaos} results and we also address the cumulative aging\index{aging!cumulative} controversy: Is the aging\index{aging} performed at temperature $T_a$ useful at different temperature $T_b$?

Specifically, the cumulative aging\index{aging!cumulative} hypothesis~\cite{jonsson:02,bert:04} considers a protocol in which the system is first suddenly quenched from high temperature to temperature $T_a < \Tc$, where it ages for time $t_a$. Then, the temperature is changed to $T_b < \Tc$ and the system evolves for a time $t_b$. The hypothesis is made that this two-steps protocol is equivalent to isothermal\index{isothermal} aging\index{aging} at temperature $T_b$ for an effective time $t^{\mathrm{eff}}_b$ defined through the equation 
\begin{equation}
\xi (t_a ; t_b ; T_a \rightarrow T_b) = \xi(t^{\mathrm{eff}}_b,T_b) \> . \labeq{cumulative_hypotesis}
\end{equation}

Fortunately, we now know how to compute the coherence length\index{coherence length} $\xi$~\cite{janus:09b,fernandez:18b} accurately (see also \refsec{finite_size_effects}). Therefore, we are able to compare the evolution of systems that have undergone either the two-steps protocol or the isothermal\index{isothermal} one, choosing the time scales to precisely match~\refeq{cumulative_hypotesis}.

Here, we will consider two symmetric two-steps proposals. On the one hand, we let age the system at temperature $T_a=0.8$ (by \textit{the system} we mean the $\NS$ samples\index{sample} and the $\NRep$ replicas\index{replica}) until it reaches coherence length\index{coherence length} $\xi(t_a)=6$. At this point, we suddenly quench to the temperature $T_b=1.0$ and let the system evolve until $\xi(t_a;t_b; T_a=0.8 \rightarrow T_b=1.0)=9$. On the other hand we perform the symmetric two-steps protocol, starting at temperature $T_a=1.0$ and cooling the system to temperature $T_b=0.8$. On the technical side, let us mention that we compute the distribution function as explained in~\refsec{procedure}.

As everywhere else in this chapter, we will compute the overlaps\index{overlap} between configurations\index{configuration} at two temperatures $T_1<T_2$. However, we shall need to introduce notation in order to specify the protocol. 

We will put the superscript \textit{iso} on a temperature to indicate that the system has always been at that temperature (\textit{iso} is a short for isothermal\index{isothermal}). For instance, the notation $\Tiso_1 = 0.8$ will mean that we are computing the overlap\index{overlap} between a system that has always been at temperature $0.8$ and a system that currently is at some higher temperature. Instead, $\Tiso_2 = 0.8$ means that the overlap\index{overlap} is computed between our isothermal\index{isothermal} system and a system which currently is at some lower temperature.

We will put the superscript $T_a \to T_b$ on a temperature to indicate that the system has followed the two-steps protocol from temperature $T_a$ to temperature $T_b$. Notice that the current temperature of the system is $T_b$. For instance, the notation $X(\Tiso_1=0.8,T^{0.8 \to 1.0}_2=1.0)$ indicates that we are computing overlaps\index{overlap} between the configurations\index{configuration} of a system that has always been at temperature $0.8$ and those of a system that has followed the two-steps protocol from temperature $T_a=0.8$ to $T_b=1.0$.

We organize the obtained results into two subsections. The first one (see~\refsubsec{no-equivalence}) captures the asymmetric behavior when the system is cooled or heated. The second one (see~\refsubsec{auxiliary}) takes advantage of an auxiliary temperature to study the effect of the two-steps protocol on the \gls{TC}\index{temperature chaos} phenomenon.

\subsection{On the asymmetry between cooling and heating}\labsubsec{no-equivalence}

\begin{figure}[!h]
  \centering
  \includegraphics[width=1.0\textwidth]{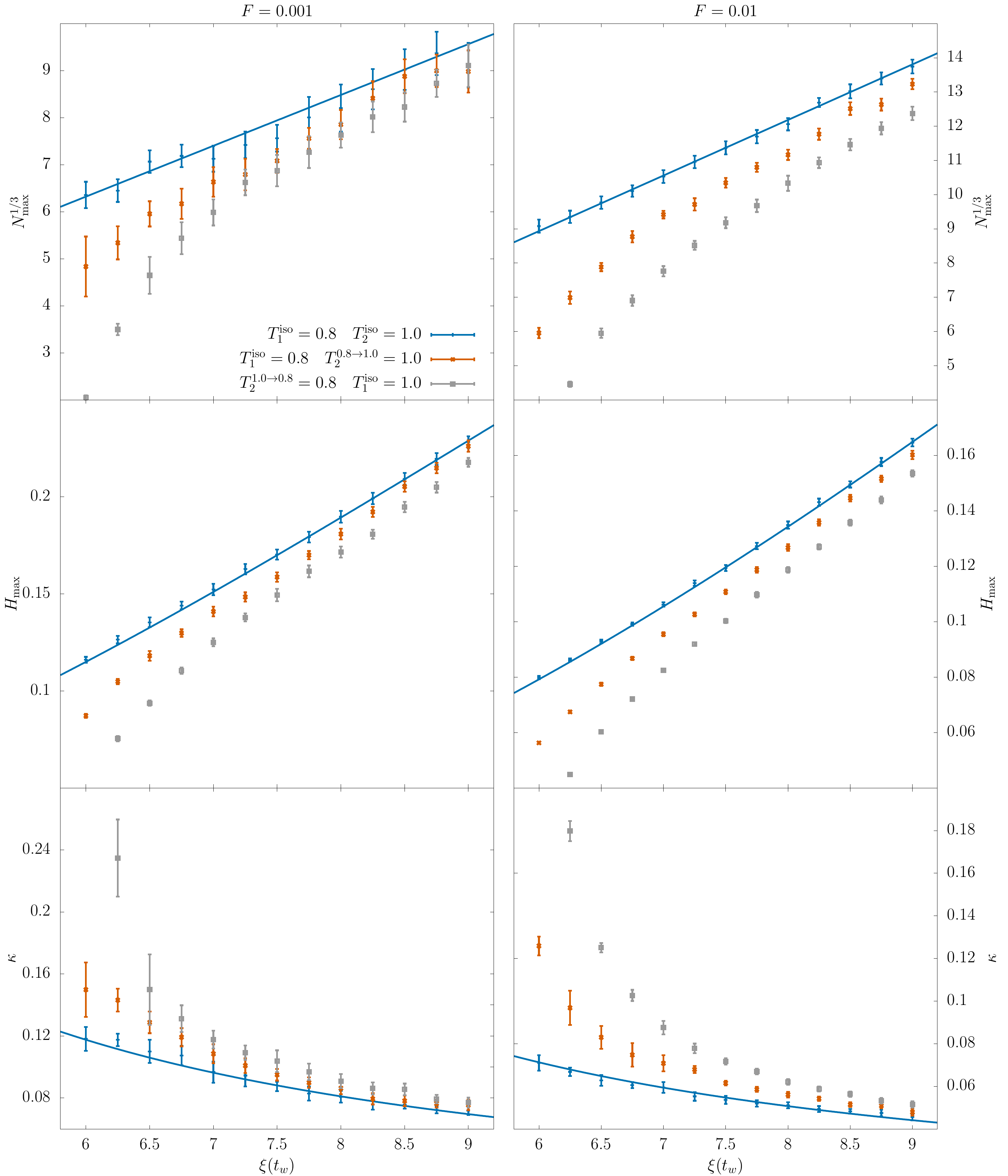}
  \caption[\textbf{Comparison between isothermal and two-step protocol.}]{\textbf{Comparison of the main parameters in the \boldmath $1-X$ curves for the two-steps protocol and the isothermal\index{isothermal} protocol. The two protocols approach to each other for large $\xi$.} Position of the peak (up panels), height of the peak (middle panels) and curvature (down panels) against $\xi(t_w)$ are plotted for the reference curve $\Tiso_1=0.8$, $\Tiso_2=1.0$ and for the temperature-varying\index{temperature-varying protocol} simulations at two different probability levels $F=0.001$ (left) and $F=0.01$ (right). Heating and cooling two-steps protocols are not equivalent.}
 \labfig{xi_noaux}
\end{figure}

We analyze the \gls{TC}\index{temperature chaos} phenomenon through the same three observables used in~\refsec{results}: the peak position, the peak height, and the (inverse) peak width (i.e. the curvature). Each of those quantities are extracted from one of the following overlap\index{overlap} combination. First we have the reference curve $\Tiso_1=0.8$ and $\Tiso_2=1.0$. Next we have the so called \textit{cooling} curve $T_1^{1.0 \to 0.8} = 0.8$ and $\Tiso_2=1.0$. Finally we have the so called \textit{heating} curve $\Tiso_1=0.8$ and $T_2^{0.8 \to 1.0} = 1.0$.

The prediction of the cumulative aging\index{aging!cumulative}, see~\refeq{cumulative_hypotesis}, is that the three curves, namely the reference, the \textit{heating} and the \textit{cooling} curves, should coincide. Indeed, we choose the times in such a way that the coherence length\index{coherence length} is the same for all the four systems involved: $\Tiso=0.8$, $\Tiso=1.0$, $T^{0.8 \to 1.0} = 1.0$ and $T^{1.0 \to 0.8} = 0.8$. Instead, our data in~\reffig{xi_noaux} show that cumulative aging\index{aging!cumulative} holds only in an approximated way. In fact, both the \textit{heating} and the \textit{cooling} curves significantly differ from the reference curve at coherence length\index{coherence length} $\xi \approx 6$, although the three curves tend to merge as $\xi$ grows. Nevertheless, we notice that the \textit{cooling} curve does not converge to the other two curves, at least in the simulated range of $\xi$. In this sense, there exist an asymmetry between the \textit{cooling} curve and the \textit{heating} curve. Furthermore, by comparing the left and right panels in~\reffig{xi_noaux}, we conclude that the larger the probability $F$, the stronger the asymmetry.

Finally, the reader may wonder about the possibility of computing overlaps\index{overlap} between configurations\index{configuration} at temperature $\Tiso_1=0.8$ and $T_2^{1.0 \to 0.8} = 0.8$ or configurations\index{configuration} at temperature $\Tiso_1=1.0$ and $T_2^{0.8 \to 1.0} = 1.0$. Indeed, we have attempted to compute those overlaps\index{overlap} but we found very large values of the chaotic parameter that made impossible the characterization of the peak.

\subsection{Auxiliary-temperature analysis}\labsubsec{auxiliary}
In this subsection, we study the behavior of the two-steps protocol simulations by comparing them with the most ordered configurations\index{configuration} we have, namely, the simulations of temperature $\Tiso=0.625$. We establish two reference curves $\Tiso_1=0.625$, $\Tiso_2=0.8$ and $\Tiso_1=0.625$, $\Tiso_2=1.0$ and we compare them with the curves $\Tiso_1=0.625$, $T^{1.0 \to 0.8}=0.8$, and $\Tiso_1=0.625$, $T^{0.8 \to 1.0}=1.0$.

\begin{figure}[!h]
  \centering
  \includegraphics[width=1.0\textwidth]{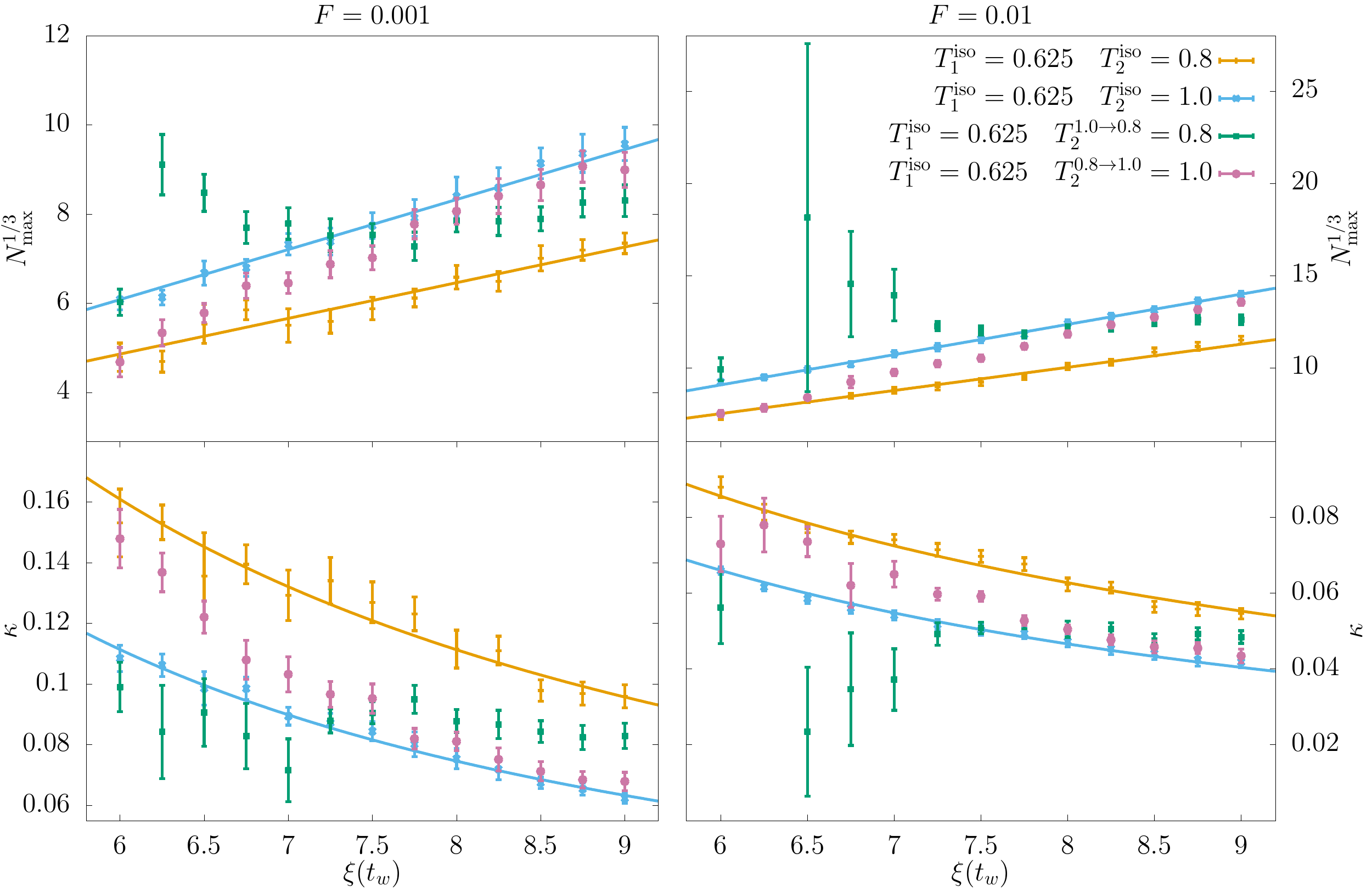}	
  \caption[\textbf{Temperature-varying\index{temperature-varying protocol} simulations of the peak position and curvature exhibit non-symmetric behavior when symmetric cooling and heating protocols are applied.}]{\textbf{Temperature-varying\index{temperature-varying protocol} simulations of the peak position and curvature exhibit non-symmetric behavior when symmetric cooling and heating protocols are applied.} The peak position (upper panels) and the curvature (lower panels) are plotted against the coherence length\index{coherence length} $\xi(t_w)$. We use a constant auxiliary temperature $\Tiso_1=0.625$ combined with two fixed-temperature simulations as the reference curves ($\Tiso_2=0.8$ and $\Tiso_2=1.0$) and the constant auxiliary temperature $\Tiso_1=0.625$ combined with two temperature-varying\index{temperature-varying protocol} simulations as the studied curves ($T^{0.8 \to 1.0}_2=1.0$ and $T^{1.0 \to 0.8}_2=0.8$). We compute all the curves at the probability levels $F=0.001$ (left panels) and $F=0.01$ (right panels).}
 \labfig{Nr_width_xi_changeT}
\end{figure}

We can observe the migration, as the coherence length\index{coherence length} grows, of the curve $\Tiso_1=0.625$, $T^{0.8 \to 1.0}_2=1.0$ from the reference curve $\Tiso_1=0.625$ $\Tiso_2=0.8$ to the reference curve $\Tiso_1=0.625$ $\Tiso_2=1.0$ by increasing the position of the peak (\reffig{Nr_width_xi_changeT} upper panels) and the curvature (\reffig{Nr_width_xi_changeT} lower panels). However, the opposite temperature-change, namely the curve corresponding to $\Tiso_1=0.625$ $T^{1.0 \to 0.8}_2=0.8$, does not lead to the symmetric behavior. The position of the peak (respectively, the curvature) does not decrease (respectively, increase) when cooling the temperature to merge with the reference curve $\Tiso_1=0.625$, $\Tiso_2=0.8$. Thus, the optimal scale to observe \gls{TC}\index{temperature chaos} is a monotonically increasing function of the coherence length\index{coherence length} while the curvature seems a monotonically decreasing function of the coherence length\index{coherence length}.

Although we do not know the behavior of the peak position and the peak curvature for values of the coherence length\index{coherence length} higher than those simulated, one could expect, as a naive ansatz, that the curve relative to the \textit{cooling} protocol would merge with its corresponding reference curve ($\Tiso_1=0.625$ $\Tiso_2=0.8$) when the coherence length\index{coherence length} is increased and then it would behave as the isothermal\index{isothermal} simulation.

\begin{figure}[!h]
  \centering
  \includegraphics[width=1.0\textwidth]{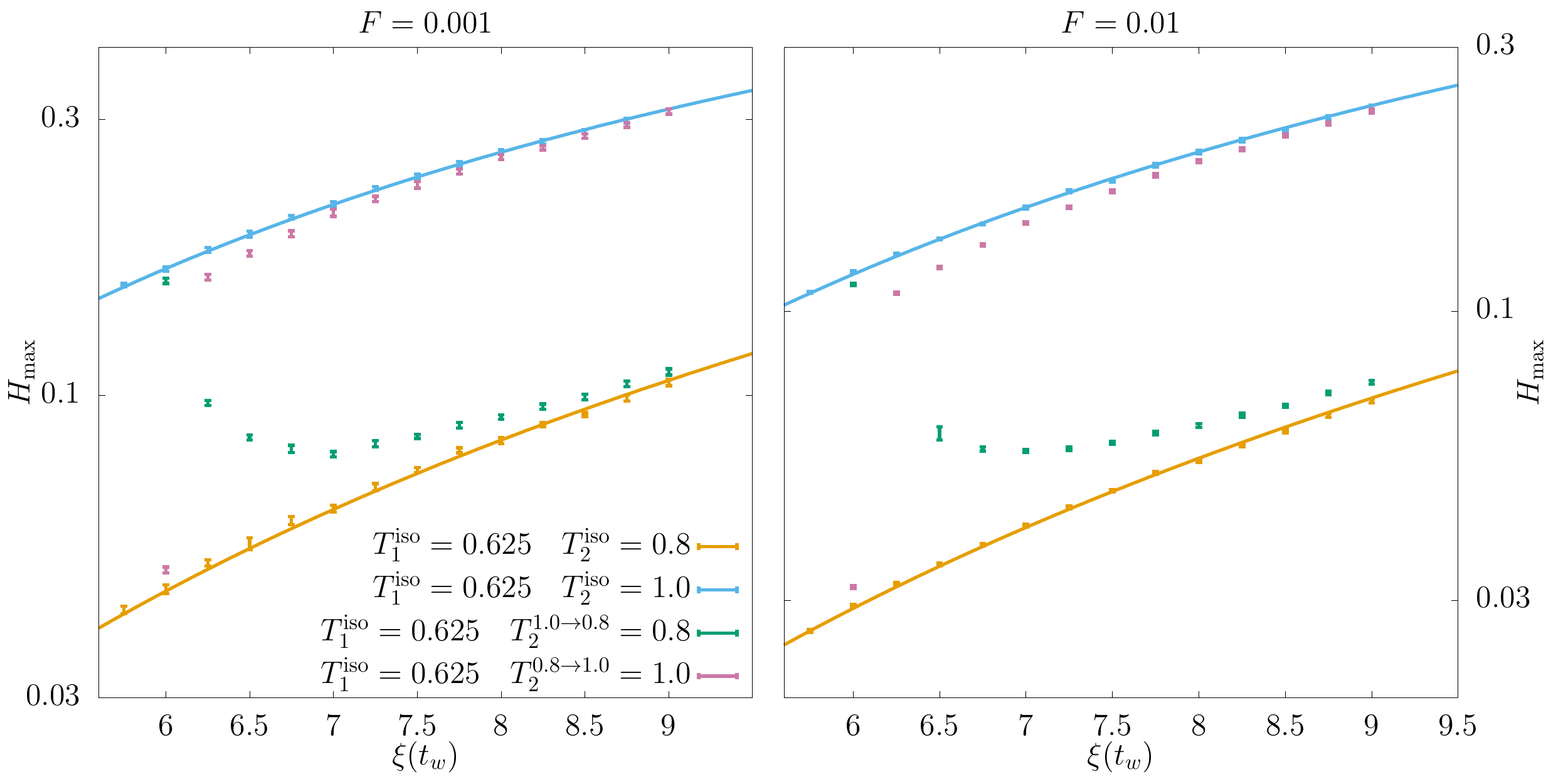}	
  \caption[\textbf{Peak height of cooled two-steps protocol simulations with an auxiliary temperature converges slower than the symmetric heating protocol.}]{\textbf{Peak height of cooled two-steps protocol simulations with an auxiliary temperature converges slower than the symmetric heating protocol.} The height of the peak is plotted against the coherence length\index{coherence length} $\xi(t_w)$ for fixed-temperature simulations and for two-steps protocol simulations with an auxiliary temperature $\Tiso_1=0.625$ at the probability levels $F=0.001$ (left panel) and $F=0.01$ (right panel).}
 \labfig{fmax_xi_changeT}
\end{figure}

As for the peak height (\reffig{fmax_xi_changeT}), we may appreciate both studied curves approaching the reference curves. However, the same information provided in the previous subsection (\refsubsec{no-equivalence}) emerges here: the curve with the \textit{cooling} protocol needs a higher coherence length\index{coherence length} to converge to the reference curve. 

Recalling the fact that the height of the peak is related to the strength of \gls{TC}\index{temperature chaos}, the system tends to erase the memory of its previous relaxation\index{relaxation} state and exhibits the same amount of \gls{TC}\index{temperature chaos} as it could be expected from its current pair of temperatures. Nonetheless, the erasing process is not complete in the cooling protocol and a small amount of information of the previous state lasts in the system, at least for the simulated coherence lengths\index{coherence length}.

Let us conclude this section with a small summary of the results. The reader might imagine a \gls{TC}\index{temperature chaos} movie in which each frame corresponds to a function that represents $1-X(F,T_1,T_2,\xi,r)$ against the size of the studied region $N_r^{1/3}$, similar to those appearing in~\reffig{Xvsr_examples}. When both temperatures $T_1$ and $T_2$ follow an isothermal\index{isothermal} protocol, the function will, monotonically, increase the height of its peak, will move to the right (i.e. higher values of the sphere size), and will become \textit{flatter} (i.e. the curvature decreases). 

The same situation is expected to happen (qualitatively) when $T_2$ is suddenly heated following the two-steps protocol. However, in this situation, the increase of the peak position, the peak height, and the decrease of the curvature \textit{accelerates} (i.e. requires shorter values of coherence length\index{coherence length}) compared with the $T_2$-isothermal\index{isothermal} movie.

Finally, when $T_2$ is suddenly cooled following the two-steps protocol, the height of the peak will decrease, while the position of the peak and the curvature will \textit{freeze}. The naive expectation is that the \textit{freezing} of the position of the curve and the curvature will end at higher values of the coherence length\index{coherence length} than the simulated ones and then, the curve will behave as the isothermal\index{isothermal} one.

\addpart{Conclusions}
\chapter{Conclusions} \labch{conclusions}

\setlength\epigraphwidth{.5\textwidth}
\epigraph{\textit{Si el pasado y el presente\\
Se reflejan y no mienten\\
Tengo que hacer algo \\
por mi porvenir.}}{-- Evaristo Páramos, \textit{Ya no quiero ser yo}}

The traditional picture of a solitary theoretical physicist, working alone with paper and pencil as her only tools, has evolved through the years. Instead, computational research has been an important part of the development of physics for the last decades and the discoveries are often performed by groups of physicists working together. However, although the increase of computational power is not new, it has been only recently that we can collect, process, and analyze a huge amount of data at affordable times.

The increase of computational power has particularly benefited theoretical physics. Specifically for spin glasses, the development of special-purpose\index{special-purpose computer} hardware has allowed simulating systems up to the experimental time-scale. This thesis is another iteration in the development of a field that takes the general path of science and embraces the interplay between experiments, theory, and computing as a fruitful relation to make relevant advances.

Throughout this thesis, we have studied the spin glasses from a numerical point of view. The study of the metastate\index{metastate} in~\refch{metastate} has been focused on addressing a theoretical problem through the first construction of the equilibrium metastate\index{metastate} in numerical simulations. In~\refch{aging_rate} and \refch{mpemba} the off-equilibrium dynamics of spin glasses has been studied. In the former, we introduce the coherence length\index{coherence length} as the most relevant quantity characterizing the aging\index{aging} state of a spin glass, and we solve a numerical discrepancy in the aging\index{aging!rate} rate between experiments and previous numerical simulations. In the latter, we discuss the Mpemba\index{Mpemba effect} effect, which is a memory\index{memory effects} effect that takes place in the off-equilibrium dynamics, providing a good example of how the coherence length\index{coherence length} governs many out-of-equilibrium phenomena. The final part of the thesis, \refch{Introduction_chaos}, \refch{equilibrium_chaos} and \refch{out-eq_chaos}, is devoted to introduce and analyze the Temperature Chaos\index{temperature chaos} phenomenon in spin glasses. Actually, we tackle this problem from the equilibrium point of view (\refch{equilibrium_chaos}) and we also observe and characterize the phenomenon in the off-equilibrium dynamics (\refch{out-eq_chaos}).

In this chapter we outline the main results of this thesis, revisiting all the parts and summarizing the relevant messages.

\section{A brief thought about the importance of the data}
This thesis is mainly focused on the numerical study of spin glasses. One idea that is central throughout this thesis is the fundamental importance of high-quality data. The reader may wonder what could we say in~\refch{aging_rate} without our precise estimation of the coherence length\index{coherence length}, or which conclusions could we extract from~\refch{out-eq_chaos} without our accurate estimations of the chaotic parameter. None of these works would be possible without our high-quality data.

This section aims to emphasize the role of the special-purpose\index{special-purpose computer} FPGA-based\index{FPGA} hardware, Janus\index{Janus} II, in this thesis. It is usual in the physicist work to spend a lot of time by using (or developing) sophisticated statistical methods to improve the quality of the data, and, it is actually, a very important task. Nevertheless, if we can combine the statistical methods with powerful hardware, we can, indeed, obtain data at the forefront of the field.

Specifically, in this thesis, we have taken advantage (in some works) of the largest spin-glass simulation in off-equilibrium dynamics. We have simulated an Edwards-Anderson\index{Edwards-Anderson!model} model with Ising\index{Ising} spins, for a lattice size $L=160$ for a modest number of samples\index{sample} $\NS=16$, but a huge number of replicas\index{replica} $\NRep=512$ (a choice that turned to be fundamental). The simulations have been performed up to temperatures well deep in the spin-glass phase\index{phase!low-temperature/spin-glass} (for instance, $T=0.625$) to unprecedented long times.

It is worthy to mention also the fundamental role of other computers such as the Madrid's Cluster in the UCM and the supercomputer Cierzo in BIFI. They have made possible the analysis of the data in this thesis through thousands of hours of computational time.

\section{Conclusions on the metastate} \labsec{conclusion_metastate}
In its origins, the Replica Symmetry Breaking\index{replica!symmetry breaking (RSB)} theory aimed to explain the nature of the low-temperature phase\index{phase!low-temperature/spin-glass} in spin glasses assuming infinite-size systems. However, some mathematical procedures in that development were ill-defined. In particular, for disordered\index{disorder!systems} systems, the thermodynamic limit\index{thermodynamic limit} $L \to \infty$ for a Gibbs state may not exist. This problem is originated from a phenomenon known as \textit{chaotic size dependence}. Mathematical physics offered a new approach in the context of disordered\index{disorder!systems} systems and brought a solution to this problem: \textit{the metastate}\index{metastate}. This concept is a generalization of the concept of Gibbs states. Nonetheless, this discussion used to be limited to the theoretical work, without any numerical or experimental counterpart.

In~\refch{metastate} we have shown that the state of the art in numerical simulations allows the construction of the Aizenman-Wehr metastate\index{metastate!Aizenman-Wehr}. Indeed, our numerical data suggest that the $1 \ll W \ll R \ll L$ limit required from the construction of this metastate\index{metastate!Aizenman-Wehr} (see~\refsubsec{aw_metastate}) can be relaxed to $W/R \approx 0.75$ and $R/L \approx 0.75$ without changing substantially the thermodynamic physical behavior.

The main quantitative result of our work is the numerical computation of the exponent $\zeta$. According to Read~\cite{read:14}, it is possible to partially discriminate between the competing pictures trying to describe the nature of the spin-glass phase\index{phase!low-temperature/spin-glass} in\footnote{Recall that $d_{\mathrm{L}}$ and $d_{\mathrm{U}}$ are the lower and the upper critical dimension\index{critical dimension!lower}\index{critical dimension!upper} respectively.} $d_{\mathrm{L}} < d < d_{\mathrm{U}}$ by computing this exponent $\zeta$. Indeed, $\zeta$ is related to the number of different states that can be measured in a system of size $W$. This number behaves as $\log n_{\mathrm{states}} \sim W^{d -\zeta}$. Therefore, $\zeta < d$ would lead, for $W \to \infty$, to a metastate\index{metastate} of infinitely many states.

Here, we find $\zeta = 2.3 (3)$. We compare our result to previous estimations of this exponent, computed in different contexts, and we summarize our knowledge about the behavior of this exponent as a function of the dimension $d$ in~\refsec{relating_numerical_theory_metastate}.

All the numerical evidence strongly supports the existence of a spin-glass metastate\index{metastate!dispersed} dispersed over infinitely many states for $d=3$. This result probably holds for $d>d_{\mathrm{L}}$. Our findings are incompatible with the droplet\index{droplet!picture} model, while they are compatible with both the chaotic pair picture and the Replica Symmetry Breaking\index{replica!symmetry breaking (RSB)} scenario.

The results concerning this work are published in~\cite{billoire:17}.

\section{Conclusions on the study of the aging rate}
The off-equilibrium dynamics in spin glasses is of glaring importance since the experiments are always conducted out-of-equilibrium. Under these conditions, domains\index{magnetic domain} of correlated spins start to grow at the microscopical level. The linear size of these domains\index{magnetic domain} is the so-called \textit{coherence length\index{coherence length}} $\xi$. The \textit{aging\index{aging!rate} rate} $z(T)$, which is none but the variation rate of the free-energy\index{free energy!barrier} barriers with the logarithm of the coherence length\index{coherence length} $\xi$ (see \refsec{how_can_study_aging}), is directly related to the growth of the coherence length\index{coherence length} with the time. Previous numerical evidences pointed to a dependency of the form $\xi \sim \tw^{1/z(T)}$. However, numerical discrepancies in the determination of $z(T)$ were recently found between experiments~\cite{zhai:17} and numerical simulations~\cite{janus:08,lulli:16}.

In~\refch{aging_rate} we have taken advantage of the previously mentioned simulations performed in Janus\index{Janus} II to study the growth of the coherence length\index{coherence length} in the glassy phase\index{phase!low-temperature/spin-glass} and solved this discrepancy. Specifically, we found that the growth of the coherence length\index{coherence length} is controlled by a time-dependent aging\index{aging!rate} rate $z(T,\xi(\tw))$. 

In this work, we have described the dynamics as governed by a crossover between a critical and a low-temperature fixed point. The characterization of that crossover has allowed us to quantitatively model the growth of the aging\index{aging!rate} rate. In particular, we have considered two different ans\"atze for that growth, and we have found that, for the convergent one [\refeq{convergent_ansatz}], the computation of the aging\index{aging!rate} rate is consistent with the most recent experimental measures~\cite{zhai:17}.

Besides, we find clear evidence of non-coarsening dynamics at the experimental scale and find that temperatures $T \lesssim 0.7$ are free of critical effects and therefore safe for numerical studies of the spin-glass phase\index{phase!low-temperature/spin-glass}.

The results concerning this work are published in~\cite{janus:18}.

\section{Conclusions on the Mpemba effect}
Consider two beakers of water that are identical to each other except for the fact that one is hotter than the other. If we put both of them in contact with a thermal reservoir (for example, a freezer) at some temperature lower than the freezing point of the water, under some circumstances, it can be observed that the initially hotter water freezes faster than the colder one. This phenomenon is known as the Mpemba\index{Mpemba effect} effect~\cite{mpemba:69}.

In~\refch{mpemba} we have shown that the Mpemba\index{Mpemba effect} effect is present in spin glasses, where it appears as an intrinsically non-equilibrium process, ruled by the spin-glass coherence length\index{coherence length} $\xi$.

First, we have identified the relevant quantities to mimic the effect in spin glasses, namely the energy-density\index{energy!density} (as the role of the temperature in the classical Mpemba effect), and the Monte\index{Monte Carlo} Carlo time. However, the introduction of the coherence length\index{coherence length} as a hidden quantity ruling the process turned out to be fundamental.

Indeed, we have provided the first explanation of this phenomenon (in the spin-glass context), by using the relation between the energy density\index{energy!density} and the coherence length\index{coherence length} [\refeq{energy_coherence_length_relation}] to characterize the effect. Although this description is approximate, it is accurate enough to describe the Mpemba\index{Mpemba effect} effect.

Our results explain how the most natural experimental setup (prepare two identical systems at $T_1,T_2>T_\text{c}$ with an identical protocol, then quench them) can fail to observe the effect. Indeed, for spin glasses at least, a different starting $\xi$ is required. 

Finally, we have investigated the inverse Mpemba\index{Mpemba effect} effect (\refsec{inverse_mpemba}). In the spin-glass phase\index{phase!low-temperature/spin-glass}, the inverse Mpemba\index{Mpemba effect} effect is completely symmetrical to the classical Mpemba\index{Mpemba effect} effect. However, we found that above the critical\index{critical temperature} temperature $\Tc$ the inverse Mpemba\index{Mpemba effect} effect is strongly suppressed because our description of \refeq{energy_coherence_length_relation} is not valid for $T> \Tc$.

The Mpemba\index{Mpemba effect} effect is peculiar among the many memory\index{memory effects} effects present in spin glasses. Indeed, this phenomenon can be studied through quantities, such as the energy density\index{energy!density}, which are just measured at one-time scale (rather than the usual two times~\cite{young:98,jonason:98,janus:17b}). However, our setup poses an experimental challenge, because we are not aware of any measurement of the non-equilibrium temperature associated with the magnetic degrees of freedom\index{degree of freedom}. Perhaps one could adapt the strategy of Ref.~\cite{grigera:99}, connecting dielectric susceptibility\index{susceptibility} and polarization noise in glycerol, to measurements of high-frequency electrical noise in spin glasses~\cite{israeloff:89}.

Our investigation of the Mpemba\index{Mpemba effect} effect offers as well a new perspective into an important problem, namely the study of the glassy coherence length\index{coherence length} in supercooled liquids and other glass formers~\cite{cavagna:09}. Indeed, the identification of the right correlation function to study experimentally (or numerically) is still an open problem. Spin glasses are unique in the
general context of the glass transition\index{phase transition}, in both senses: we know which correlation functions should be computed microscopically~\cite{edwards:75,edwards:76}, while accurate experimental determinations of the coherence length\index{coherence length} have been obtained~\cite{guchhait:17}.

The results concerning this work are published in~\cite{janus:19}.

\section{Conclusions on the equilibrium Temperature Chaos}
In a spin glass, the Temperature\index{temperature chaos} Chaos phenomenon refers to the complete reorganization of the Boltzmann\index{Boltzmann!weight} weights that determines the frequency with which each configuration\index{configuration} of spins will appear, upon an arbitrary small change in the temperature $T$.

In~\refch{equilibrium_chaos} we have studied the Temperature Chaos\index{temperature chaos} phenomenon in equilibrium simulations and have proposed an efficient variational method\index{variational method} to estimate the elusive exponential autocorrelation time\index{autocorrelation time!exponential} of a Monte\index{Monte Carlo} Carlo Markov\index{Markov chain} chain, specific to the case of a Parallel Tempering\index{parallel!tempering} simulation.

This variational method\index{variational method} takes into account three parameters and performs a maximization of the estimation of the integrated autocorrelation time\index{autocorrelation time!integrated} in the phase-space\index{phase space} of these parameters. Since the exponential autocorrelation time\index{autocorrelation time!exponential} is an upper bound of the parameter-dependent integrated autocorrelation time\index{autocorrelation time!integrated}, our procedure leads to robust estimations.

In addition, we have studied the scaling properties of the probability distribution of the autocorrelation time, obtained using the proposed variational\index{variational method} approach. In particular, we have shown that scaling holds for lattices of sizes $L \geq 24$, consistently with previous studies using effective potentials.

The presence of Temperature Chaos\index{temperature chaos} is related to the poor performance of the Parallel Tempering simulations in the spin-glass phase\index{phase!low-temperature/spin-glass}. Then, the exponential autocorrelation time\index{autocorrelation time!exponential} provides us a \textit{dynamic} characterization of this phenomenon.

In this work, we have also characterized the Temperature Chaos\index{temperature chaos} from a \textit{static} point of view by studying the equilibrium configurations\index{configuration} of the system. The observable that allows us to quantitatively study the Temperature Chaos\index{temperature chaos} is the so-called \textit{chaotic parameter}. The empirical observation of the most chaotic samples\index{sample} (from the dynamical point of view), led to the construction of observables derived from the chaotic parameter to characterize the Temperature Chaos\index{temperature chaos} (see \cite{fernandez:13,fernandez:16}). In our work, we introduce a new observable to characterize the phenomenon.

Finally, we have checked the statistical correlations between these \textit{static} chaotic indicators and the dynamical correlation times. The introduction of the new \textit{static} indicator has improved previous results in the correlations.

The results concerning this work are published in~\cite{billoire:18}.

\section{Conclusions on the Temperature Chaos in off-equilibrium dynamics}
In~\refch{out-eq_chaos} we have studied an interesting phenomenon in off-equilibrium dynamics that closely mimics Temperature Chaos\index{temperature chaos}. Indeed, as we have just discussed in the previous section, Temperature Chaos\index{temperature chaos} is defined as an equilibrium phenomenon. Therefore, studying it in a non-equilibrium system is an open challenge that we have addressed here. 

First, we tried a naive approach to non-equilibrium Temperature Chaos\index{temperature chaos} (see \refsec{average_killed_chaos_signal}), which found only an exceedingly small chaotic signal. Fortunately, the statics-dynamics equivalence\index{statics-dynamics equivalence}~\cite{barrat:01,janus:08b,janus:10b,janus:17} combined with the rare events analysis performed in equilibrium simulations, see~\cite{fernandez:13}, provide the crucial insight to approach the problem. Specifically, the statics-dynamics equivalence allows us to relate the non-equilibrium dynamics of a spin glass (of infinite size) with a finite coherence length\index{coherence length} $\xi(\tw)$, with small samples\index{sample} of size $L\sim\xi(\tw)$ which can be equilibrated.

Our numerical protocol considers spherical-like regions of radius $r$. We focus on the probability distribution function of the chaotic parameter as computed over the spheres. We find that only the spheres in the tail of the distribution exhibit a strong Temperature Chaos\index{temperature chaos}.

Choosing a suitable length scale $r$ for the spherical-like regions turns out to be instrumental in the study of dynamic Temperature Chaos\index{temperature chaos}. This optimal length scale is proportional to the coherence length\index{coherence length}. However, our data mildly suggests that the importance of choosing exactly the correct $r$ becomes less critical in the $\xi \to \infty$ limit.

A striking link emerges between the dynamic and the static faces of the Temperature Chaos\index{temperature chaos} phenomenon. Indeed, we find a characteristic length scale $\xi^*(T_2-T_1,F)$ at which the crossover between the weak chaos and the strong chaos regimes occurs. The physical meaning of the characteristic length $\xi^*$ suggests that the equilibrium chaotic length $\ell_c$~\cite{fisher:86,bray:87b} is its equilibrium counterpart. In fact, both quantities depend on $T_2-T_1$ in the same way and the exponent\footnote{The reader should be warned that this exponent $\zeta$ is completely different from the exponent $\zeta$ in \refsec{conclusion_metastate}. We follow here the usual notation in the literature and denote both of them with the same letter despite their completely different meanings.} $\zeta$, controlling the temperature-dependence of $\ell_c$ turns out to be equal to $\zeta_{\mathrm{NE}}$, obtained from the off-equilibrium estimation, at the two-$\sigma$ level. We regard this coincidence as new and important evidence for the statics-dynamics equivalence\index{statics-dynamics equivalence}.

In the second part of the work we have performed temperature-varying\index{temperature-varying protocol} simulations to address the cumulative aging\index{aging!cumulative} problem~\cite{jonsson:02,bert:04,komori:00,berthier:02,picco:01,takayama:02,maiorano:05,jimenez:05}. We have found small but clear violations of cumulative aging\index{aging!cumulative}, which are stronger upon \textit{cooling} than upon \textit{heating}. Although both protocols display a memory-erasing process, the \textit{cooling} process shows better memory, i.e. larger coherence-lengths are needed to lose the memory of the previous state at the higher temperature. 

The results concerning this work are published in~\cite{janus:21}.

\chapter{Conclusiones} \labch{conclusiones_castellano}
La visión tradicional de un físico teórico solitario, trabajando sólo con papel y lápiz como sus únicas herramientas, ha ido evolucionando a lo largo de los años. En su lugar, la investigación computacional ha sido una parte importante del desarrollo de la física en las últimas décadas y los descubrimientos son normalmente hechos por grupos de físicos trabajando juntos. Sin embargo, aunque el incremento de la capacidad computacional no es nuevo, ha sido recientemente cuando hemos conseguido recolectar, procesar y analizar una gran cantidad de datos en tiempos razonables.

La física teórica ha sido una de las grandes beneficiadas de este incremento de la capacidad computacional. Si nos centramos en los vidrios de espín, el desarrollo de hardware dedicado ha permitido simular sistemas hasta las escalas de tiempo experimentales. Esta tesis es otra iteración en el desarrollo de un campo que ha tomado el camino general de la ciencia y ha abrazado la interacción entre experimentos, teoría y computación como una relación fructífera para hacer avances relevantes.

A lo largo de esta tesis, hemos estudiado los vidrios de espín desde un punto de vista numérico. El estudio del metaestado en el Capítulo 2 ha estado enfocado a abordar un problema teórico a través de la primera construcción del metaestado en equilibrio en simulaciones numéricas. En los capítulos 3 y 4, hemos estudiado la dinámica fuera del equilibrio de los vidrios de espín. En el Capítulo 3, hemos introducido la longitud de coherencia como la cantidad más relevante para caracterizar el estado de envejecimiento de un vidrio de espín y hemos resuelto una discrepancia numérica en el ratio de envejecimiento entre experimentos y simulaciones numéricas. En el capítulo 4, hemos discutido el efecto Mpemba, que es un efecto de memoria que tiene lugar en la dinámica fuera del equilibrio, dándonos un buen ejemplo de cómo la longitud de coherencia gobierna una multitud de fenómenos en este régimen. La parte final de esta tesis, los capítulos 5, 6 y 7, están dedicados a introducir y analizar el fenómeno del Caos en Temperatura en los vidrios de espín. De hecho, abordamos este problema desde un punto de vista de equilibrio (Capítulo 6) y también caracterizamos este fenómeno en la dinámica de fuera del equilibrio (Capítulo 7).

En este capítulo, resaltamos los principales resultados de esta tesis, revisitando todas las partes de la misma y resumiendo los mensajes más relevantes.

\section{Una breve reflexión sobre la importancia de los datos}
Esta tesis está principalmente enfocada al estudio numérico de los vidrios de espín. Una idea que es central a lo largo de esta tesis es la importancia fundamental de los datos de alta calidad. El lector puede preguntarse qué podríamos haber dicho en el Capítulo 3 sin una estimación precisa de la longitud de coherencia, o qué conclusiones podríamos sacar del Capítulo 7 sin una estimación precisa del parámetro caótico. Ninguno de estos trabajos sería posible sin nuestros datos de alta calidad.

Esta sección tiene por objetivo enfatizar el rol del hardware dedicado basado en FPGA Janus II en esta tesis.  Es usual en el trabajo de un físico dedicar mucho tiempo a usar (o desarrollar) métodos estadísticos sofisticados para mejorar la calidad de los datos y esta es una tarea muy importante. No obstante, si podemos combinar éstos métodos estadísticos con un hardware potente, podremos obtener datos en la vanguardia del campo.

En concreto, en esta tesis hemos tenido acceso (para varios trabajos) a la simulación más grande en vidrios de espín fuera del equilibrio. Hemos simulado un modelo de Edwards-Anderson con espines de Ising para una red de tamaño $L=160$, para un modesto número de samples $\NS=16$, pero para un gran número de réplicas $\NRep=512$ (una elección que ha resultado ser fundamental). Las simulaciones han sido llevadas a cabo a temperaturas bastante bajas, en la fase vítrea (por ejemplo $T=0.625$) hasta tiempos muy largos, sin precedentes.

También cabe destacar el papel fundamental de otros ordenadores como el Cluster de Madrid de la UCM y el superordenador Cierzo en el BIFI. Estos ordenadores han hecho posible el análisis de los datos de esta tesis a través de miles de horas de tiempo computacional.

\section{Conclusiones sobre el metaestado}
En sus orígenes, la teoría de \textit{Replica Symmetry Breaking} tenía como objetivo explicar la fase de bajas temperaturas en los vidrios de espín asumiendo sistemas de tamaño infinito. Sin embargo, algunos procedimientos matemáticos involucrados en la teoría estaban mal definidos. En concreto, para sistemas desordenados, el límite termodinámico $L\to\infty$ para un estado de Gibbs puede no existir. Este problema está originado por un fenómeno conocido como la \textit{dependencia caótica con el tamaño}. La irrupción de la física matemática en el contexto de los sistemas desordenados trajo la solución a este problema: el metaestado. Este concepto es una generalización del concepto de estado de Gibbs. No obstante, esta discusión solía estar limitada al trabajo teórico, sin una contraparte numérica o experimental.

En el capítulo 2 hemos demostrado que el estado del arte en simulaciones numéricas permite la construcción del metaestado de Aizenman-Wehr~\cite{aizenman:90}. De hecho, nuestros datos numéricos sugieren que el límite $1 \ll W \ll R \ll L$ requerido para la construcción del metaestado puede relajarse a $W/R \approx 0.75$ y $R/L \approx 0.75$ sin que haya cambios sustanciales en el comportamiento físico del mismo.

El principal resultado cuantitativo de nuestro trabajo es la computación numérica del exponente $\zeta$. Según Read \cite{read:14}, es posible discriminar parcialmente entre las teorías contrapuestas que tratan de describir la naturaleza de la fase vítrea entre la dimensión crítica inferior $d_{\mathrm{L}}$ y la dimensión crítica superior $d_{\mathrm{U}}$, calculando este exponente $\zeta$. De hecho, $\zeta$ está relacionado con el número de estados distintos que pueden ser medidos en un sistema de tamaño $W$. Este número se comporta como $\log n_{\mathrm{states}} \sim W^{d -\zeta}$. Por lo tanto, si $\zeta < d$ tendríamos que, para el límite $W \to \infty$, el metaestado tendría un número infinito de estados.

Nosotros encontramos $\zeta = 2.3(3)$. Comparamos nuestro resultado numérico con estimaciones previas de este exponente, calculadas en contextos distintos, y resumimos todo nuestro conocimiento sobre el comportamiento de dicho exponente como función de la dimensión $d$ en la sección 2.8.

Todas las evidencias numéricas apoyan firmemente la existencia de un metaestado disperso en vidrios de espín, formado por infinitos estados, en $d=3$. Este resultado, probablemente es válido para todo $d>d_{\mathrm{L}}$. Nuestros resultados son incompatibles con el modelo \textit{droplet}, mientras que son compatibles tanto con el modelo \textit{chaotic pair} como con el modelo \textit{replica symmetry breaking}.

Los resultados de este trabajo están publicados en~\cite{billoire:17}.

\section{Conclusiones sobre el estudio del ratio de envejecimiento}
La dinámica fuera del equilibrio tiene una importancia palmaria en los vidrios de espín puesto que los experimentos siempre son llevados a cabo en estas condiciones. En estos experimentos, dominios de espines correlacionados empiezan a crecer a nivel microscópico. La longitud de estos dominios es conocida como \textit{longitud de coherencia} $\xi$. La \textit{tasa de envejecimiento} $z(T)$, que no es más que la tasa de variación de las barreras de energía libre con el logaritmo de la longitud de coherencia $\xi$ (véase la Sección 3.2), está directamente relacionada con el crecimiento de la longitud de coherencia con el tiempo. La evidencia numérica señala que el comportamiento de este crecimiento con el tiempo es, esencialmente, $\xi \sim \tw^{1/z(T)}$. Sin embargo, discrepancias numéricas en la determinación de $z(T)$ han sido encontradas recientemente entre experimentos~\cite{zhai:17} y simulaciones numéricas~\cite{janus:08,lulli:16}.

En el capítulo 3 hemos aprovechado las simulaciones previamente mencionadas, llevadas a cabo por Janus II, para estudiar el crecimiento de la longitud de coherencia en la fase vítrea. En concreto, hemos descubierto que el crecimiento de la longitud de coherencia está controlado por un ratio de envejecimiento que depende del tiempo $z(T, \xi(\tw))$.

En este trabajo hemos descrito la dinámica mediante el cruce entre la influencia del punto fijo a temperatura crítica y el punto fijo a temperatura 0. La caracterización de ese cruce nos ha permitido modelizar cuantitativamente el crecimiento del ratio de envejecimiento. En concreto, hemos considerado dos hipótesis para ese crecimiento y hemos encontrado que la hipótesis convergente [\refeq{convergent_ansatz}] es consistente con el ratio de envejecimiento obtenido por las medidas experimentales más recientes~\cite{zhai:17}.

Además, hemos encontrado claras evidencias de dinámica \textit{non-coarsening} a la escala de tiempo experimental y hemos encontrado que las temperaturas $T \lesssim 0.7$ están libres de efectos críticos y, por lo tanto, son seguras para el estudio numérico.

Los resultados correspondientes a este capítulo están publicados en~\cite{janus:18}.

\section{Conclusiones sobre el efecto Mpemba}
Considere dos recipientes de agua que son idénticos entre ellos con la única excepción de que el agua contenida en uno de ellos está más caliente que la del otro recipiente. Si ponemos ambos recipientes en contacto con un baño térmico (por ejemplo, un congelador) a una temperatura por debajo del punto de congelación del agua, bajo ciertas circunstancias, puede observarse que el agua inicialmente más caliente se congela antes que la fría. Este fenómeno es conocido como el efecto Mpemba \cite{mpemba:69}.

En el Capítulo 4 hemos demostrado que el efecto Mpemba está presente en los vidrios de espín, donde es un proceso que ocurre fuera del equilibrio y que está gobernado por la longitud de coherencia $\xi$.

En primer lugar, hemos identificado los observables relevantes para imitar el efecto Mpemba clásico en los vidrios de espín. Esos observables han sido la densidad de energía, cumpliendo el rol de la temperatura en el efecto Mpemba clásico, y el tiempo de Monte Carlo. Sin embargo, la introducción de la longitud de coherencia como el observable oculto que gobierna el proceso ha resultado ser fundamental.

Hemos dado una primera explicación de este fenómeno en los vidrios de espín usando la relación entre la densidad de energía y la longitud de coherencia [\refeq{energy_coherence_length_relation}] para caracterizar dicho fenómeno. Aunque la descripción dada es aproximada, es suficientemente precisa para caracterizar el efecto Mpemba.

Nuestros resultados explican cómo la configuración experimental más habitual (preparar dos sistemas idénticos a $T_1,T_2 > \Tc$ con un protocolo idéntico y después enfriarlos) puede fallar para ver este efecto. De hecho, al menos para los vidrios de espín, diferentes longitudes de coherencia $\xi$ iniciales son necesarias.

Finalmente, hemos investigado el efecto Mpemba inverso. En la fase vítrea, el efecto Mpemba inverso es completamente simétrico respecto del efecto Mpemba clásico. Sin embargo, hemos encontrado que por encima de la temperatura crítica $\Tc$, el efecto Mpemba inverso se ve fuertemente menguado debido a que la descripción de \refeq{energy_coherence_length_relation} no es válida para $T>\Tc$.

El efecto Mpemba es peculiar dentro de los muchos efectos de memoria presentes en los vidrios de espín. De hecho, este fenómeno puede estudiarse a través de cantidades como la densidad de energía que requieren una sola escala temporal para ser medidas (al contrario que las dos escalas temporales que suelen requerir estos fenómenos para ser estudiados y caracterizados~\cite{young:98,jonason:98,janus:17b}). Sin embargo, nuestra configuración para el experimento numérico plantea un desafío experimental, dado que no tenemos noticia de ninguna medida de temperatura fuera del equilibrio asociada con los grados de libertad magnéticos. Quizás podría adaptarse la estrategia de~\cite{grigera:99}, que conecta la susceptibilidad dieléctrica y el ruido de polarización en glicerol, con las mediciones de ruido eléctrico de alta frecuencia en vidrios de espín~\cite{israeloff:89}.

Nuestro estudio del efecto Mpemba inverso sugiere una vía experimental más fácil, donde los sistemas son calentados, en lugar de enfriados. En este caso, aunque la respuesta de la energía es muy pequeña, el proceso va acompañado de un efecto de memoria dramático en la longitud de coherencia. Esta cantidad tiene una evolución temporal no monótona al calentarse desde la fase vítrea a la fase paramagnética, antes de converger a la curva maestra (isoterma). Además puede medirse con las técnicas experimentales actuales.

Nuestra investigación del efecto Mpemba ofrece también una nueva perspectiva sobre un problema importante, a saber, el estudio de la longitud de coherencia vítrea en líquidos sobreenfriados y otros formadores de vidrio~\cite{cavagna:09}. De hecho, la identificación de la función de correlación correcta para el estudio experimental (o numérico) sigue siendo un problema abierto. Los vidrios de espín son únicos en el contexto general de la transición vítrea: sabemos qué funciones de correlación deben calcularse microscópicamente~\cite{edwards:75,edwards:76} y se han obtenido determinaciones experimentales precisas de la longitud de coherencia~\cite{guchhait:17}.

Los resultados relacionados con este trabajo se han publicado en~\cite{janus:19}.

\section{Conclusiones sobre el Caos en Temperatura en equilibrio}
En un vidrio de espín, el efecto del \textit{caos en temperatura} es la completa reorganización de los pesos de Boltzmann que determinan la frecuencia con la que cada configuración de espines aparece, debido a un cambio arbitrariamente pequeño en la temperatura $T$.

En el capítulo 6 hemos estudiado el fenómeno del caos en temperatura en simulaciones de equilibrio y hemos propuesto un eficiente método variacional para estimar la elusiva cantidad del tiempo de autocorrelación exponencial en una cadena de Markov, específicamente para el caso del \textit{Parallel Tempering}.

Este método variacional toma en cuenta tres parámetros y realiza una maximización de la estimación del tiempo de autocorrelación integrado en el espacio de fases de dichos parámetros. Puesto que el tiempo de autocorrelación exponencial es una cota superior del tiempo de autocorrelación integrado (que depende de los parámetros anteriormente mencionados), nuestro procedimiento es capaz de dar estimaciones robustas.

Además, hemos estudiado propiedades de escalado de la distribución de probabilidad del tiempo de autocorrelación obtenido a través del método variacional. En concreto, hemos demostrado que el escalado se mantiene para redes de tamaños $L \geq 24$. Este resultado es consistente con resultados previos que usaban potenciales efectivos.

La presencia del caos en temperatura está relacionada con el pobre desempeño del algoritmo de \textit{Parallel Tempering} en la fase vítrea. Por lo tanto, el tiempo de autocorrelación exponencial constituye una caracterización dinámica de este fenómeno.

En este trabajo, también hemos caracterizado el caos en temperatura desde un punto de vista estático mediante el estudio de configuraciones de equilibrio del sistema. El observable que nos permite estudiar cuantitativamente el caos en temperatura es el parámetro caótico. La observación experimental de las \textit{samples} más caóticas (desde un punto de vista dinámico) llevó a la construcción de cantidades derivadas del parámetro caótico para estudiar el caos en temperatura (véase \cite{fernandez:13,fernandez:16}). En nuestro trabajo, hemos introducido un nuevo observable para caracterizar este fenómeno.

Finalmente, hemos comprobado las correlaciones estadísticas entre los estimadores estáticos del caos y los estimadores dinámicos. La introducción del nuevo estimador estático del caos ha mejorado considerablemente la correlación con los estimadores dinámicos con respecto a los estudios previos.

Los resultados relacionados con este trabajo están publicados en~\cite{billoire:18}.

\section{Conclusiones sobre el Caos en Temperatura fuera del equilibrio}
En el Capítulo 7 hemos estudiado el interesante fenómeno del caos en temperatura en dinámica fuera del equilibrio que imita de cerca al caos en temperatura en equilibrio. De hecho, como hemos visto en la sección previa, el caos en temperatura está definido como un fenómeno de equilibrio y, por lo tanto, el estudio de este fenómeno fuera del equilibrio ha sido históricamente un desafío abierto que nosotros trataremos aquí.

En primer lugar, intentamos una metodología \textit{naive} para explicar el fenómeno (véase sección 7.2), con la cual encontramos un caos extremadamente débil. Afortunadamente, la correspondencia estática-dinámica ~\cite{barrat:01,janus:08b,janus:10b,janus:17} junto con el análisis de eventos raros llevado a cabo en simulaciones de equilibrio~\cite{fernandez:13}, nos proporcionaron el enfoque correcto para abordar el problema. En concreto, la equivalencia estática-dinámica nos permite relacionar la dinámica fuera del equilibrio de un vidrio de espín (de tamaño infinito) con una longitud de coherencia $\xi(\tw)$, con pequeñas \textit{samples} de tamaño $L\sim\xi(\tw)$ que pueden ser llevadas al equilibrio.

Nuestro protocolo numérico considera regiones quasi-esféricas de radio $r$. En este trabajo, nos centramos en la función densidad de probabilidad del parámetro caótico calculado para dichas esferas. Encontramos que sólo para las esferas en la cola de la distribución puede observarse un caos en temperatura fuerte.

Escoger una escala de longitud $r$ adecuada para las regiones esféricas ha resultado ser fundamental para estudiar el caos en temperatura. Esta escala de longitud óptima es proporcional a la longitud de coherencia. Merece la pena resaltar que nuestros datos sugieren que la importancia de escoger la escala de longitud $r$ correcta es menos crítica para el límite $\xi \to \infty$.

Una conexión llamativa surge entre la dinámica y la estática en lo que al caos en temperatura se refiere. Encontramos una longitud característica $\xi^*(T_2-T_1,F)$ que marca el límite entre un régimen de caos débil y un régimen de caos fuerte. El significado físico de esta longitud característica $\xi^*$ sugiere que la llamada \textit{longitud caótica de equilibrio} $\ell_c$~\cite{fisher:86,bray:87b} podría ser su contrapartida en los sistemas en equilibrio. De hecho, ambas cantidades dependen de $T_2-T_1$ de la misma forma, a través del exponente $\zeta$\footnote{Advertimos al lector que este exponente $\zeta$ es completamente diferente al exponente $\zeta$ de la Sección 8.2. Nosotros seguimos la notación habitual en la literatura, denotando ambas cantidades con la misma letra a pesar de sus significados completamente distintos.}, que ha resultado ser el mismo para el equilibrio, y para fuera del equilibrio $\zeta_{\mathrm{NE}}$ dentro del nivel de dos desviaciones estándar. Contemplamos esta coincidencia como una nueva e importante evidencia de la equivalencia estática-dinámica.

En la segunda parte de este trabajo hemos llevado a cabo simulaciones con protocolos de cambios de temperatura para referir el problema del \textit{cummulative aging}~\cite{jonsson:02,bert:04,komori:00,berthier:02,picco:01,takayama:02,maiorano:05,jimenez:05}. Hemos encontrado pequeñas pero claras violaciones al \textit{cummulative aging} que son más fuertes cuando se enfría el sistema que cuando se calienta. Aunque ambos protocolos (calentar y enfriar), muestran procesos de borrado de memoria, el protocolo de enfriamiento muestra una mejor memoria, es decir, se necesitan tiempos de coherencia más largos para borrar la memoria del estado previo a mayor temperatura.

Los resultados relativos a este trabajo están publicados en~\cite{janus:21}.

\appendix 

\addpart{Appendices}
\setchapterstyle{lines}

\chapter{Statistical tools} \labch{AP_statistics}
\setlength\epigraphwidth{.5\textwidth}
\epigraph{\textit{Most people use statistics like a drunk man uses a lamppost; more for support than illumination.}}{-- Andrew Lang}

In this appendix, we expose the main statistical tools that we have used throughout this thesis. The huge amount of data (often correlated) and the non-linear nature of many quantities in our analysis are the main reasons for our interest in non-traditional statistics. We will discuss each of the methods used in this thesis and we provide examples.

\section{Estimating error bars} \labsec{estimating_errorbars}
Let us consider a random variable $x$ which follows a \gls{pdf} with mean $\mu_x$ and variance $\sigma_x$. The common situation in both, numerical and real experiments, is that the \gls{pdf} of $x$ is completely unknown for us. Thus, we can only \textit{sample} the \gls{pdf} with a finite number $N$ of measurements of the variable $x$: $\{x_1,x_2,\dots,x_N\}$. The usual estimators for $\mu_x$ and $\sigma_x$ are the arithmetic mean of the data $m(x)$ and the sample variance $s^2(x)$
\begin{equation}
m(x) = \dfrac{1}{N} \sum_{i=1}^N x_i \quad , \quad s^2(x) = \dfrac{1}{N-1} \sum_{i=1}^N \left[x_i - m(x)\right]^2 \, .
\end{equation}

These estimators are perfectly suitable for random variables and a linear combination of them. However, it is usual to be interested in a non-linear combination of random variables where the previous simple method does not work. In this case, we need a more advanced analysis in order to estimate the mean and the variance.

The traditional method is the well-known \textit{error propagation}. This method is based on Taylor expansions and is unbiased (up to arbitrary precision, determined by the number of terms we retain in the Taylor expansion). Nonetheless, it is dependent on the specific expression of the non-linear quantity studied and, therefore, a different computation (involving derivatives) has to be performed for each particular case.

Here, we introduce two alternatives that consist of different strategies to resample the data and compute the error bars\index{error bars}: the Jackknife\index{Jackknife} method and the Bootstrap\index{Bootstrap} method.

\subsection{The Jackknife method} \labsubsec{jk_method}
Just as before, suppose we have $N$ independent measures $\{x_1,x_2,\dots,x_N\}$. We define the Jackknife\index{Jackknife} variables as
\begin{equation}
\xjk_i = m(x) + \dfrac{1}{N-1} \left[ m(x) - x_i \right] = \dfrac{1}{N-1} \sum_{j \neq i} x_j \, . \labeq{jackknife_variable_definition}
\end{equation}
Now, suppose we have a set of random variables $x_{\alpha}$, $x_{\beta}$, $x_{\gamma}$, $\dots$ (each one with $N$ independent measures $x_{\alpha,i}$, $x_{\beta,i}$, $\dots$ with $i=1,\dots,N$) and a non-linear combination of them $f(x_{\alpha},x_{\beta},x_{\gamma},\dots)$. We define the Jackknife\index{Jackknife} variable (or Jackknife\index{Jackknife} block) of that non-linear function as
\begin{equation}
\fjk_i = f(\xjk_{\alpha,i},\xjk_{\beta,i},\xjk_{\gamma,i},\dots) \, .
\end{equation}
Our aim is to estimate $\mu_f=f(\mu_{x_{\alpha}},\mu_{x_{\beta}},\dots)$ and the standard deviation of the quantity $f$, $\sigma_f$. To estimate that quantity, we introduce the mean of the Jackknife\index{Jackknife} blocks
\begin{equation}
m(\fjk) = \dfrac{1}{N} \sum_{i=1}^N \fjk_i \, .
\end{equation}
For non-linear quantities is easy to show~\cite{young:12} that there is a bias in the estimation of $f(\mu_{x_{\alpha}},\mu_{x_{\beta}},\dots)$ with the quantity $m(\fjk)$. Indeed, the elimination of the leading bias leads to the following estimator $f_{\mathrm{JK,estimator}}$ of $f(\mu_{x_{\alpha}},\mu_{x_{\beta}},\dots)$:
\begin{equation}
f_{\mathrm{JK,estimator}} = N f\left(m(x_{\alpha}),m(x_{\beta}),\dots\right) - (N-1) m(\fjk) \, .
\end{equation}

In this case, the bias is of order $1/N^2$. If we just take the quantity $m(\fjk)$ as an estimator of $f(\mu_{x_{\alpha}},\mu_{x_{\beta}},\dots)$ the bias is of order $1/N$. As we will discuss below, this is not a big problem since the bias in the statistical error is of order $1/\sqrt{N}$ which is much bigger than $1/N$ for large $N$.

We can also define the Jackknife\index{Jackknife} variance as
\begin{equation}
s^2(\fjk) = m\left((\fjk)^2\right) - \left[m(\fjk)\right]^2 \, .
\end{equation}

After some easy mathematical steps~\cite{young:12}, we can reach an expression for the estimator $\sigma_{\mathrm{JK,estimator}}$ of the variance $\sigma_f$
\begin{equation}
\sigma^2_{\mathrm{JK,estimator}} = (N-1) s^2(\fjk) \, .
\end{equation}
This is the great success of the Jackknife\index{Jackknife} method, the estimator of the variance can be obtained from the raw data, and no derivatives are needed to be computed. Hence, the error bars\index{error bars} of our desired quantity, that usually corresponds to the standard deviation of $f(\mu_{x_{\alpha}},\mu_{x_{\beta}},\dots)$, would be 
\begin{equation}
\sigma_{\mathrm{JK,estimator}} = \sqrt{N-1} \, s(\fjk) = \sqrt{N-1} \left[  m\left((\fjk \right)^2) - \left[m(\fjk)\right]^2 \right]^{1/2} \, . \labeq{errorbar_estimator}
\end{equation}

The reader can found an example of a specific application of this method in~\refsec{procedure}. 

Another advantage of this method is that the Jackknife\index{Jackknife} blocks can be easily propagated to perform further computations. This propagation usually involves non-linear operations, fortunately, we have just discussed that the Jackknife\index{Jackknife} method can deal with them. We illustrate our statement with one example.

\subsubsection{Computing the error of the coherence length}
As we have said, the coherence length\index{coherence length} is fundamental in order to characterize the off-equilibrium dynamics. Let us sketch the procedure to compute its error bars\index{error bars}.

First, we recall that our off-equilibrium simulation consists of an \gls{EA}\index{Edwards-Anderson!model} model with Ising\index{Ising} spins for which we have simulated $\NS=16$ different samples\index{sample} and, for each sample\index{sample}, $512$ replicas.

Before explaining the method, we briefly recall the definition of the estimator of the coherence length\index{coherence length} $\xi_{12}(t)$ from the four-point\index{correlation function!four point} correlation function $C_4(T,r,t)$
\begin{align}
C_4(T,r,t) & =  \overline{\braket{q^{\sigma,\tau}_{\vec{x}}(t)q^{\sigma,\tau}_{\vec{x}+\vec{r}}(t)}} \, , \\
I_k(t) & = \int_0^{\infty} r^k C_4(T,r,t) \dd r \, , \\
\xi_{12}(t) & = \dfrac{I_{2}(t)}{I_1(t)} \, .
\end{align}

We start by computing $C_4(T,r,t)$. The Jackknife\index{Jackknife} method will allow us to estimate the disorder\index{disorder!average} average $\overline{(\cdots)}$. For each sample\index{sample} $\mathcal{J}_i$, we compute
\begin{equation}
C_4^{\mathcal{J}_i}(T,r,t) = \braket{q^{\sigma,\tau}_{\vec{x}}(t)q^{\sigma,\tau}_{\vec{x}+\vec{r}}(t)}_{\mathcal{J}_i} \, ,
\end{equation}
where the label $\mathcal{J}_i$ stresses the fact that the quantity is sample-dependent. Then we can build the Jackknife\index{Jackknife} blocks for this quantity\footnote{The details on the computation of $ C_{4,i}^{\mathrm{JK}}(T,r,t)$ can be found in~\refsec{finite_size_effects}.}
\begin{equation}
C_{4,i}^{\mathrm{JK}}(T,r,t) = \dfrac{1}{\NS-1} \sum_{i \neq j} C_4^{\mathcal{J}_j}(T,r,t)
\end{equation}

If we were interested only in computing the four-point\index{correlation function!four point} correlation function, we would simply estimate the errors of $C_4(T,r,t)$ by using~\refeq{errorbar_estimator}. However, we are interested in the quantity $\xi_{12}(t)$. The \textit{propagation} of the Jackknife\index{Jackknife} method to derived quantities is quite simple. We now compute $I_k(t)$ [in our case we are interested in $I_1(t)$ and $I_2(t)$] for each Jackknife\index{Jackknife} block of the four-point\index{correlation function!four point} correlation function. Therefore, we will obtain
\begin{equation}
I^{\mathrm{JK}}_{k,i}(t) = \int_0^{\infty} r^k C_{4,i}^{\mathrm{JK}}(T,r,t) \dd r \, .
\end{equation}
Finally we can, again, propagate the Jackknife\index{Jackknife} block by computing
\begin{equation}
\xi^{\mathrm{JK}}_{12,i}(t) = \dfrac{I^{\mathrm{JK}}_{2,i}(t)}{I^{\mathrm{JK}}_{1,i}(t)} \, .
\end{equation}

In the end, we will have $\NS$ Jackknife\index{Jackknife} blocks and we can estimate the error bars\index{error bars} of $\xi_{12}(t)$ by using~\refeq{errorbar_estimator}.

\subsection{The Bootstrap method} \labsubsec{bootstrap_method}
Although most of the time we use the Jackknife\index{Jackknife} method, there exists an alternative that also takes advantage of resampling the data: the Bootstrap\index{Bootstrap} method. 

Let us consider again that we have a set of $N$ points $\{x_1,x_2,\dots,x_N\}$. In the Jackknife\index{Jackknife} method, we build $N$ blocks from all the points of the original data except one. On the contrary, in the Bootstrap\index{Bootstrap} method, we build $\Nb \neq N$ data sets, each one containing $N$ points picked at random from the original data. The probability of selecting one point of the original data is $1/N$ and does not depend on whether it has been selected before.

For a given data set $k$, with $k$ running from 1 to $\Nb$, we can define the mean of the points inside this data set as
\begin{equation}
x^{B}_k = \dfrac{1}{N} \sum_{i=1}^N n^{B}_{i,k} x_i \, ,
\end{equation}
where $n^{B}_{i,k}$ is the number of times that the point $x_i$ have been selected for the data set $k$. The mean of all the $x^{B}_k$ over the $\Nb$ data sets is
\begin{equation}
m\left(x^{B}\right) = \dfrac{1}{\Nb} \sum_{k=1}^{\Nb} x_k^{B} \, . \labeq{def_mean_boots}
\end{equation}

It can be shown~\cite{young:12}, that for $\Nb \to \infty$ the value of $m\left(x^{B}\right)$ corresponds to the value of $m(x)$, which is an unbiased estimator of $\mu_x$. The variance of $x^{B}$ is
\begin{equation}
s^2\left(x^{B}\right) = m \left( (x^{B})^2\right) - \left[ m\left( x^{B} \right) \right]^2 \, .
\end{equation}

To address the more interesting scenario, suppose again that we have a set of random variables $x_{\alpha}$, $x_{\beta}$, $x_{\gamma}$, $\dots$ (each one with $N$ independent measures $x_{\alpha,i}$, $x_{\beta,i}$, $\dots$ with $i=1,\dots,N$) and a non-linear combination of them $f(x_{\alpha},x_{\beta},x_{\gamma},\dots)$.  We are again interested in estimating $\mu_f= f(\mu_{x_{\alpha}},\mu_{x_{\beta}},\dots)$ and its standard deviation $\sigma_f$. The Bootstrap\index{Bootstrap} variable will be
\begin{equation}
\fb_k = f(x^{B}_{\alpha,k},x^{B}_{\beta,k},x^{B}_{\gamma,k},\dots) \, . \labeq{bootstrap_variable}
\end{equation}

Similarly to the Jackknife\index{Jackknife} case, the mean and the variance of $\fb_k$ would be
\begin{equation}
m\left( \fb \right) = \dfrac{1}{\Nb} \sum_{k=1}^{\Nb} \fb_{k} \, ,
\end{equation}
\begin{equation}
s^2 \left( \fb \right) = m\left( \left(\fb \right)^2 \right)- \left[ m\left(\fb\right) \right]^2 \, .
\end{equation}

Again, the bias in $m\left( \fb \right)$ is of order $1/N$, much smaller than the statistical error. Indeed, the Bootstrap\index{Bootstrap} estimator $\sigma_{\mathrm{B,estimator}}$ of the standard deviation $\sigma_f$ is
\begin{equation}
\sigma_{\mathrm{B,estimator}} = \sqrt{\dfrac{N}{N-1}} \left[ m\left( \left(\fb \right)^2 \right)- \left[ m\left(\fb\right) \right]^2 \right]^{1/2} \, . \labeq{standard_deviation_bootstrap}
\end{equation}
In the $N\to \infty $ limit, the prefactor tends to $1$.

It is worthy to note that a high enough number of data sets $\Nb$ is needed to ``achieve'' the $\Nb \to \infty$ limit in~\refeq{def_mean_boots}. How large should be $\Nb$ to achieve that limit? We have not performed a systematic study but the relation $\Nb = 10N$ seems to be safe. Indeed, our use of the Bootstrap\index{Bootstrap} method is limited and the computations involved do not require large computational times, but a much more detailed study should be done for a different situation.

As a final note, the reader may wonder why we would use the Bootstrap method if we can use the Jackknife method that involves less computations and provide similar results. The Jackknife method assumes a Gaussian behavior of the variables $x_{\alpha}$, $x_{\beta}$, $x_{\gamma}$, $\dots$, on the contrary, the Bootstrap method allows us to access to the probability distribution of $f(x_{\alpha},x_{\beta},x_{\gamma},\dots)$ in those cases in which $x_{\alpha}$ does not follow a Gaussian distribution. 

\subsubsection{Computing the Pearson coefficient}
Let us consider a set of $N$ pairs of independent measures $\{(x_1,y_1),(x_2,y_2),\dots,(x_N,y_N)\}$. In some situations, we want to test the linear dependence between the pairs of measures. To test that dependence, the proper tool is the Pearson coefficient $r$ which is between $-1$ and $1$. The extreme values are clear, $r=-1$ corresponds to a complete linear-anticorrelation between $x$ and $y$ and the value $r=1$ corresponds to a perfect correlation between $x$ and $y$.

The Pearson coefficient is defined as
\begin{equation}
r = \dfrac{\sum_{i=1}^N x_i y_i - N m(x) m(y)}{(n-1) \sqrt{m(x^2) - [m(x)]^2} \sqrt{m(y^2) - [m(y)]^2}} \, , \labeq{pearson_coefficient}
\end{equation}
where $m(x)=(1/N)\sum_i x_i$.

The reader may find one practical example of this computation in \refch{equilibrium_chaos}. Indeed, the Bootstrap\index{Bootstrap} estimation of the error for this quantity is very simple. From the complete cloud of $N$ points, we randomly (with uniform probability) select $N$ pairs with repetition (i.e. picking more than once the same pair is allowed). We repeat this procedure $\Nb$ times and, for each of the $\Nb$ clouds of $N$ points, we compute the Pearson coefficient $r$. 

The correspondence between each of the $r_k$ computed values (with $k$ running from 1 to $\Nb$) and the values of $\fb_k$ appearing in~\refeq{bootstrap_variable} is evident. Then, the estimation of the error bars\index{error bars} is straightforward, we have just to apply~\refeq{standard_deviation_bootstrap}.

\section{Improving the statistics} \labsec{improving_statistics}
Here, we explain how the correlation between variables (in a controlled situation) can reduce the error bars\index{error bars} of the quantities we are interested in. We take advantage of the \textit{control\index{control variate} variates} for that purpose. The reader may find further information in~\cite{fernandez:09c,ross:14}.

\subsection{Control Variates}
The starting point is the same of~\refsec{estimating_errorbars}, consider a random variable $x$ which follows a \gls{pdf} with mean $\mu_x$ and standard deviation $\sigma_x$ and a set $N$ measurements of the variable $x$: $\{x_1,x_2,\dots,x_N\}$. Since we only have a finite number of measurements, we can only aspire to estimate the mean and the standard deviation through the arithmetic mean $m(x)$ and the sample standard deviation $s(x)$ (or equivalently, the sample variance $s^2(x)$)
\begin{equation}
m(x) = \dfrac{1}{N} \sum_{i=1}^N x_i \quad , \quad s^2(x) = \dfrac{1}{N-1} \sum_{i=1}^N \left[x_i - m(x)\right]^2 \, .
\end{equation}

In order to apply the control\index{control variate} variates method, we need another variable $y$ (the control\index{control variate} variate) from which we know exactly its mean $\mu_y$. Hence, we define a new variable $z$
\begin{equation}
z= x + c(y-\mu_y) \, , \labeq{cv_new_variable}
\end{equation}
where $c \in \mathbb{R}$ is a constant. It is clear that 
\begin{equation}
m(z) = m\left(x + c(y-\mu_y) \right) = m(x) + c(m(y) - \mu_y) = m(x) \, .
\end{equation}
Therefore, $m(z)$ is an unbiased estimator of $\mu_x$. The situation is more interesting for the variance
\begin{equation}
s^2(z) = s^2 \left( x + c(y -\mu_y)\right) = s^2(x) + c^2 s^2(y) + 2c \Cov(x,y) \, ,
\end{equation}
being $\Cov(x,y)=m(xy) - m(x)m(y)$ the covariance between the random variables $x$ and $y$. As long as we are interested in reduce the variance of $z$ we can look for the constant value $c$ minimizing $s^2(z)$
\begin{equation}
\dfrac{\partial s^2(z)}{\partial c} = 0 \implies c^* = - \dfrac{\Cov(x,y)}{s^2(y)} \, ,
\end{equation}
where $c^*$ is the value of the constant $c$ that minimizes the variance of $s^2(z)$. Then, the variance of $z$ will be
\begin{equation}
s^2(z) = s^2\left(x+c^*(y-\mu_y)\right) = s^2(x) - \dfrac{\left[ \Cov(x,y) \right]^2}{s^2(y)} \, .
\end{equation}
The plan for reducing the error bars\index{error bars} and improve the statistics is clear: we have to find a quantity $y$ with the smallest possible variance and the biggest possible covariance with $x$.

\subsubsection{Improving the energy statistics}
One practical example of the control\index{control variate} variates has been already mentioned in \refsec{numerical_simulation_mpemba}, here, we provide details of our particular implementation. To put the reader in context, we recall here that we are interested in computing the energy-density\index{energy!density} $e(t,\mathcal{J})$ to study the Mpemba\index{Mpemba effect} effect.

The definition of the energy\index{energy!density} density in our three-dimensional lattice of linear size $L$ is
\begin{equation}
e(t,\mathcal{J}) = \dfrac{1}{L^3} \braket{\mathcal{H}_{\mathcal{J}}(t)} \quad , \quad e(t) = \overline{e(t,\mathcal{J})} \, .
\end{equation}

In order to apply the above-exposed method, we need to introduce a control\index{control variate} variate with a large covariance with the energy-density\index{energy!density}. Our proposal for the control\index{control variate} variate is the following. For each of the $\NS=16$ samples\index{sample} simulated, we took a fully random initial configuration\index{configuration}, placed suddenly at $T=0.7$. We perform an isothermal\index{isothermal} simulation for $t=2^{21}$ Monte\index{Monte Carlo} Carlo steps, and we store in our disk those configurations\index{configuration} that fulfills $t = \Floor{2^{n/8}}$ with $n$ integer. Then, we choose all the configurations\index{configuration} with times $\mathcal{C} \, \in \, [2^{19},2^{21}]$, the number of selected times is $N_{\mathcal{C}}$. Then, we compute
\begin{equation}
\ecvj = \dfrac{1}{L^3 N_{\mathcal{C}}}  \sum_{t \in \mathcal{C}} \braket{\mathcal{H}_{\mathcal{J}}(t)} \, . \labeq{control_variate}
\end{equation}

The alert reader may wonder why we choose this quantity $\ecvj$ if we do not know the real value of its mean $\mu_{\ecvj}$. At the first sight, it seems that we have made no improvements. However, although we do not know $\mu_{\ecvj}$ we can estimate it with great accuracy. Indeed, we perform a separated simulation with $1.6 \cdot 10^5$ samples\index{sample} and two replicas for each sample\index{sample} (i.e. the energy\index{energy!density} was averaged over $3.2 \cdot 10^5$ systems). Thus, the estimation of $\mu_{\ecvj}$ would have negligible error compared with $e(t)$.

Then, the final estimates of the internal energy\index{energy!density} is
\begin{equation}
\tilde{e}(t,\mathcal{J}) = e(t,{\mathcal{J}}) - \left[ \ecvj - \mu_{\ecvj} \right] \quad \, \quad \tilde{e}(t) = \overline{\tilde{e}(t,\mathcal{J})} \, . 
\end{equation}

The benefits of using this control\index{control variate} variate are obvious from~\reffig{control_variate}. Moreover, the importance of the high covariance between the studied quantity [in our case, the energy-density $e(t,\mathcal{J})$] and the control\index{control variate} variate $\ecvj$ is evident. In the three simulations showed in~\reffig{control_variate}, the larger improvement in the error estimation is achieved for $T=0.7$ for a coherence length\index{coherence length} $\xi(\tw)$ with $\tw$ roughly corresponding to $\tw \in [2^{19},2^{21}]$. Because our control\index{control variate} variate has been computed with great accuracy for those conditions, it is logical that we achieve the best performance in the method for that case, although, improvements are also achieved for different temperatures (see for example $T=0.9$).

\begin{figure}[h!]
\centering
\includegraphics[width=0.8\columnwidth]{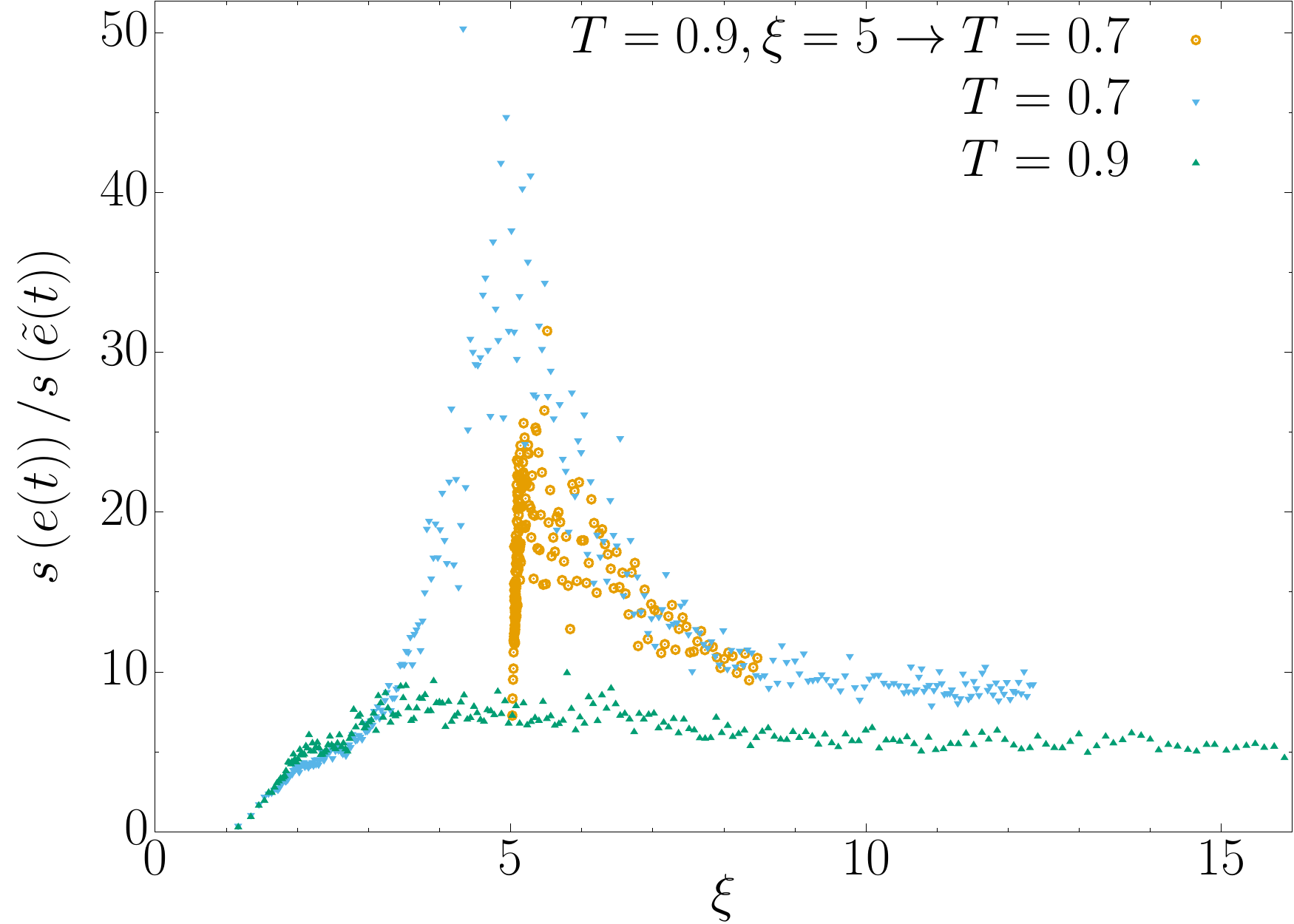}
\caption[\textbf{Improving the accuracy with control variates.}]{\textbf{Improving the accuracy with control\index{control variate} variates.} The figure shows the ratio of statistical errors, as a function of $\xi(t)$, for the naive $e(t)$ and improved $\tilde{e}(t)$ estimates of the energy density. The data shown correspond to three different relaxations\index{relaxation}. Two of them are isothermal\index{isothermal} relaxations\index{relaxation} starting from $\xi=0$ at $t=0$. The third relaxation\index{relaxation} corresponds to the preparation starting at $(T=0.9,\xi=5)$ which is quenched to $T=0.7$ at $t=0$. The error reduction is largest for the isothermal\index{isothermal} relaxation\index{relaxation} at $T=0.7$ and $2^{19}\leq t\leq 2^{21}$, of course (after all, this is the temperature and time region defining the control\index{control variate} variate), but the error reduction is also very significant at other times and temperatures.}
\labfig{control_variate}
\end{figure}

\setchapterstyle{lines}
\chapter{The aging rate. Technical details.} \labch{AP_technical_details_aging}
In this appendix, we expose the technical details that are fundamental to understand the procedure followed in~\refch{aging_rate} but that would blur the mean message of the chapter and would make reading difficult.

In~\refsec{Nr_aging} we explore the error reduction for the four-point\index{correlation function!four point} correlation function when a high number of replicas\index{replica} is simulated. We perform exhaustive checks to our $L=160$ system to be sure that our results are not affected by finite-size effects\index{finite-size effects} in~\refsec{finite_size_effects}. Then, in~\refsec{Josephson_length}, we give additional details of the crossover between the $T=\Tc$ fixed point and the $T=0$ one, which is controlled by the Josephson length. Finally, we discuss the choices of the parameters in our fits in~\refsec{parameters_aging}.

\section{Error reduction for high number of replicas}\labsec{Nr_aging}
As mentioned in~\refch{aging_rate}, the choice of the number of replicas\index{replica} ($\NRep$) and samples\index{sample} ($\NS$) was taken with the aim of improving the estimation of observables related to \gls{TC}\index{temperature chaos} in future work, where it is important to maximize the number of possible overlaps\index{overlap} (pairs of replicas\index{replica}) $N_\text{ov}=\NRep(\NRep - 1)/2$.

Unexpectedly, this has led to a dramatic increase in precision.~\reffig{error-NR} shows the reduction of the statistical error in the correlation function $C_4$ as a function of $1/\sqrt{\Nov}$. Moreover, this effect is enhanced as $r$ increases, which leads to a qualitative improvement in the computation of the $I_k(T,r,\tw)$ integrals (\reffig{error-r2Cr}).

The reader should be alert that, in view of the above results, a previous study can be done for any quantity which involves disorder\index{disorder!average} averages and thermal averages. The general form of the variance of such a quantity would be
\begin{equation}
\sigma^2 (\NRep,\NS) = \left[ \sigma_{\mathrm{S}}^2 + \sigma_{\mathrm{R}}^2 \left( \dfrac{2}{\NRep (\NRep-1)}\right)^x \right] \dfrac{1}{\NS} \, ,
\end{equation}
where $\sigma_{\mathrm{S}}^2$ and $\sigma_{\mathrm{R}}^2$ are the sample\index{sample} and thermal contributions to the variance respectively and $x$ is the exponent controlling the error reduction due to the number of overlaps\index{overlap} $\Nov$ which should be between $0.5$ and $1$.

A careful analysis in the context of thin-film \gls{SG}s can be found in \cite{fernandez:19b}. The great advantage of this analysis is that we can choose, for a given computational effort $E=\NRep \NS$, the number of replicas\index{replica} (or equivalently, of samples\index{sample}) that minimizes the error instead of assuming that the maximum possible number of samples\index{sample} should be simulated (as it was usual).

\begin{figure}[h!]
\includegraphics[width=0.8\linewidth]{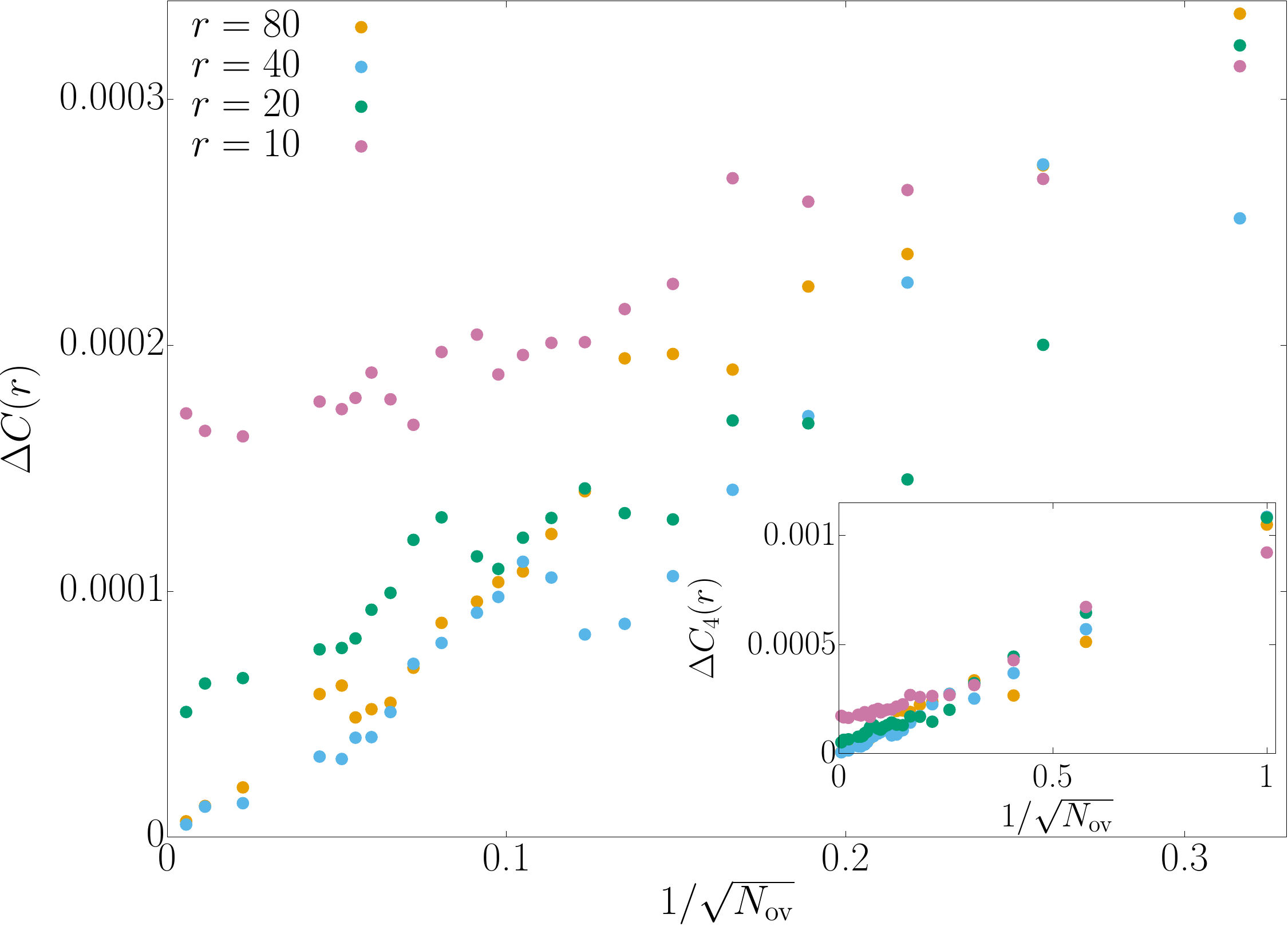}
\caption[\textbf{Reduction of statistical error with the number of overlaps.}]{\textbf{Reduction of statistical error with the number of overlaps.\index{overlap}} Rapid reduction in the statistical error for the overlap\index{overlap} autocorrelation function $C_4(T,r,\tw)$ with increasing number of replicas\index{replica} $N_\text{R}$. We plot $\Delta C(r)$ for $T=0.7$ and $\tw=2^{32}$ and several values of $r$ as a function of the number of possible overlaps\index{overlap} [$N_\text{ov} = N_\text{R}(N_\text{R}-1)/2$]. For large values of $r$ the error is essentially linear in $1/N_\text{R}$. The inset shows the whole range from $N_\text{R}=2$ (the minimum to define $C_4$), while the main plot is a close up of the large-$N_\text{R}$ sector. The simulations reported in this paper have $N_\text{R}=256$.}
\labfig{error-NR}
\end{figure}

\begin{figure}[h!]
\includegraphics[width=0.7\linewidth]{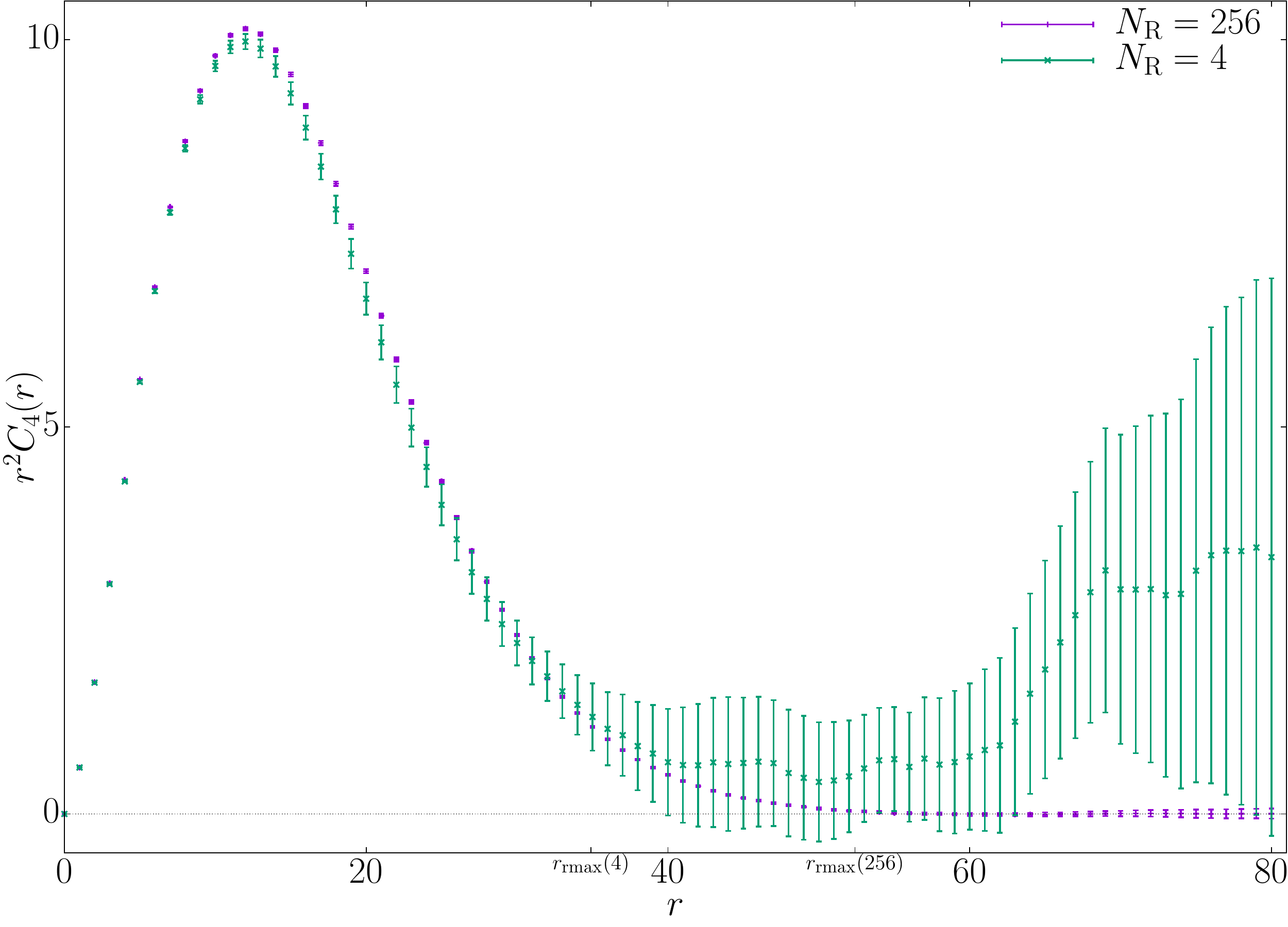}
\caption[\textbf{Improving the determination of \boldmath $C_4(T,r,\tw)$.}]{\textbf{Improving the determination of \boldmath $C_4(T,r,\tw)$.} Comparison of $r^2 C_4(r)$, function whose integral is used to estimate the coherence length\index{coherence length}, as computed with $4$ and $256$ replicas\index{replica} and the same number $N_\text{S}=16$ of samples\index{sample} ($T=0.7, \tw=2^{34}$). We have marked on the $x$ axis the resulting self-consistent cutoffs (see \refsec{finite_size_effects}) used to estimate the $I_k$ integrals in both cases.}
\labfig{error-r2Cr}
\end{figure}

\section{Controlling finite-size effects} \labsec{finite_size_effects}
In order to obtain an estimate for $\xi_{k,k+1}(T,\tw) = I_{k+1}(T,\tw)/I_k(T,\tw)$ we need to compute the integrals \begin{equation}
I_k(T,\tw) = \int_0^\infty r^k C_4(T,r,\tw)\dd r \, . \labeq{I_SI} 
\end{equation}
As discussed in~\cite{janus:09b}, the main difficulty  in this computation is handling the large-$r$ tail where relative errors [$\Delta C_4(T,r,\tw) / C_4(T,r,\tw)$] are big. Our main goal in this part is to minimize the statistical errors and to check for finite-size effects\index{finite-size effects} (which will appear when $\xi/L$ becomes relatively large).

To compute those $I_k(T,\tw)$, we proceed in a similar way as in~\cite{janus:09b,fernandez:18b}. We separate the integration range in two parts. From $0$ to $r_{\mathrm{cutoff}}$, we compute the numerical integral of $C_4(T,r,\tw)$. From $r_{\mathrm{cutoff}}$ to infinity, we estimate the contribution of the tail with a smooth extrapolating function $F(r) \sim r^{-\vartheta} f\bigl(r/\xi)$.

In short, the procedure is
\begin{enumerate}
\item Obtain $F(r)$ with fits of $C_4$ in a self-consistent region $[r_\text{min},r_\text{max}]$  where the signal-to-noise ratio is still good. 
\item Integrate $C_4$ numerically up to some cutoff and add the analytical integral of $F(r)$  beyond the
cutoff to estimate the tail contribution.
\end{enumerate}
The selection of the function $F(r)$ can be done in several ways, here, we propose two particular functions. First, we consider
\begin{equation}
F_1(r) = A_1 r^{-\vartheta} \ee^{-(r/\xi)^{\beta_1}} \, . \labeq{F1}
\end{equation}
Where $\vartheta$ is the replicon\index{replicon} exponent and we fit for $A_1,\beta_1$ and $\xi$. This analytical form is motivated by the fact that this is the simplest choice that avoids a pole singularity in the Fourier transform of $C_4(T,r,\tw)$ at finite $\tw$ and has been used in several previous works, for example, see~\cite{janus:09b}. In order to check for finite-size effects\index{finite-size effects}, we also consider a second function $F_2(r)$ resulting from fits that include the first-image term:
\begin{equation}
F_2^*(r) = A_2 \left[ \frac{ \ee^{-(r/\xi)^{\beta_2}}}{r^\vartheta} + \frac{ \ee^{-((L-r)/\xi)^{\beta_2}}}{(L-r)^\vartheta}\right],
\end{equation}
so we have a second extrapolating function
\begin{equation}\labeq{F2}
F_2(r) = A_2 r^{-\vartheta} \ee^{-(r/\xi)^{\beta_2}}.
\end{equation}
For these fits we used $\vartheta=0.35$. However, this value has very little effect on the final computation of $\xi_{k,k+1}$. We have checked this by recomputing the integrals with $\vartheta = 0 $ and $\vartheta = 1+\eta\approx 0.61$ 
(prediction of the droplet\index{droplet!picture} theory and influence of the $T=T_\mathrm{c}$ fixed point respectively). The different choices of $\vartheta$  led to a systematic effect smaller than $20 \%$ of the error bars\index{error bars} in the worst case. 

Once we have our two extrapolating functions $F_1$ and $F_2$, we can combine them with the $C_4$ data in several ways:
\begin{eqnarray}
I^1_k &=& \int_0^{r_{\mathrm{max}}} \dd r~r^k C_4(T,r,\tw) + \int_{r_{\mathrm{max}}}^\infty \dd r~r^k F_1(r) \, , \\
I^2_k &=& \int_0^{r_{\mathrm{max}}} \dd r~r^k C_4(T,r,\tw) + \int_{r_{\mathrm{max}}}^\infty \dd r~r^k F_2(r) \, , \\
I^3_k &=& \int_0^{r_{\mathrm{min}}} \dd r~r^k C_4(T,r,\tw) + \int_{r_{\mathrm{min}}}^\infty \dd r~r^k F_2(r) \, . \labeq{I3}
\end{eqnarray}

The difference between $I^2_k$ and $I^1_k$ is always under $1\%$ in the error, so choosing between them has no effect in any computation. In contrast, $I^1_k$ and $I^3_k$ present measurable differences for long $\tw$, at least for our highest temperatures, where the faster dynamics allows us to reach higher values of $\xi_{12}/L$. As a (very conservative) cutoff we have discarded all the $\tw$ where $|I^1-I^3|$ is larger than $20\%$ of the error bar\index{error bars}, thus obtaining a $\xi_{12}^\text{max}(T)$ below which we are sure we do not have any finite-size effects\index{finite-size effects} in our $L=160$ systems. The reader can find the values in~\reftab{selected_ximax}.

\begin{table}[h!]
\centering
\begin{tabular}{cccccccc}
\toprule
\toprule
$T$ & $0.55$ & $0.625$ & $0.7$ & $0.8$ & $0.9$ & $1.0$ & $1.1$  \\
$\xi_\mathrm{max}$ & - & - & - & - & $18.1$ & $17.3$ & $17$ \\
\bottomrule
\end{tabular}
\caption[\textbf{Cutoff values of \boldmath $\xi_{12}^\mathrm{max}(T)$.}]{\textbf{Cutoff values of \boldmath $\xi_{12}^\mathrm{max}(T)$.} Cutoff values of $\xi_{12}^\mathrm{max}(T)$ below which we are sure that no finite-size effects\index{finite-size effects} are present in our $L=160$ lattices. For $T<0.9$ the growth of $\xi_{12}$ is very slow and we never reach the cutoff value.}
\labtab{selected_ximax}
\end{table}

As a final check that our data is not affected by finite-size effects\index{finite-size effects}, we have compared our $\xi_{12}(T,\tw)$ with that of~\cite{fernandez:15}. This reference considers shorter simulations but with $L=256$ and $50$ samples\index{sample}. As shown in~\reffig{xi256}, the $L=160$ and $L=256$ data coincide even  beyond our cutoff $\xi_{12}^\text{max}$.

\begin{figure}[h!]
\includegraphics[width=0.7\linewidth]{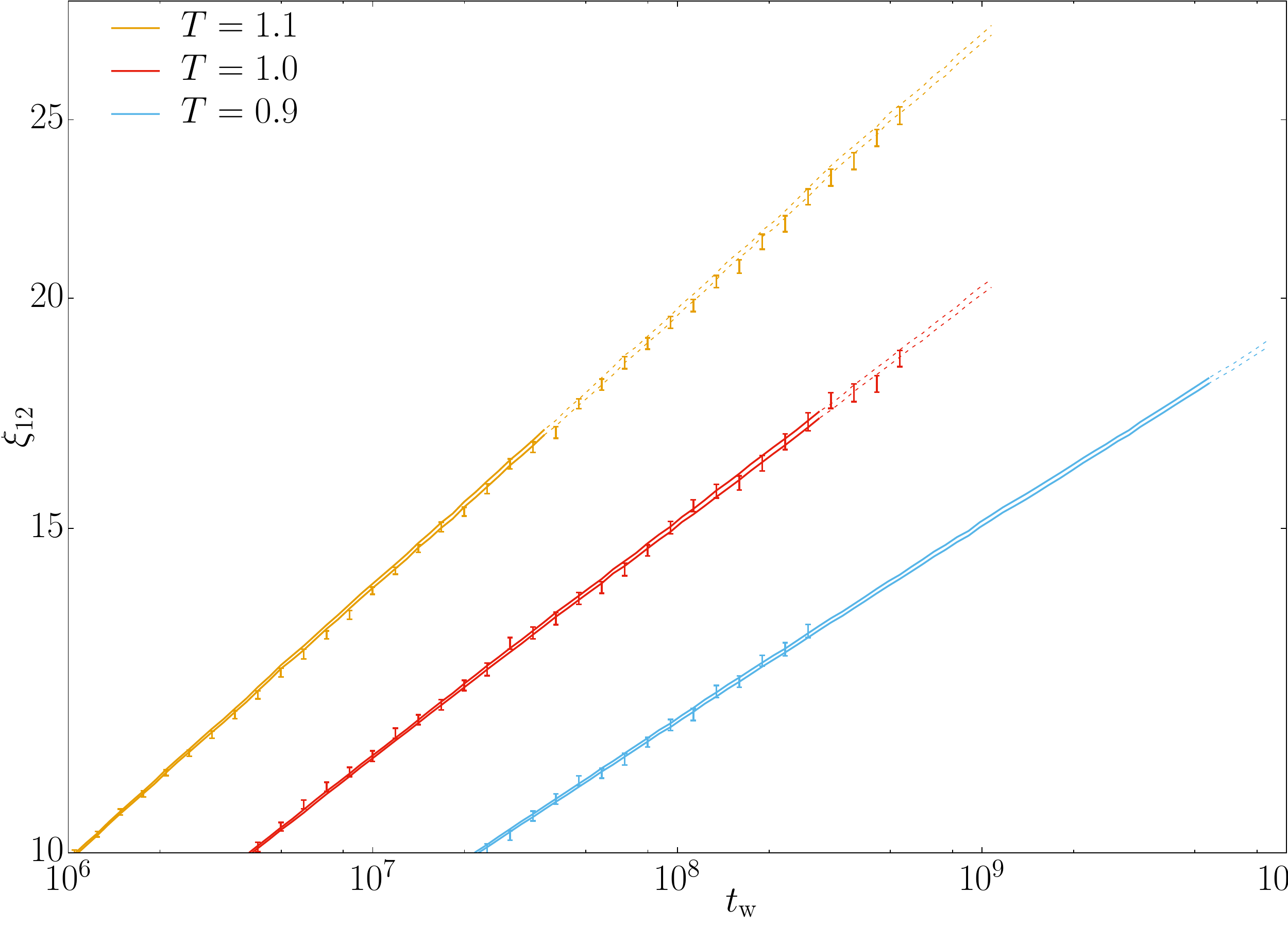}
\caption[\textbf{Comparision of \boldmath $\xi_{12}$.}]{\textbf{Comparision of \boldmath $\xi_{12}$.} Comparison between the $\xi_{12}(T,\tw)$ computed in $L=160$ lattices with $N_\text{S}=16$ samples\index{sample} and $N_\text{R}=256$ replicas\index{replica}, and that of~\cite{fernandez:15} ($L=256$; $N_\text{S}=50$; $N_\text{R}=4$  for $T=0.9,1.0$ and $N_\text{R}=8$ for $T=1.1$). For the $L=160$ simulations, we plot with two parallel lines the error interval for $\xi_{12}(T,\tw)$. Only the \tw range depicted with  continuous lines is used in the paper, the  extension with dashed lines represents the discarded times with $\xi_{12}>\xi_{12}^\text{max}$ (see~\reftab{selected_ximax}). These curves were generated with the $I^3_k$ estimator for the  integrals~\refeq{I3}. The values from the $L=256$ simulations are plotted with conventional error bars\index{error bars}. Notice that both curves are compatible even beyond this cutoff.}
\labfig{xi256}
\end{figure}

\section{Josephson length} \labsec{Josephson_length}
The Josephson length $\ell_J$ is expected to grow as $\ell_J(T) \sim (\Tc - T)^{-\nu}$ with $\nu = 2.56(4)$~\cite{janus:13} for temperatures close to $\Tc$. Scaling corrections are expected for lower temperatures
\begin{equation}
\ell_J(T) = (\Tc - T)^{-\nu} \left[ a_0 + a_1(\Tc-T)^{\nu} + a_2(\Tc-T)^{\omega \nu}   \right] \, , \labeq{josephson_length}
\end{equation}
where $a_0$, $a_1$ and $a_2$ are coefficients chosen to perform the best collapse in \reffig{vartheta_collapse} aging\index{aging} and $\omega=1.12(10)$~\cite{janus:13}.

Assuming $\xi(T,\tw) \gg \ell_J(T)$, the crossover from the fixed point at $T=\Tc$ and the fixed point at $T=0$ is affecting our basic quantity $C_4(T,r,\tw)$ in the following way
\begin{equation}
C_4(T,r,\tw) \sim
\begin{cases}
  \displaystyle \dfrac{1}{r^{D-2+\eta}}\,,  & r\ll \ell_\text{J}(T)\,, \\[3mm]
  \displaystyle \dfrac{\ell_\text{J}^\vartheta}{\ell_\text{J}^{D-2+\eta}} \dfrac{1}{r^\vartheta} f(r/\xi)\,, & r \gg \ell_\text{J}(T) \,.
\end{cases} \labeq{C4_josephson}
\end{equation}
The prefactor $\ell_J^{\vartheta}/\ell_\text{J}^{D-2+\eta}$ is fixed by the condition that the two asymptotic expansions in $r$ connect smoothly at $r=\ell_J$.

From this expression, we arrive at an asymptotic expansion for the $I_k$ integrals
\begin{equation}
I_k = \int_0^{\infty} r^k C_4(T,r,\tw) \dd r = \dfrac{F_k}{\ell_\text{J}^{D-2+\eta}} \left(\dfrac{\xi}{\ell_\text{J}}\right)^{k+1-\vartheta} \left[ 1 + a_k\left(\dfrac{\xi}{\ell_\text{J}}\right)^{k+1-\vartheta}+\ldots\right]\,, \labeq{Ik_expansion}
\end{equation}
where $F_k$ and $a_k$ are amplitudes.

Finally, we need to eliminate the unknown $\xi$ in favor of the computable $\xi_{12}$,
\begin{equation}
\xi_{12}(T,\xi) = \dfrac{F_2}{F_1} \xi \left[ 1+ a_1'\left(\dfrac{\xi}{\ell_\text{J}}\right)^{2-\vartheta} + a_2\left(\dfrac{\xi}{\ell_\text{J}}\right)^{3-\vartheta}+\ldots\right]\,,
\end{equation}
where $a_1'$ considers contributions both from the numerator ($-a_1$) and  from the denominator.  The easiest way to obtain $\vartheta$ is to study the evolution of $\log I_2$ as a function of $\log \xi$. However, we have to settle for using $\log\xi_{12}$ as independent variable (see~\reffig{I2xi}).

We can define an effective $\vartheta(T,\xi_{12})$ as
\begin{equation}
\vartheta(T,\xi_{12}) = 3 - \dfrac{\dd \log I_2(T,\xi_{12})}{\dd \log \xi_{12}} = \vartheta + b_2 \left(\dfrac{\xi_{12}}{\ell_\text{J}}\right)^{\vartheta-2}+b_3 \left(\dfrac{\xi_{12}}{\ell_\text{J}}\right)^{\vartheta-3}+\ldots\, . \labeq{fit}
\end{equation}
To estimate this derivative for a given $\xi_{12}^*$, we  fit $\log I_2$ to a quadratic polynomial in $\log \xi_{12}$  in a $[0.75\xi_{12}^*,1.25\xi_{12}^*]$ window. We then take the derivative of this polynomial at $\xi^*$. The procedure, as well as the wiggles in the resulting values of $\vartheta$ due to the extreme data correlation (see~\reffig{replicon_sin_reescalar}), may remind the reader of Fig.~1 in~\cite{janus:08b}.

We have computed a fit to the first two terms in~\refeq{fit} in the range $0\leq \ell_\text{J}/\xi_{12}\leq 0.33$, resulting in the value of $\vartheta\approx 0.30$ reported in the main text.

The previous analysis solves the problem of the crossover between the $T=\Tc$ and $T=0$ fixed points. However, in the framework of the droplet\index{droplet!picture} picture, one would also need to consider corrections to scaling at the $T=0$ fixed point. This is precisely what the droplet\index{droplet!picture} fit in the main text to $\vartheta(x) \simeq Cx^\zeta$ does.

\begin{figure}
\includegraphics[width=0.8\linewidth]{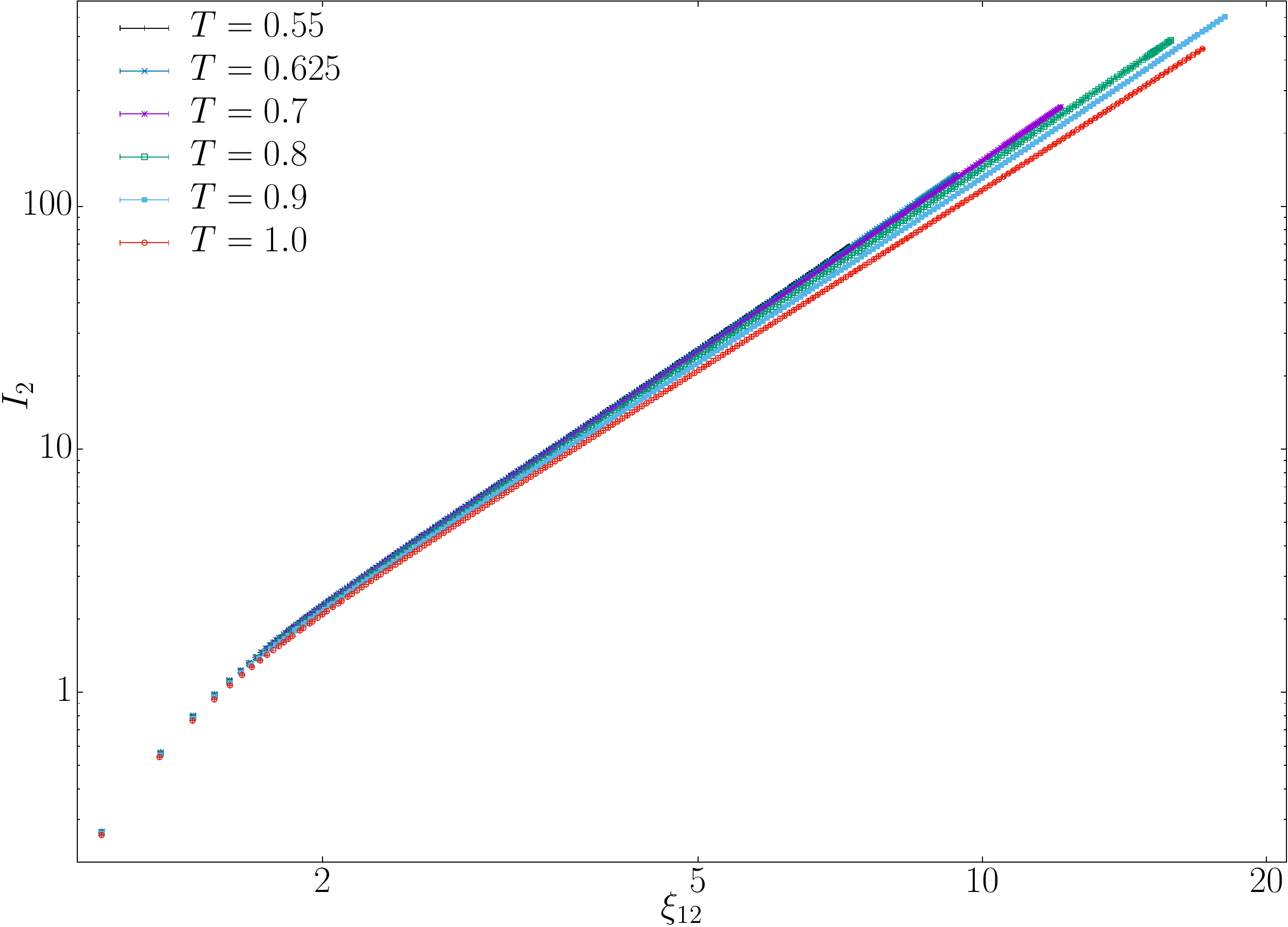}
\caption[\textbf{Integral \boldmath $I_2$ as a function of $\xi_{12}$ in a logarithmic scale, for all our $T<\Tc$ temperatures.}]{\textbf{Integral \boldmath $I_2$ as a function of $\xi_{12}$ in a logarithmic scale, for all our $T<\Tc$ temperatures.} We use the numerical derivative of this curve to compute the replicon\index{replicon} exponent $\vartheta$.}
\labfig{I2xi}
\end{figure}

\begin{figure}
\includegraphics[width=0.8\linewidth]{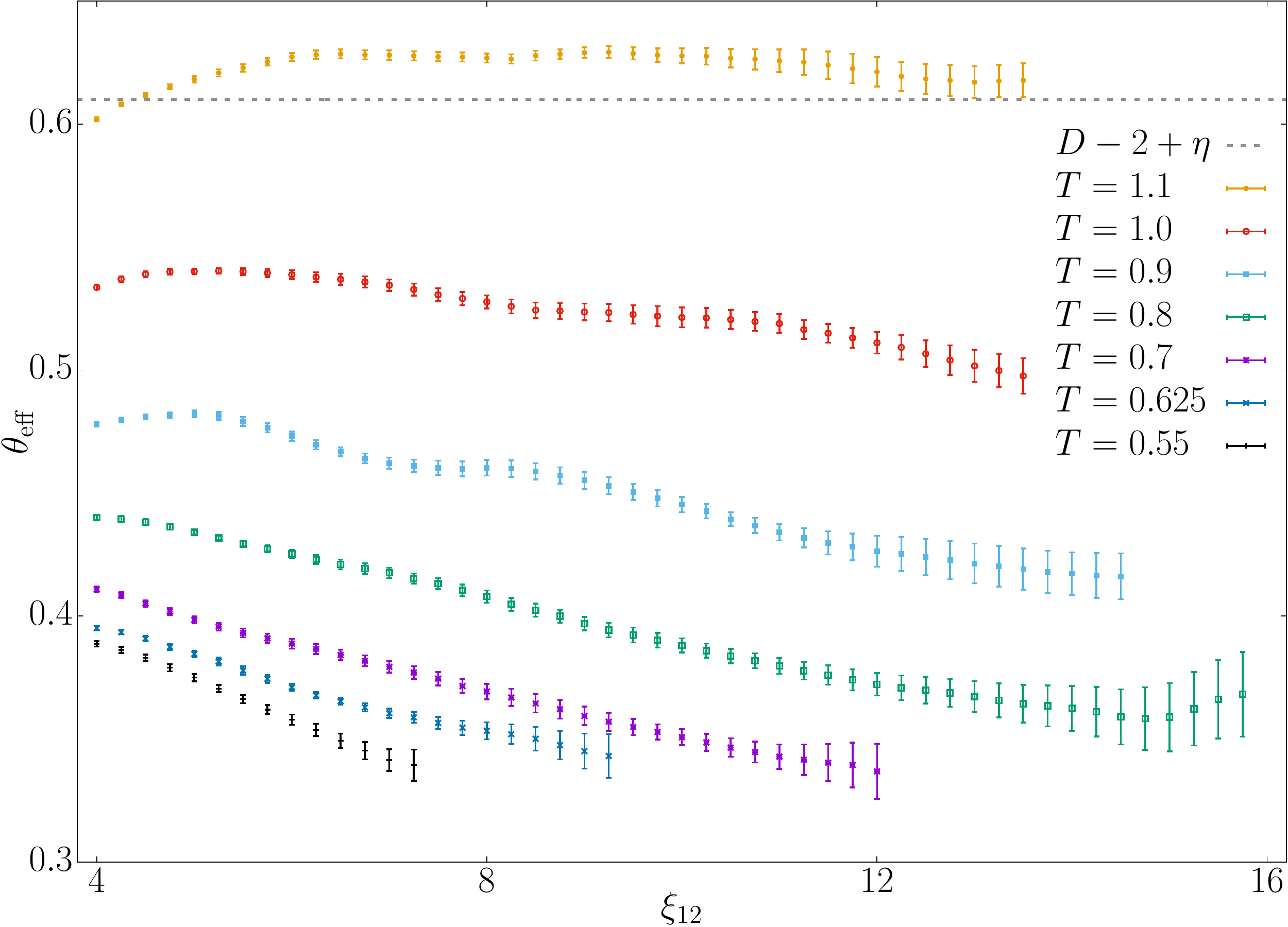}
\caption[\textbf{The replicon \boldmath $\vartheta(T,\xi_{12})$.}]{\textbf{The replicon\index{replicon} \boldmath $\vartheta(T,\xi_{12})$.} Value of the replicon\index{replicon} exponent $\vartheta(T,\xi_{12})$ computed from a numerical derivative of $\log I_2$ as a function of $\log \xi_{12}$, nicely illustrating the crossover between the $T=\Tc$ and  $T=0$ fixed points.}
\labfig{replicon_sin_reescalar}
\end{figure}

\section{Parameter choices in our fits}\labsec{parameters_aging}
We will discuss separately the choice of $\xi_{\mathrm{min}}$ for different
temperatures and the choice of the value of $\omega$.

\subsection[Selection of $\xi_{12}^{\mathrm{min}}$ for each temperature]{Selection of \boldmath $\xi_{12}^{\mathrm{min}}$ for each temperature}
We have reported fits of our data to three different functional forms

\begin{align}
\log \tw&= a_0(T) + a_1(T) \log \xi_{12} + a_2(T) \log^2 \xi_{12},\\
\log t_\mathrm{w} &= C_1(T) + z_\infty(T) \log \xi_{12} + C_2(T) \xi_{12}^{-\omega} \, , \labeq{RSB}\\
\log t_\mathrm{w} &= D_1(T) + z_\mathrm{c} \log \xi_{12} + D_2(T) \xi_{12}^{\psi} \, .
\labeq{Bouchaud}
\end{align}

In these fits we have used $z_\text{c}=6.69$ and $\omega=0.35$ ($T<\Tc$), $\omega=1.12$ ($T=\Tc$), as discussed in the~\refch{aging_rate}. Full results for the fits to~\refeq{RSB} and~\refeq{Bouchaud} can be seen in~\reftab{RSB_omega_0.35} and~\reftab{Saclay_6.69}, for different fitting ranges. We include for both cases the extrapolated values of $z(T,\xi)$ for the experimental scale (as explained in the~\refch{aging_rate} we use both $\xi_{12}=38$ and $\xi_{12}=76$) and for~\refeq{RSB} also the value of the $\xi\to\infty$ aging\index{aging!rate} rate $z_\infty$.

In order to make the choice of the fitting range, we have followed two criteria. Firstly we require the parameters of the fit to be stable inside the error when we increase $\xi_{12}^\mathrm{min}$. Secondly, we impose $\xi_\mathrm{min}$ to be monotonically increasing in $T$ (with the exception of \Tc, which has different behavior).~\reftab{selected_ximin} shows our final choices for $\xi_{12}^\text{min}(T)$, which is the same for all three fits.

\begin{table}[h!]
\centering
\begin{tabular}{cccccccc}
\toprule
\toprule
$T$ & $0.55$ & $0.625$ & $0.7$ & $0.8$ & $0.9$ & $1.0$ & $1.1$  \\
$\xi_{12}^\mathrm{min}$ & $4$ & $5$ & $6$ & $8$ & $8$ & $9$ & $5$ \\
\bottomrule
\end{tabular}
\caption{Values of $\xi_{12}^\text{min}(T)$ determining the common fitting range $\xi_{12}\geq\xi_{12}^\text{min}$ for our three different fits of $\log\tw$ as a function of $\log \xi_{12}$.}
\labtab{selected_ximin}
\end{table}

\subsection[Selection of $\omega$]{Selection of \boldmath $\omega$}
For our most important result, namely the extrapolation of the aging\index{aging!rate} rate to the experimental scale of $\xi_{12}=38,76$, we have repeated our fits with our upper and lower bounds for $\omega=\vartheta(\xi_\text{films})$ (\gls{RSB}\index{replica!symmetry breaking (RSB)} and droplet\index{droplet!picture} extrapolations, respectively).  The results are completely compatible, as we can see in~\reftab{omega_xi38}.

\begin{table}[h!]
\begin{tabular}{lccccc}
\toprule
\toprule
& \multicolumn{2}{c}{$z(T,\xi_{12}=38)$} & & \multicolumn{2}{c}{$z(T,\xi_{12}=76)$} \\ 
       & $\omega = 0.35$    & \multicolumn{1}{c}{$\omega = 0.28$} & 
       & $\omega = 0.35$    & \multicolumn{1}{c}{$\omega = 0.25$} \\ 
\hline
$T=0.55$  & 19.80(20) & 20.08(22) &  & 20.75(24) & 21.41(27)                    \\
$T=0.625$ & 16.90(19)  & 17.07(20)& & 17.69(24)  & 18.13(27)                   \\
$T=0.7$   & 14.81(15) & 14.93(16)&  & 15.54(19) & 15.87(21)                   \\
$T=0.8$   & 12.73(22) & 12.81(23)&  & 13.47(30) & 13.71(32)                    \\
$T=0.9$   & 10.55(25) & 10.61(26)&  & 11.11(34) & 11.28(37)                   \\
$T=1.0$   & 8.63(32) & 8.68(33)  &  & 8.98(44) & 9.02(42)                  \\ 
\bottomrule
\end{tabular}
\caption[\textbf{Estimations of the aging rate.}]{\textbf{Estimations of the aging rate.} Comparison of our estimates of the experimental aging\index{aging!rate} rate $z(T,\xi_{12}=\xi_\text{films})$ for $\xi_\text{films}=38$ and $\xi_\text{films}=76$ using our lower and upper bounds for $\omega=\vartheta(\xi_\text{films})$. The choice of $\omega$ is immaterial, since even in the worst case (lowest temperatures for $\xi_\text{films}=76$) there is  only a two-sigma difference.}
\labtab{omega_xi38}
\end{table}

\begin{table}[h!]
\resizebox{\textwidth}{!}{\begin{tabular}{lcccccccc}
\toprule
\toprule
& & $\xi_{\mathrm{min}}$= 3.5 & $\xi_{\mathrm{min}}$= 4 & $\xi_{\mathrm{min}}$= 5 & $\xi_{\mathrm{min}}$= 6 & $\xi_{\mathrm{min}}$= 7 & $\xi_{\mathrm{min}}$= 8 & $\xi_{\mathrm{min}}$= 9 \\ \hline
\multirow{4}{*}{{$T=0.55$}}&
$z_\infty$ & 23.61(28)   & \bfseries 24.22(40)   & 25.30(86)   & 22.9(31)   & & &   \\ 
&$z (\xi\!=\!38)$ & 19.49(15)   &  \bfseries 19.80(20)   & 20.32(41)  & 19.2(14)   & & &   \\
&$z (\xi\!=\!76)$ & 20.38(18)   &  \bfseries 20.75(24)   & 21.39(51)   & 20.0(18)   & & &   \\
&$\chi^2$/dof &  $40(17)/133$ &  \bfseries 10.2(54)/111 & 3.0(12)/73 & 1.71(76)/40  &  &  & \\ 
\hline
\multirow{4}{*}{{$T=0.625$}}&
$z_\infty$ & 19.85(17)   & 20.26(23)   &  \bfseries 20.60(41)   & 20.16(84)   &  &  &  \\ 
&$z (\xi\!=\!38)$ & 16.538(91)   & 16.74(12)   &  \bfseries 16.90(19)   & 16.71(37)   &  &  &  \\
&$z (\xi\!=\!76)$ & 17.25(11)   & 17.50(14)   &  \bfseries 17.69(24)   & 17.45(47)   &  &  &  \\
&$\chi^2$/dof & 81(34)/167
 &  18(10)/147
 &  \bfseries  8.3(21)/114
&  5.1(19)/86
&  &  &  \\ \hline
\multirow{4}{*}{{$T=0.7$}}&
$z_\infty$ & 17.04(18)   & 17.23(21)   & 17.61(27)   &  \bfseries 18.23(35)   & 18.63(62)   &  &  \\ 
&$z (\xi\!=\!38)$ & 14.295(88)   & 14.38(11)   & 14.55(13)   &  \bfseries 14.81(15)   & 14.96(25)&  &  \\
&$z (\xi\!=\!76)$ & 14.89(11)   & 15.00(13)   & 15.21(16)   &  \bfseries 15.54(19)   & 15.75(32)   &  &  \\ 
&$\chi^2$/dof & 116(40)/190 & 66(36)/173 & 33(24)/144 &  \bfseries 9.3(84)/119 & 4.9(21)/98
 &  &  \\ \hline
\multirow{4}{*}{{$T=0.8$}}&
$z_\infty$ & 13.76(15)   & 14.06(19) & 14.53(26)   & 15.19(35)   & 15.68(42)   &  \bfseries 16.18(58)   & 16.55(78)   \\ 
& $z (\xi\!=\!38)$ & 11.787(73)   & 11.921(89)   & 12.11(12)   & 12.37(15)   & 12.55(17)   & \bfseries  12.73(22)   & 12.85(28)   \\
& $z (\xi\!=\!76)$ & 12.211(93)   & 12.38(11)   & 12.63(15)   & 12.98(19)   & 13.23(23)   & \bfseries  13.47(30)   & 13.65(39)   \\ 
& $\chi^2$/dof & 351(104)/185 & 188(72)/170& 93(41)/146& 27(16)/125& 12.4(82)/107
& \bfseries  6.2(31)/91
& 4.9(21)/77
\\ \hline
\multirow{4}{*}{{$T=0.9$}}&
$z_\infty$ & 11.00(13)  & 11.29(18)   & 11.54(24)   & 11.80(31)   & 12.55(41)   & \bfseries  13.16(68)   & 12.3(13)   \\ 
& $z (\xi\!=\!38)$ & 9.748(65)   & 9.883(93)   & 9.98(11)   & 10.08(13)   & 10.34(16)   & \bfseries  10.55(25)   & 10.33(41)   \\
& $z (\xi\!=\!76)$ & 10.017(82)   & 10.18(11)   & 10.32(14)   & 10.45(17)   & 10.82(21)   &  \bfseries 11.11(34)   & 10.80(60)   \\ 
& $\chi^2$/dof & 310(150)/165 & 129(64)/152 & 79(44)/131 & 63(35)/113 & 22(13)/98
 & \bfseries  5.9(21)/84
 & 6.4(79)/72
 \\ \hline
\multirow{4}{*}{{$T=1.0$}}&
$z_\infty$ & 8.60(11)   & 8.69(15)   & 8.83(20)   & 8.97(53)   & 9.29(45)   & 9.36(46)   &  \bfseries 10.28(89)   \\
& $z (\xi\!=\!38)$ & 8.041(59)   & 8.080(73)   & 8.132(93)   & 8.21(22)   & 8.27(18)   & 8.34(17)   &  \bfseries 8.63(32)   \\ 
& $z (\xi\!=\!76)$ & 8.162(74) & 8.210(86) & 8.28(11) & 8.38(29) & 8.46(24) & 8.57(24) & \bfseries  8.98(44)   \\ 
& $\chi^2$/dof & 43(30)/137
 & 27(18)/126
 & 16(13)/107
 & 12(25)/91
 & 10.4(91)/78
 & 8.4(60)/66
 &  \bfseries 2.9(21)/55
\\ \hline
\multirow{4}{*}{{$T=1.1$}}&
$z_\infty$ & 6.672(44) & 6.671(41) & \bfseries  6.689(63) & 6.751(84) & 6.80(12) & 7.00(18) & 7.02(21)  \\
& $z (\xi\!=\!38)$ & 6.682(32) & 6.673(41) & \bfseries  6.694(50) & 6.732(68) & 6.77(10) & 6.92(14) & 6.94(16)   \\
& $z (\xi\!=\!76)$ & 6.677(33) & 6.671(41) & \bfseries  6.691(54) & 6.742(72) & 6.79(11) & 6.96(16) & 6.99(16)   \\
& $\chi^2$/dof & 32(19)/119
 & 31(20)/109
 &  \bfseries 26(16)/92
 & 19(10)/78
 & 16.8(74)/66
 & 5.9(20)/55
 & 6.3(27)/46 \\ 
 \bottomrule
\end{tabular}}
\caption[\textbf{Parameters for the convergent ansatz.}]{\textbf{Parameters for the convergent ansatz.}Parameters of the fits to~\refeq{RSB} for different fitting ranges $\xi_{12}\geq\xi_{12}^\text{min}$.  We use $\omega = 0.35$ ($\omega = 1.12$ for $T = T_\mathrm{c}$). The fitting range that we choose for our final values is highlighted in boldface.}
\labtab{RSB_omega_0.35}
\end{table}

\begin{table}[h!]
\resizebox{\textwidth}{!}{\begin{tabular}{lcccccccc}
\toprule
\toprule
& & $\xi_{\mathrm{min}}$= 3.5 & $\xi_{\mathrm{min}}$= 4 & $\xi_{\mathrm{min}}$= 5 & $\xi_{\mathrm{min}}$= 6 & $\xi_{\mathrm{min}}$= 7 & $\xi_{\mathrm{min}}$= 8 & $\xi_{\mathrm{min}}$= 9 \\ \hline

\multirow{4}{*}{{$T=0.55$}}& 
$z(\xi\!=\!38)$ & 24.07(41)   & \bfseries24.25(55)  & 24.6(11)& 24.7(81)  &  &  & \\
&$z(\xi\!=\!76)$ & 28.86(69)   & 2\bfseries9.18(95)  & 29.9(19)   & 30(15)  &  &  & \\
& $G(T)$ & 13.78(65)   & \bfseries13.45(92) & 12.8(18)  & 18(13) &  &  &  \\ 
& $\psi$ & 0.3512(92)   & \bfseries0.355(21)   & 0.372(33)   & 0.29(24)  &  &  & \\
& $\chi^2$/dof & 13.3(47)/133
 & \bfseries6.8(20)/111
 & 3.2(15)/73
 & 1.7(27)/40
 &  &  &  \\ \hline
\multirow{4}{*}{{$T=0.625$}}& 
 $z(\xi\!=\!38)$ & 19.73(22)   & 19.72(28)   & \bfseries19.36(45)   & 18.53(77)  &  &  & \\
& $z(\xi\!=\!76)$ & 23.33(38)   & 23.31(49)   & \bfseries22.66(79)   & 21.1(13)  &  &  & \\
& $G(T)$ & 10.36(37)   & 10.39(52)   & \bfseries11.3(11)   & 14.0(30)  &  &  & \\
& $\psi$ & 0.354(14)   & 0.352(12)   & \bfseries0.334(21)   & 0.290(39)  &  &  &  \\
& $\chi^2$/dof & 19(10)/167
 & 15(10)/147
 & \bfseries8.5(33)/114
 & 4.5(14)/86
 &  &  &  \\ \hline
\multirow{4}{*}{{$T=0.7$}}& 
$z(\xi\!=\!38)$ & 16.58(22)   & 16.44(23)   & 16.35(27)   &\bfseries 16.51(32)   & 16.55(52) &  &   \\ 
& $z(\xi\!=\!76)$ & 19.40(37)   & 19.14(40)   & 18.98(47)   &\bfseries 19.29(58)   & 19.4(10)  &  &  \\
& $G(T)$ & 7.32(34)   & 7.63(43)   & 7.84(59)   &\bfseries 7.41(75)   & 7.3(14)   &  & \\ 
& $\psi$ & 0.364(13)   & 0.354(12)   & 0.354(24)   &\bfseries 0.358(23)   & 0.360(39)   &  & \\
& $\chi^2$/dof & 49(38)/190
 & 28(20)/173
 & 10.5(83)/144
 & \bfseries6.3(31)/119
 & 5.7(29)/98
 &  &  \\ \hline
\multirow{4}{*}{{$T=0.8$}}& 
$z(\xi\!=\!38)$ & 13.37(18)   & 13.39(21) & 13.45(25) & 13.68(31) & 13.80(35) & \bfseries13.94(45) & 14.1(17)   \\
& $z(\xi\!=\!76)$ & 15.44(33) & 15.48(37) & 15.60(46) & 16.06(59)   & 16.31(68)   & \bfseries16.59(93)   & 17.1(38)   \\ 
& $G(T)$ & 4.16(25)   & 4.13(29)   & 4.01(38)   & 3.57(43)   & 3.36(48)   & \bfseries3.13(66)   & 3.0(19)   \\
& $\psi$ & 0.392(13) & 0.390(19) & 0.395(18)   & 0.421(27) & 0.443(31) & \bfseries0.447(50) & 0.46(18) \\ 
& $\chi^2$/dof & 31(19)/185
 & 29(19)/170
 & 22(17)/146
 & 10.0(60)/125
 & 7.5(34)/107
 & \bfseries5.5(21)/91
 & 5(11)/77
 \\ \hline
\multirow{4}{*}{{$T=0.9$}}& 
$z(\xi\!=\!38)$ & 10.76(17) & 10.86(21) & 10.82(24) & 10.86(27) & 11.23(35) & \bfseries11.49(54) & 11.13(56) \\
& $z(\xi\!=\!76)$ & 12.12(30) & 12.31(39) & 12.24(45) & 12.31(53) & 13.12(73) & \bfseries13.7(12)   & 12.9(12)  \\
& $G(T)$ & 2.15(19) & 2.01(23) & 2.07(31) & 2.00(39) & 1.52(31)   & \bfseries1.18(45)   & 1.67(77)  \\
& $\psi$ & 0.417(23)  & 0.430(34) & 0.431(28) & 0.427(41) & 0.490(51) & \bfseries0.546(88) & 0.47(10) \\
& $\chi^2$/dof & 68(44)/165
 & 46(25)/152
 & 41(25)/131
 & 38(24)/113
 & 17(10)/98
 & \bfseries8.7(45)/84
 & 4.8(25)/72
 \\ \hline
\multirow{4}{*}{{$T=1.0$}}& 
$z(\xi\!=\!38)$ & 8.53(15) & 8.54(18) & 8.55(20) & 8.56(30) & 8.59(60) & 8.74(72) &\bfseries 9.22(18)  \\
& $z(\xi\!=\!76)$ & 9.19(27) & 9.20(33) & 9.22(39) & 9.25(58) & 9.3(12) & 9.7(16) &\bfseries 10.9(45)  \\
& $G(T)$ & 0.85(14) & 0.84(18) & 0.83(22) & 0.79(40) & 0.7(10)   & 0.5(10) &\bfseries 1.4(19)  \\
& $\psi$ & 0.440(36)   & 0.441(51)   & 0.444(64) & 0.45(10) & 0.49(25) & 0.55(30)   & \bfseries0.34(71) \\ 
& $\chi^2$/dof & 12.6(95)/137
 & 12.1(90)/126
 & 10.0(87)/107
 & 9.3(83)/91
 & 8.2(97)/78
 & 07(11)/66
 & \bfseries11(11)/55
 \\ \hline
\multirow{4}{*}{{$T=1.1$}}& 
$z(\xi\!=\!38)$ & 6.684(12)   & 6.682(14)   &\bfseries 6.684(11)   & 6.672(31)   & 6.683(32)& 6.694(31)& 6.712(41)   \\ 
& $z(\xi\!=\!76)$ & 6.684(13)   & 6.681(11)   & \bfseries 6.682(14)   & 6.674(41)   & 6.684(41)& 6.692(31)& 6.721(42)  \\ 
& $G(T)$ & 1.9(10)   & 1.71(91) & \bfseries0.4(27) & 0.02(64)  & 0.0(26)   & 1.5(26)   & 1.2(10)   \\ 
& $\psi$ & -0.0030(49)   & 0.0037(68)   &\bfseries 0.03(15)   & 0.29(42)   & 0.37(55)   & 0.002(16)   & 0.023(51)   \\
& $\chi^2$/dof & 34(20)/119
 & 33(19)/109
 &\bfseries 27(18)/92
 & 25(18)/78
 & 23(15)/66
 & 21(14)/55
 & 11.0(26)/46
 \\ 
 \bottomrule
\end{tabular}}
\caption[\textbf{Parameters for the divergent ansatz.}]{\textbf{Parameters for the divergent ansatz.}Parameters of the fits to ~\refeq{Bouchaud} for different fitting ranges $\xi_{12}\geq\xi_{12}^\text{min}$.  We use $z_\text{c} = 6.69$. The fitting range that we choose for our final values is highlighted in boldface.}
\labtab{Saclay_6.69}
\end{table}

\setchapterstyle{lines}
\chapter{Parallel Tempering. Technical details.} \labch{AP_PT}

\section{On the selection of relevant parameters of the simulation} \labsec{selection_parameters}
In this section, we try to clarify some of the choices made for the parameters of the simulation that can be obscure from the reader's point of view.

\subsection{Setup-independence of the results}
At the beginning of the chapter, we assured that our results do not depend on the particular setup of our simulations, motivated by the metastate\index{metastate} study exposed in~\refch{metastate}. Now, it is time to quantitatively justify our statement.

If we focus on the relevant observable from the dynamical point of view, $\tintf=\tau$ we can check in~\reffig{selection_samples_dinamica} that the internal disorder\index{disorder} is affecting almost nothing the probability distribution of our observable. The average over the outer disorder\index{disorder} (which we can call the metastate\index{metastate} average, as introduced in~\refch{metastate}) dramatically reduces the fluctuations due to the internal disorder\index{disorder}.

The same discussion is straightforwardly applicable to the chaos integral in~\reffig{selection_samples_estatica}.

\begin{figure}[h!]
\centering
\includegraphics[width=0.7\columnwidth]{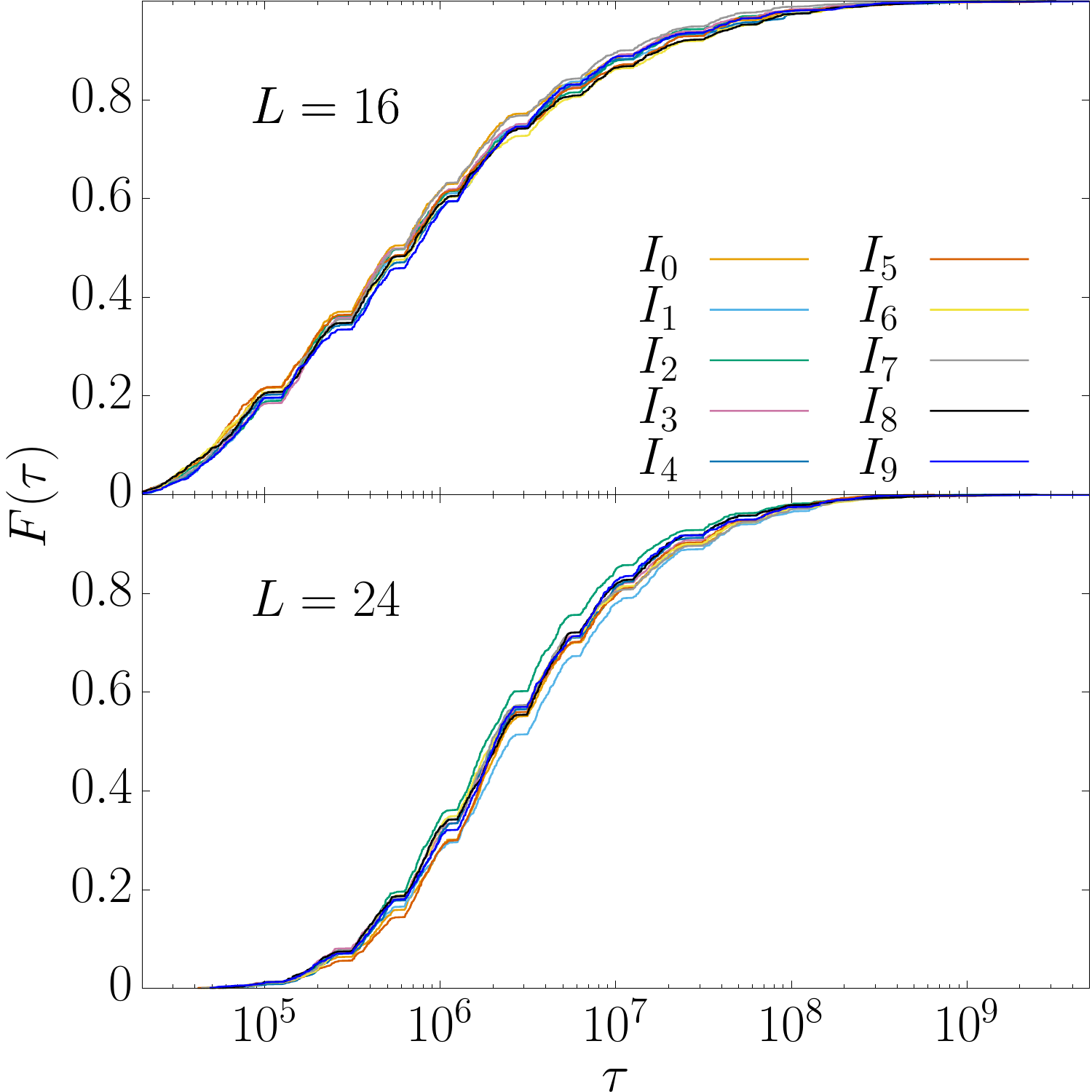}
\caption[\textbf{The integrated time is independent on the simulation setup.}]{\textbf{The integrated time is independent on the simulation setup.} Empirical probability distribution function of $\tau$ represented for the $10$ inner samples\index{sample} separately.  $L=16$ case (top) and $L=24$ case (bottom). Averaging over the metastate\index{metastate} (i.e. the outer samples\index{sample}) with fixed inner couplings\index{couplings} strongly reduces the fluctuations between the inner samples\index{sample}.}
\labfig{selection_samples_dinamica}
\end{figure}

\begin{figure}[h!]
\centering
\includegraphics[width=0.7\columnwidth]{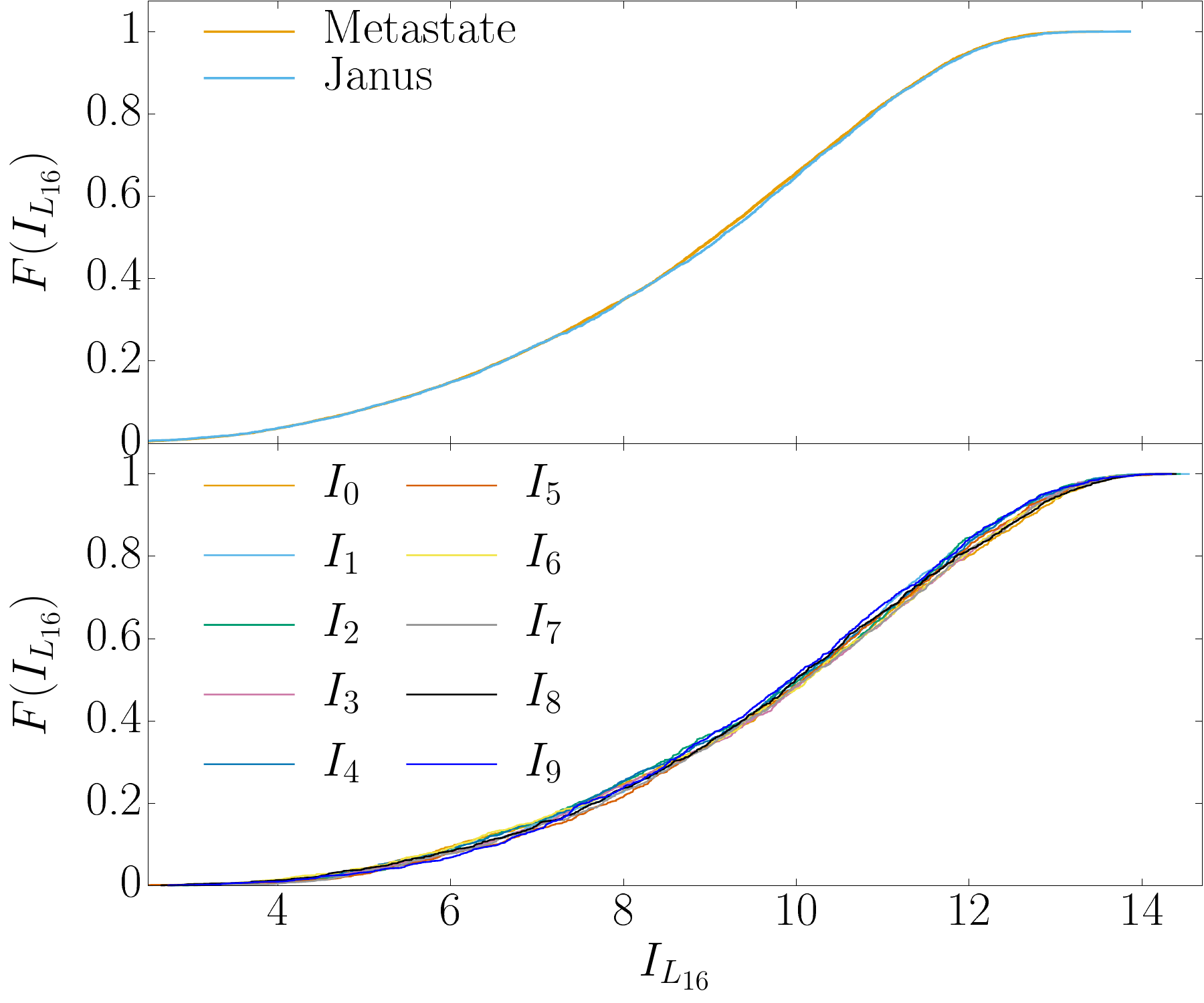}
\caption[\textbf{The chaos integral is independent on the simulation setup.}]{\textbf{The chaos integral is independent on the simulation setup.} Empirical probability distribution function of the integrated chaotic parameter. \textbf{Top} We compare the distribution (labeled as "Metastate"\index{metastate}) obtained with our particular choice of samples\index{sample} with the distribution obtained from $4000$ fully independent samples\index{sample} (data from Janus\index{Janus}). \textbf{Bottom:} Distributions obtained for the $10$ inner samples\index{sample} plotted separately. Averaging over the metastate\index{metastate} (over the outer couplings\index{couplings}) strongly reduces the fluctuations between the inner samples\index{sample}.}
\labfig{selection_samples_estatica}
\end{figure}

\subsection{Temperature-mesh's choice}
Our main results are mainly concerned about the $L=16$ and $L=24$ simulations. The reader may find it surprising that we have two simulations for the $L=16$ case, one with $\Tmin=0.479$ and the other one with $\Tmin=0.698$. The latter is obvious since is the $\Tmin$ needed to compare with the other lattice sizes in our scaling result of~\refsec{scaling_eq_chaos}. However, the choice of $\Tmin=0.479$ is not capricious. We carefully choose this $\Tmin$ with the aim of having similar $\tint$ for the most chaotic samples\index{sample} in both simulations ($L=24$, $N=24$, $\Tmin=0.698$) and ($L=16$, $N=16$, $\Tmin=0.479$). This is shown in~\reffig{comparison_dynamics_L24_L16}.

\begin{figure}[h!]
\centering
\includegraphics[width=0.7\columnwidth]{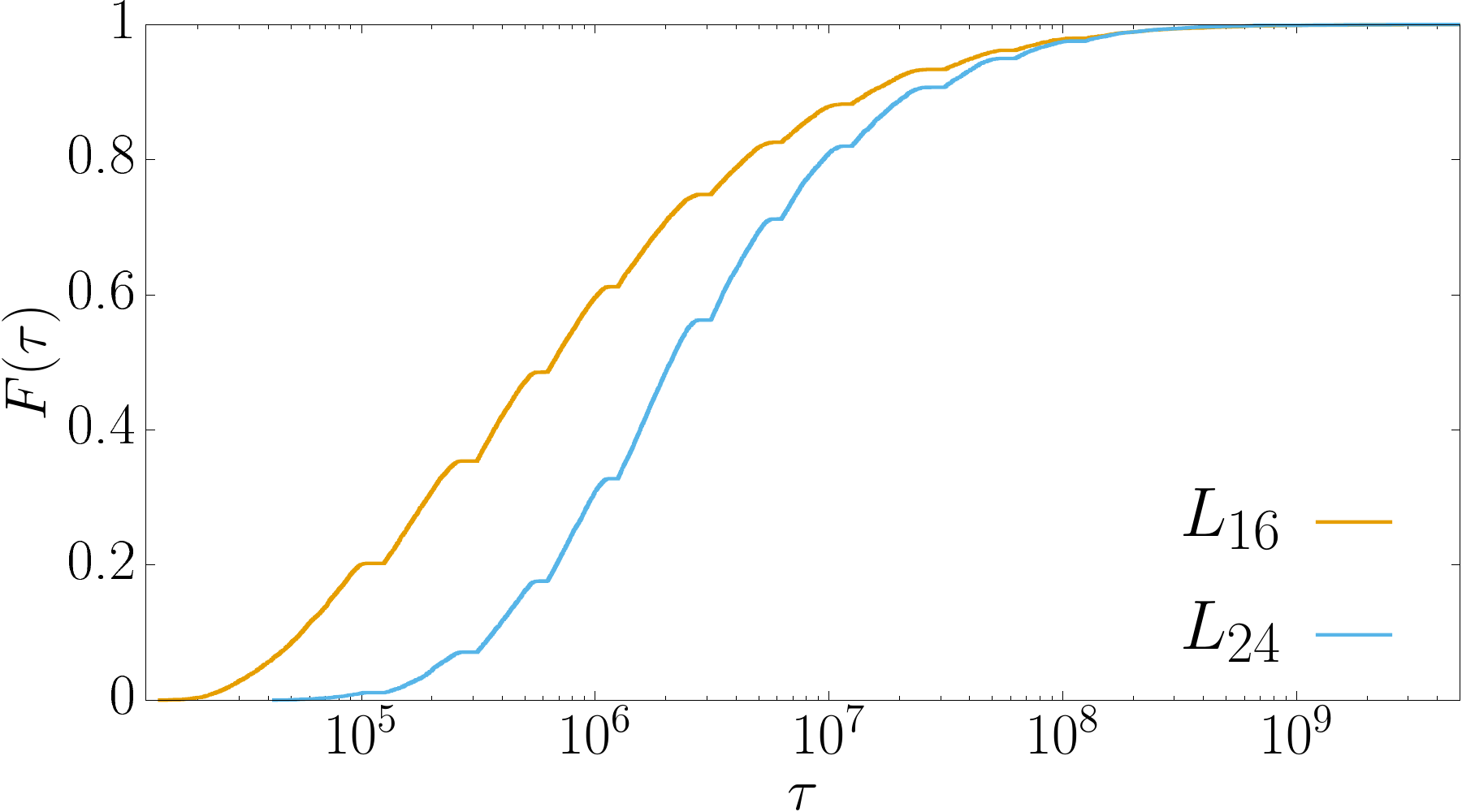}
\caption[\textbf{Empirical probability distribution function of $\mathbf{\tau}$.}]{\textbf{Empirical probability distribution function of $\mathbf{\tau}$.} Comparison of results for the simulations ($L=24$,$T_\mathrm{min}=0.698$) and ($L=16$,$T_\mathrm{min}=0.479$). Note that at the high-end of very difficult samples\index{sample}, these two simulations are similarly challenging.}
\labfig{comparison_dynamics_L24_L16}
\end{figure}

\section{Thermalizing with Parallel Tempering} \labsec{thermalizing_PT}
The reader used to deal with thermalization\index{thermalization} protocols might find surprising that we focus on the temperature-random-walk\index{random walk!temperature} to establish our particular thermalization\index{thermalization} protocol. This appendix is devoted to justify our decision (which is not pioneer and can be find in other works, see for example~\cite{janus:10}).

We have introduced in~\refsubsec{Monte_Carlo} several properties and theorems of the Markov\index{Markov chain} chains. Specifically, there exists one theorem~\cite{sokal:97} that assures, for an aperiodic\index{aperiodicity} Markov\index{Markov chain} chain that fulfills the balance\index{balance condition} condition, that the time evolution of the Markov\index{Markov chain} chain eventually reaches a stationary distribution (which is unique).

In the case of \gls{PT}, that stationary distribution appears in~\refeq{PT_eq_distribution} but we recall it here for the reader's convenience.
\begin{equation}
P_{\mathrm{eq}}(\{ \pi, \{s^{(\alpha)}\}_{\alpha=1}^N\}) = \dfrac{1}{N!}\prod_{\alpha=1}^N \dfrac{\exp \left[-\beta_{\alpha} \mathcal{H}(\{s^{\pi^{-1}(\alpha)}\})\right]}{Z_{\beta_{\alpha}}} \, , 
\end{equation}
In the previous expression, $N$ is the number of temperatures in the temperature-mesh of our \gls{PT} method, the index $\alpha$ labels a given clone and, here, $\pi(\alpha)$ is the permutation of the clone $\alpha$ at a given time and $\pi^{-1}$ is its inverse.

This equilibrium configuration\index{configuration} for the \gls{PT} is telling us two main things. On the one hand, the marginal probability for the clones permutation is uniform as the factor $1/N!$ implies. On the other hand if we focus on the marginal probability for the spins of one clone at temperature $T_{\alpha'}$, namely $\pi(\alpha) = \alpha'$, we find that the equilibrium probability at a given temperature is
\begin{equation}
P_{\mathrm{eq}}(\{\{s^{(\alpha)}\} | \pi(\alpha)=\alpha' ) = \dfrac{\exp \left[-\beta_{\alpha'} \mathcal{H}(\{s^{\alpha}\})\right]}{Z_{\beta_{\alpha'}}} \, , 
\end{equation}
which is none but the Boltzmann-Gibbs\index{Boltzmann!-Gibbs distribution} distribution.

The message is clear, when the random-walks\index{random walk} equilibrate, the Boltzmann-Gibbs\index{Boltzmann!-Gibbs distribution} distribution is reached by all the $N$ clones simultaneously.

The theorem ensuring the equilibration in the Markov\index{Markov chain} chains is fundamental because it legitimates its use to study equilibrated systems. However, that theorem only assures that we will reach, at some point, our desired equilibrium distribution, but it says nothing about the convergence to that limit. We have succinctly explained in~\refsec{time_scales_eq_chaos} the usual way to study this convergence through the computation of time autocorrelation functions\index{correlation function!time auto-} $\hat{C}_f(t)$. 

Nevertheless, the reader familiarized with those methods may be used to build that time autocorrelation functions\index{correlation function!time auto-} from spin-dependent functions instead of focusing on the temperature random-walk\index{random walk!temperature} of \gls{PT}. Actually, this assumption is customary rather than necessary. All the statements done concerning the time autocorrelation function\index{correlation function!time auto-} $\hat{C}_f(t)$ are valid from any function $f$ related to the Markovian\index{Markovian} dynamics. The reader can check that no special condition, restricting the function $f$, was imposed in~\refsec{time_scales_eq_chaos}. In our particular case, it is really convenient to study the random-walk\index{random walk!temperature} in the temperature. We expose here an example to illustrate our statements.

We shall consider an $L=24$ sample\index{sample} chosen from the simulated samples\index{sample} with $N=24$ and $T_{\min}=0.698$ (see~\refsec{numerical_simulations_eq_chaos}). Specifically, the chosen sample\index{sample} will be more difficult (i.e. will have a larger autocorrelation time) than the $90\%$ of the simulated samples\index{sample}.

Now, we consider two different \gls{PT} setups. The first one corresponds to the setup used in the simulations, namely a temperature mesh with $N=24$ temperatures and $T_{\min}=0.698$. The second one will consist only of the four lowest temperatures: $T_1=T_{\min}=0.698$, $T_2 = 0.735$, $T_3=0.771$ and $T_4=0.808$.

The $N=24$ setup is found to need $2\cdot10^9$ Metropolis sweeps to meet our thermalization\index{thermalization} criteria and we decided to run both \gls{PT} setups for that number of Metropolis sweeps. We know that the $N=24$ system will thermalize and the natural expectation for the $N=4$ system is that it will not reach our thermalization\index{thermalization} criteria. The reason for that expectation is that, while the standard simulation will spend around $1/2$ of the total time above the critical\index{critical temperature} temperature, the truncated simulation ($N=4$) will not. 

Actually, we know that, for the highest temperature in the (standard) mesh, $T=1.6$, the exponential autocorrelation time\index{autocorrelation time!exponential} for Metropolis dynamics is about $10^4$ lattice sweeps~\cite{ogielski:85}, which corresponds to $1000$ times more lattice sweeps that each clone will spend at the highest temperature ($2\cdot 10^9/24 \approx 8 \cdot 10^7$). On the contrary, for the truncated system with $N=4$ the highest temperature is still in the \gls{SG} phase\index{phase!low-temperature/spin-glass} and the time that each clone spends at the highest temperature is not enough to decorrelate the spins.

Now, we need a function $f$ depending on the spin configurations\index{configuration} in order to compare with our function $f$ depending on the temperature random-walk\index{random walk!temperature} of the \gls{PT}. Using the fact that we have already equilibrated this sample\index{sample}, we have selected randomly four equilibrium spin configurations\index{configuration} at our lowest temperature $T_{\min}=0.698$. We denote those configurations\index{configuration} as $\{ \tau_{x,a}\}$ with $a=$1, 2, 3, 4. Our four proposed functions (for each clone $\alpha$) $f_{a,\alpha}$ will be the time-dependent overlap\index{overlap} between the spin configurations\index{configuration} of the clone $\alpha$, and each of the four selected equilibrium configurations\index{configuration} $\{ \tau_{x,a} \}$
\begin{equation}
f_{a,\alpha} = q_{a,\alpha}(t) = \dfrac{1}{L^3} \sum_x \tau_{x,a} s_x^{(\alpha)}(t) \, . \labeq{spin-dependent_observable}
\end{equation}

We compute those overlaps\index{overlap} for a set of $10$ new standard simulations (i.e. $N=24$) and for the truncated simulations $N=4$. Because we are particularly interested in the thermalization\index{thermalization} process, we measure $q_{a,\alpha}(t)$ very often (every $5\cdot 10^4$ Metropolis sweeps\footnote{This suppose a total of $40000$ measure from $t=0$ to $t=2\cdot 10^9$ Metropolis sweeps.}). 

The global spin-flip symmetry of the \gls{EA}\index{Edwards-Anderson!Hamiltonian} Hamiltonian\index{Hamiltonian} implies that the equilibrium distribution for $q_{a,\alpha}$ is symmetric under overlap\index{overlap} inversions $q_{a,\alpha} \leftrightarrow -q_{a,\alpha}$. It is important to check this symmetry, since it is believed that the largest dynamical barriers are related to global spin-flips~\cite{billoire:01}.

Once we simulate both setups, we notice that the truncated simulation is not able to thermalize within the time span of our simulations. To illustrate our findings, we show~\reffig{historia_q}.

\begin{figure}[t!]
\includegraphics[width=0.8\textwidth]{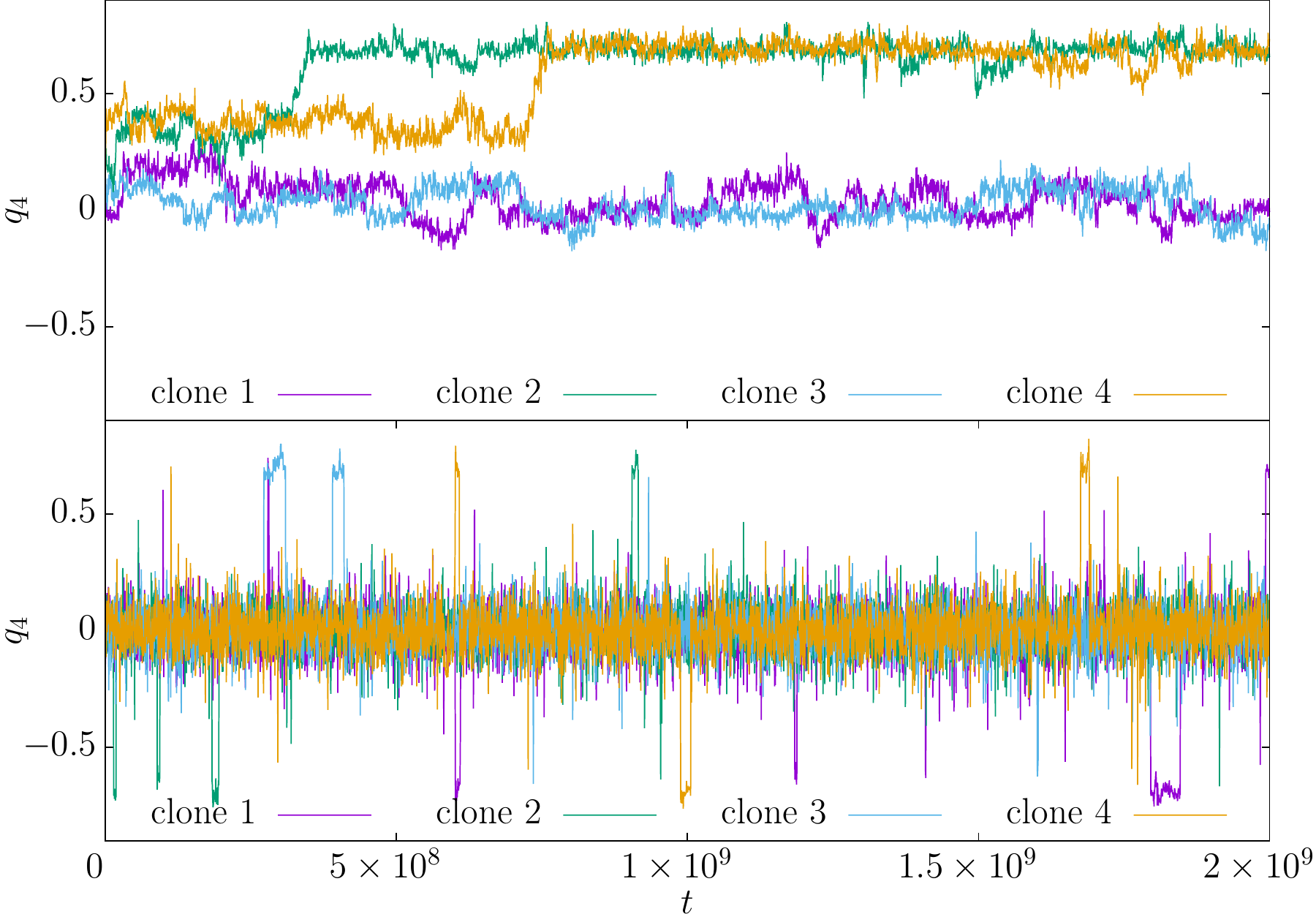}
\caption[\textbf{Overlap history. Exploring the thermalization process.}]{\textbf{Overlap\index{overlap} history. Exploring the thermalization\index{thermalization} process.}\textbf{ Top:} Monte\index{Monte Carlo} Carlo history for the overlap\index{overlap} $q_4(t)$. Note that our simulation time is too short to expose the symmetry $q_4 \leftrightarrow -q_4$. As a consequence, we know for sure that thermal equilibrium has not been reached for the truncated simulation. \textbf{Bottom:} as in the top panel, for the first four clones in one of our standard simulations with $N = 24$ temperatures (there were 10, completely independent, standard simulations). For each clone, the overlap\index{overlap} $q_4(t)$ changes sign many times along the simulation (as it is to be expected for a well-equilibrated simulation). Note that, with small probability, each clone reaches a state where $|q_4| \sim 0.8$. These events, which are not observed for the other three overlaps\index{overlap} $q_a$ with $a=$1, 2, 3, make it particularly interesting to study the dynamics of $q_4$.}
\labfig{historia_q}
\end{figure}

In the top panel of this figure, we represent the time-dependent overlap\index{overlap} $q_4(t)$ for each of the four clones in the truncated simulation. Furthermore, in the bottom panel, we represent the same quantity for the clones corresponding to the four lowest temperatures in the standard simulation. We clearly observed that nor the overlap\index{overlap} symmetry neither the clone equivalence is respected in the truncated simulation. On the contrary, the standard simulation with $N=24$ displays the expected overlap\index{overlap} symmetry. The Monte\index{Monte Carlo} Carlo histories (in the standard simulation) for $q_{a,\alpha}$ with $a =$ 1, 2, 3 (not shown) are symmetric as well. However, only $q_4$ uncovers a state that arises with small probability, characterized by $|q_4| \sim 0.8$. This feature suggests that $q_4$ is the most interesting overlap\index{overlap} to look at.

It is clear, from the above discussion, that thermalization\index{thermalization} is not reached. But we have focused on the overlaps\index{overlap} $q_a$ with $a=$ 1, 2, 3, 4. One could wonder if looking at the functions based on the temperature random-walk\index{random walk!temperature} of the \gls{PT} we would infer a different result, after all, the truncated simulation is composed only of $4$ copies of the system and one could think that it should not be so difficult to equilibrate the clone permutation. The answer is negative as can be seen in~\reffig{histogramas}. The fact that the spins are out of equilibrium makes it also impossible to equilibrate the clone permutations.

\begin{figure}[t!]
\includegraphics[width=0.8\textwidth]{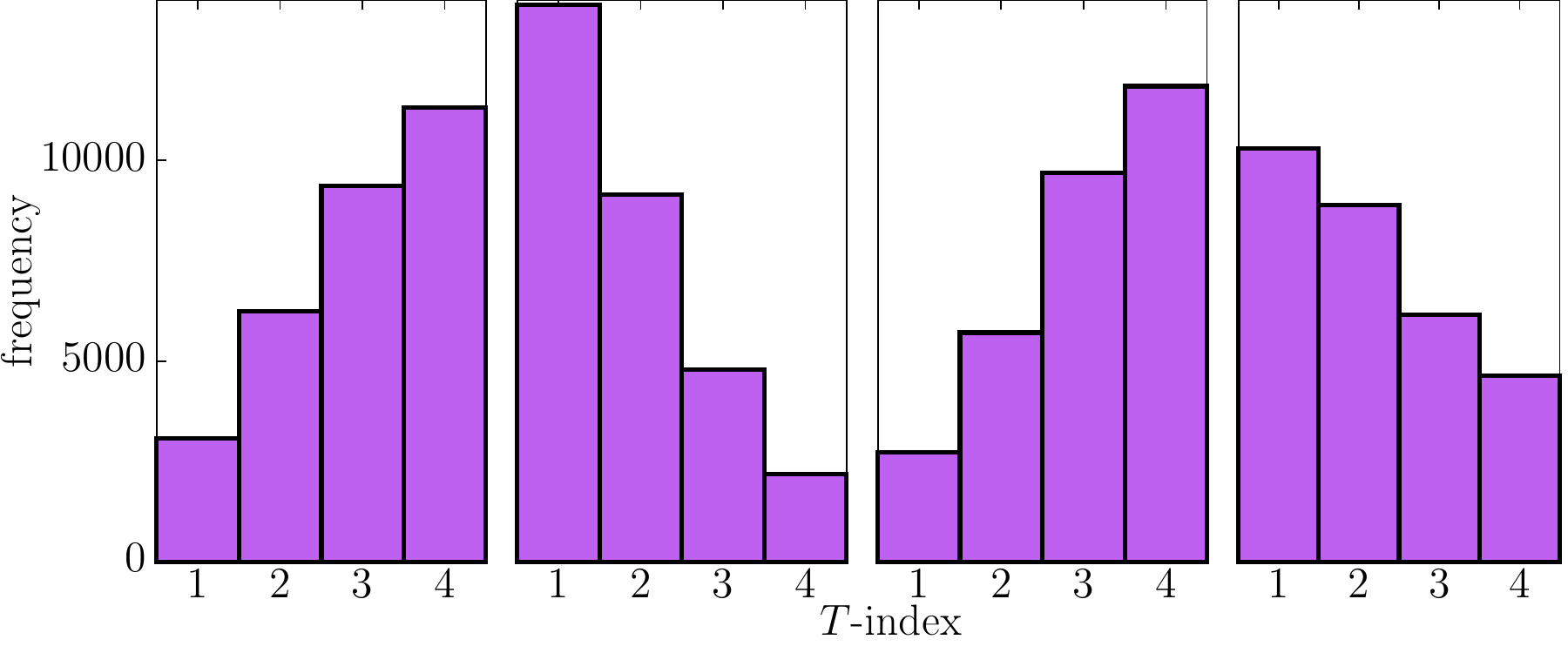}
\caption[\textbf{Time spent in each temperature by the four clones of the truncated Parallel Tempering simulation.}]{\textbf{Time spent in each temperature by the four clones of the truncated  Parallel Tempering simulation.} For each of the four clones in the truncated simulation, we show the histogram of temperature (i.e. the number of times that each clone can be found at $T_1$, $T_2$, \dots). The temperature state was sampled every $5\times 10^4$ Metropolis sweeps (per clone). Had the simulation equilibrated, we would have expected the occupation histograms to be uniform.}
\labfig{histogramas}
\end{figure}

Now, we have exemplified that if the equilibrium is not reached by spin-related quantities it would neither be reached by clone-permutations-related quantities but, are the autocorrelation timescales of the spin-related quantities and the clone-permutations-related quantities equivalent?

To answer this question we focus on the standard setup with $N=24$, which is the only one that equilibrates and, therefore, the only one in which we can compute the equilibrium autocorrelation functions\index{correlation function!time auto-}. In~\reffig{correlaciones} we plot the time autocorrelation functions\index{correlation function!time auto-} for the four quantities $q_a$ with $a=$ 1, 2, 3, 4, and also the time autocorrelation function\index{correlation function!time auto-} for the temperature random-walk\index{random walk!temperature} (labeled as $T$-correlation), specifically the piecewise linear function with $T^* = T_3$ (see~\refch{equilibrium_chaos}).

\begin{figure}[t!]
\includegraphics[width=0.8\textwidth]{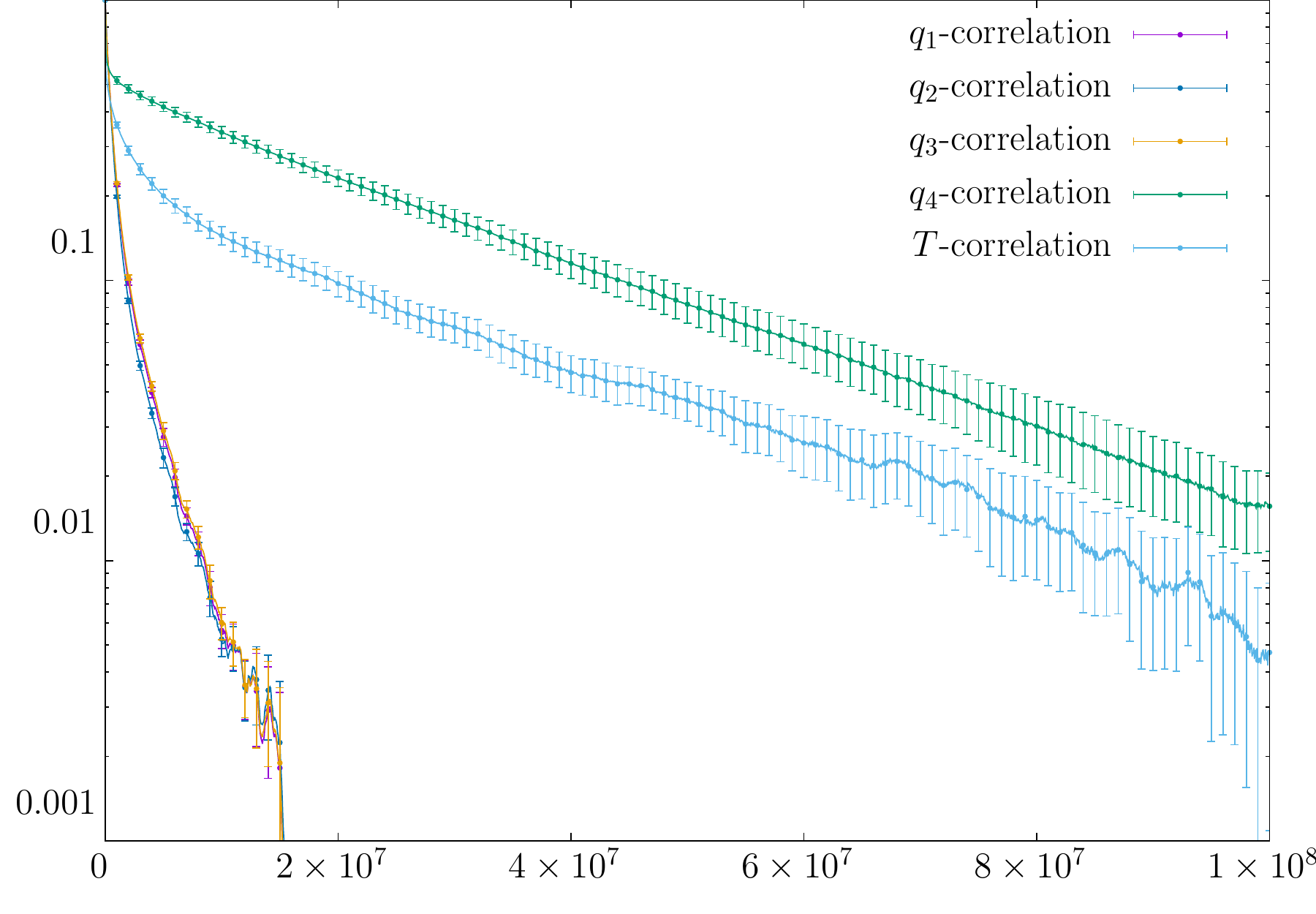}
\caption[\textbf{Equilibrium time dependent correlation functions, as computed from the standard simulation with \boldmath $N_T=24$.} ]{\textbf{Equilibrium time dependent correlation functions, as computed from the standard simulation with \boldmath $N_T=24$.}  We consider five observables, one related to temperature (computed from the piece-wise linear function with $T^*=T_3$, see \refch{equilibrium_chaos}), and the overlaps\index{overlap} $q_a$ with $a=1,2,3,4$. The fact that the $T$ and $q_4$ correlations become parallel in this semi-logarithmic scale indicates that we are safely computing the exponential auto-correlation time (which is independent of the observable). Instead, the $q_{a=1,2,3}$ correlations do not become parallel to the other curves, at least not within the range we can measure, which probably indicates that the amplitudes $A_{n=1,q_{a=1,2,3}}$, see~\refeq{autocorrelation_function_decomposition}, are much smaller for these observables.}
\labfig{correlaciones}
\end{figure}

We observe that the long-time behavior of both the $q_4$-correlation function and the $T$-correlation function are equivalent. Recall that this long-time behavior is giving us the exponential autocorrelation time\index{autocorrelation time!exponential} (which is independent of the particular quantity we are studying). This is indicating us that we are safely computing the exponential autocorrelation time\index{autocorrelation time!exponential} and, furthermore, that both timescales are equivalent. Specifically, and measuring time in Metropolis sweeps, we find $\texp = 3.0(4)\cdot 10^{7}$ from $q_4$, while we find the fully compatible value $\texp = 3.1(6) \cdot 10^7$ from the $T$ random-walk\index{random walk!temperature}. 

A final warning should be made in this respect. The reader might wonder, since we have shown the equivalence of both methods, if it is not simpler to study $q_4$ rather than the clone permutations. The answer is negative. In order to discover the $q_4$-clone-permutation equivalence we had to thermalize the system and find the equilibrium configuration\index{configuration} $\{\tau_{x,4}\}$. Moreover, we randomly pick four different configurations\index{configuration} and only one of them was useful. It is not guaranteed that one can identify an interesting overlap\index{overlap} by randomly picking a small number of equilibrated configurations\index{configuration}.

\setchapterstyle{lines}
\chapter{Temperature Chaos. Technical details.} \labch{AP_technical_details_out-eq_chaos}

\section{Our procedure to obtain the distribution functions} \labsec{procedure}
Here, we explain the details in our computation of the distribution functions $F(X,T_1,T_2,\xi,r)$ or, rather, the quantity we really compute, namely its inverse $X(F,T_1,T_2,\xi,r)$. First, in~\refsubsec{parameters} we provide the relevant parameters for the construction of the distribution functions. Next, in~\refsubsec{construction} we explain how we compute the chaotic parameter for a given sphere and a given number of replicas\index{replica} $\NRep$. In~\refsubsec{sample_to_sample_fluctuations} we explain our computation of $X(F,T_1,T_2,\xi,r)$ for a given $\NRep$, including our procedure for the computation of the error bars\index{error bars}. Finally, in~\refsubsec{extrapolation} we explain the process of the extrapolation to $\NRep \to \infty$.

\subsection{The parameters in our computation} \labsubsec{parameters}
The computation of $X^{s,r}_{T_1,T_2}(\xi)$, namely the chaotic parameter for a given sphere $s$ of radius $r$, recall~\refsubsec{observables-locales}, is specified by five parameters: two temperatures $T_1$ and $T_2$, the radius $r$, the coherence length\index{coherence length} $\xi$ and the number of replicas\index{replica} used to estimate the thermal noise $\NRep$. Our choice of the parameters has been the following:

\begin{itemize}
\item \textbf{Temperatures:} we impose $T_1<T_2$ with $T_1 \in \{0.625,0.7,0.8\}$ and $T_2 \in \{ 0.7,0.8,0.9,1.0\}$.
\item \textbf{Radius:} integer values of $r$ from $r=1$ to $r=15$ are simulated.
\item \textbf{Coherence lengths\index{coherence length}:}  from $\xi_{\min} = 3$ to $\xi_{\max,T_1}$ with intervals $\Delta_\xi=0.25$ where $\xi_{\max,T_1}$ is the maximum $\xi$ simulated for temperature $T_1$\footnote{This condition is imposed by~\refeq{el_reloj_doble}. Indeed, we need $\xi$ to be reached at both temperatures $T_1$ and $T_2$. So, as long as $\xi_{\max}$ increases with temperature (recall~\reftab{xi_max}), the maximum $\xi$ has to be $\xi_{\max,T_1}$}.
\item \textbf{Number of replicas\index{replica}:} in order to quantify the thermal noise, we have computed $X^{s,r}_{T_1,T_2}(\xi)$ by using different number of replicas\index{replica} $\NRep \in \{ 16,32,64,128,256,512\}$ (see below).
\end{itemize}

\subsection{On the computation of $X^{s,r}_{T_1,T_2}(\xi)$}\labsubsec{construction}
The reader will recall from~\refeq{def_chaotic_parameter} that the computation of $X^{s,r}_{T_1,T_2}(\xi)$ requires an infinite number of replicas\index{replica}. Unfortunately we only have $512$ replicas\index{replica} at our disposal. Our choice has been to produce different estimates of $X^{s,r}_{T_1,T_2}(\xi)$ by varying the number of replicas\index{replica} $\NRep$. Specifically, our procedure has been the following:

\begin{enumerate}
\item For each $\NRep < \NRep^{\max}$ we order randomly the $\NRep^{\max}$ replicas\index{replica} and divide them in $\NRep^{\max}/\NRep$ groups of $\NRep$ replicas\index{replica}.
\item In this way, we get $\NRep^{\max}/\NRep$ independent estimates of $X^{s,r}_{T_1,T_2}(\xi)$.
\item In order to eliminate the effect of the initial permutation of the $\NRep^{\max}$ replicas\index{replica}, we repeat this procedure $10$ times for all $\NRep < \NRep^{\max}$.
\end{enumerate}

In a nutshell, for every sphere of radius $r$ we obtain $\NTherm(\NRep)$ estimates of $X^{s,r}_{T_1,T_2}(\xi)$ where
\begin{equation}
\NTherm(\NRep<\NRep^{\max}) = 10 \times \dfrac{\NRep^{\max}}{\NRep} \, ,
\end{equation}
or
\begin{equation}
\NTherm(\NRep=\NRep^{\max}) = 1 \, .
\end{equation}

\subsection{The computation of $X(F,T_1,T_2,\xi,r)$}\labsubsec{sample_to_sample_fluctuations}
Given that, for every radius $r$, we have computed the chaotic parameter for $8000$ different spheres in each sample\index{sample}, we shall fix the probability level $F$ as
\begin{equation}
F_i=\dfrac{i}{8000} \, , \quad i=1,2,\ldots,7999 \, .
\end{equation}
Errors have been computed with the Jackknife method (see~\refsubsec{jk_method}). Let us briefly describe our procedure:

\begin{enumerate}
\item \textbf{Our estimate of $X(F,T_1,T_2,\xi,r)$ for each $\NRep$.}
	\begin{enumerate}
	\item We group together the $\NS \times 8000 \times \NTherm(\NRep)$ estimates of $X^{s,r}_{T_1,T_2}(\xi)$.
	\item We sort them in increasing order.
	\item Our estimate of the central values of $X(F=F_i,T_1,T_2,\xi,r)$ is the $X^{s,r}$ that occupies the position $F_i \times \NS \times 8000 \times \NRep$ in the sorted list. Note that our choice of $F_i$ ensures that $F_i \times \NS \times 8000 \times \NRep$ is an integer.
	\end{enumerate}

\item \textbf{Computation of the fluctuations.}\\
	For the $j$-th Jackknife block, $j=1,2,\ldots,\NS$, we proceed as follows:
	\begin{enumerate}
	\item We group together the $(\NS-1) \times 8000 \times \NTherm(\NRep)$ estimates of $X^{s,r}_{T_1,T_2}(\xi)$, obtained excluding the sample\index{sample} $j$.
	\item We sort them in increasing order.
	\item Our estimate of $X(F=F_i,T_1,T_2,\xi,r)$ for the $j$ Jackknife block, is the $X^{s,r}$ that occupies the position $F_i \times (\NS-1) \times 8000 \times \NRep$ in the sorted list.
	\end{enumerate}
	The errors $X(F,T_1,T_2,\xi,r)$ are computed through the Jackknife blocks using the standard formula (see e.g.~\cite{amit:05} and \refch{AP_statistics}).
\end{enumerate}
Examples of these computations can be seen in~\reffig{linear_quadratic} and~\reffig{exponente_libre}.

\subsection{Extrapolation to infinite number of replicas}\labsubsec{extrapolation}

The computation of the thermal expectation values necessary to calculate the chaotic parameter, see~\refeq{def_chaotic_parameter}, requires an infinite number of replicas\index{replica}. As the maximum number of replicas\index{replica} in our simulation is $\NRep^{\max}=512$, we need an extrapolation to $\NRep \to \infty$. 

Fixing $T_1$, $T_2$, $\xi$, $r$ and $F$ we find a smooth behavior of $X(F,T_1,T_2,\xi,r)$ with $\NRep$ (see~\reffig{linear_quadratic} and~\reffig{exponente_libre}). This smooth evolution with $\NRep$ makes feasible the extrapolation to the $\NRep \to \infty$ limit. Our main ansatz for the extrapolation is
\begin{equation}
X_{\NRep} = X_{\infty} + \dfrac{A}{\NRep} \>\>\> , \labeq{extrapolacion_lineal}
\end{equation}
where $A$ is an amplitude, $X_{\NRep}$ is a short hand for $X(F,T_1,T_2,\xi,r;\NRep)$ and also a short hand for $X_\infty=X(F,T_1,T_2,\xi,r; \NRep=\infty)$. As a check for the linear ansatz in~\refeq{extrapolacion_lineal}, we consider two alternative functional forms for the extrapolation:
\begin{equation}
X_{\NRep} = X_{\infty} + \dfrac{B}{\NRep} + \dfrac{C}{\NRep^2} \>\>\> , \labeq{extrapolacion_cuadratica}
\end{equation}
\begin{equation}
X_{\NRep} = X_{\infty} + \dfrac{D}{\NRep^\gamma} \>\>\> , \labeq{extrapolacion_libre}
\end{equation}
where $B$, $C$ and $D$ are amplitudes and $\gamma$ is a free exponent. We perform independent fits to~\refeq{extrapolacion_lineal},~\refeq{extrapolacion_cuadratica} and~\refeq{extrapolacion_libre} for every value of the parameters $(F,T_1,T_2,\xi,r)$. We reject fits with a diagonal $\chi^2/\text{d.o.f.}\geq 1.1$.\index{degree of freedom} Errors in $X_{\infty}$ are computed from the fluctuations of the Jackknife blocks. Indeed, we perform separated fits for each Jackknife block (the fitting procedure consists in minimizing the diagonal $\chi^2$, see~\cite{yllanes:11}).

\begin{figure}[!h]
  \centering
  \includegraphics[width=0.9\textwidth]{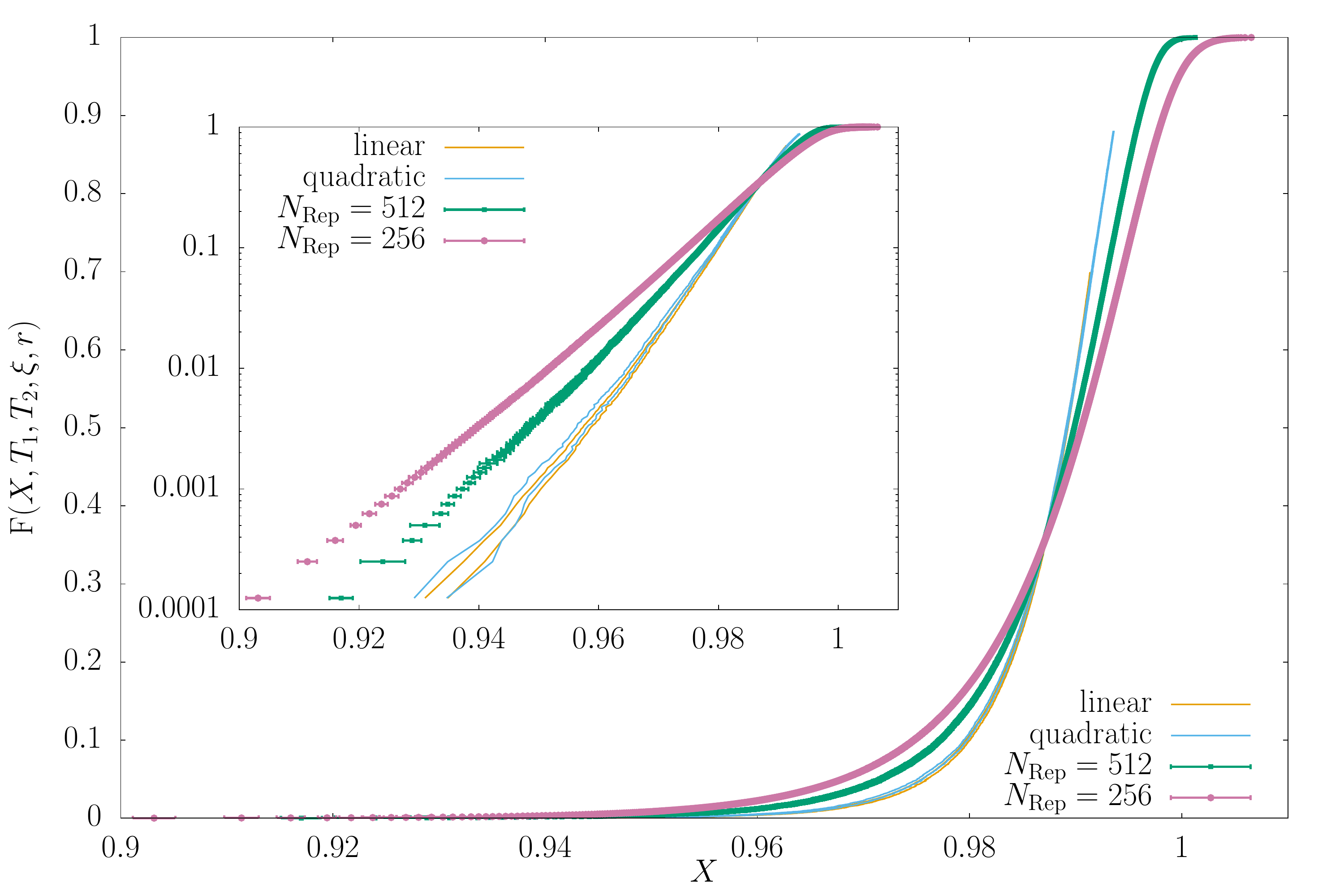}
  \caption[\textbf{Equivalence of linear and quadratic extrapolations.}]{\textbf{Linear and quadratic extrapolations,~\refeq{extrapolacion_lineal} and~\refeq{extrapolacion_cuadratica}, turn out to be equivalent for the tail of the distribution function.} The continuous lines are the linear (golden curves) and quadratic (blue curves) extrapolations to $\NRep \to \infty$ for $F(X,T_1,T_2,\xi,r)$ as a function of $X$. The data shown correspond to the case $T_1=0.7$, $T_2=0.8$, $\xi=11$ and $r=8$. The two curves shown for each extrapolation correspond to the central value plus or minus the standard error. We show horizontal error bars\index{error bars} because we are computing the inverse distribution function $X(F,T_1,T_2,\xi,r)$. We only show extrapolated data when $\chi^2/\text{d.o.f.} < 1.1$\index{degree of freedom} in the fits to~\refeq{extrapolacion_lineal} or to~\refeq{extrapolacion_cuadratica}. For comparison, we also plot the data corresponding to $\NRep=512$ and $\NRep=256$ (yellow and blue dots respectively) \textbf{Inset:} As in the main plot, but with the vertical axis in log-scale.}
\labfig{linear_quadratic}
\end{figure} 

\begin{figure}[!h]
  \centering
  \includegraphics[width=0.9\textwidth]{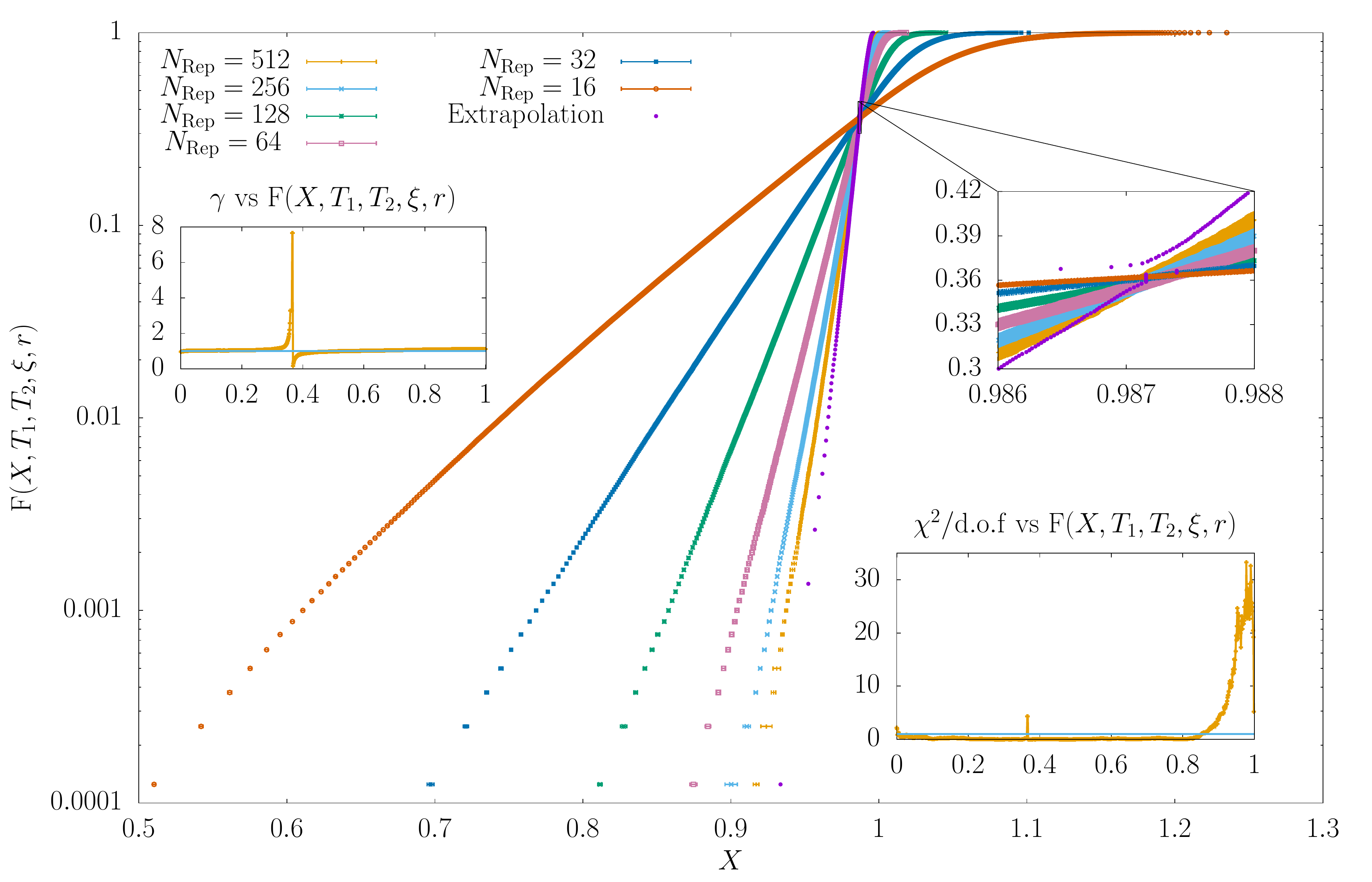}
  \caption[\textbf{Free-exponent extrapolation.}]{\textbf{The exponent $\gamma$ in~\refeq{extrapolacion_libre}, remains close to one when it becomes a fit parameter.} The distribution function $F(X,T_1,T_2,\xi,r)$ is plotted as a function of $X$ for $\NRep=\{512, 256, 128 ,64, 32, 16\}$ together with their extrapolation to $\NRep \to \infty$ as obtained from a fit to~\refeq{extrapolacion_libre}. The data shown correspond to $T_1=0.7$, $T_2=0.8$, $\xi=11$ and $r=8$. In order not to clutter the figure, we do not show error bars\index{error bars} in $\NRep \to \infty$ extrapolation. \textbf{Left inset:} exponent $\gamma$, which is plotted against the probability $F$, remains close to $\gamma=1$ for all $F$, with the exception of the unstable behavior at $F \approx 0.35$, where curves for different $\NRep$ cross (see also top right inset). \textbf{Bottom right inset:} $\chi^2$ per degree of freedom\index{degree of freedom} is plotted against $F$. The blue line corresponds to $\chi^2/\text{d.o.f.} = 1$.\index{degree of freedom} \textbf{Top right inset:} Zoom of the main plot, emphasizing the crossing region at $F \approx 0.35$. Note that at that particle value of $F$ data show almost no dependence with $\NRep$, which makes unstable the fit to~\refeq{extrapolacion_libre}.}
 \labfig{exponente_libre}
\end{figure}

As a first check, we compare the linear and the quadratic extrapolations (see~\reffig{linear_quadratic} for an illustrative example). The figure shows that even for our largest $\NRep$, namely $\NRep=256$, and $\NRep=512$, we are still far from the extrapolation to the $\NRep \to \infty$ limit. Fortunately, the linear and the quadratic extrapolations provide compatible results in our region of interest, i.e. the tail of the distribution function. We remark that the consistency condition $\chi^2/\text{d.o.f.} < 1.1$\index{degree of freedom} is met in a larger range for the quadratic extrapolation ($F<0.9$) than in the linear extrapolation ($F<0.7$). However, because both coincide in the low-$F$ range we are interested in, we have kept the simpler linear extrapolation.

Our second check in~\refeq{extrapolacion_libre} seeks the natural exponent $\gamma$ for the extrapolation, which is a fitting parameter. We have found, see~\reffig{exponente_libre}, that the consistency condition $\chi^2 / \mathrm{d.o.f.}<1.1$\index{degree of freedom} is met for $F<0.85$. Fortunately, $\gamma$ turns out to be very close to the value $\gamma=1$, with the exception of the instability in the crossing region around $F\approx 0.35$. 

In summary, the quadratic and the free-exponent extrapolations support our choice of~\refeq{extrapolacion_lineal} as the preferred form for the $\NRep \to \infty$ extrapolation.

\section{On the most convenient variable to characterize the sphere size}\labsec{cambio_de_r}

\begin{figure}[!h]
  \centering
  \includegraphics[width=0.48\textwidth]{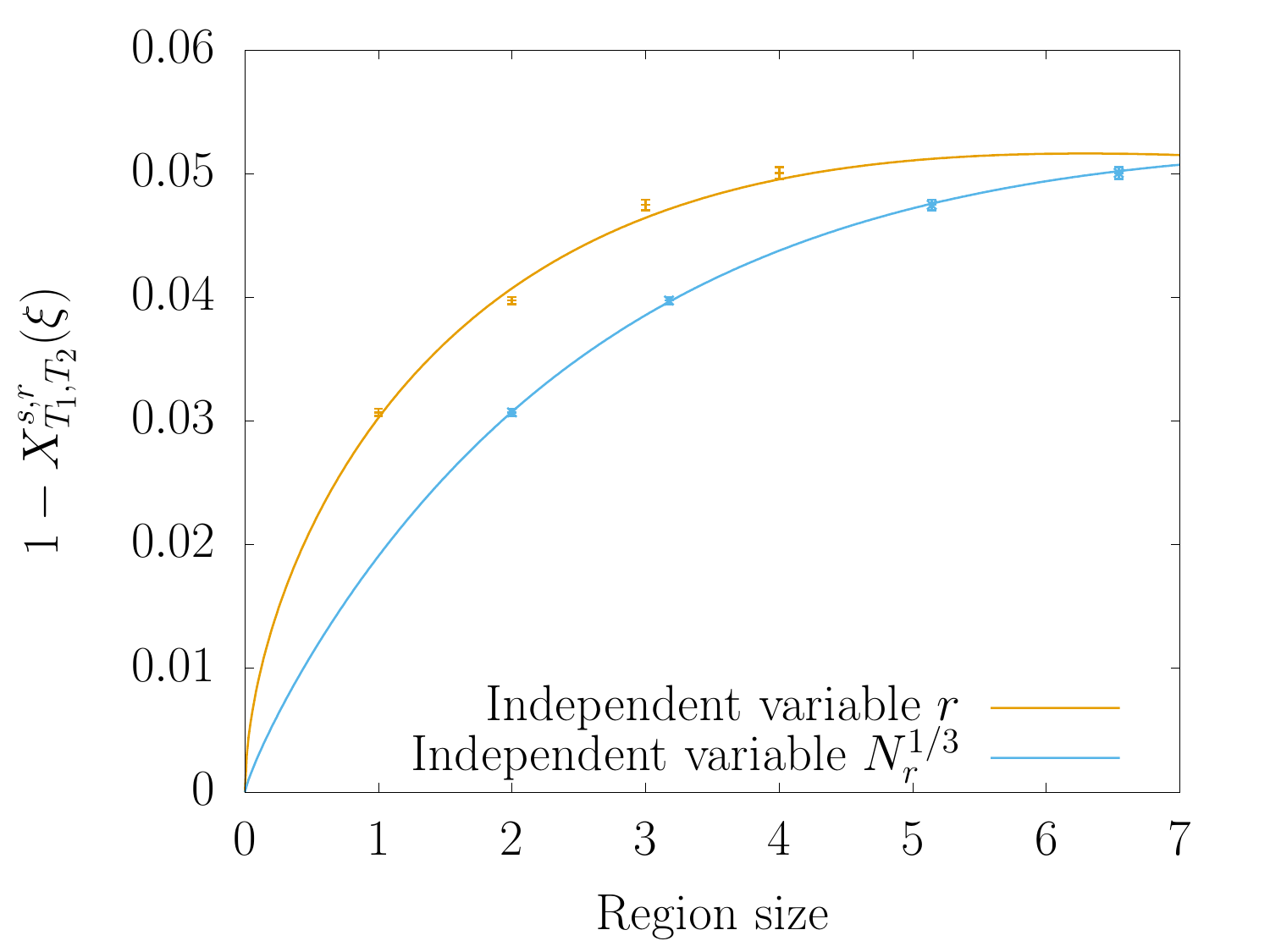}
  \includegraphics[width=0.48\textwidth]{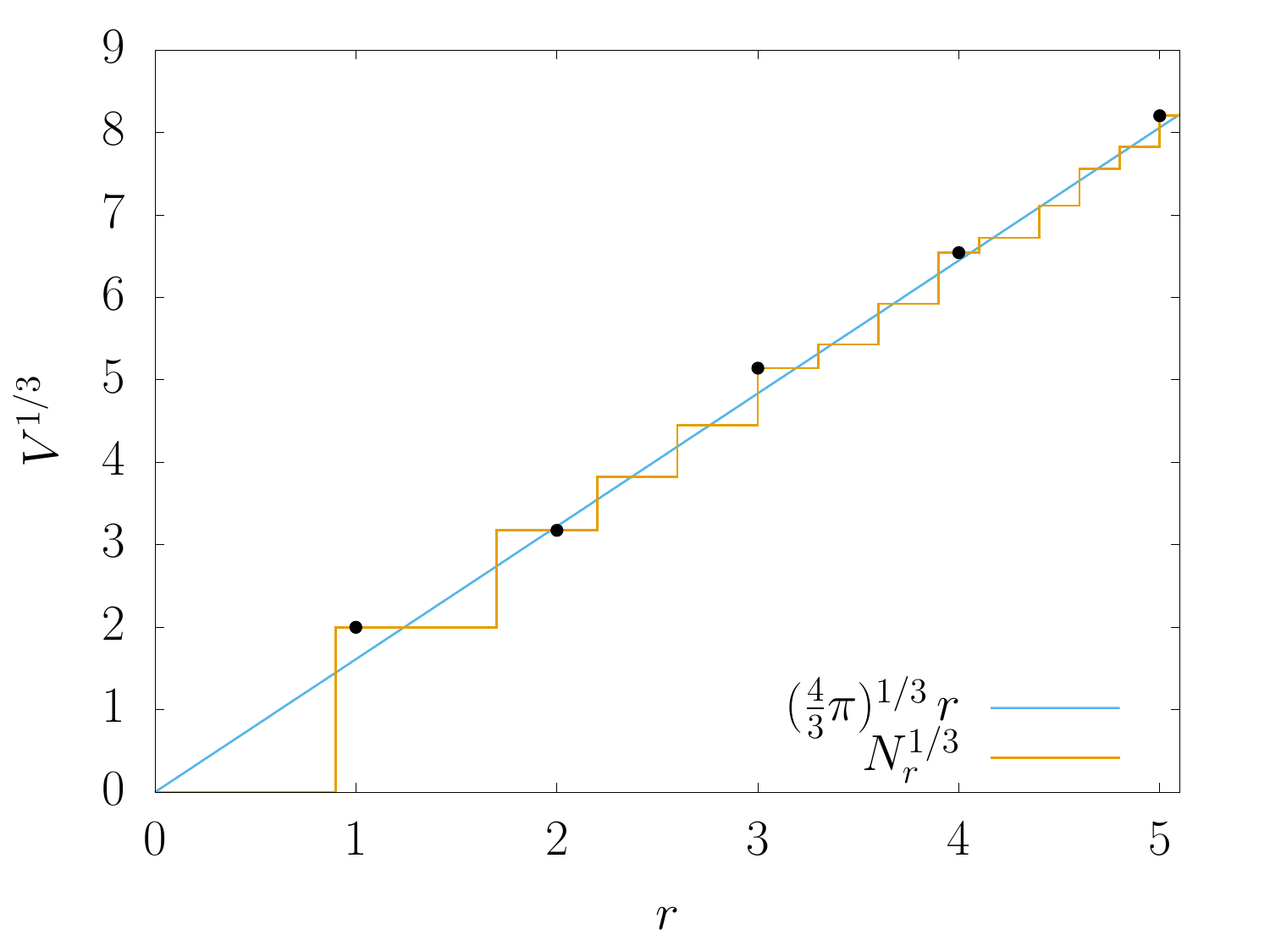}	
  \caption[\textbf{$N_r^{1/3}$ postulates as a better variable to describe short length scales.}]{\textbf{$N_r^{1/3}$ postulates as a better variable to describe short length scales.} \textbf{Left:} complementary of \gls{TC}\index{temperature chaos} $1-X^{s,r}_{T_1,T_2}(\xi)$ against the region size for the two discussed independent variables, namely $N_r^{1/3}$ and the radius $r$. The continuous lines are fits to~\refeq{functional_form} taking as variables $z=r$ (golden curve) and $z=N_r^{1/3}$ (blue curve). The shown data correspond to $T_1=0.7$, $T_2=0.9$, $F=0.01$ and $\xi=7$. We zoom the region of small spheres, where both independent variables most differ. \textbf{Right:} the cubic root of the volume of a sphere (blue curve) is plotted as a function of the radius of the sphere $r$. The golden curve is $N_r^{1/3}$, namely the cubic root of the number of lattice points contained in a sphere of radius $r$, centered at a node of the dual lattice corresponding to our cubic lattice. Values of $N_r^{1/3}$ corresponding to integer $r$ are highlighted as black dots.}
 \labfig{Nr_vs_r}
\end{figure}

Here we explain our rationale for choosing the cubic root of the number of spins contained in the sphere $N_r^{1/3}$, rather than its radius $r$, to characterize the size of the spheres considered in our analysis.

We asked ourselves this question because our first attempt to fit the peaks of $1-X$ to~\refeq{functional_form} by using as an independent variable the radius of the spheres $r$ failed. Indeed, see~\reffig{Nr_vs_r} left, $1-X$ is not a smooth function of $r$. After some reflection, we realized that the number of lattice points in our spheres is not a smooth function of $r$ either (see~\reffig{Nr_vs_r} right). At that point, it was only natural to trade $r$ with $N_r^{1/3}$ as the independent variable. In fact, see~\reffig{Nr_vs_r} left, the new independent variable $N_r^{1/3}$ solved our problem of fitting to~\refeq{functional_form}. Of course, the difference between both independent variables becomes immaterial for very large spheres.

\section{Difficulties in peak characterization for the weak temperature chaos regime}\labsec{peak_characterization}

\begin{figure}[!h]
  \centering
  \includegraphics[width=1.0\textwidth]{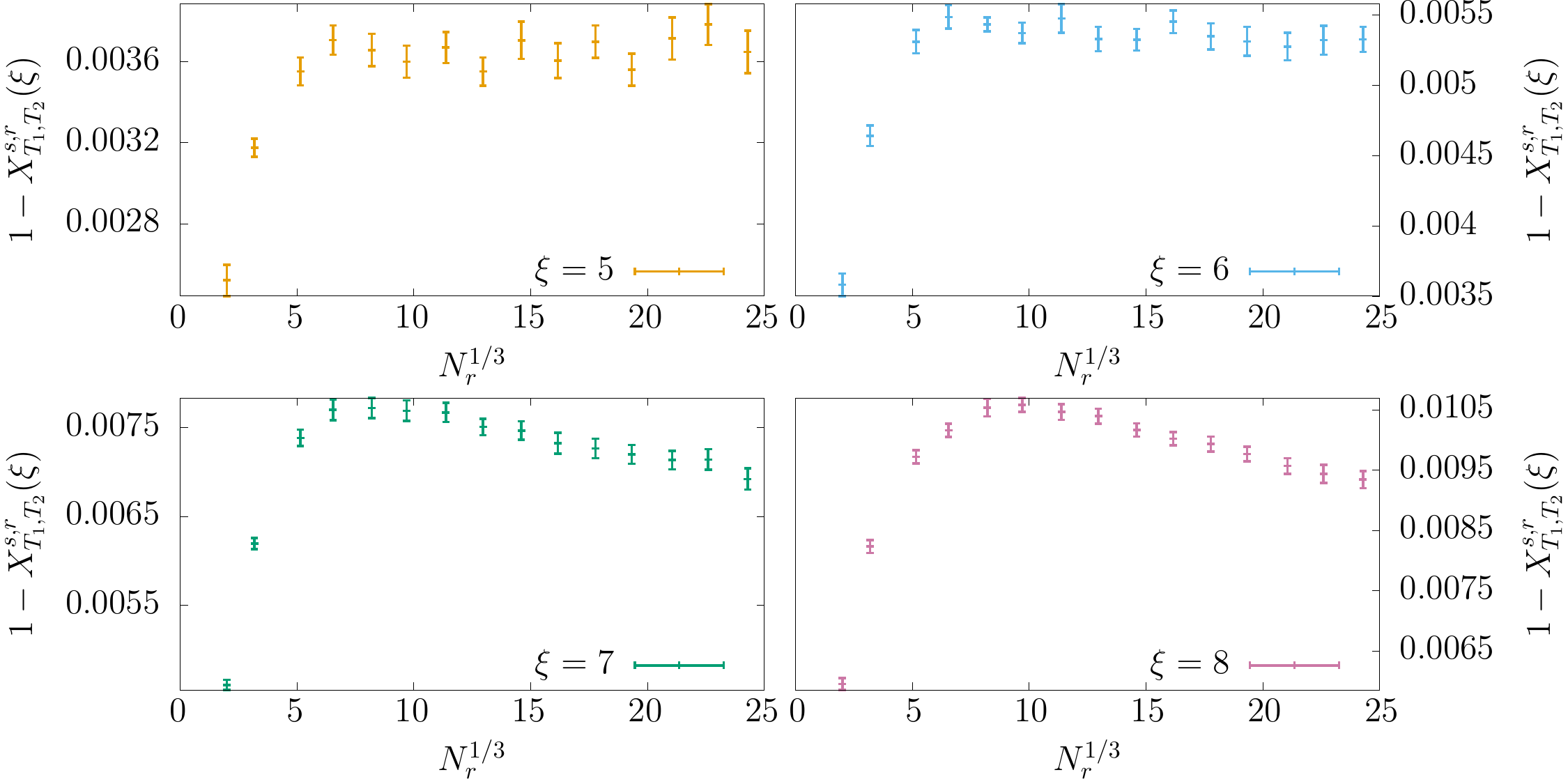}
  \caption[\textbf{Too weak a temperature chaos makes it difficult to characterize the peak.}]{\textbf{Too weak a temperature chaos makes it difficult to characterize the peak.} The complementary of the chaotic parameter $1-X^{s,r}_{T_1,T_2}(\xi)$ is represented against $N_r^{1/3}$ for the temperatures $T_1=0.625$ and $T_2=0.7$, the probability $F=0.01$ and different values of $\xi$. A quick growth for small $N_r^{1/(3}$ followed by a plateau is observed in all the plots.}
 \labfig{T062507_F01}
\end{figure}

Here we illustrate the difficulty of characterizing the peak of the function $f(z)$ defined in~\refeq{functional_form} when \gls{TC}\index{temperature chaos} is extremely weak. Indeed, as~\reffig{T062507_F01} shows, the size of the error bars makes\index{error bars} data almost compatible with a plateau (rather than a peak). Moreover, mind the vertical scale in~\reffig{T062507_F01}, \gls{TC}\index{temperature chaos} is almost nil, which suggests that this set of parameters ($T_1=0.625,T_2=0.7,F=0.01$) is not suitable to study \gls{TC}\index{temperature chaos}. Consequently, we have decided to exclude from the analysis the data obtained with the temperatures $T_1=0.625$ and $T_2=0.7$ and the probability $F=0.01$.

\setchapterstyle{lines}
\chapter{Multispin Coding} \labch{AP_multispin_coding}
\setlength\epigraphwidth{.7\textwidth}
\epigraph{\textit{Premature optimization is the root of all evil.}}{-- Donald Knuth, \textit{The Art of Computer Programming} }

Parallel computing\index{parallel!computing} has turned out to be essential in scientific computing to achieve the computational power that we enjoy nowadays. The tasks that usually tackle the computational hardware do not need a completely sequential work-flow. Instead, many computational tasks can be cut up into independent smaller tasks that can be performed at the same time.

The evolution of the hardware through the years has reflected this fact. From the most general-purpose hardware, the \gls{CPU}, to the (originally) game-oriented hardware, the \gls{GPU}, the mainstream-design tends to increase the independent cores with great benefits for the parallel\index{parallel!computation} computation. Of course, there exists a lot of hardware capable of performing parallel tasks. In addition to the above-mentioned \gls{CPU} and \gls{GPU}, the \gls{FPGA}\index{FPGA} stands as a great example of parallel hardware.

In this appendix, we focus on the \gls{CPU} parallel\index{parallel!computation} computation. Specifically, we will focus on the streaming extensions proposed for the first time by Intel with the MultiMedia eXtension (MMX)~\cite{yu:97} and that was subsequently improved until the most recent iteration of this technology, the Advanced Vector Extensions (AVX, AVX-2, and AVX-512)~\cite{intel_avx}. This technology allows one-clock-cycle boolean\index{boolean} operations of registers of 128 or 256 bits, i.e. we can perform 128 or 256 boolean\index{boolean} operations at the same time in one cycle of the \gls{CPU}'s clock.

We first introduce the Multispin Coding\index{Multispin Coding} as a general concept in~\refsec{what_is_multispin_coding} and then we explain with great detail one implementation that takes advantage of the \gls{CPU} streaming extensions AVX in \refsec{overlap_spheres_multispin_coding}. Last, we explain how we use Multispin Coding\index{Multispin Coding} for more general and complex simulation programs in \refsec{musa_musi_multispin_coding}.

\section{What is the Multispin Coding?} \labsec{what_is_multispin_coding}
The \gls{MSC}\index{Multispin Coding}~\cite{friedberg:70,jacobs:81} is a method that is born out of the necessity of performing simulations with limited computational resources. The basic idea is that, for Ising\index{Ising} spins, we are wasting a lot of memory and computational resources if we encode each spin as an integer number. Indeed, in a 32-bit processor\footnote{Actually, nowadays the standard of 64-bit processors is imposing.} we can encode a variable with $2^{32}$ possible values, and an Ising\index{Ising} spin needs only $2$ values to be encoded.

For binary variables, the solution is clear. By using an integer, we can store $32$ spins at the same time. However, our problem is not the memory but the performance. Our main task is not to be more efficient by storing data, but by running our algorithm. Here is where the streaming extensions (specifically we will focus on the family AVX) take center stage.

The new Intel and AMD processors are able to execute the AVX set of instructions, which allows us to perform one-clock-cycle instructions for registers of $128$ or $256$ bits. We are interested in perform boolean\index{boolean} operations like the AND boolean\index{boolean} operation \\
\begin{center}
\begin{tabular}{|c|c|c|c|c|}
\hline
0 & 0 & 1 & $\cdots$ & 1 \\
\hline
\end{tabular} $\&$ \begin{tabular}{|c|c|c|c|c|}
\hline
1 & 0 & 1 & $\cdots$ & 0 \\
\hline
\end{tabular} $=$ \begin{tabular}{|c|c|c|c|c|}
\hline
0 & 0 & 1 & $\cdots$ & 0\\
\hline
\end{tabular} ,
\end{center}
but also in performing rotations of the bits in our registers
\begin{center}
\begin{tabular}{|c|c|c|c|c|c|c|c|} 
\hline
$s_0$ & $s_1$ & $s_2$ & $s_3$ & $s_4$ & $s_5$ & $s_6$ & $s_7$ \\
\hline
\end{tabular} $\to$ \begin{tabular}{|c|c|c|c|c|c|c|c|}
\hline
$s_7$ & $s_0$ & $s_1$ & $s_2$ & $s_3$ & $s_4$ & $s_5$ & $s_6$ \\
\hline
\end{tabular}
\end{center}

All the operations available for each one of the instructions sets can be found on the official page of Intel (\href{https://software.intel.com/sites/landingpage/IntrinsicsGuide/}{Intel Intrinsics Guide}).

Now, we know how to efficiently compute a large number of boolean\index{boolean} operations but, how can we use this in a real situation? The next two sections are devoted to explain specific implementations in programs developed during this thesis. All the programs showed here have been coded in the programming language C.

\section{An easy example: computing overlaps inside spheres} \labsec{overlap_spheres_multispin_coding}
We start with a simple implementation of \gls{MSC}\index{Multispin Coding}. The program that we describe here was coded to compute the chaotic parameter of a given sphere (see~\refch{out-eq_chaos}). We will briefly describe the program and focus on the implementation of the \gls{MSC}.

In the program, we receive as an input two $L=160$ three-dimensional cubic lattices for each of the $512$ replicas\index{replica}, where the nodes correspond to the spins and the edges to the couplings\index{couplings}. One of the lattices has been simulated with a thermal reservoir at temperature $T_1$ and the other one with $T_2$, both at a time $\tw$ such that the coherence length\index{coherence length} of both systems $\xi_1=\xi_2=\xi(\tw)$. 

Once we have stored the lattices, we select (randomly) 8000 centers in the dual lattice\footnote{The dual lattice of a cubic lattice with \gls{PBC}\index{boundary conditions!periodic} is another cubic lattice of the same size, and with \gls{PBC}\index{boundary conditions!periodic} as well. The nodes of the dual lattice are the centers of the elementary cells of the original lattice. See \refch{out-eq_chaos}.} and we build for each center a sphere of radius $r$. The number of spins inside the sphere is $N_r$, for example, the smallest sphere $r=1$ has $N_r=8$ spins inside.

We want to compute the chaotic parameter of the sphere, defined in \refeq{def_chaotic_parameter} and repeated here for the reader's convenience
\begin{equation} 
X^{s,r}_{T_1,T_2}(\xi) = \dfrac{\langle [q_{T_1,T_2}^{s,r}(\xi)]^2\rangle_T}{\sqrt{\langle[q_{T_1,T_1}^{s,r}(\xi)]^2\rangle_T \,\langle[q_{T_2,T_2}^{s,r}(\xi)]^2\rangle_T}} \, . 
\end{equation} 

The square of the overlaps\index{overlap} $[q_{T_1,T_2}^{s,r}(\xi)]^2$ has to be averaged over the thermal noise. As long as we are in an out-of-equilibrium simulation, our estimation of that thermal average would be an average over the replicas\index{replica}. Therefore, if we focus now on a given sphere, we have a total of $N_r \times \NRep$ number of spins at each temperature. Furthermore, we have to repeat this procedure for the 8000 spheres, for different sizes of spheres $r$, for different coherence lengths\index{coherence length} $\xi(\tw)$ and for different pairs of temperatures $T_1$ and $T_2$. 

Just to put the reader in context, if we fix $T_1$, $T_2$ and $\xi$, we have to perform $2.5526747136 \cdot 10^{14}$ overlaps\index{overlap}. This is, indeed, a huge amount of computations that are completely independent of each other. Hence, we can greatly benefit from the \gls{MSC} in this situation. 

First, to each sphere, we associate a vector of pairs of $256$-bit registers. To this purpose, we coded an specific function \textit{void fill\_sphere(int, int*, int, int)}
\begin{lstlisting}[language=C,style=mystyle]

#define NR_PIECES 2 //The number of words of 256 bits
typedef __m256i MY_WORD; //MY_WORD is a register of 256 bits (see Intel Intrinsics for further information)

MY_WORD replicas[NR_PIECES][V],replicas2[NR_PIECES][V]; //Full lattice for temperature T1 and T2 respectively
MY_WORD sphere_replicas[NR_PIECES][V], sphere_replicas2[NR_PIECES][V]; //Spins inside the sphere for the replica 1 and replica 2.

void fill_sphere(int size,int* sphere_index, int size_word, int temperature_flag){
  int is; //loop variable to run over the spins of the sphere
  int i512; //loop variables to run over the two words of 256-bits
  
  //Auxiliar variables  
  MY_WORD* aux1[NR_PIECES];
  MY_WORD* aux2[NR_PIECES];
  
  //The value of the flag is arbitrary, but this specific set of values
  //allows us to iterate over it in a loop
  
  for(i512=0;i512<NR_PIECES;i512++){
    if(temperature_flag==0){ //Means T1T1
      aux1[i512]=replicas[i512];
      aux2[i512]=replicas[i512];
    }else if(temperature_flag==1){ //Means T2T2
      aux1[i512]=replicas2[i512];
      aux2[i512]=replicas2[i512];
    }else if(temperature_flag==2){ //Means T1T2
      aux1[i512]=replicas[i512];
      aux2[i512]=replicas2[i512];
    }else{ //There are no more options
      print_and_exit("Temperature flag cannot be differente from 0,1 or 2 -> temperature_flag = %d \n", temperature_flag);
    }
  }

  if(size_word==512){
    for(is=0;is<size;is++){
      for(i512=0;i512<NR_PIECES;i512++){
		sphere_replicas[i512][is]=aux1[i512][sphere_index[is]];
		sphere_replicas2[i512][is]=aux2[i512][sphere_index[is]];
      }
    }
  }
}
\end{lstlisting}
This function is really easy, it only associates each spin of the sphere (\textit{sphere\_replicas} and \textit{sphere\_replicas2}) with its corresponding value of the full lattice. Now, we have built our spheres. The variable \textit{temperature\_flag} is selecting the proper temperature of the replicas\index{replica} to construct $[q_{T_1,T_1}^{s,r}(\xi)]^2$, $[q_{T_2,T_2}^{s,r}(\xi)]^2$ or $[q_{T_1,T_2}^{s,r}(\xi)]^2$.

From now on, we focus on the $T_1 \neq T_2$ case for the sake of simplicity, but for the $T_1=T_2$ case, the situation is almost the same (below, the reader can find a minor discrepancy). The next step is to compute the overlap\index{overlap} between the $512$ replicas\index{replica}. The code for this operation, using the AVX instructions, is really easy:
\begin{lstlisting}[language=C,style=mystyle]
MY_WORD q[NR_PIECES][V]; // Matrix (2 x V) of registers of size 256-bits
int i512,j; //Loop variables

for (j=0;j<sphere_size;j++){
   for(i512=0;i512<NR_PIECES;i512++){
      q[i512][j]=_mm256_xor_si256(sphere_replicas[i512][j],sphere_replicas2[i512][j]);
   }
}

\end{lstlisting}

Now, the variable \textit{q[NR\_PIECES][V]} contains, for each site inside the sphere, a set of $512$ overlaps\index{overlap}. We can imagine it as a $(512 \times n_{\mathrm{sph}})$ matrix, where $n_{\mathrm{sph}}$ is the number of spins in the sphere. To compute $[q_{T_1,T_2}^{s,r}(\xi)]^2$ we fix one of the $512$ bits of the register and compute the square of the boolean\index{boolean} sum over the $n_{\mathrm{sph}}$ elements. At the end of this process, we will have computed $512$ values of $[q_{T_1,T_2}^{s,r}(\xi)]^2$.

However, we can compute much more values of $[q_{T_1,T_2}^{s,r}(\xi)]^2$. Indeed, we can rotate the 512 replica-spins\index{replica} in one of the spheres (as done in the example of~\refsec{what_is_multispin_coding}) and compute again the overlap\index{overlap}. This code makes the rotations of the bits

\begin{lstlisting}[language=C,style=mystyle]
typedef union {
    __m256i sse;
    uint64_t vec[4];
} VectorUnion64;

void rota(int size, MY_WORD *sphere1, MY_WORD *sphere2)
{

  VectorUnion64 w[2];
  uint64_t resto[8];

  int i,iw; //Loop variables
  
  for (i=0;i<size;i++){
     w[0].sse=sphere1[i];
     w[1].sse=sphere2[i];
     for (iw=0;iw<8;iw++){
        resto[iw]=w[iw/4].vec[iw%4]&1;
	    w[iw/4].vec[iw%4]>>=1;
     }
    
     for (iw=0;iw<8;iw++){
	    w[iw/4].vec[iw%4]|=resto[(iw+1)&7]<<63;
     }
     sphere1[i]=w[0].sse;
     sphere2[i]=w[1].sse;
  }
}
\end{lstlisting}

This rotation process can be repeated $511$ times until we reach again the initial disposition of spins. Therefore, we compute a total of $512^2$ values of $[q_{T_1,T_2}^{s,r}(\xi)]^2$ that will allow us to estimate the thermal average $\langle [q_{T_1,T_2}^{s,r}(\xi)]^2\rangle_T$. Here is where the small discrepancy between the $T_1 \neq T_2$ and the $T_1 = T_2$ cases appears. For the $T_1=T_2$ case, we can compute only $512(512-1)/2$ overlaps\index{overlap}.

With this simple code, we are able to reduce by a factor $512$ the total number of operations we have to compute. We have considered explaining with detail this simple case because it is very illustrative with a relatively small amount of code and technical details. Nonetheless, the same idea sketched here can be extended to more complex simulations as we will discuss in the next section.

\section{Spin glass simulations by using Multispin Coding} \labsec{musa_musi_multispin_coding}

The equilibrium simulations performed for the works appearing in \refch{metastate} and \refch{equilibrium_chaos} required a huge numerical effort. Because those simulations were performed in \gls{CPU}s, we could take advantage of the \gls{MSC}\index{Multispin Coding} method to save a great amount of time. Here, we explain two different optimizations that are based on the \gls{MSC}\index{Multispin Coding} technique.

\subsection{Multisample Multispin Coding} \labsubsec{multisample_multspin_coding_appendix}
As said in~\refch{equilibrium_chaos}, is known for around 20 years that we can perform the Metropolis update of a single spin by using a sequence of boolean\index{boolean} operations \cite{newman:99}. Thus, we can take advantage of the \gls{MSC}\index{Multispin Coding} technique to simulate several independent \gls{EA}\index{Edwards-Anderson!model} models with Ising\index{Ising} spins and $\pm 1$ couplings\index{couplings}. This method is widely used in computational physics \cite{newman:99,leuzzi:08,fernandez:09f,banos:12,janus:13,manssen:15,lulli:15,billoire:17,billoire:18,gonzalez-adalid-pemartin:19,fernandez:19} and it is known as \gls{MUSA}\index{Multispin Coding!Multisample}. Because \gls{MUSA} is well known (see e.g. Appendix B.1 in~\cite{seoane:13}) we will only briefly describe it here.

The idea is to take a set of $N$ independent systems (in our particular case, $N$ different samples\index{sample}) that can be simulated in a parallel\index{parallel!computation} way and encode, for each site of the lattice, the spins of the $N$ systems in a bit-register of size $N$. Then, all the computations can be done by performing boolean\index{boolean} operations and rotations of spins, just like we have exemplified in \refsec{overlap_spheres_multispin_coding}. In our particular case, with the available hardware (Intel Xeon E5-2680 and AMD Opteron Processor 6272 processors), we found that the $128$ bits version was the most efficient one.

However, as introduced in~\refch{equilibrium_chaos}, this method is not efficient for all the samples\index{sample}. Indeed, if only a few of the 128 samples\index{sample} coded in a computer word are not yet thermalized, continuing the simulation of the already equilibrated samples\index{sample} is a waste of computer time. In \reffig{all_L_prob_tau} one can understand that the width of the autocorrelation time distribution increases with $L$, which makes this problem more important for large sizes. For those samples\index{sample}, we need an extra level of optimization.

\subsection{Multisite Multispin Coding} \labsubsec{musi}
The \gls{MUSI}\index{Multispin Coding!Multisite} method provides a solution for the problem of slow equilibration mentioned at the end of \refsubsec{multisample_multspin_coding_appendix}. In this case, we found that the $256$ version was more efficient. The idea is to use the 256 bits in a computer word to code 256 distinct spins of a single replica\index{replica} of a single sample\index{sample}~\cite{fernandez:15}. In this way, we execute the Metropolis algorithm in $L^3/256$ steps. The major problem of the \gls{MUSI}\index{Multispin Coding!Multisite} method is the generation of random numbers. Because we are performing $256$ updates of spins at the same time, we need $256$ independent random numbers. In our realization of the \gls{MUSI}\index{Multispin Coding!Multisite} we circumvent this problem as explained in~\cite{fernandez:15}.

The geometry used to encode the lattice with $L=16$ is the same that appears in~\cite{fernandez:15}. Here, we briefly recall it.

The physical lattice of Cartesian coordinates $0\leq x,y,z<L$ is mapped to a \emph{super-spin} lattice. Each super-spin is coded in a 256-bits computer word (of course, the 256 bits correspond to 256 physical spins which are updated in parallel). The crucial requirement is that spins that are nearest-neighbors in the physical lattice are coded into nearest-neighbors super-spins. In particular, our super-spins are placed at the nodes of a cubic lattice with the geometry of a parallelepiped of dimensions $L_x=L_y=L/8$, and $L_z=L/4$. The relation between physical coordinates $(x,y,z)$ and the coordinates in the super-spin lattice $(i_x,i_y,i_z)$ is
\begin{eqnarray}
x&=& b_x L_x + i_x\,,\ 0\leq i_x<L_x\,,\ 0\leq b_x < 8\,,\nonumber\\\labeq{MSC-lattice}
y&=& b_y L_y + i_y\,,\ 0\leq i_y<L_y\,,\ 0\leq b_y < 8\,,\\
z&=& b_z L_z + i_z\,,\ 0\leq i_z<L_z\,,\ 0\leq b_z < 4\,.\nonumber
\end{eqnarray}
In this way, exactly 256 sites in the physical lattice are given the same super-spin coordinates $(i_x,i_y,i_z)$. We distinguish between them by means of the bit index:
\begin{equation}
i_b=64 b_z+8b_y+b_x\,,\ 0\leq i_b\leq 255\,.
\end{equation}
Since we have to simulate $N_T$ independent system copies in our \gls{PT} simulation (see~\refch{equilibrium_chaos} and \refsec{thermalizing_PT}), we simply carry out successively the simulation of the $N_T$ systems. 

The alert reader will note that the above geometric construction is very anisotropic (we start with a cube, but end-up with a parallelepiped). Fortunately, this unsightly feature can be easily fixed by noticing that the single-cubic lattice is bipartite. Indeed the lattice splits into the \emph{even} and \emph{odd} sub-lattices according to the parity of $x+y+z$. The two sub-lattices contain $L^3/2$ sites. Furthermore, odd spins interact only with even spins and vice versa. It follows that the update ordering is irrelevant, provided that our full-lattice sweep first updates all the (say) odd sites and next to all the even sites. Now, provided that $L_x$, $L_y$ and $L_z$ are all \emph{even}, the parity of $x+y+z$ and $i_x+i_y+i_z$ coincide.  This implies that all the spins coded in a single super-spin share the same parity, making irrelevant the super-spin lattice asymmetry. For $L=16$ one finds that $L_x=L_y=2$ and $L_z=4$, the three of them even numbers, and hence the above geometric construction works smoothly. 

Despite the success for $L=16$, we found that changes in the geometry were needed in order to correctly simulate the $L=24$ system because one has $L_x=L_y=3$ and $L_z=6$ which implies that the super-spin lattice cannot be split into even and odd sub-lattices.

Our solution consisted in introducing \emph{logical} super-spins of $512$ physical spins, that were later on coded into two computer words of 256 bits each. The geometrical correspondence was ($L_x=L_y=L_z=L/8$)
\begin{eqnarray}
x&=& \tilde b_x L_x + j_x\,,\ 0\leq j_x<L_x\,,\ 0\leq \tilde b_x < 8\,,\nonumber\\\labeq{MSC-lattice-L24}
y&=& \tilde b_y L_y + j_y\,,\ 0\leq j_y<L_y\,,\ 0\leq \tilde b_y < 8\,,\\
z&=& \tilde b_z L_z + j_z\,,\ 0\leq j_z<L_z\,,\ 0\leq \tilde b_z < 8\,.\nonumber
\end{eqnarray}
In this way, exactly 512 sites in the physical lattice are given the same super-spin coordinates $(j_x,j_y,j_z)$. We distinguish between them by means of the bit index:
\begin{equation}
j_b=64 \tilde b_z+8\tilde b_y+\tilde b_x\,,\ 0\leq i_b\leq 511\,.
\end{equation}
Now, the crucial observation is that (because $L_x=L_y=L_z=3$ for $L=24$) the parity of $x+y+z$ coincides with that of $j_x+j_y+j_z$ if (and only if) the parity of $\tilde b_x+\tilde b_y+\tilde b_z$ is even. In other words, given super-spin coordinates $(j_x,j_y,j_z)$, the 512 spins coded in the super-spin split into 256 even spins and 256 odd spins. Because same-parity spins are guaranteed to be mutually non-interacting, we decided to code the 256 bits with the same parity in the same computer word, with the corresponding bit index being the integer part of $j_b/2$.


\backmatter 
\setchapterstyle{plain} 



\printbibliography[heading=bibintoc, title=Bibliography] 



%
%

\cleardoublepage
\phantomsection
\addcontentsline{toc}{chapter}{\listfigurename}
\begingroup 

\setlength{\textheight}{23cm} 

\etocstandarddisplaystyle 
\etocstandardlines 

\listoffigures 

\let\cleardoublepage\bigskip
\let\clearpage\bigskip

\phantomsection
\addcontentsline{toc}{chapter}{\listtablename}
\listoftables 

\endgroup


\setglossarystyle{listgroup} 
\printglossary[title=Acronyms, toctitle=Acronyms] 



\printindex 





\end{document}